\documentclass{ws-rv9x6}
\usepackage{subfigure}   
\usepackage{ws-rv-thm}   
\usepackage{ws-rv-van}   

\usepackage{amssymb}
\usepackage{mathbbol}
\usepackage{amsfonts}
\usepackage{mathrsfs}
\usepackage{epsfig,bm,dcolumn}
\usepackage{graphicx}
\usepackage{color}
\usepackage{amsmath}
\usepackage{dcolumn}
\usepackage{overpic}
\usepackage{hyperref}
\usepackage{slashed}

\usepackage{siunitx}

\makeindex

\begin{document}

\chapter{Chiral Magnetic Effect in Heavy Ion Collisions: \\ 
The Present and Future \label{ch_cme}}

\author[DK]{Dmitri E. Kharzeev \footnote{dmitri.kharzeev@stonybrook.edu}}
\address{Center for Nuclear Theory, Department of Physics and Astronomy,
Stony Brook University, Stony Brook, New York 11794-3800, USA}
\address{Department of Physics, Brookhaven National Laboratory Upton, New York 11973-5000, USA}

\author[JL]{Jinfeng Liao \footnote{liaoji@indiana.edu}}
\address{Physics Department and Center for Exploration of Energy and Matter,	Indiana University, 2401 N Milo B. Sampson Lane, \\ Bloomington, IN 47408, USA}

\author[PT]{Prithwish Tribedy \footnote{ptribedy@bnl.gov}}
\address{Department of Physics, Brookhaven National Laboratory Upton, New York 11973-5000, USA}



\begin{abstract}
The chiral magnetic effect (CME) is a collective quantum phenomenon that arises from the interplay between gauge field topology and fermion chiral anomaly,  encompassing a wide range of physical systems from semimetals to quark-gluon plasma. 
This review, with a focus on CME and related effects in heavy ion collisions, aims to provide an introductory discussion on its conceptual foundation and measurement methodology, a timely update on the present status in terms of experimental findings and theoretical progress, as well as an outlook into the open problems and future developments.   
\end{abstract}

\markboth{Chiral Magnetic Effect in Heavy Ion Collisions: The Present and Future}{Dmitri E. Kharzeev, Jinfeng Liao, Prithwish Tribedy} 

\body

\tableofcontents


\section{Introduction }

The chiral magnetic effect (CME) is the generation of an electric current parallel to an external magnetic field in the presence of an imbalance between the densities of left- and right-handed fermions~\cite{Kharzeev:2004ey,Kharzeev:2007jp,Fukushima:2008xe}, see \cite{Kharzeev:2015znc,Landsteiner:2016led,Kharzeev:2020jxw,Zhao:2019hta,Li:2020dwr,Huang:2015oca,Hattori:2016emy,Liu:2020ymh} for earlier reviews. The CME is a collective quantum phenomenon induced by the anomalous breaking of chiral invariance in relativistic field theory, the chiral anomaly. It is important to note that CME is a non-equilibrium phenomenon which is absent if the chirality is broken by the Hamiltonian of the system (as proposed early \cite{vilenkin1980equilibrium}) -- the chiral imbalance has to be a property of the state. Because the chiral charge is not conserved as a consequence of chiral anomaly, a configuration with non-zero density of chiral charge is an excited unstable state. In the presence of an external magnetic field, the decay of this unstable chirally imbalanced state proceeds through the generation of electric current.

The CME was originally proposed to uncover topologically nontrivial transitionss in the QCD vacuum.  These transitions (instantons, sphalerons, ...) are believed to define many properties of the physical world, including nearly all of the mass of the physical Universe. This is because these transitions are accompanied by the flip of quark's chirality, and thus induce the mass terms. However these chirality-violating transitions have never been directly detected in an experiment, because quarks are confined, and we usually have no way of directly detecting their chirality. This is where the CME can provide a unique way of detecting the domains with chirality imbalance induced by topological configurations -- in the presence of a background magnetic field, such domains will generate an electric current. Since electric current is a conserved quantity, transformation from quarks to hadrons should not destroy the resulting electric charge asymmetry. Therefore a topological fluctuation in the QCD vacuum should manifest itself as a fluctuation in the electric current carried by quarks, which can be detected in experiment.

\subsection{Chiral anomaly} 

Consider a system of charged fermions and antifermions subjected to a backgroud magnetic field. 
In classical physics, charged particles experience a Lorentz force ${\bf F}_m = e\ {\bf v} \times {\bf B}$ in the presence of an external background magnetic field ${\bf B}$. If the projection of their velocities on the direction of the magnetic field is equal to zero, the magnetic field results in the charged particle motion along closed cyclotron orbits with ${\bf \Omega} = {\bf \nabla} \times {\bf v} \neq 0$, but ${\bf v} \cdot {\bf \Omega} = 0$. Even if the charge has a non-zero velocity component along ${\bf \Omega}$, but there is no external force directed along ${\bf B}$, we can always choose a frame in which ${\bf v} \cdot {\bf B} = {\bf v} \cdot {\bf \Omega} = 0$, i.e. the motion of the charge is not helical. This is not true if there is a force applied along the direction of ${\bf B}$, e.g. an electric force ${\bf F}_e = e\ {\bf E}$ resulting from an electric field ${\bf E}$ parallel to ${\bf B}$ - then the motion is helical in any inertial frame.

In quantum theory, charged particles occupy quantized Landau levels. For massless fermions the lowest Landau level (LLL) is chiral and has a zero energy -- qualitatively, this happens due to a cancellation between a positive kinetic energy of the electron and a negative Zeeman energy of the interaction between magnetic field and spin. Therefore, on the LLL the spins of positive (negative) fermions are aligned along (against) the direction of magnetic field. All excited Landau levels are degenerate in spin, and are thus not chiral. 

More formally, the zero energy of the chiral LLL can be seen as a consequence of the Atiyah-Singer index theorem \cite{Atiyah:1968mp} that relates the analytical index of Dirac operator to its topological index. In other words, it relates the number of zero modes of the Dirac operator on a manifold $M$ to the topology of this manifold. The analytical index of Dirac operator $D$ is given by the difference in the numbers of zero energy modes with right ($\nu_+$) and left ($\nu_+$) chirality:
\begin{equation}
{\rm ind}\ D = {\rm dim\ ker} D^+ - {\rm dim\ ker} D^- = \nu_+ - \nu_- ,
\end{equation}
where $\ker D$ is the subspace spanned by the kernel of the operator $D$, i.e. the subspace of states that obey $D^+ \psi = 0$, or $D^- \psi = 0$. 

For a two dimensional manifold $M$, the topological index of this operator is equal to $\frac{1}{2\pi} \int_M {\rm tr}\ {\bf F}$, and Atiyah-Singer index theorem states that
\begin{equation}\label{asi}
\nu_+ - \nu_- = \frac{1}{2\pi} \int_M {\rm Tr}\ {\bf F} .
\end{equation}
Performing analytical continuation to Euclidean ($x,y$) space (with ${\bf B}$ along the ${\bf z}$ axis), we thus find that the number of chiral zero fermion modes is given by the total magnetic flux through the system \cite{Aharonov:1978gb}. For positive fermions with charge $e>0$, we have no left-handed modes, $\nu_-=0$ and the number of right-handed chiral modes from (\ref{asi}) is given by
\begin{equation}
\nu_+ = \frac{e \Phi}{2 \pi} ,
\end{equation}
which is the number of LLLs in the transverse plane; we have included an explicit dependence on the electric charge $e$. For negative fermions $\nu_+=0$ and $\nu_- = \frac{e \Phi}{2 \pi}$.

Let us assume for simplicity that the electric charge chemical potential is equal to zero, $\mu = 0$. In this case it is clear that $\nu_+ = \nu_-$, and the system possesses zero chirality. Let us now turn on an external electric field ${\bf E} \parallel {\bf B}$. The dynamics of fermions on the LLL is $(1+1)$ dimensional along the direction of ${\bf B}$, so we can apply the index theorem (\ref{asi}) to the $(z,t)$ manifold; for positive fermions of $"+"$ chirality
\begin{equation}\label{asi-ep}
\nu_+ = \frac{1}{2\pi} \int dz dt\ e E, \ \ \ \nu_- = 0,
\end{equation}
and for negative fermions of $"-"$ chirality
\begin{equation}\label{asi-en}
\nu_+ = 0, \ \ \ \nu_- = - \frac{1}{2\pi} \int dz dt\ e E .
\end{equation}
These relations can be qualitatively understood from a {\it seemingly} classical argument \cite{Nielsen:1983rb}: the positive charges are accelerated by the Lorentz force along the electric field ${\bf E}$, and thus acquire Fermi-momentum $p_F^+ = e E t$. The density of states in one spatial dimension is $p_F/(2 \pi)$, so the total number of positive fermions with positive chirality is $\nu_+ = 1/(2\pi) \int dz dt\ e E$, in accord with (\ref{asi-ep}). The same argument applied to negative fermions explains the relation (\ref{asi-en}). While the notion of acceleration by Lorentz force is classical, in assuming that it increases the Fermi momentum, we have made an implicit 
assumption that there is an infinite tower of states in the vacuum that are accelerated by the Lorentz force. This tower of states does not exist in classical theory; however it is a crucial ingredient of the (quantum) Dirac theory. 

In $(3+1)$ dimensions, multiplying the density of states in the longitudinal direction $p_F/(2 \pi)$ by the density of states $eB/(2 \pi)$ in the transverse direction, we find from (\ref{asi-ep}) and (\ref{asi-en}) 
\begin{equation}\label{asi3}
\nu_+ - \nu_- = 2 \times \frac{e^2}{4\pi^2} \int d^2x\ dz\ dt\ {\bf E\cdot B}  = \frac{e^2}{2\pi^2} \int d^2x\ dz\ dt\ {\bf E\cdot B},
\end{equation}
where the factor of 2 is due to the contributions of positive and negative fermions. This relation represents the Atiyah-Singer theorem for $U(1)$ theory in $(3+1)$ dimensions, so we could use it directly instead of relying on dimensional reduction of the LLL dynamics. The quantity that appears on the r.h.s. of (\ref{asi3}) is the derivative of the Chern-Simons three-form. 

The relation (\ref{asi3}) can also be written in differential form in terms of the axial current
\begin{equation}
J_\mu^5 = {\bar \psi} \gamma_\mu \gamma^5 \psi = J_\mu^+ - J_\mu^-
\end{equation}
as \cite{Adler:1969gk,Bell:1969ts} 
\begin{equation}\label{an-d}
\partial^\mu J_\mu^5 =  \frac{e^2}{2\pi^2}\ {\bf E\cdot B}.
\end{equation}
The equation (\ref{an-d}) expresses the chiral anomaly, i.e. non-conservation of axial current. It is an operator relation. In particular, we can use it to evaluate the matrix element of transition from a pseudoscalar mesonic excitation of the Dirac vacuum (a neutral pion) into two photons. This can be done by using on the l.h.s. of (\ref{an-d}) the Partial Conservation of Axial Current (PCAC) relation that amounts to replacing the divergence of axial current by the interpolating pion field $\varphi$
\begin{equation}
\partial^\mu J_\mu^5 \simeq F_\pi M_\pi^2\ \varphi,
\end{equation}
where $F_\pi$ and $M_\pi$ are the pion decay constant and mass. 
Taking the matrix element between the vacuum and the two-photon states $\langle 0| \partial^\mu J_\mu^5 |\gamma \gamma \rangle$ then yields the decay width of $\pi^0 \to \gamma  \gamma$ decay, which is a hallmark of the chiral anomaly. However,  chiral anomaly has much broader  implications when the classical gauge fields are involved, as we will now discuss.

\subsection{Chiral magnetic effect}

Let us now show that the chiral anomaly  implies the existence of a non-dissipative electric current in parallel electric and magnetic fields. Indeed, the (vector) electric current 
\begin{equation} \label{eq_intro_jv}
J_\mu = {\bar \psi} \gamma_\mu  \psi = J_\mu^+ + J_\mu^-
\end{equation}
contains equal contributions from positive charge, positive chirality fermions flowing along the direction of $\bf E$ (which we assume to be parallel to $\bf B$), and negative charge, negative chirality fermions flowing in the direction opposite to $\bf E$:
\begin{equation}\label{cur-a}
J_z = 2 \times \frac{e^2}{4\pi^2} {\bf E\cdot B}\ t  = \frac{e^2}{2\pi^2} {\bf E\cdot B}\ t .
\end{equation}
In constant electric and magnetic fields, this current grows linearly in time -- this means that the conductivity $\sigma$ defined by $J = \sigma E$ becomes divergent, and resistivity $\rho = 1/\sigma$ vanishes. Therefore the current (\ref{cur-a}) is non-dissipative, similarly to what happens in superconductors!

We can also write down the relation (\ref{cur-a}) in terms of the chemical potentials $\mu_+ = p_F^+$ and $\mu_- = p_F^-$ for right- and left-handed fermions, which for the massless case are given by the corresponding Fermi momenta $p_F^+ = e E t$ and $p_F^- = - e E t$. It is useful to define the {\it chiral chemical potential} 
\begin{equation}
\mu_5 \equiv \frac{1}{2} \left( \mu_+ - \mu_- \right) 
\end{equation}
related to the density of chiral charge $\rho_5 = J_0^5$; for small $\mu_5$, it is proportional to $\rho_5$, $\mu_5 \simeq \chi^{-1} \rho_5$ where $\chi$ is the chiral susceptibility.
The relation (\ref{cur-a}) then becomes the CME equation \cite{Fukushima:2008xe}
\begin{equation}\label{eq_intro_cme}
{\bf J} = \frac{e^2}{2\pi^2}\ \mu_5\ {\bf B} .
\end{equation}
It shall be noted that by examining the axial vector current $J_\mu^A  = J_\mu^+ - J_\mu^-$ in addition to the vector current in (\ref{eq_intro_jv}), one arrives at another interesting effect known as the chiral separation effect (CSE) \cite{son2004quantum,son2008axial}: 
\begin{equation}\label{eq_intro_cse}
{\bf J}_A = \frac{e^2}{2\pi^2}\ \mu\ {\bf B} .
\end{equation}
where $\mu= \frac{1}{2} \left( \mu_+ + \mu_- \right)$ is the usual chemical potential corresponding to the vector charge density. 

It is important to point out that unlike a usual chemical potential, the chiral chemical potential $\mu_5$ does not correspond to a conserved quantity -- on the contrary, the non-conservation of chiral charge due to chiral anomaly is necessary for the  Chiral Magnetic Effect (CME) described by (\ref{eq_intro_cme}) to exist. Indeed, static magnetic field cannot perform work, so the current (\ref{eq_intro_cme}) can be powered only by a change in the chiral chemical potential. Another way to see this is to consider the power \cite{Basar:2013iaa} of  the CME current (\ref{eq_intro_cme}): $P = \int d^3x\ {\bf E}\ {\bf J} \sim \mu_5\ \int d^3x\ {\bf E}\ {\bf B}$. For a constant $\mu_5$, it can be both positive or negative, in contradiction with energy conservation. In particular, one would be able to extract energy from the ground state of the system with $\mu_5 \neq 0$! On the other hand, if $\mu_5$ is dynamically generated through the chiral anomaly (\ref{an-d}), it has the same sign as $\int d^3x\ {\bf E}\ {\bf B}$, and the electric power is always positive, as it should be. Note that in the latter case the state with $\mu_5 \neq 0$ is not the ground state of the system, and can relax to the true ground state through the anomaly by generating the CME current.

For the case of parallel ${\bf E}$ and ${\bf B}$, the CME relation (\ref{eq_intro_cme}) is a direct consequence of the Abelian chiral anomaly. However it is valid also when the chiral chemical potential is sourced by non-Abelian anomalies \cite{Fukushima:2008xe}, coupling to a time-dependent axion field \cite{Wilczek:1987mv}, or is just a consequence of some non-equilibrium dynamics \cite{Kharzeev:2016mvi}.

\begin{figure}[!hbt]
\includegraphics[width=1.0\textwidth]{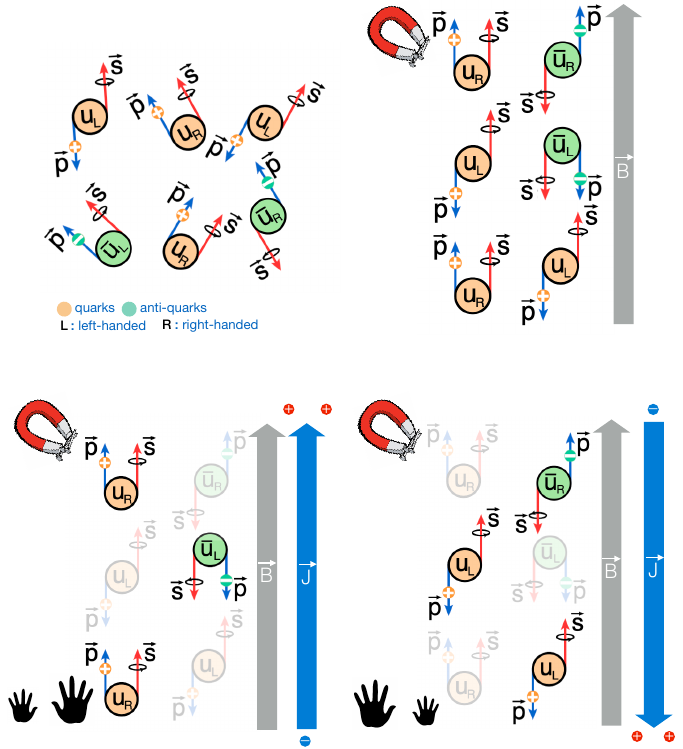}
\caption{\label{cme_cartoon} A simplified cartoon to represent the phenomenon of CME in four steps. (Upper left) Four types of chiral fermions are produced in the system: left and right handed quarks (with positive electric charge) and anti-quarks (with negative electric charge). Their spins are randomly oriented, as are their momentum directions. This does not lead to any observable effect. (Upper right) The presence of an external electromagnetic field ($\vec{B}$) aligns the spin of the quarks due to polarization, where positive quarks are aligned along $\vec{B}$ and the anti-quarks are anti-aligned. Still, nothing noticeable happens that can be detected by known experiments. (Lower left) In a situation where more right handed quarks are produced, a vector current ($\vec{J}$) will be generated along $\vec{B}$. (Lower right) In a situation where more left handed quarks are produced, a vector current ($\vec{J}$) will be generated anti-parallel to $\vec{B}$. This phenomenon of generation of $\vec{J}||\vec{B}$ or $\vec{J}||-\vec{B}$ is called the CME. This will lead to an observable effect of charge separation along $\vec{B}$.}
\end{figure}

The CME is currently under intense experimental investigation in heavy ion collisions, and these studies will be reviewed here. An illustration of the CME current generation via quarks/anti-quarks available in heavy ion collisions is shown in Fig.~\ref{cme_cartoon}.  In addition, the CME has been already observed \cite{li2016chiral} in a number of Dirac and Weyl semimetals \cite{ong2021experimental}. There is a vigorous ongoing research of the CME and related phenomena in condensed matter physics, with potential applications in quantum sensing, quantum communications and quantum computing. The discovery of the effect in condensed matter systems turns CME into a calibrated probe of QCD topology. It is of utmost importance to detect CME in heavy ion collisions -- this would mark the first direct experimental observation of topological transitions in QCD.

\clearpage

\subsection{Chiral vortical effect and other anomaly-induced phenomena}
\label{sec:1:3}

The chiral anomaly gives rise not only to the CME, but to a number of other related phenomena. In addition to the magnetic field ${\vec B}$, we can consider vorticity ${\vec \omega}$ \cite{Kharzeev:2007tn}, and in addition to the electric current we can consider also the vector baryon current \cite{Kharzeev:2010gr}. The chiral anomaly then yields 
\begin{equation}\label{cve}
    {\vec J} = \frac{N_c \mu_5}{2 \pi^2}\ \left[ {\rm tr}(VAQ) {\vec B} + {\rm tr}(VAB) 2 \mu {\vec  \omega}  \right],
\end{equation}
where $V = Q, B$ is the matrix of vector charges (either electric $Q$ or baryon $B$), $A = \mathbb{1}$ is the matrix of axial charges (here we have assumed it is the same for different quark flavors), and $\mu$ is the (vector) chemical potential. Let us consider the case of three light flavors, which is relevant when the temperature $T$ of the QCD plasma is much higher than the strange quark mass $m_s$. 

The second term in (\ref{cve}) was introduced in \cite{Kharzeev:2007tn} and termed in \cite{Kharzeev:2009fn} the {\it chiral vortical effect}.
The corresponding current has also been identified in holography by considering the fluid dynamics of R-charged black holes \cite{Erdmenger:2008rm}.
\vskip0.3cm

Using the matrix of $(u,d,s)$ quark electric charges 
$$Q={\rm diag} (2/3, -1/3, -1/3)$$ and the matrix of baryon charges $$B={\rm diag} (1/3, 1/3, 1/3),$$ one can derive the following expressions for the CME and CVE electric currents:
\begin{equation}
    J^E_{CME} = \frac{2}{3}\ \frac{N_c \mu_5}{2 \pi^2}\ {\vec B},
\end{equation}
and 
\begin{equation}
    J^E_{CVE} = 0. 
\end{equation}
Likewise, for the currents of baryon charge one gets \cite{Kharzeev:2010gr} 
\begin{equation}
    J^B_{CME} = 0,
\end{equation}
and 
\begin{equation}
    J^B_{CVE} = \frac{N_c \mu_5 \mu}{\pi^2}\ {\vec \omega}. 
\end{equation}
\vskip0.3cm
These results suggest that the magnetic field is effective in separating the electric charge relative to the reaction plane, but not the baryon number. On the other hand, vorticity is effective in separating baryon number, but not the electric charge. This observation may have important consequences for the experiment. Indeed, it appears that the induced magnetic field falls off faster at higher collision energies - this is because the initial magnetic field is provided mostly by spectator protons, and at high energies the resulting pulse of magnetic field is Lorentz-contracted. 

However, the vorticity field falls off much slower, due to the fluid dynamics of QCD plasma characterized by a small shear viscosity-to-entropy ratio. It is thus likely that at the LHC energies the electric charge separation driven by CME would be very small, but the baryon charge separation driven by CVE would be significant. This seems consistent with the recent results from the ALICE experiment at the LHC reported at Quark Matter 2023 \cite{wang2023search}.

Note that at RHIC energies both the magnetic field and the vorticity are expected to contribute, so there should be both separation of electric charge and baryon number. In particular, there is an evidence for strong vorticity effects at lower collision energies at RHIC. According to the theory predictions reviewed above, the vorticity should lead to baryon number separation, especially at the energies of the beam energy scan at RHIC. The predictions for the baryon-electric charge correlations and the corresponding observables have recently been developed \cite{Frenklakh:2024jff}.

The same $VAA$ structure of the anomalous vertex also gives rise to the so-called axial separation and axial vortical effects. They may be relevant for the interpretation of polarization of $\Lambda$ hyperons and vector mesons observed at RHIC and the LHC, since the expectation value of the axial current results in spin polarization of chiral fermions. Additionally, interesting transport effects in chiral materials induced by an electric field have also been found~\cite{Huang:2013iia,Jiang:2014ura}. A more detailed discussion of this interesting problem is, however, outside the scope of this review.


\section{Chiral Magnetic Effect in Heavy Ion Collisions}

\subsection{CME in Quark-Gluon Plasma }

For the CME in Eq.(\ref{eq_intro_cme}) to occur,  one needs both a net axial charge (as quantified by its corresponding chiral chemical potential $\mu_5$) for the fermions and  a strong magnetic field $\mathbf{B}$. Fortunately we have both conditions available in the environment created by heavy ion collisions. 

Let us first discuss the axial charge for the light quarks, which originates from the gluonic topological configurations such as instantons and sphalerons. The essence of such topological configurations is the tunneling transitions across energy barriers between the topologically distinct vacuum sectors of a non-Abelian gauge theory like  Quantum Chromodynamics (QCD), which are  characterized by different Chern-Simons numbers. In doing so, the gluon fields themselves  ``twist'' topologically  around spacetime boundaries and the twist can be described by the so-called {\em topological winding numbers}, defined as:  
\begin{eqnarray}  \label{eq_qw}
Q_w = \int d^4 x \,  q(x) = \int d^4 x \left[- \frac{g^2\epsilon^{\mu\nu\rho\sigma}}{32\pi^2}\text{Tr} \left\{G_{\mu\nu}G_{\rho\sigma} \right\}\right ] \,\,\, 
\end{eqnarray}
where the integrand $q(x)$ is the local topological charge density of a given gauge field configuration described by  field strength tensor $G_{\mu\nu} (x)$ with $g$ being the corresponding gauge coupling constant. 
Despite their significance,   the topological configurations are difficult to find  experimentally. A direct detection in a laboratory experiment  could substantially advance our understanding of the underlying tunneling mechanism. A concrete proposal~\cite{Kharzeev:1998kz,Kharzeev:2001ev,Kharzeev:2004ey,Voloshin:2004vk} toward this goal  is to look for the parity-odd ``bubbles'' (i.e. local domains) arising from the topological transitions of QCD gluon fields. Specifically, these bubbles could occur in the hot quark-gluon plasma (QGP) created by relativistic heavy ion collisions. The parity-odd nature of such a bubble can be quantified by the macroscopic {\em chirality} $N_5$ generated for the light quarks in the bubble. Indeed, this is enforced by the  chiral anomaly relation between the chirality of  light flavor quarks and the topology of gluon fields: 
\begin{eqnarray}  \label{eq_anomaly_local} 
 \partial_\mu J^\mu_5 &=& 2q(x) =- \frac{g^2}{16\pi^2} \epsilon^{\mu\nu\rho\sigma} \text{Tr}\left\{G_{\mu\nu}G_{\rho\sigma} \right\} \,\, , \\
N_5 &\equiv& N_R - N_L = 2 Q_w \label{eq_anomaly_global}   \,\, .
\end{eqnarray}
In the equations above $J^\mu_5$ is the local chiral or axial current for each quark flavor while  Eq.(\ref{eq_anomaly_global}) is the spacetime-integrated version of Eq.(\ref{eq_anomaly_local}), with $N_R$ and $N_L$ being the number of right-handed (RH) and left-handed (LH) quarks. This latter equation has its deep mathematical roots in the celebrated Atiyah-Singer index theorem and physically means that each topological winding generates two units of net chirality per flavor of light quarks. Therefore, measuring the net chirality of a QGP provides a unique way of directly accessing the fluctuations of the gluon field  topological windings  in heavy ion collision experiments. 

\begin{figure}[!hbt]
\begin{center}
\includegraphics[height=0.3\textwidth]{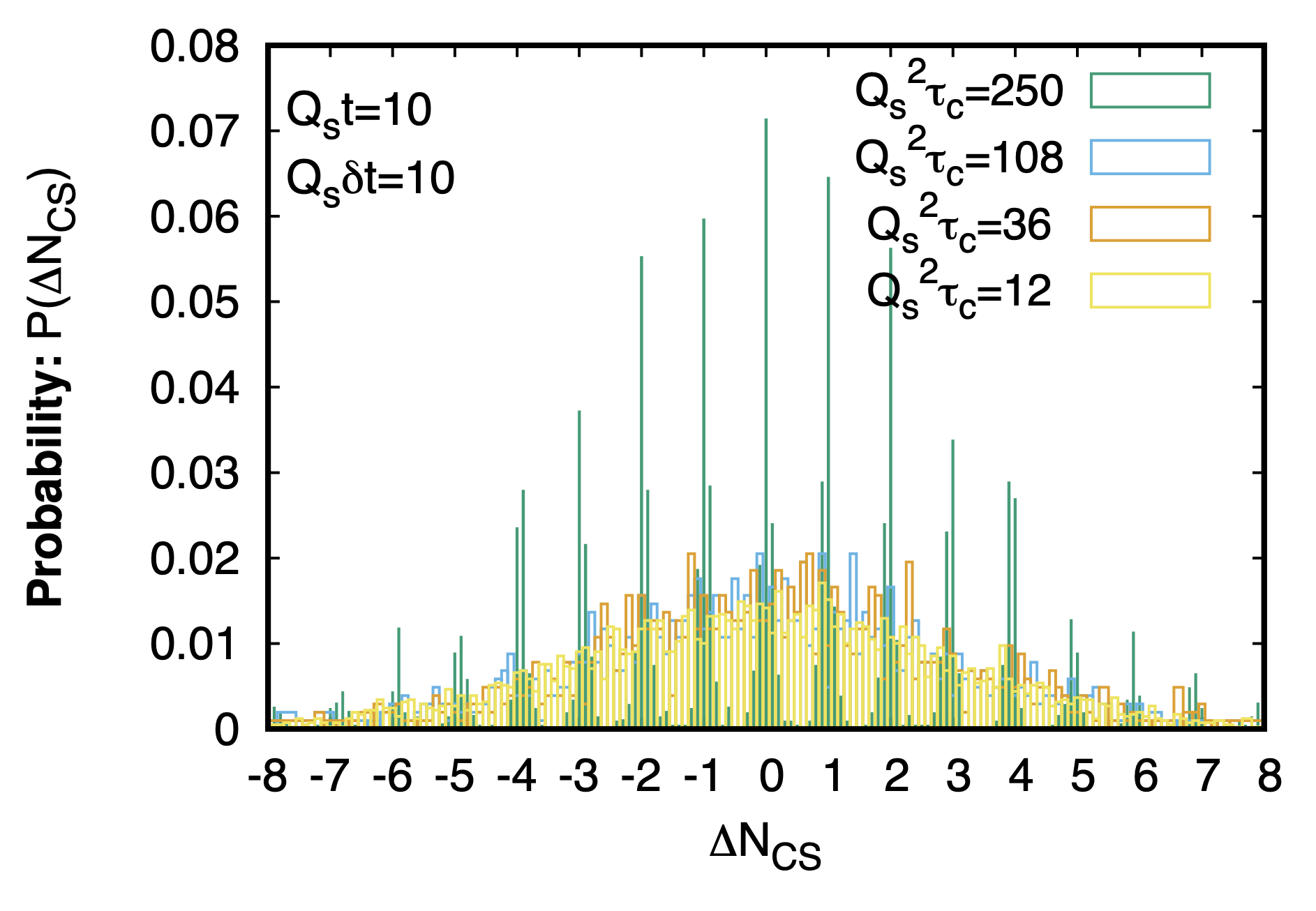} \quad 
\includegraphics[height=0.3\textwidth]{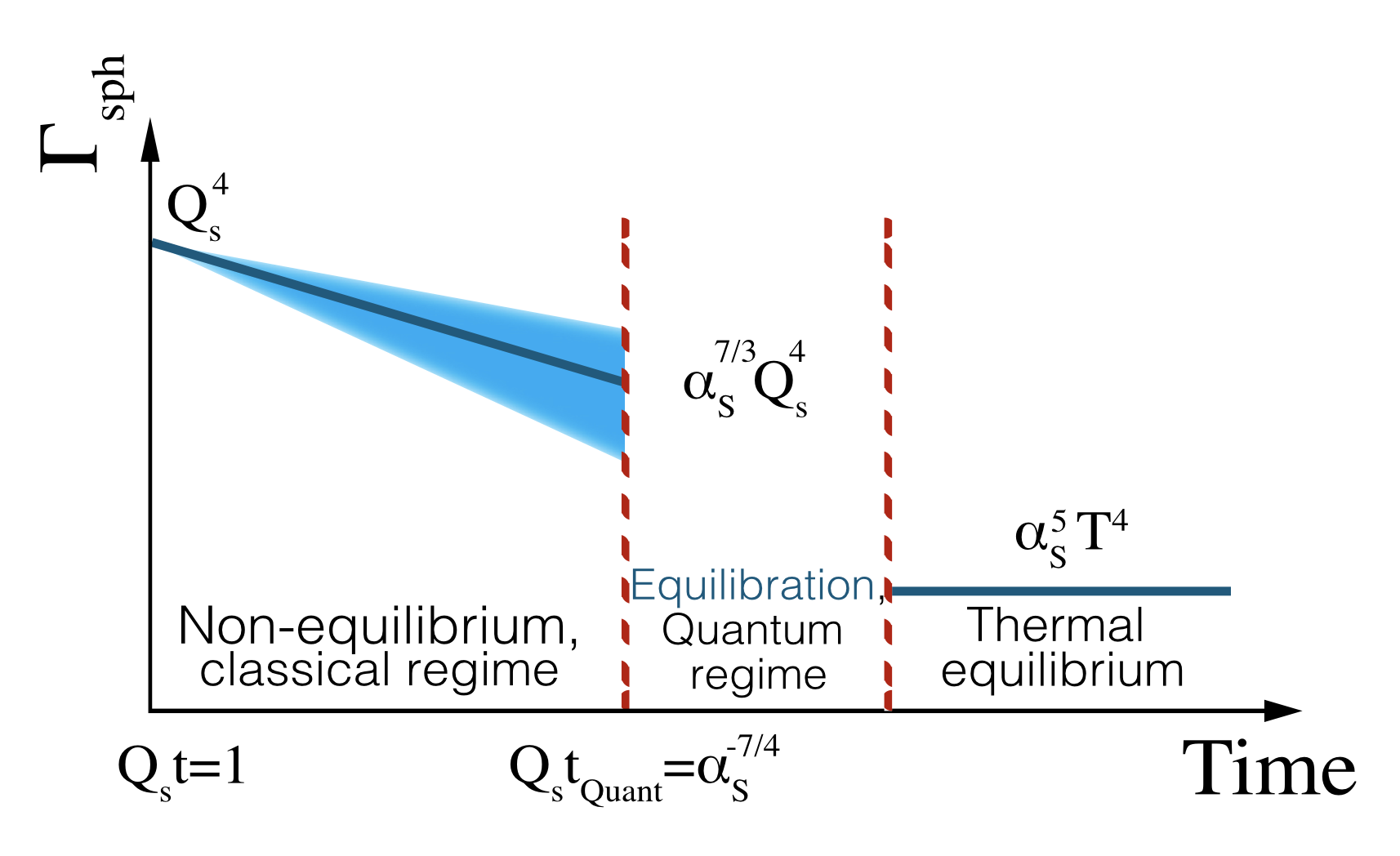}
\caption{ Results from classical statistical simulations on a real-time lattice for  
the probability distribution of guage field Chern-Simons number diffusion (left) and 
 the temporal evolution of the sphaleron transition rate (right) in the glasma. See \cite{Mace:2016svc} for details. 
\label{fig-Ncs}
}
\end{center}
\end{figure}
\begin{figure}[!hbt]
\begin{center}
\includegraphics[height=0.4\textwidth]{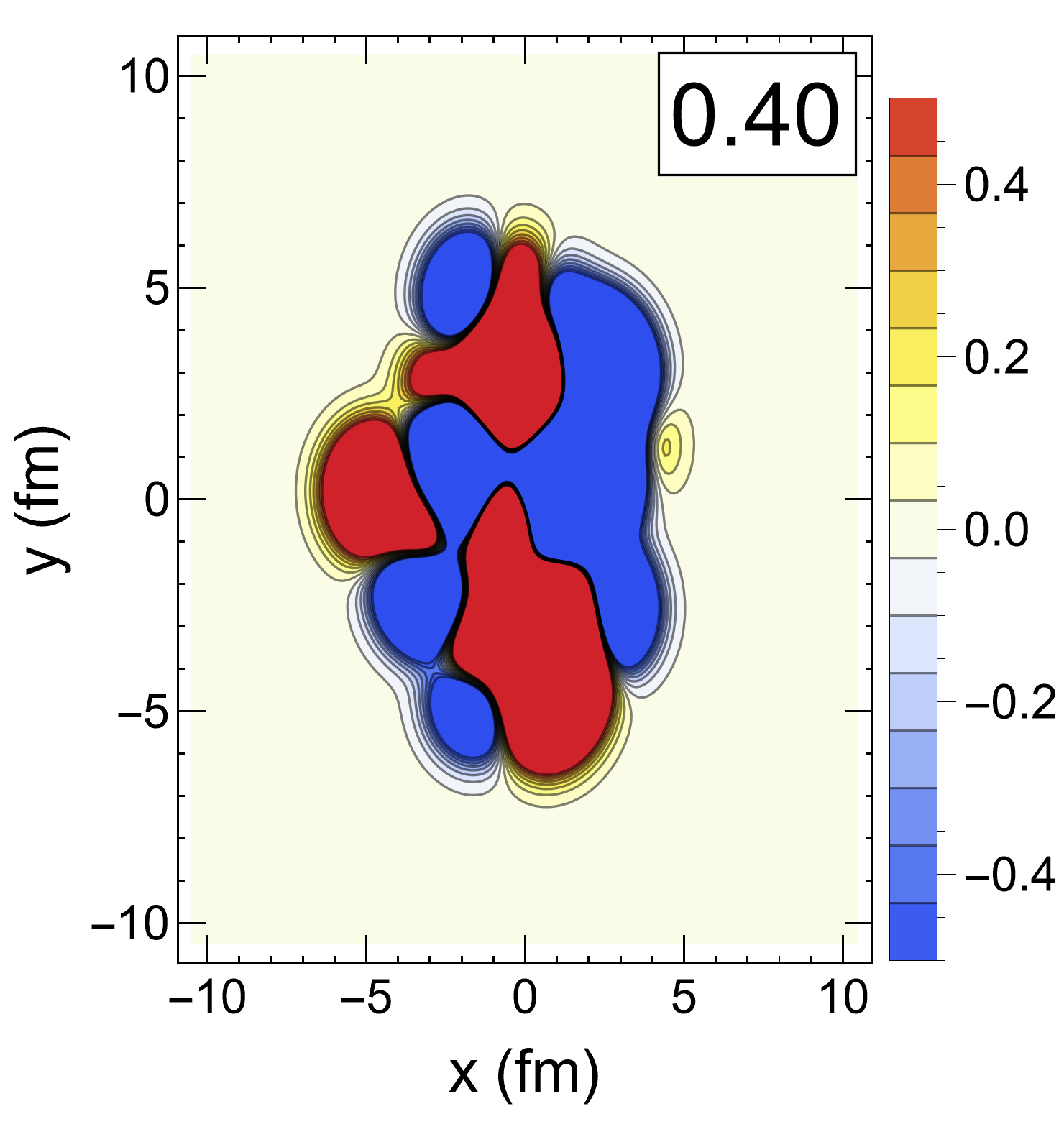} 
\qquad 
\includegraphics[height=0.4\textwidth]{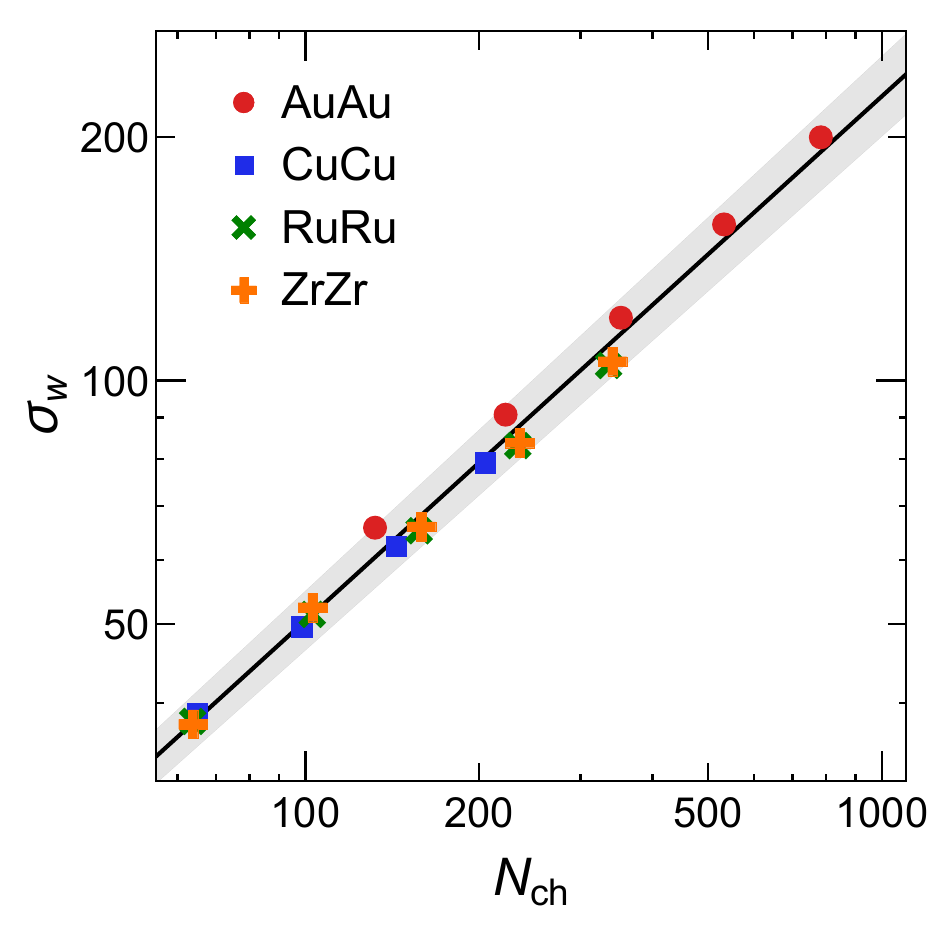}
\caption{Stochastic hydrodynamic simulation results based on glasma-type initial conditions  for an example of initial axial charge density profile from event-by-event (left) and the event-averaged variance of total initial axial charges as a function of charged particle multiplicity (right). See \cite{Huang:2021bhj} for details. 
\label{fig-Q5}
}
\end{center}
\end{figure}

In a typical heavy ion collision, the fireball possesses considerable initial axial charge $N_5$ from the random   topological fluctuations of the strong initial color fields. This has been  demonstrated by phenomenological simulations based on the so-called glasma framework~\cite{Mueller:2016ven,Mace:2016svc,Mace:2016shq,Mace:2017wcl,Lappi:2017skr}, see e.g. an example shown in Fig.~\ref{fig-Ncs}.  A quantitative estimate of  the axial charge initial condition   would be important but has proved challenging. 
Event-by-event simulations of axial charge initial conditions have been developed:  see e.g. Fig.~\ref{fig-Q5}~\cite{Huang:2021bhj,Hirono:2014oda}. These are 
based on phenomenological models of topological fluctuations generated during glasma evolution at the early stage.  
Given a nonzero initial $N_5$, there is also the  question of its subsequent relaxation  toward a vanishing equilibrium value in the QGP. This is because the axial charge is not a strictly conserved charge: it can fluctuate to be nonzero but also dissipates toward zero, featuring the competition between fluctuations and dissipations. Both finite quark masses and gluonic topological fluctuations contribute to the relaxation rate for the  random flipping of individual quark chirality. Realistic estimates including both gluonic and mass contributions to axial charge relaxation~\cite{Guo:2016nnq,Hou:2017szz,Lin:2018nxj,Liang:2020sgr} suggest that the QGP should be able to maintain its finite chirality for considerable time. The relatively long lifetime of axial charges is also supported by recent studies with real-time lattice simulations. Nevertheless, it is important to include the stochastic dynamics for a realistic description of the axial charge evolution. Recent calculations in~\cite{Huang:2021bhj} have shown that such an effect would reduce the final state charge-dependent correlations from the CME transport by a factor of two and therefore should be accounted for.   

The net chirality for light quarks in itself is, however, also challenging to detect due to the fact that the QGP born from collisions would expand, cool down, and eventually transition into a low temperature hadron phase where the spontaneous breaking of chiral symmetry makes the net chirality unobservable. Fortunately, there is a way out by virtue of the CME transport~\cite{Kharzeev:2007jp,Fukushima:2008xe}. In the context of a QGP, one expects  an electric current $\mathbf{J}$ induced by the CME along an external magnetic field $\mathbf{B}$ to be:  
\begin{eqnarray}  \label{eq_cme} 
  \mathbf{J} &=&  N_c \left (\sum_{f} Q_f^2 \right) \frac{ e^2}{2\pi^2} \mu_5  \mathbf{B}  \, .
\end{eqnarray} 
where $N_c=3$ is the number of color and the sum is over $N_f$ flavors of light quarks with electric charge factor $Q_f$, respectively. 
The $\mu_5$ is a chiral chemical potential that quantifies the net chirality $N_5$.  In the context of heavy ion collisions, the CME current (\ref{eq_cme}) leads to a charge separation in the quark-gluon plasma that results in a specific hadron emission pattern and can be measured via charge-dependent azimuthal correlations~\cite{Voloshin:2004vk}.  
In short, there is a promising pathway for experimental probes  of gauge field topology in heavy ion collisions: the winding number $Q_w$ of gluon fields $\Rightarrow$ net chirality of quarks 
$\Rightarrow$ CME current $\Rightarrow$ correlation observables.

\begin{figure}[!hbt]
\begin{center}
\includegraphics[height=0.35\textwidth]{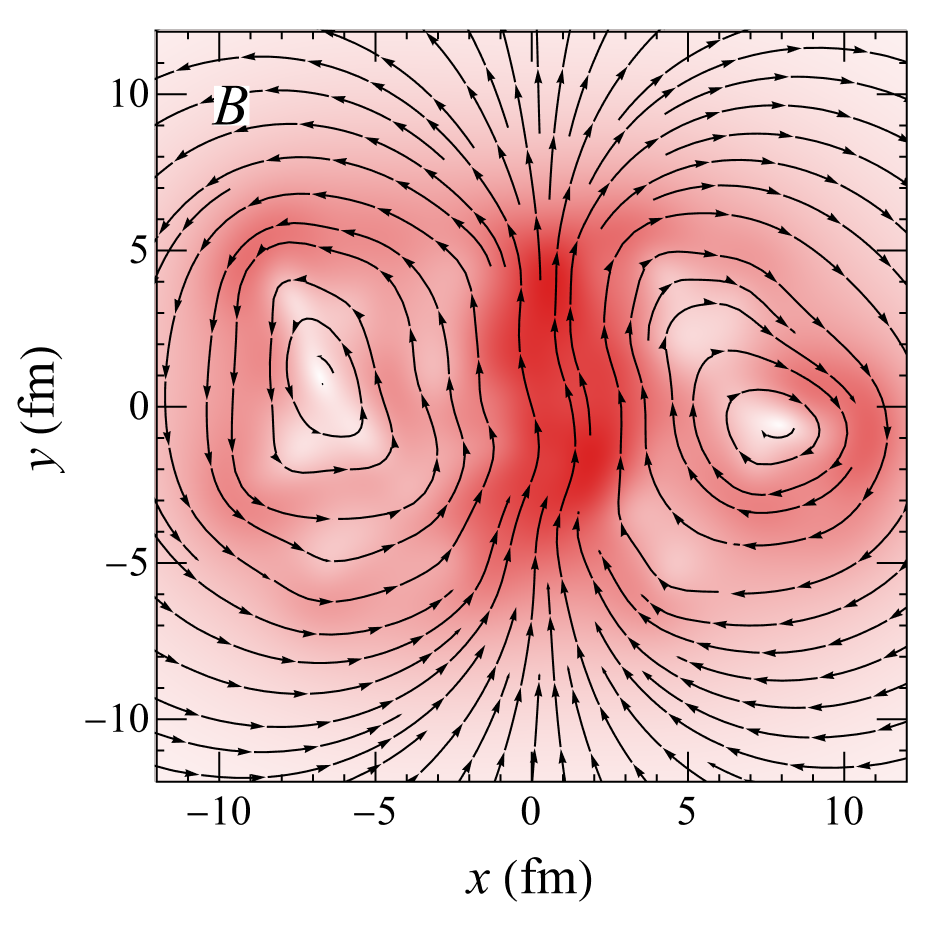}\quad
\includegraphics[height=0.35\textwidth]{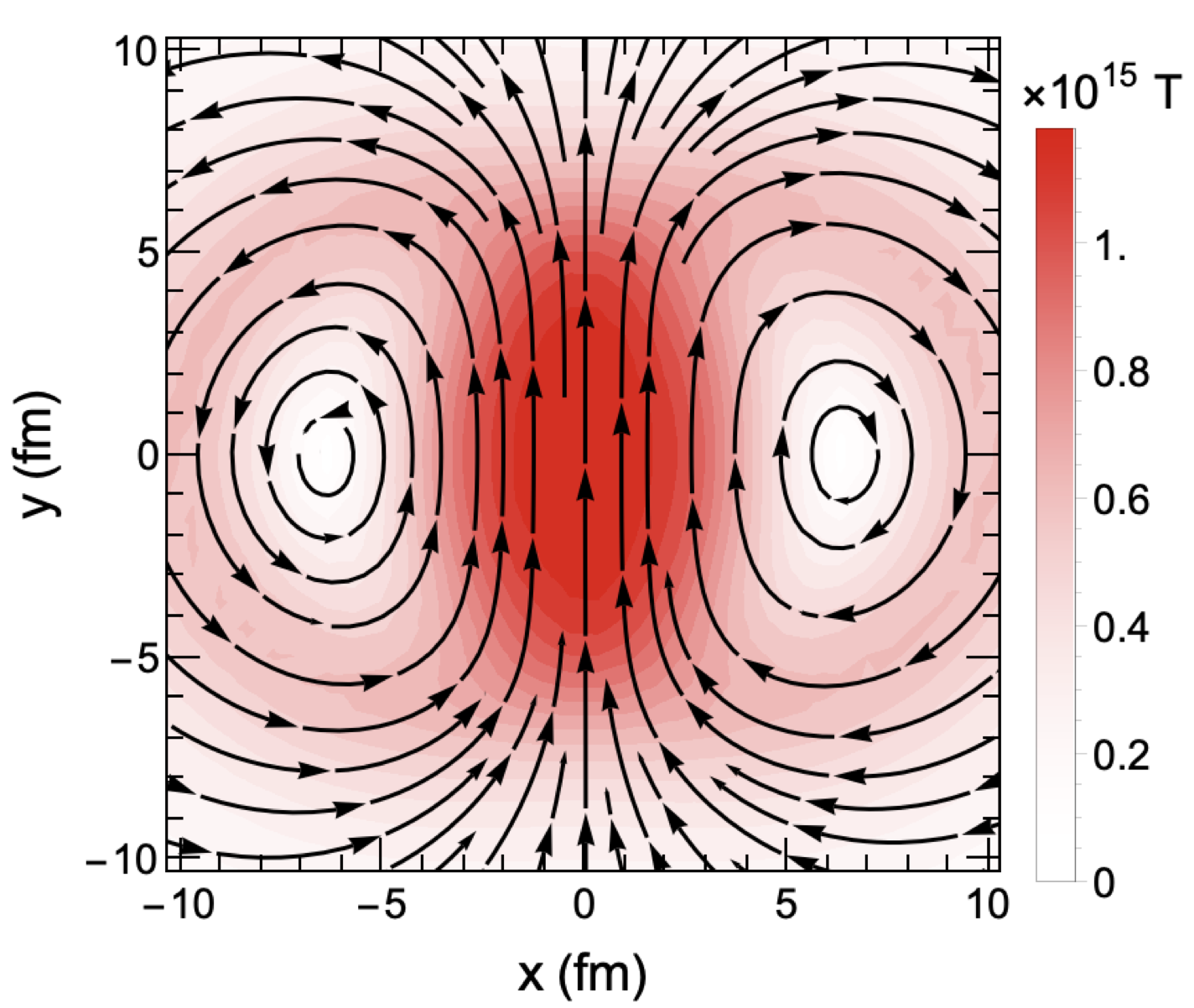}\quad
\includegraphics[height=0.35\textwidth]{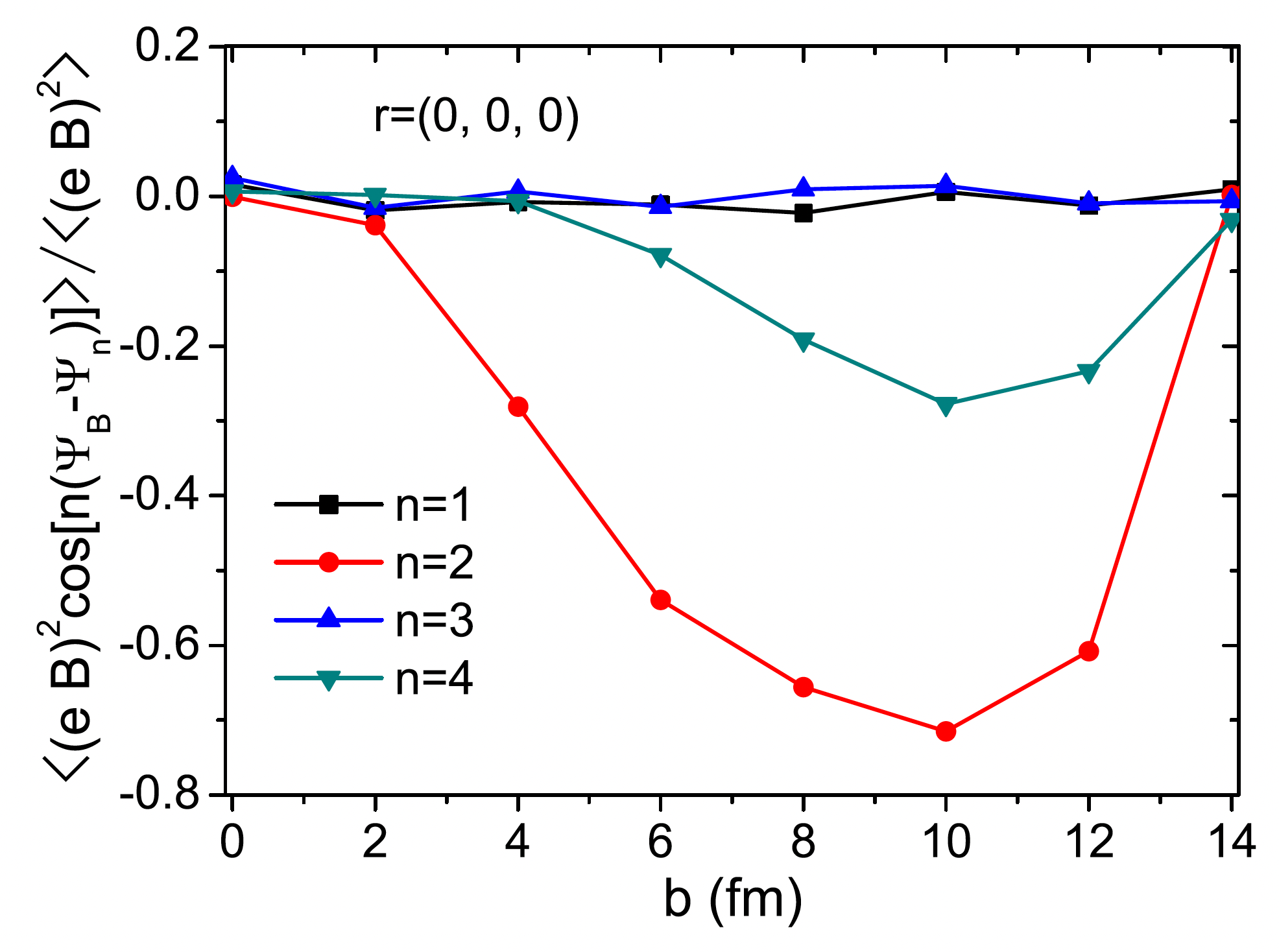}
\caption{Event-by-event simulation results for the spatial distributions of initial magnetic field strength and orientation, with a single event profile shown in the upper left panel and event-averaged smooth profile shown in the upper right panel~\cite{Shi:2019wzi}. The lower panel shows the azimuthal correlations between the magnetic field orientation and various harmonic participant planes, see \cite{Bloczynski:2012en} for details.  
\label{fig-B-field}
}
\end{center}
\end{figure}

The other key element is the magnetic field $\bf{B}$. Heavy ion collisions create an environment with an extreme magnetic field -- at least at very early times -- which arise from the fast-moving, highly-charged nuclei. A simple estimate gives $|e\mathbf{B}|\sim \frac{\alpha_{EM} Z\gamma b}{R_A^2} \sim m_\pi^2$ at the center point between two colliding nuclei upon initial impact.    Given such a large magnetic field and a chiral QGP, one expects the CME to occur. However, for a quantitative understanding of possible CME signals, two crucial factors need to be understood: its azimuthal orientation as well as its time duration. A randomly oriented magnetic field which is not correlated to any other observable such as flow prevents the CME, even if present, to be measured in experiment. A magnetic field which, although very strong initially, decays too fast would lead to a small and   undetectable signal~\cite{Shi:2017cpu}. 

As first shown in  \cite{Bloczynski:2012en}, strong fluctuations of the initial protons in the colliding nuclei   bring significant fluctuations to the azimuthal orientation of the $\mathbf{B}$ field relative to the bulk matter geometry.  Fortunately one can use simulations to quantify the azimuthal correlations between the magnetic field and various geometric orientations (e.g. reaction plane, elliptic and triangular participant planes) in the collision.   Such magnetic field fluctuations  turn out to be useful features for experimental analysis,  by comparing relevant charge-dependent correlations measured with respect to the reaction plane as well as elliptic and triangular event planes, see 
e.g. discussions in \cite{Bzdak:2019pkr,Wang:2018ygc,Li:2020dwr}.

\begin{figure}[!hbt]
\begin{center}
\includegraphics[height=0.4\textwidth]{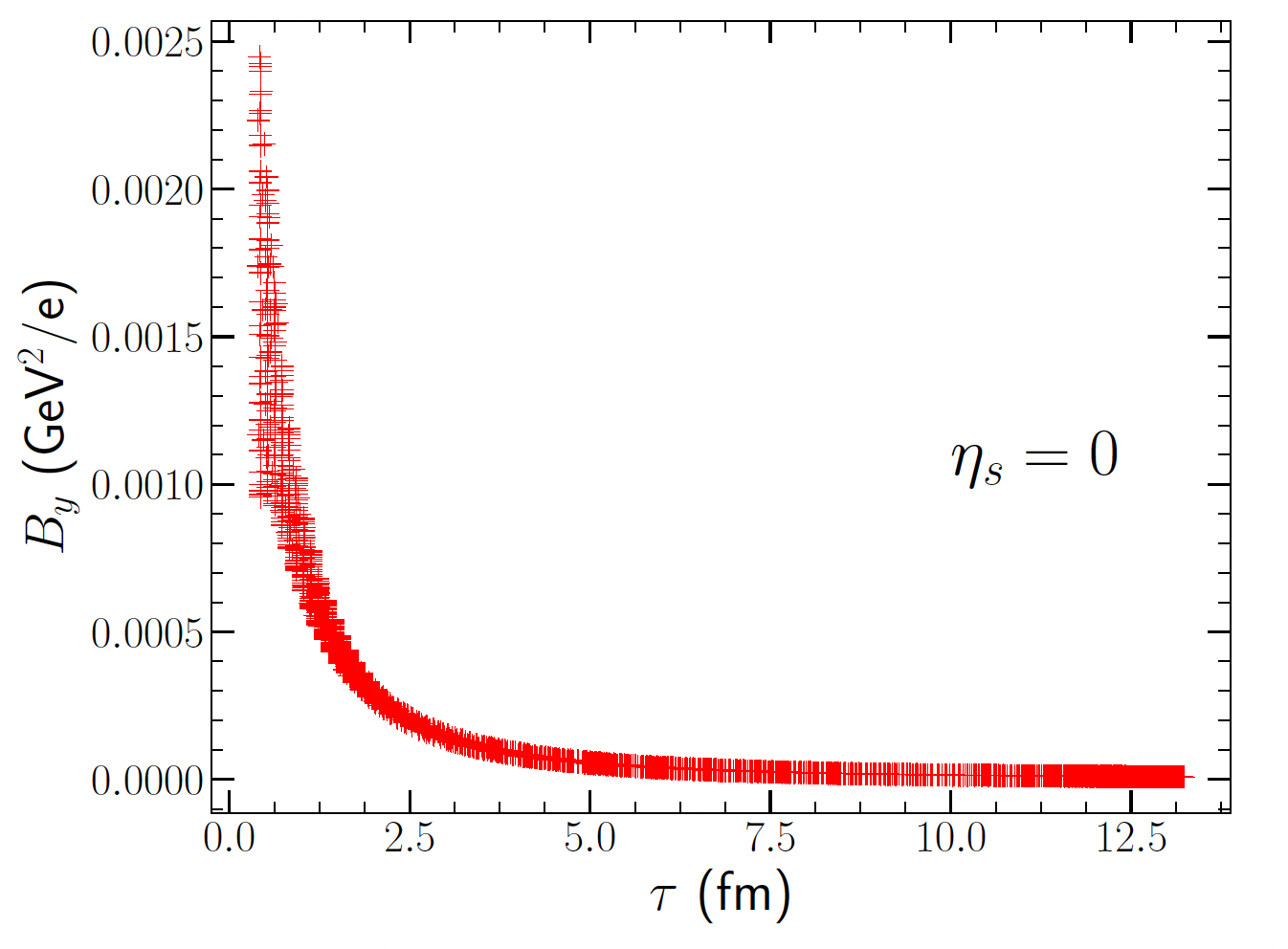}\quad
\includegraphics[height=0.4\textwidth]{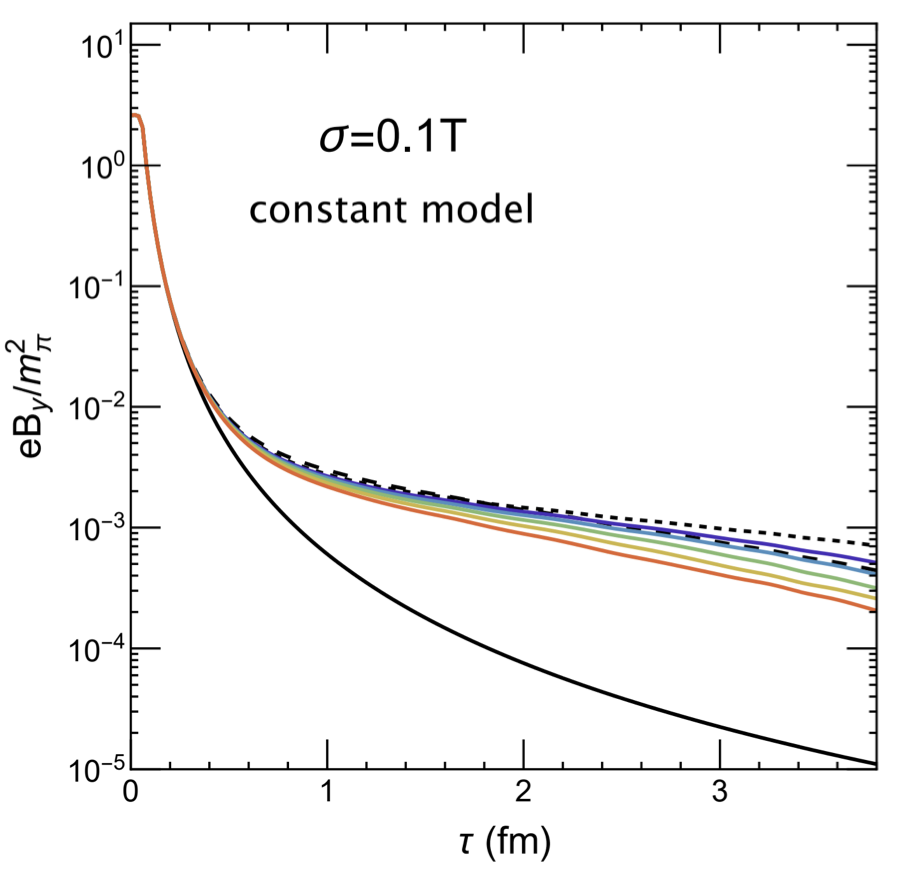}
\caption{Examples of dynamical magnetic field evolution with time in the hydrodynamic medium created by heavy ion collisions, with results from \cite{Gursoy:2018yai} (left) and from \cite{Huang:2022qdn} (right). 
\label{fig-B-time} 
}
\end{center}
\end{figure}

 The strong initial magnetic field   rapidly decays over a short period of time due to the departure of spectator protons down the collision pipeline. However, it has been proposed that the electric conducting current of the dense partonic medium in response to the $\mathbf{B}$ field decrease could significantly alter its time dependence. 
The dynamical evolution of the residue in-medium magnetic field in the mid-rapidity region is a very challenging problem to solve. 
Many attempts with varied degrees of rigor and approximations have been  made~\cite{McLerran:2013hla,Tuchin:2015oka,Inghirami:2016iru,Inghirami:2019mkc,Gursoy:2018yai,Roy:2017yvg,Pu:2016ayh,Muller:2018ibh,Guo:2019joy,Guo:2019mgh}. While it is  qualitatively expected that the medium effect could help increase the lifetime of $\mathbf{B}$ field, a quantitative answer is still lacking. Progress has been made on the framework of a  magneto-hydrodynamic (MHD) approach and  phenomenological simulations have been performed~\cite{Inghirami:2016iru,Inghirami:2019mkc,Hernandez:2017mch,Denicol:2018rbw,Denicol:2019iyh,Shokri:2018qcu,Siddique:2019gqh,Hattori:2022hyo,Wang:2023imu}.  However  the QGP may not have a large enough electrical  conductivity to be in an ideal MHD regime.   Another perhaps more  realistic approach aims to solve the in-medium Maxwell's equations in an  expanding and conducting fluid while neglecting the feedback of the $\mathbf{B}$ field on the medium's bulk evolution~\cite{Gursoy:2018yai,Huang:2022qdn}. See e.g.  examples in Fig.~\ref{fig-B-time}. In both approaches, an enhancement of the magnetic field as compared with the vacuum case is clearly observed.  The current conclusion is that the in-medium magnetic field lifetime sensitively depends on both the early time pre-equilibrium contributions and the precise value of conductivity for the hydrodynamic medium.    Additionally,   there are  interesting studies on other effects  induced by strong magnetic fields which could  be measured to help extract/constrain the in-medium $\mathbf{B}$ field in heavy ion collisions~\cite{Inghirami:2019mkc,Gursoy:2018yai,Guo:2015nsa,Muller:2018ibh,Guo:2019joy,Guo:2019mgh,Xu:2020sui,Aaboud:2018eph,Adam:2018tdm,Zha:2018tlq,Zha:2018ywo,Klein:2018fmp}.


\subsection{CME signature  in heavy ion collisions  }

 \begin{figure}[!hbt]
\begin{center}
\includegraphics[height=0.35\textwidth]{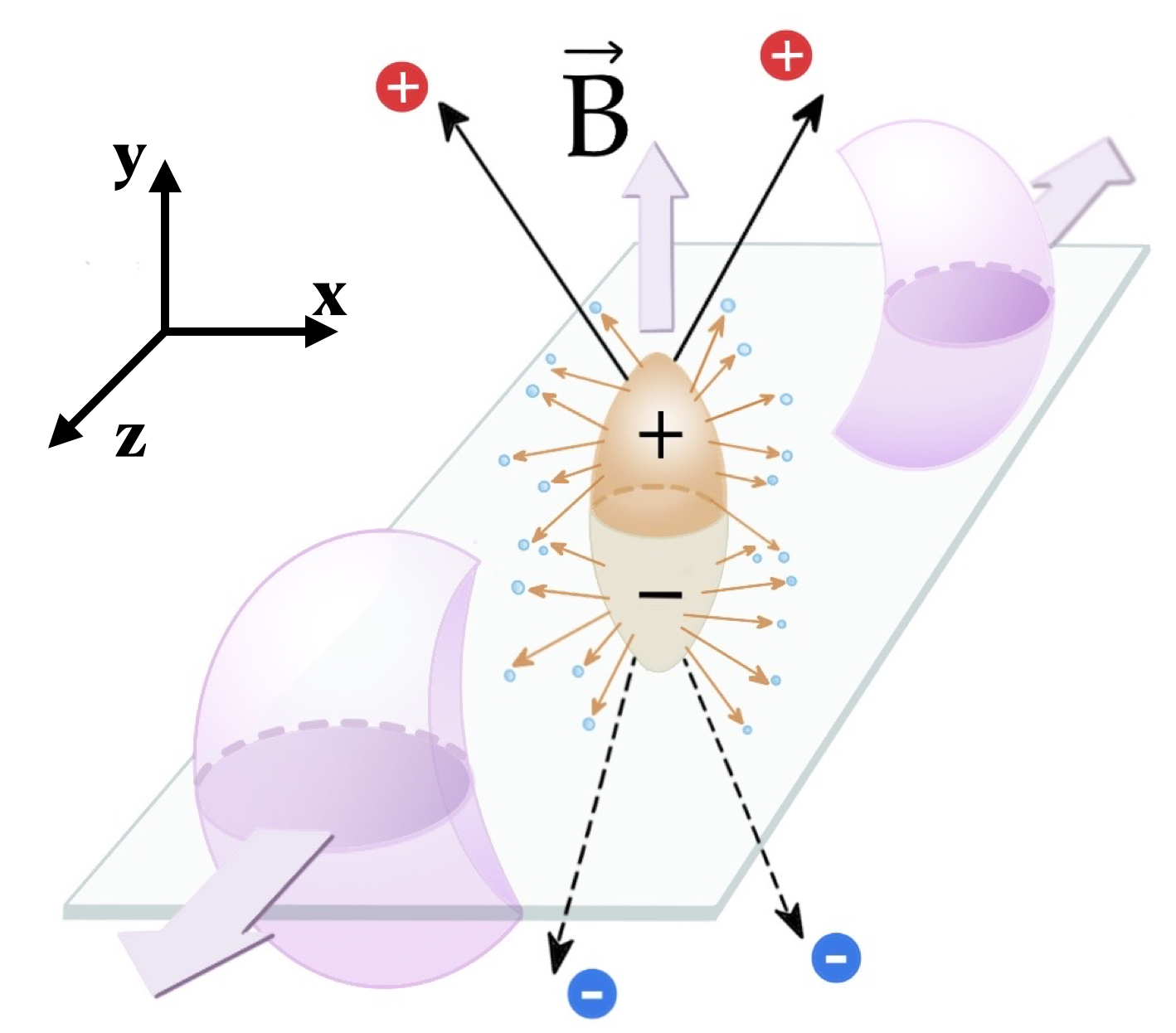}\qquad
\includegraphics[height=0.35\textwidth]{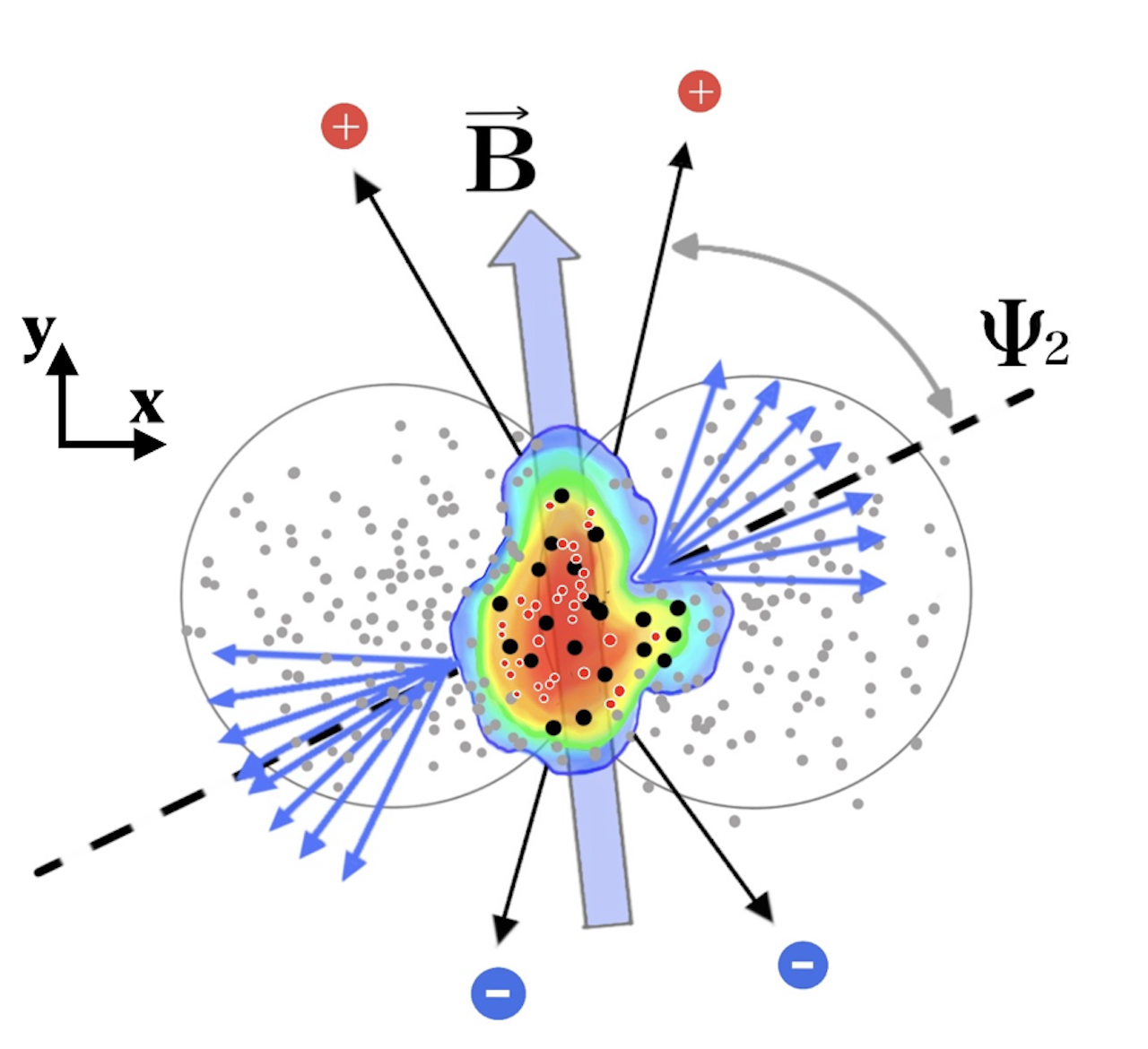}
\caption{ The 3-dimensional (left) and 2-dimensional (right) illustrations of the charge separation induced by chiral magnetic effect across the reaction plane. (Reproduced from \cite{Kharzeev:2020jxw}.)  
\label{fig-signal}
}
\end{center}
\end{figure}

As we have seen in the previous subsection, the CME is a phenomenon that causes an electric current along the magnetic field due to an imbalance of left and right handed fermions. On the experimental side, The CME-induced transport is expected to result in a dipole-like charge separation  along $\mathbf{B}$ field direction~\cite{Kharzeev:2004ey}, as illustrated in Fig.~\ref{fig-signal}, which could be measured by charge asymmetry in two-particle azimuthal correlations~\cite{Voloshin:2004vk}. Extensive searches have been carried out over the past decade  to look for its traces by STAR at the Relativistic Heavy Ion Collider (RHIC) as well as by ALICE and CMS at the Large Hadron Collider (LHC) with a variety of observables~\cite{Voloshin:2004vk,Xu:2017qfs,Zhao:2017nfq,Voloshin:2018qsm,Magdy:2017yje,Magdy:2018lwk,Tang:2019pbl}.  While showing encouraging hints of the CME, particularly over the RHIC beam energy range in both the AuAu collision system and the more recent isobar collision systems~\cite{Kharzeev:2020jxw,Kharzeev:2019zgg,Skokov:2016yrj,Voloshin:2010ut,Adam:2019fbq}, the interpretation of these data however remains inconclusive  due to significant background contamination. See more in-depth  discussions  in e.g. \cite{Bzdak:2019pkr,Kharzeev:2015znc,Li:2020dwr,Zhao:2019hta,Bzdak:2012ia}.

\subsubsection{How to detect the CME?}

To detect the CME in heavy ion collisions, we need to keep a few things in mind. First, any observable we devise must be averaged over many detected particles and events. Second, we need to measure a phenomenon that is both directional and parity odd, which may vanish after averaging unless done properly. Therefore, it is crucial to get a handle on the directionality, in this case, the direction of the magnetic field in each event. Finally, even if the direction of the magnetic field is known, the signal itself can flip from event to event. This is because the CME-driven current could either align with or oppose the magnetic field, depending on the handedness imbalance. Let's break this down step by step. 


The CME involves the preferential emission of charged particles along a specific direction. Therefore, any observable designed to detect the CME will likely involve measuring the angles of emitted particles and analyzing their azimuthal distribution. In order to quantify various modes of collective motion of the medium formed in hadronic and heavy-ion collisions, the azimuthal distribution of final-state particles can be Fourier-decomposed as
\begin{equation}
    \frac{dN_{\alpha}}{d\phi} \approx \frac{N_\alpha}{2\pi} \left[1 + 2v_{1,\alpha}\cos(\phi) + 2a_{1,\alpha}\sin(\phi) + 2v_{2,\alpha}\cos(2\phi) + \cdots \right]\,
\label{equ:Fourier_expansion}
\end{equation}
where, various coefficients ($v_{n,\alpha}, a_{n,\alpha}$) are moments of the angular distribution of particles. To construct an observable for detecting the CME, we will explore whether any of these coefficients can be utilized for our purpose. We want to measure the separation of electric charges along the magnetic field direction. To do this, we need to find the magnetic field direction in each event. Conveniently, one can use the directed or elliptic flow coefficients $v_1$ and $v_2$ for this purpose. These coefficients are defined as $v_2=\left<\cos(2\phi-2\Psi_{2})\right>$ and $v_1=\left<\cos(\phi-\Psi_{1})\right>$, where $\phi$ is the particle azimuthal angle. The brackets $\left<\cdots\right>$ indicates an average over all particles in the event and then over all events. The flow angles $\Psi_1$ and $\Psi_2$ are estimates for the reaction plane angle $\Psi_{RP}$, which is the plane constructed from the impact parameter vector and the collision axis. The magnetic field direction is approximately perpendicular to the reaction plane, according to theoretical models.

\begin{figure}
    \centering
    \includegraphics[width=0.8\textwidth]{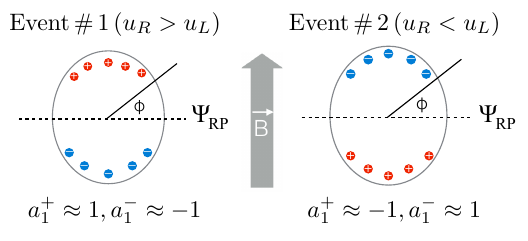}
    \caption{A cartoon to illustrate the basic idea of how the CME can be measured. In an ideal scenario, the CME causes charge separation across the reaction plane $\Psi_{RP}$, which is perpendicular to the magnetic field $\vec{B}$ direction. This charge separation means that in one event (Event \#1), positive charges will be emitted at azimuthal angles $\phi=\pi/2$ while negative charges will be along $\phi=-\pi/2$, if there are more right-handed quarks than left-handed ($u_R>u_L$). The quantity $a_1$ defined in Eq.\protect\ref{eq_a1_observables} $a_1^\pm=\left<\sin(\phi^\pm - \Psi_{RP})\right>$ can detect this angular pattern of emission, leading to $a_1^+=1$, $a_1^-=-1$ for the distribution for a positive or a negative particle, respectively. This makes $a_1$ an observable sensitive to the CME and the difference in the sign of $a_1^+$ and $a_1^-$ can be a signature of the CME. The limitation of $a_1$ is as follows: in the next event (Event \#2), the imbalance can be opposite, leading to $u_L<u_R$, causing $a_1^+=-1$, $a_1^-=+1$. Averaging over many events, one will find $\left<a_1^+\right>=\left< a_1^-\right>=0$. Therefore, an observable constructed out of the variance $\left<a_1^2\right>$ called $\gamma$ is necessary to detect signatures of CME.} 
    \label{fig:cme_a1_observable}
\end{figure}

Once we have a handle on the magnetic field direction, the next step is to detect if there is charge separation along that direction -- in other words, charge separation perpendicular to $\Psi_{RP}$. One option is to use the quantity
\begin{equation}
a_1=\left<\sin(\phi^\alpha - \Psi_{RP})\right>,    
\label{eq_a1_observables}
\end{equation}
 where $\phi^{\alpha}$ is the azimuthal angle of a particle with $\alpha=+,-$. This quantity is based on the idea that the magnetic field is perpendicular to $\Psi_{RP}$, and the CME will cause a positive particle to be emitted perpendicular to $\Psi_{RP}$ with $\phi^+-\Psi_{RP}=\pi/2$, and a negative particle to be emitted opposite to $\Psi_{RP}$ with $\phi^--\Psi_{RP}=-\pi/2$. Therefore, we can compare $a^+=\sin(\phi^+-\Psi_{RP})=1$ and $a^-=
 \sin(\phi^--\Psi_{RP})=-1$, which are expected to be different. However, there is a problem.

We must not forget to  account for the fact that the CME current can reverse its direction in different events, depending on the excess of left-handed or right-handed fermions. See Fig.\ref{fig:cme_a1_observable}.
This is a local parity-violating effect. Therefore, we cannot use a single charge particle and calculate a quantity like $\Delta a_1=a_1^+-a_1^-$ where $a_1^+=\left<\sin(\phi^+- \Psi_{RP})\right>$ and $a_1^-=\left<\sin(\phi^--\Psi_{RP})\right>$, where $\phi^+$ and $\phi^-$ are the azimuthal angles of positive and negative particles and $\Psi_{RP}$ is the reaction plane angle. These quantities are expected to average to zero over many events leading to $\left<a_1^+\right> = \left<a_1^-\right> = 0$ by the symmetry of the problem. Any signature of $\left<a_1^+ \right>\ne \left<a_1^-\right>$ would suggest the presence of a net global violation of parity which is not allowed in QCD. This was demonstrated by the STAR collaboration in Ref~\cite{STAR:2013ksd}.

\begin{figure}
    \centering
    \includegraphics{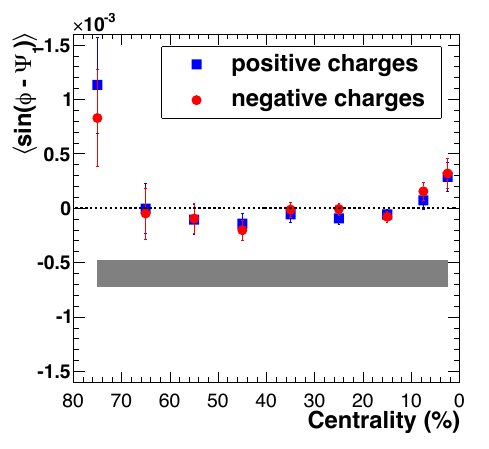}
    \caption{The figure taken from Ref.~\cite{STAR:2013ksd} shows the average value of $a_1=\left<sin(\phi-\Psi_{1})\right>$ for positively and negatively charged particles as a function of centrality in Au+Au collisions at $\sqrt{s_{NN}}=$ 200 GeV measured by the STAR collaboration. The systematic uncertainty for both charge types is estimated by comparing correlations from opposite pseudorapidity regions and is indicated by the width of the shaded area below the data points. The figure indicates that there is no significant difference between positively and negatively charged particles.}
    \label{fig:a1_variable}
\end{figure}

On Fig.\ref{fig:a1_variable} one can see the measurements of $\left<\sin(\phi-\Psi_{1})\right>$, which was  measured separately for positive and negative particles. Here $\Psi_1$  is the first order event plane constructed using the directed flow of the spectator neutrons using the RHIC zero-degree-calorimeter (ZDC)~\cite{STAR:2005btp}  and was used as a proxy for $\Psi_{RP}$. Azimuthal angle, $\phi$, is measured using the Time-Projection Chamber (TPC). The results show that $a_1^+=a_1^-$ within uncertainties ($\Delta a_1$=0) for various centralities (that is a proxy for impact parameter or violence of collision in other words) of Au+Au collisions at $\sqrt{s_{_{NN}}}=200$ GeV. The absolute values of $a_1^+$ and $a_1^-$ have small but statistically insignificant deviation from zero which is charge independent and the origin of which is not clearly understood. For example, it could be that the moments of the angular distributions also suffer from various background effects that are not charge dependent but can lead to non-zero values -- we need to eliminate them. It will be more clear when we discuss them in the next section, for now, let us ignore that fact that $a_1\ne0$ and only focus on the fact that $\Delta a_1=0$. Why does this happen? Is there some cancellation due to averaging going on like what is shown on Fig.\ref{fig:cme_a1_observable}.

The observation of $\Delta a_1$=0 in Fig.\ref{fig:a1_variable} is consistent with the expectations that $a_1$ may still be sensitive to CME driven charge separation and the observation of $\Delta a_1$=0 is just a consequence of the cancellation of CME current that flips event-by-event. Regardless of whether $\Delta a_1$=0 is indicative of such conclusion, it is evident that the observables $a_1$ or $\Delta a_1$ cannot be used to measure charge separation in an experiment. A possible solution to this problem is to move from moments of the distribution of a single particle to a pair of particles. So, instead of measuring $a_1$ we measure a quantity that is equivalent to  the variance of $a_1$, i.e. equivalent to measuring $\left<a_1^2\right>$. This naturally leads to the following way to overcome the cancellation due to flipping:
\begin{equation}
\gamma^{\alpha,\beta}_{\perp}=\langle \sin(\phi^{\alpha}-\Psi_{RP})\sin(\phi^{\beta}-\Psi_{RP}) \rangle,
\label{eq_gamma_perp_observable}
\end{equation}
where $\alpha=\pm$ and $\beta=\pm$ refer to the electric charges of the two particles.The use of the subscript $\perp$ is intentional; this is to remind ourselves that we are referring to emission perpendicular to $\Psi_{RP}$. To measure charge separation, we now need to consider four different quantities $\gamma^{+,-}_\perp$, $\gamma^{-,+}_\perp$, $\gamma^{+,+}_\perp$, $\gamma^{-,-}_\perp$ -- let us break them down one by one.

The two combinations of opposite sign pairs $\gamma^{+,-}_\perp$, $\gamma^{-,+}_\perp$ are equivalent but measure in which direction/angle a positive particle is emitted with a reference to a negative particle, or vice-versa. For every pair of particles with opposite signs, the goal is to achieve a maximum signal when the positive particle is emitted perpendicularly to a reference plane ($\Psi_{RP}$) and the negative particle is emitted in the opposite direction. 
In the following we show that unlike $a_1$ this quantity (equivalent to $\left<a_1^2\right>$) will not cancel after event averaging. 

Let us consider an idealized scenario of CME to understand this better. If in one event $\phi^{+}-\Psi_{RP}=\pi/2$ for a positive particle, we expect  $\phi^{-}-\Psi_{RP}=-\pi/2$ for a negative particle  -- this is consistent with the CME given an initial excess of right handed and positively charged fermions. The result will be
\begin{equation}
   \sin(\phi^{+}-\Psi_{RP})\sin(\phi^{-}-\Psi_{RP})=1\times -1=-1. 
\end{equation}
If in the next event, the handedness is flipped, i.e., then there will be more left-handed fermions. In this event one expects the positive particle to emit at $-\pi/2$, while the negative particle to emit at $\pi/2$. The quantity of interest will be 
 \begin{equation}
      \sin(\phi^{+}-\Psi_{RP})\sin(\phi^{-}-\Psi_{RP})=1\times -1=-1
 \end{equation}
 Therefore, averaged over two events, according to Eq.\ref{eq_gamma_perp_observable},  $\gamma^{+,-}_{\perp}=-1$. The result is same if we had considered $\gamma^{-,+}_{\perp}=-1$. One can further average the two as :
 \begin{equation}
  \gamma^{os}_{\perp}=\frac{1}{2} \left(\gamma^{+,-}_{\perp}+\gamma^{-,+}_{\perp}\right),    
 \end{equation}
 the super-script ``os" refers to opposite-sign pairs ($+-$, or $-+$). Clearly, in this example, $\gamma^{os}_{\perp}=-1\ne 0$. In other words, the flipping of handedness imbalance did not average $\gamma^{+,-}$ to zero.

Now let us go to the two remaining quantities of interest combining two same sign pairs, i.e. $\gamma^{+,+}_\perp$ or $\gamma^{-,-}_\perp$, combining them together we get
\begin{equation}
\gamma^{ss}_\perp=\frac{1}{2}(\gamma^{+,+}_\perp+\gamma^{-,-}_\perp)
\end{equation}
Using the same logic as before we will get the following results in the first event where $\phi^{+}-\Psi_{RP}=\pi/2$ and $\phi^{-}-\Psi_{RP}=-\pi/2$, the quantity of interest is:
\begin{eqnarray}
\sin(\phi^{+}-\Psi_{RP})\sin(\phi^{+}-\Psi_{RP})=1, \nonumber \\    
\sin(\phi^{-}-\Psi_{RP})\sin(\phi^{-}-\Psi_{RP})=1.
\label{eq_sin_phipm}
\end{eqnarray}
 If in the second event, $\phi^{+}-\Psi_{RP}=-\pi/2$ and $\phi^{-}-\Psi_{RP}=\pi/2$ due to flipping of handedness, the result will still be the same as Eq.\ref{eq_sin_phipm}, giving us $\gamma^{ss}_\perp=1$. The immediate contrast between $\gamma^{os}_\perp=-1$ and $\gamma^{ss}_\perp=1$ already indicates that we can use this variable to detect the charge separation effect, which is the expected signature of the CME.

 Now that we have understood how $\gamma^{ss}_\perp$ and $\gamma^{os}_\perp$ work, let us ask a few questions. In an experiment after averaging over many events we will get a number. How are we going to interpret this number? One possibility is we take the difference $\gamma^{ss}_\perp- \gamma^{os}_\perp$. But then what should we compare these number to?  In the above case, we considered an ideal case, where we expected the positive particles emitted at an angle  $\phi^{+}-\Psi_{RP}=\pi/2$ while negative particles emitted at angles $\phi^{-}-\Psi_{RP}=-\pi/2$. But in the real collisions, conditions are far from reality. The strength of $\gamma^{\alpha, \beta}_\perp$ due to the CME is unknown, it is predicted to be small (we will discuss about the magnitudes later) as the CME is a phenomenon driven by quantum fluctuations. We expect $0\le\gamma^{\alpha,\beta}_\perp\le1$. Like any other measurement it is expected that the signal can easily be overwhelmed by various sources of background. An experimental baseline of $\gamma^{\alpha, \beta}_\perp$ is desirable. We can start to think about possible sources of background as follows. 
The implementation of $\gamma^{\alpha, \beta}_\perp$ as an observable is much better than $a_1$ because former one takes care of the parity-odd feature of the phenomenon. But then by design $\gamma^{\alpha, \beta}_\perp$ becomes susceptible to parity-even processes that are not related to CME. 

If $\gamma^{os}_\perp\ne\gamma^{ss}_\perp$ then the difference would indicate a separation of charge with respect to the reaction plane. This is something that we expect to happen due to the CME, see figure~\ref{fig:cme_a1_observable}. The problem is that there will certainly be background processes that can affect $\gamma^{os}_\perp - \gamma^{ss}_\perp$ and so we need a baseline for reference that is not sensitive to the CME. 
%
%
One solution is to create such a baseline. What does a baseline mean? It refers to an observable that has similar features, constructed in a similar way as $\gamma_\perp$ but will not have the sensitivity to CME. This can be done by exploiting one feature of CME which is that the phenomenon leads to charge separation across $\Psi_{RP}$ (or along $\vec{B}$). Therefore one naive expectation is that CME will not lead to charge separation across along $\Psi_{RP}$ (or perpendicular to $\vec{B}$. Therefore a possible experimental baseline observable of $\gamma_\perp$ will be $\gamma_{||}$. This quantity is defined in a way similar to eq.\ref{eq_gamma_perp_observable} but instead of sine one uses cosine to measure charge separation along $\Psi_{RP}$ as
\begin{equation}
    \gamma^{\alpha,\beta}_{||}=\langle \cos(\phi^{\alpha}-\Psi_{RP})\cos(\phi^{\beta}-\Psi_{RP}) \rangle. 
\label{eq_gamma_parallel_observable}
\end{equation}
One can therefore definitely measure $\gamma^{os}_{||}$ and $\gamma^{ss}_{||}$. The question is what this observable measures. It measures charge separation along $\Psi_{RP}$ that is not due to CME by construction, because there is no B-field component parallel to $\Psi_{RP}$. Therefore, these observables are purely by construction measures of charge separation driven solely by non-CME backgrounds. They can be compared to $\gamma^{os}_{\perp}$ and $\gamma^{ss}_{\perp}$.

\begin{figure}
    \centering
    \includegraphics[width=0.8\textwidth]{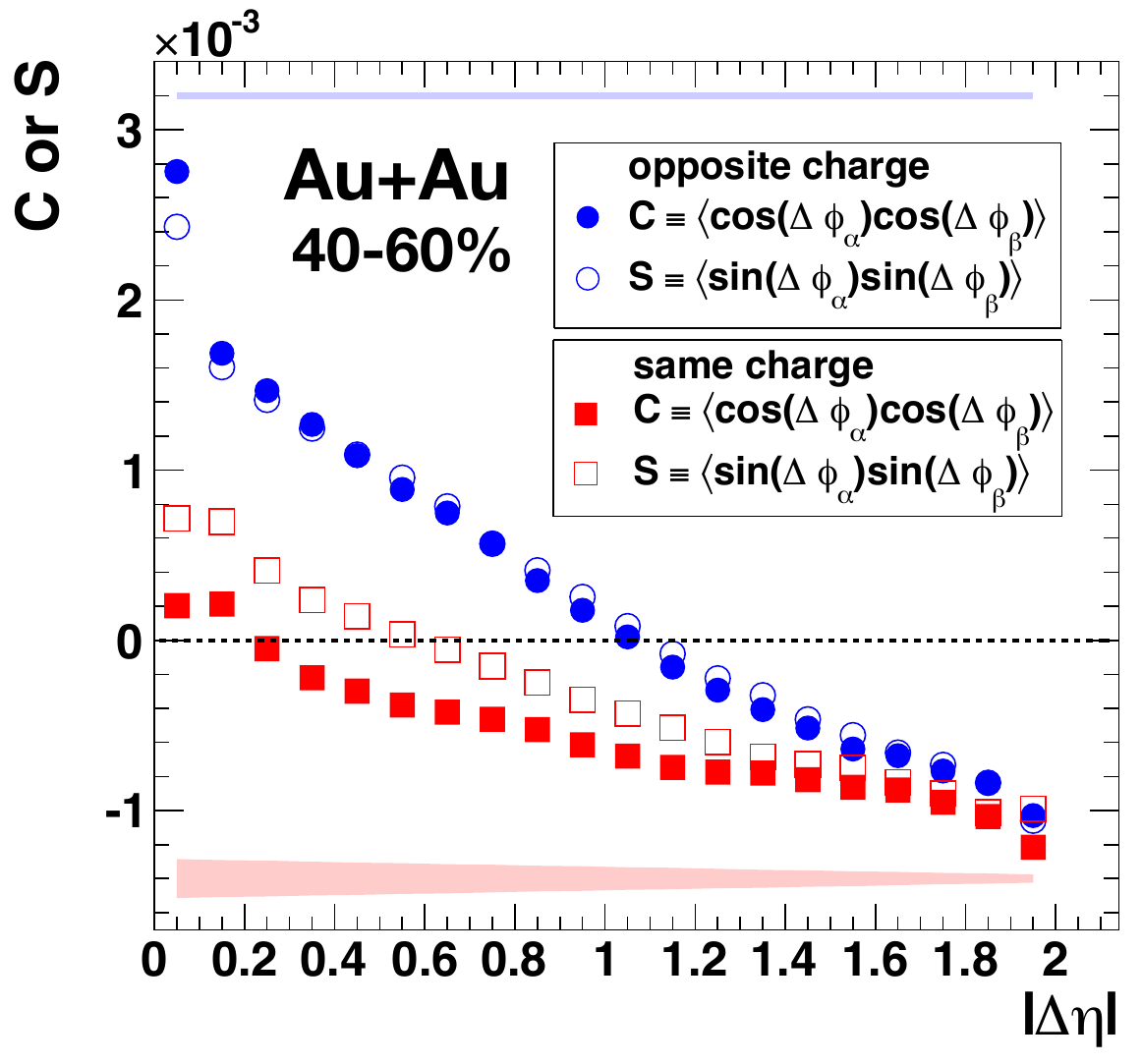}
    \caption{Three-point correlations of out-of-plane and in-plane components, denoted by $\gamma^{ss}_\perp$, $\gamma^{os}_\perp$ and $\gamma^{ss}_{||}$, $\gamma^{os}_{||}$, respectively, for 40-60\% Au+Au collisions at $\sqrt{s_{_{NN}}}$=200 GeV. The correlations are shown for different ranges of the average pseudorapidity of the detector acceptance, where $\langle \eta \rangle = (\eta_\alpha + \eta_\beta)/2$. The shaded bands represent the systematic errors, which only apply to the difference between the out-of-plane and in-plane components.}
    \label{fig:cme_components_observables}
\end{figure}

Fig.\ref{fig:cme_components_observables} shows the measurements of individual $\gamma^{os}_{\perp}$, $\gamma^{ss}_{\perp}$, $\gamma^{os}_{||}$ and $\gamma^{ss}_{||}$ performed by the STAR experiment~\cite{STAR:2013ksd}. The measurements are presented in Au+Au collisions in 40-60\% centrality with respect to the relative pseudorapidity $|\Delta \eta|$ between the pairs of particles $\alpha$ and $\beta$. For the time being, we can ignore the quantity on the x-axis. The relative pseudorapidity dependence of charge separation will be important later. The notations on the plots are a bit different from what we introduced in the previously. On the plots, the notations $\Delta\phi_{\alpha,\beta}$ refers to $\phi^{\alpha,\beta}-\Psi_{RP}$.

The first observation is that the observables for opposite charges, represented by circles, differ from those for same-sign charges shown by squares. This difference suggests that there is a non-zero charge separation across or along $\Psi_{RP}$, which is a step forward for experimental detection of CME. This looks much more promising over the $\left<a_1\right>$ results shown in Fig.\ref{fig:a1_variable}, which yielded no difference between positive and negative charges. The $\gamma$ observables, on the other hand, show non-zero difference between opposite and same sign charges.

The solid and open circles show the measurements of $\gamma^{os}_{||}=C=\left<\cos(\Delta\phi_\alpha)\cos(\Delta\phi_\beta)\right>$ and $\gamma^{os}_{\perp}=S=\left<\sin(\Delta\phi_\alpha)\sin(\Delta\phi_\beta)\right>$. In other words, the charge separation along and perpendicular to $\Psi_{RP}$. Two things to notice here.  First of all, it is interesting to note that they are very similar, except for the highest values of $|\Delta\eta|$. 
At first glance, this result may seem disappointing. We might expect that a pair of opposite charges would be more correlated perpendicular to $\Psi_{RP}$ due to CME. In other words, we might expect that $\gamma^{os}_{\perp}=S$ would be larger than $\gamma^{os}_{||}=C$. The latter should only reflect background effects, right? Does this mean that we do not see any evidence of CME? Or could it be that the background correlations are so strong that they mask the CME signal? Clearly, we need more information to figure out what is happening. Another observation is that the magnitudes of these observables are of the order of $10^-3$, much smaller than unity, as we discussed in the ideal scenario of CME. However, this is not surprising, as in a real experiment the signal strength can be much lower than the ideal expectations. 
 
The two other correlators, shown by the solid and open squares on the same plot, reveal a significant difference. They represent the measurements of $\gamma^{ss}_{||}=C=\left<\cos(\Delta\phi_\alpha)\cos(\Delta\phi_\beta)\right>$ and $\gamma^{ss}_{\perp}=S=\left<\sin(\Delta\phi_\alpha)\sin(\Delta\phi_\beta)\right>$. These two correlators indicate that the correlated emission of a pair of same-sign particles is different when they are perpendicular or parallel to the $\Psi_{RP}$. This is an interesting finding that could be consistent with the expectations of CME. We can refer to the cartoon of Fig.\ref{fig:cme_a1_observable} and recall that CME indeed causes same-sign particles to move together perpendicular to the $\Psi_{RP}$. However, this may require more evidence to confirm such conclusions. Why do we also see even correlated emission of same-sign particles along (parallel to) the $\Psi_{RP}$? Is this due to some non-CME background?

The experimental measurements in Fig.\ref{fig:cme_components_observables} suggest some possible conclusions. The $\gamma_\perp$ and $\gamma_{||}$ observables are better than $\left< a_1\right>$ for detecting charge separation, as they show a clear difference between same-sign and opposite-sign pairs. However, the magnitude of $\gamma_\perp$ and $\gamma_{||}$ are much smaller than what one would expect from ideal CME scenarios. Moreover, there could be other sources of correlations that are not related to CME, which need to be carefully studied. To further investigate the charge separation along and perpendicular to the $\Psi_{RP}$, it is useful to define a new observable as the difference:
\begin{equation}
\gamma = \gamma_{||} - \gamma_{\perp}.
\end{equation}
\begin{figure}[htb]
    \centering
    \includegraphics[width=0.8\textwidth]{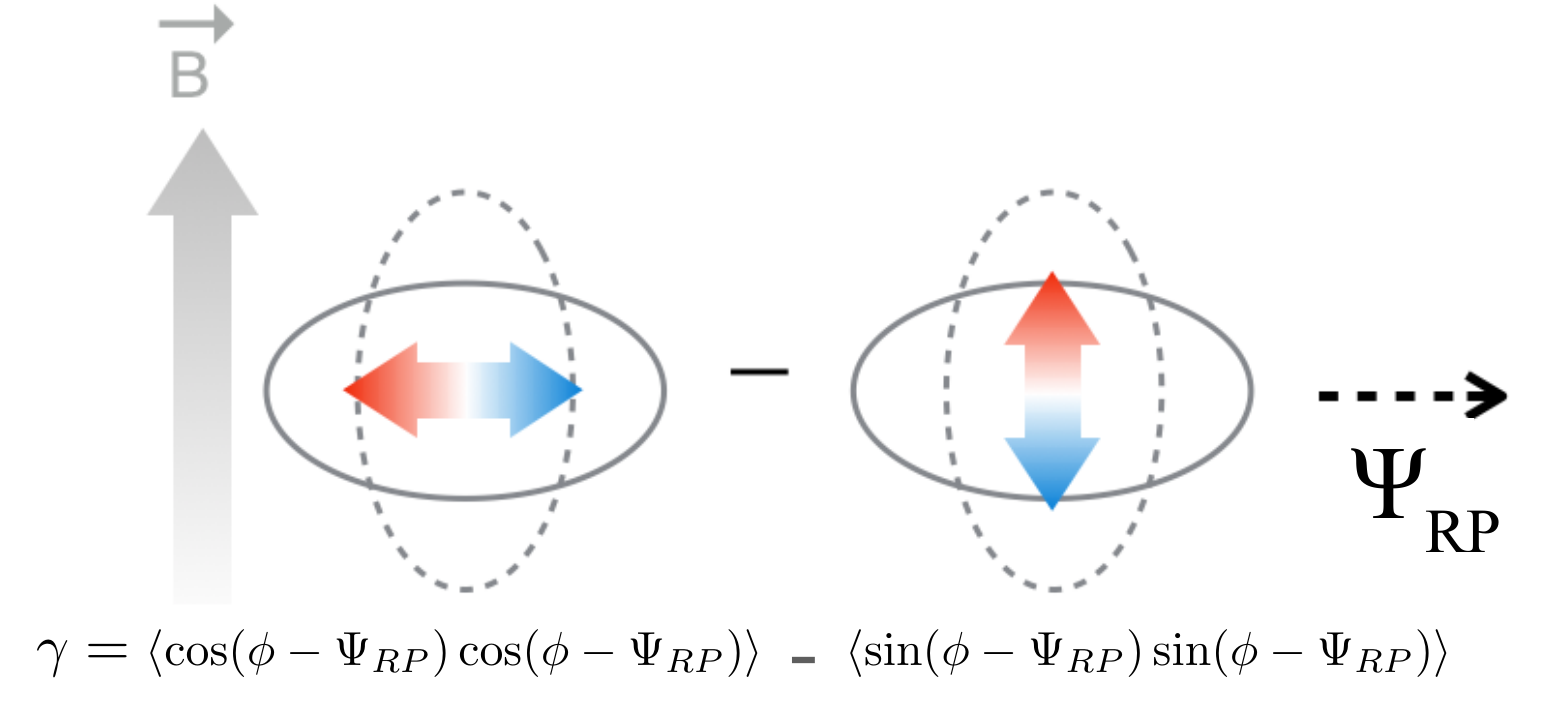}
    \caption{A possible cartoon representation of the $\gamma$-correlator introduced in Ref~\cite{Voloshin:2004vk} is designed to measure the difference between the correlated emission of a pair of particles ($\alpha$ and $\beta$) along and perpendicular to the reaction plane $\Psi_{RP}$, as accounted for by the cosine and the sine terms. In an ideal scenario, CME is expected to lead to such a correlated emission along $\vec{B}$, which is expected to be perpendicular to $\Psi_{RP}$ -- this effect will be captured by the sine-term. However, the introduction of the cosine term is a clever design suited to provide a data-driven baseline. CME cannot lead to correlated emission of a pair parallel to $\Psi_{RP}$, therefore the measurement of the cosine term must come from non-CME sources and serve as an experimental baseline. Choosing $\alpha$ and $\beta$ as same charge or opposite charge provides us with two measurements of $\gamma^{SS}$ and $\gamma^{OS}$ -- together they provide us more information about charge separation across $\Psi_{RP}$. There is in fact one flaw to this clever design, which is discussed in the texts.}
    \label{fig:gamma_correlator_observables}
\end{figure}

This is called the $\gamma$-correlator and was first proposed in Ref~\cite{Voloshin:2004vk}. It can be rewritten as:
\begin{equation}
    \gamma^{\alpha\beta} = \left<\cos(\phi_\alpha-\Psi_{RP})\cos(\phi_\beta-\Psi_{RP})\right>-\left<\sin(\phi_\alpha-\Psi_{RP})\sin(\phi_\beta-\Psi_{RP})\right>. 
    \label{eq:gamma_correlation_terms_observables}
\end{equation}
Or in a more compact representation of 
\begin{equation}
    \gamma^{\alpha\beta} = \left<\cos(\phi_\alpha + \phi_\beta - 2 \Psi_{RP} \right>,  
\label{eq:gamma_correlator_observable}
\end{equation}
which can be represented by the cartoon of Fig.~\ref{fig:gamma_correlator_observables}. 

It is important to pause for a moment and look at Eq.~\ref{eq:gamma_correlator_observable} and appreciate that this is indeed a clever design to search for a complex phenomenon such as CME. Let us therefore recap how we got here. We started with an ideal picture of CME separating charge along $\vec{B}$ or perpendicular to $\Psi_{RP}$ in Fig.~\ref{fig:a1_variable}. We therefore came up with an intuitive picture and introduced Eq.~\ref{eq_a1_observables} $a_1=\sin(\phi^\alpha-\Psi_{RP})$ that measures the single-particle asymmetry. It was a good start, but the problem was that it led to vanishing $\left<a_1^+\right>-\left<a_1^-\right>=0$, after averaging over configurations. Therefore, in Eq.~\ref{eq_gamma_perp_observable}, we introduced $\gamma_\perp^{\alpha,\beta}=\left<\sin(\phi^\alpha-\Psi_{RP})\sin(\phi^\beta-\Psi_{RP})\right>$, which basically measures the variance of $a_1$, $\left<a_1^2\right>$, and solves the issue of null result. Finally, the last piece of the puzzle was to introduce an equivalent term $\gamma_{||}^{\alpha,\beta}$ by replacing the sine-terms with cosine in $\gamma_{\perp}$. Since CME-driven signal is in $\vec{B}$ direction, it cannot impact correlations perpendicular to $\vec{B}$, so $\gamma_{||}^{\alpha,\beta}=\left<\cos(\phi^\alpha-\Psi_{RP})\cos(\phi^\beta-\Psi_{RP})\right>$ is a good experimental baseline that will capture everything that is going on not from CME. The $\gamma$-correlator, $\gamma=\gamma_{||}-\gamma_{\perp}$ in Eq.\ref{eq:gamma_correlator_observable}, is a compact way of capturing the aforementioned effects. Depending on the combination of $\alpha,\beta$ (charge of particles) the correlator can be measured of the same-sign pairs, i.e. $\gamma^{SS}$ or for the opposite sign pairs $\gamma^{OS}$. They are defined as 
\begin{eqnarray}
    \gamma^{SS} &=& \frac{1}{2} (\gamma^{++} + \gamma^{--}) \\
    \gamma^{OS} &=& \frac{1}{2} (\gamma^{+-} + \gamma^{-+}). \\
        \label{eq_gamma_SS_OS_observables}
\end{eqnarray}
The expectation is that in the ideal scenario of CME after event averaging $\gamma^{OS}$ will be different from $\gamma^{SS}$. In fact, a possible CME-sensitive observable of choice could be
\begin{equation}
    \Delta \gamma = \gamma^{OS} - \gamma^{SS}.
    \label{eq_deltagamma_observables}
\end{equation}
One can go back to the scenario described in the previous section in connection to the cartoon of Fig.\ref{fig:cme_a1_observable}, but this time one needs to consider a pair of particles. If we have CME, it will cause two positive particles to be emitted perpendicular to $\Psi_{RP}$ with $\phi^+-\Psi_{RP}=\pi/2$, and two negative particles to be emitted opposite to $\Psi_{RP}$ with $\phi^--\Psi_{RP}=-\pi/2$. This will lead to $\gamma^{OS}=1/2(\cos(\pi/2 - \pi/2) + \cos(-\pi/2 + \pi/2))=1$ and $\gamma^{SS}=1/2(\cos(\pi/2 + \pi/2) + \cos(-\pi/2 - \pi/2))=-1$. Therefore, $\gamma^{OS} \ne \gamma^{SS}$, or in other words, $\Delta\gamma>0$. One gets the same results for the scenario if the direction of positive and negative particles are flipped across $\Psi_{RP}$ as shown in Fig.\ref{fig:cme_a1_observable}.

There is however, a flaw in the design of this correlator which was already realized in the original paper ~\cite{Voloshin:2004vk} where the $\gamma$-correlator was introduced. This is related to the fact that background sources may not be the same in the parallel and the perpendicular direction of $\Psi_{RP}$. Using $\gamma_{||}$ as a baseline for background of non-CME origin for $\gamma_{\perp}$ assuming is not correct. We discuss this in the next section in detail. 

Before concluding the discussion of observables, let's us consider another way to combining $\gamma_{||}$ and $\gamma_{\perp}$. Unlike Eq.\ref{eq:gamma_correlation_terms_observables} what if one adds them instead, this gives rise to a different observable known as: 
\begin{equation}
    \delta = \gamma_{||} + \gamma_{\perp}, 
\end{equation}
which leads to
\begin{equation}
       \delta^{\alpha\beta} = \left<\cos(\phi_\alpha-\Psi_{RP})\cos(\phi_\beta-\Psi_{RP})\right>+\left<\sin(\phi_\alpha-\Psi_{RP})\sin(\phi_\beta-\Psi_{RP})\right>. 
    \label{eq:delta_correlation_terms_observables}
\end{equation}
In a more compact form the $\delta$-correlator is written as
\begin{equation}
        \delta^{\alpha\beta} = \left<\cos(\phi_\alpha - \phi_\beta) \right>.
\label{eq:delta_correlator_observable}
\end{equation}
It is interesting to note that $\delta-$correlator does not include any $\Psi_{RP}$, therefore it measures the charge separation that may or may not be correlated to the reaction plane. This can be done by measuring a quantity similar to Eq.\ref{eq_deltagamma_observables} written as
\begin{equation}
    \Delta\delta = \delta^{OS}-\delta^{SS}.
    \label{eq_delta_delta_observable}
\end{equation}



\subsubsection{The first measurement by the STAR collaboration}
\begin{figure}
    \centering
    \includegraphics{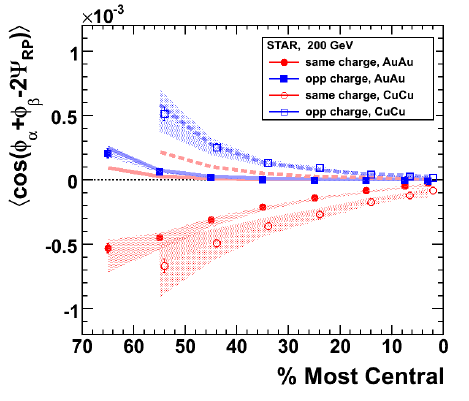}
    \caption{The first measurement of the three-particle correlator that probes the charge separation across the reaction plane in Au+Au and Cu+Cu collisions at $\sqrt{s_{_{NN}}}$= 200 GeV from Ref~\cite{STAR:2009wot}. The data points are compared with the HIJING model predictions for the background correlations from non-collective effects, shown as the thick solid (Au+Au) and dashed (Cu+Cu) lines. The shaded bands represent the uncertainty from the measurement of $v_2$, which is used to subtract the background correlations. The figure shows that the data points deviate significantly from the HIJING model, indicating the presence of charge separation, which is consistent with the chiral magnetic effect (CME) and its background contributions due to flowing resonances. The collision centrality increases from left to right.}
    \label{fig:first_measurement}
\end{figure}

Fig.\ref{fig:first_measurement} shows the first measurements from the STAR collaboration on the $\gamma$-correlator~\cite{STAR:2009wot}. The results are presented for same-sign ($\gamma^{SS}$) and opposite ($\gamma^{OS}$) pairs. The results are plotted as a function of centrality of collisions, where 0-5\% indicates the most central (most violent, smaller impact parameter and more overlap between nuclei) and 70-80\% indicates the most peripheral (least violent, larger impact parameter and less overlap between two nuclei) collisions. The measurements are presented in two collision systems: Au+Au (solid symbols) and Cu+Cu (open symbols) at the same collision energy. The harmonic plane $\Psi_2$ associated with the elliptic anisotropy determined by the particles within the central detector (within $|\eta|<1$) was used as a proxy for the $\Psi_{RP}$. Therefore, the quantity estimated in Fig.\ref{fig:first_measurement}
\begin{equation}
\gamma=\cos(\phi^\alpha + \phi^\beta - 2 \Psi_2) \
\label{eq_gamma_psi2_plane_observables}
\end{equation}

The first important observation is that in both systems, the measurements of $\gamma^{OS}$, shown by blue color points, are above zero. The measurements for $\gamma^{SS}$ are distinctly negative and therefore different from $\gamma^{OS}$. A few more observations can be made, all of which are important. The overall magnitude of all the measurements is of the order of 10$^{-4}$-10$^{-5}$, and the deviation of $\gamma^{SS}$ from zero is larger than $\gamma^{OS}$. There is an overall increase in such a deviation or the magnitude of the $\gamma-$correlator while going from central to peripheral events. The magnitude of the $\gamma$-correlator vanishes in most central events. A number of conclusions can be made from such observations.

Based on the discussions in the previous sections, the observation of different magnitude and sign of $\gamma^{OS}$ and $\gamma^{SS}$ is consistent with the expectations of CME. The magnitude of $\gamma$ going to zero from central to peripheral events is also consistent with the expectation of CME. This is because, in the scenario of CME, the strength of the signal is expected to be dependent on the correlation between the orientation of the $\vec{B}-$field, say $\Psi_B$, and the event plane $\Psi_2$ that is used for the proxy of $\Psi_{RP}$
\begin{equation}
\gamma_B = \left<\cos(2\Psi_B - 2\Psi_2)\right>.
\label{eq_gamma_B_observables}
\end{equation}
This quantity $\gamma_B$ measures the correlation of $\vec{B}-$field and $\Psi_{RP}$, and is expected to vanish in most central events where the impact parameter is small, as shown in Fig.\ref{fig-B-field}. This is because both the direction of $\vec{B}-$field and the proxy $\Psi_2$ for $\Psi_{RP}$ used for estimation of $\gamma$-correlator become random.

However, it is important to note that a number of observations can be made that are not consistent with expectations of CME. First of all, as shown in Fig.\ref{fig-B-field}, the correlation between $\vec{B}-$field and the proxy for the reaction plane $\Psi_2$ is expected to vanish also in peripheral collisions. This is because although $\vec{B}-$field will be directional, the orientation of the proxy for $\Psi_2$ is highly random. The expectation is that $\gamma_B\rightarrow 0$ for peripheral events. However, we see in Fig.\ref{fig:first_measurement} that both $\gamma^{OS}$ and $\gamma^{SS}$ continue to increase in peripheral events.

Another observation from Fig.\ref{fig-B-field} is the strength of $\gamma-$correlator in a given centrality is larger for the smaller size system Cu+Cu than that of the larger size system Au+Au. It is not obvious that  this observation is consistent with CME. If one estimates $\gamma_B$ it will be smaller in Cu+Cu than Au+Au. 

Once again, in the ideal CME scenario, the expectation was $\gamma^{OS}$ should be opposite sign but of similar magnitude compared to $\gamma_{SS}$. However, in the measurement it is pretty obvious the magnitude of $\gamma^{OS}$ is much smaller than $\gamma^{SS}$ (quantitative details are not important at this stage). This is also not consistent with the ideal expectations of CME scenario. 

Based on the above observation, STAR collaboration concluded in Ref~\cite{STAR:2009wot} that in heavy ion collisions indicate a finite non-zero charge separation across reaction plane. This could be consistent with the expectations of CME. However, many features of the data are not consistent with the ideal expectations of CME which could be driven by non-CME phenomenon that were later on known as background charge separation. In fact on Fig.\ref{fig:first_measurement} there are curves using the HIJING event generator that do not include the physics of CME. Albeit small, HIJING was able to predict some qualitative features of the data such as rising strength with centrality and collision system size ordering of $\gamma$ (Cu+Cu)$>\gamma$(Au+Au). The conclusion from the STAR collaboration was that there maybe non-CME source of charge separation, that may or maynot be correlated with the reaction plane.

\subsection{Background correlations for CME observables}

In the previous section we mention that the $\gamma$-correlator was constructed out of two components $\gamma_{||}$ and $\gamma_{\perp}$, $\gamma=\gamma_{||}-\gamma_{\perp}$. Where  $\gamma_{||}$, measures charge separation perpendicular to $\vec{B}$ (parallel to $\Psi_{RP}$), therefore entirely due to non-CME origin. Using $\gamma_{||}$ as a baseline for $\gamma_{\perp}$ assumes that the non-CME baseline is the same along and perpendicular to $\vec{B}$ is not a good assumption. If non-CME background has any correlation with the direction of $\Psi_{RP}$ (therefore not being the same along $||$ and $\perp$) this assumption fails. As a result, a major challenge that the  $\gamma$-correlator faces towards detecting signals of CME involves  large non-CME background sources that are: 1) correlated to $\Psi_{RP}$ and 2) also independent of $\Psi_{RP}$. The distinction between the two sources must be carefully noted as they are crucial to the interpretation of several key measurements performed at both RHIC and LHC. The second case of a background independent of $\Psi_{RP}$ affecting $\gamma$-correlator is very tricky. On a first thought it seems to have no as it will impact the $||$ and $\perp$ direction in an equal way. However, these $\Psi_{RP}$-independent correlations, also called nonflow correlations, impact the $\gamma$-correlator in a non-trivial way, as discussed in a later section.


\begin{figure}
    \centering
\includegraphics[width=0.6\textwidth]{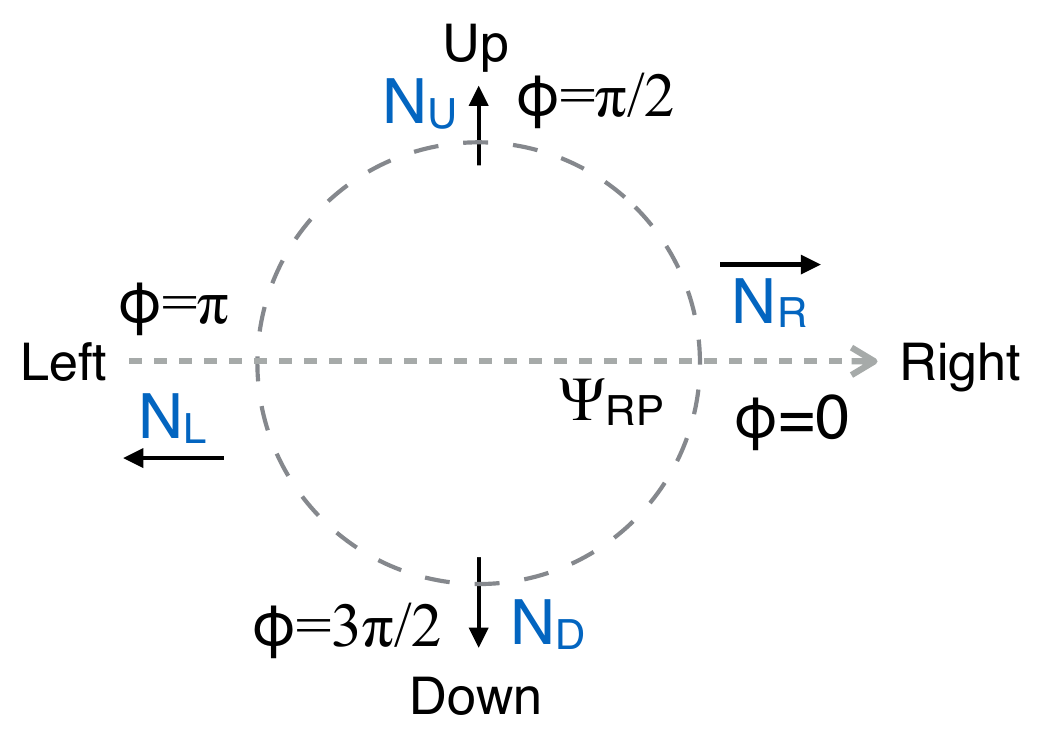}
    \caption{A simple scenario of particle emission across the reaction plane $\Psi_{RP}$ to understand the observable $\gamma$-correlator defined in Eq.~\ref{eq:gamma_correlation_terms_observables}. This scenario is ideal, i.e., in the absence of any preferential emission in any direction of particles due to CME or flow. Even in such a scenario, combinatorics due to finite number of particles lead to nonzero correlations.}
\label{fig:cme_background_zero_flow_observables}
\end{figure}

 To understand the robustness of $\gamma$ correlator let us now consider the scenario that there are no CME. By construction, the correlator should vanish. We will consider the simplified cartoon  shown in Fig.\ref{fig:cme_background_zero_flow_observables}. There are a total of $N$ particles produced in a collision which emitted such that there are $N_U$, $N_D$, $N_R$ and $N_L$ number of particles going up, down, right and left which directions on average, such directions are defined based on the value of the angles $\phi-\Psi_{RP}=0,\pi/2, \pi$ and $3\pi/2$, respectively. Let us now estimate the $\gamma$-correlator ($\gamma=\gamma_{||}-\gamma_\perp$). We first consider the perpendicular component 
\begin{eqnarray}
\gamma_\perp &=& \left\langle\sin \left(\phi_1-\Psi_{RP}\right)\sin \left(\phi_2-\Psi_{RP}\right)\right\rangle.
\end{eqnarray}
The numerator of this quantity can be written as: 
\begin{eqnarray}
\operatorname{Num}\left(\gamma_\perp \right) &=& \frac{N_U}{2}\left(N_U-1\right) \sin(\pi/2)\sin(\pi/2) \\
&+& \frac{N_D}{2}\left(N_D-1\right) \sin(3\pi/2)\sin(3\pi/2) + N_U N_D \sin(\pi/2)\sin(3\pi/2) \nonumber \\
&=& \frac{1}{2}\left[\left(N_U-N_D\right)^2-\left(N_U+N_D\right)\right], \nonumber
\end{eqnarray}
where clearly the up and down going particles will contribute and other terms will give zero. 

Similarly for the other case we will have
\begin{eqnarray}
\gamma_{||} &=& \left\langle \cos\left(\phi_1-\Psi_{RP}\right) \cos\left(\phi_2-\Psi_{RP}\right) \right\rangle,
\end{eqnarray}
since the other combination will not contribute leding to the numerator to be
\begin{eqnarray}
\operatorname{Num}\left(\gamma_{||}\right) &=& \frac{N_R\left(N_R-1\right)}{2} \cos (0)\cos(0) + \frac{N_L\left(N_L-1\right)}{2} \cos (\pi) \cos(\pi) \nonumber \\
&+& N_R N_L \cos(0) \cos(\pi) \\
&=& \frac{1}{2}\left[\left(N_R-N_L\right)^2-\left(N_R+N_L\right)\right]. \nonumber
\end{eqnarray}

The denominator of both $\gamma_\perp$ and $\gamma_{||}$ will be the same. The quantity can be written as
\begin{eqnarray}
\operatorname{Den}\left(\gamma_{||,\perp}\right) &=& \frac{1}{2}\left[N_R\left(N_R-1\right)+N_L\left(N_L-1\right)+2 N_R N_L\right. \nonumber \\
&+& N_U\left(N_U-1\right)+N_D\left(N_D-1\right)+2 N_U N_D \nonumber \\
&+& \left. 2 N_U N_R + 2 N_U N_L + 2 N_D N_R + 2 N_D N_L \right], \nonumber
\end{eqnarray}
since all combinations will contribute. In a compact form we can write
\begin{eqnarray}
    \operatorname{Den}\left(\gamma_{||,\perp}\right)  &=& \frac{1}{2}\left[\left(N_R+N_L\right)\left(N_R+N_L-1\right) +\left(N_U+N_D\right)\left(N_U+N_D-1\right) \right. \nonumber \\
    &+& \left. 2\left(N_R+N_L\right) \left(N_U+N_D\right) \right].
\end{eqnarray}
Given the above quantities established, one can estimate $\gamma=\gamma_{||}-\gamma_\perp$ and $\delta=\gamma_{||}+\gamma_\perp$. Let us know also consider another simplified quantity of single-particle elliptic harmonic anisotropy
\begin{equation}
v_2\{RP\}=\left<\cos(2\phi-2\Psi_{RP})\right>    
\label{eq_v2_RP_observables}
\end{equation}
 It will be clearer later why $v_2\{RP\}$ is important in the discussion of the background correlations for $\gamma$-correlator. As per the definition we have from cartoon of Fig.\ref{fig:cme_background_zero_flow_observables} as the numerator and denominators of $v_2\{RP\}$ to be
\begin{eqnarray}
    \operatorname{Num}(v_2) = (N_R+N_L) -(N_U+N_D), \,\, \operatorname{Den}(v_2) = (N_R+N_L) +(N_U+N_D). \nonumber \\
\end{eqnarray}
This should make sense as elliptic flow is the imbalance of particle going along vs. perpendicular to reaction plane. There is another way to estimate the elliptic anisotropy using two-particle correlations which is defined as
\begin{equation}
    v_2^2\{2\} = \left<\cos(2\phi_1 - 2\phi_2)\right>.
\label{eq_v2_2pc_observables}
\end{equation}
Note that this quantity is actually a square of the elliptic anisotropy which could be either negative or positive. According to the cartoon on Fig.\ref{fig:cme_background_zero_flow_observables} one will have the numerator of such as quantity
\begin{eqnarray}
    \operatorname{Num}(v_2^2\{2\}) &=& \frac{N_R\left(N_R-1\right)}{2} \cos (0) + \frac{N_L\left(N_L-1\right)}{2} \cos (0) + N_R N_L \cos(-2\pi)  \nonumber \\
    &+& \frac{N_U\left(N_U-1\right)}{2} \cos (0) + \frac{N_D\left(N_D-1\right)}{2} \cos (0) + N_U N_D \cos(-2\pi) \nonumber \\
&+& N_R N_U \cos(-\pi) + N_U N_L \cos(-\pi) + N_L N_D \cos(-\pi) + N_D N_R \cos(3\pi). \nonumber \\
\end{eqnarray}
In a compact form one can show the numerator of $v_2^2\{2\}$ can be expressed as
\begin{eqnarray}
    \operatorname{Num}(v_2^2\{2\}) &=& \frac{1}{2}\left[\left(N_R+N_L\right)\left(N_R+N_L-1\right) +\left(N_U+N_D\right)\left(N_U+N_D-1\right) \right. \nonumber \\
    &-& \left. 2\left(N_R+N_L\right) \left(N_U+N_D\right) \right].
\end{eqnarray}

Let us now consider a number of scenarios in the absence of CME but other phenomenon are present and affecting the preferential emission of particles. 
\subsubsection{Global momentum conservation (GMC)}
Global momentum conservation (GMC) is most simple and commonly working assumption. In this scenario we have $N_R=N_L=N_{||}$ and $N_U=N_D=N_\perp$. Let us consider a few cases of observables discussed above. This will lead to according to Eq.\ref{eq_v2_RP_observables}:
\begin{equation}
    v_2 = \frac{N_{||}-N_\perp}{N_{||}+N_\perp},\,\, \gamma = \frac{N_\perp - N_{||}}{N_{||}(2N_{||}-1) + N_\perp(2N_\perp -1) + 4N_{||}N_\perp}, \delta = \frac{\gamma}{v_2}.
\end{equation}
Which indicates the presence of non-zero $\gamma$ purely form combinatorics due to finite number of particles. Such non-zero values of $\gamma$ can be considered as background to CME. 

\subsubsection{GMC and Isotropic emission ($v_2=0$)}
In this scenario we have $(N_R=N_L)=(N_U=N_D)$. It is obvious that for this case one get $v_2\{\text{EP}\}=\gamma=0$. Which is consistent with our expectations that isotropic emission by definition ensure no charge separation. 

\subsubsection{GMC and Elliptic flow ($v_2>0$)}
In this scenario we have $(N_R=N_L) > (N_U=N_D)$. In fact, to simplify, one can take a limit of $(N_U+N_D)=N_\perp\rightarrow 0$, which indicates $v_2=1$. It also follows from the above that 
\begin{equation}
    \gamma= -\frac{1}{(2N_{||}-1)}, \, , \delta= -\frac{1}{(2N_{||}-1)}
\end{equation}
Clearly in the presence of GMC and flow, one expects an artificial effect that mimics charge separation. The effect is due to finite number of particles and vanishes when $N_\perp\rightarrow 0$. It must be noted that this effect is not charge sensitive. Although GMC and elliptic flow can lead to $\gamma\ne0$, the difference between opposite and same sign case $\Delta \gamma=\gamma^{OS}-\gamma^{SS}=0$. 

\subsubsection{GMC, elliptic flow and Local Charge Conservation (LCC)}
In this scenario of local charge conservation (LCC) we have $N_R^+=N_R^-$ and $N_L^+ = N_L^-$, i.e. the electric charge must be conserved locally. Previously, we have been dealing with charge inclusive case. We need to take one more step and define that each of the left or right going particle can now decay to a positive and a negative particle. It means $N_R = N_R^+ + N_R^-$ or $N_L = N_L^+ + N_L^-$. In this case one can also show
\begin{equation}
    \Delta \gamma = \Delta \delta \times v_2. 
\end{equation}
That indicates the combined effect of flow, GMC and LCC will lead to non-zero values of $\Delta\gamma$ that mimics CME. We discuss the case of LCC in more depth in the next subsection. 

\subsubsection{Summary of the simple picture of background expectation}

\begin{figure}[!hbt]
\begin{center}
\includegraphics[height=0.3\textwidth]{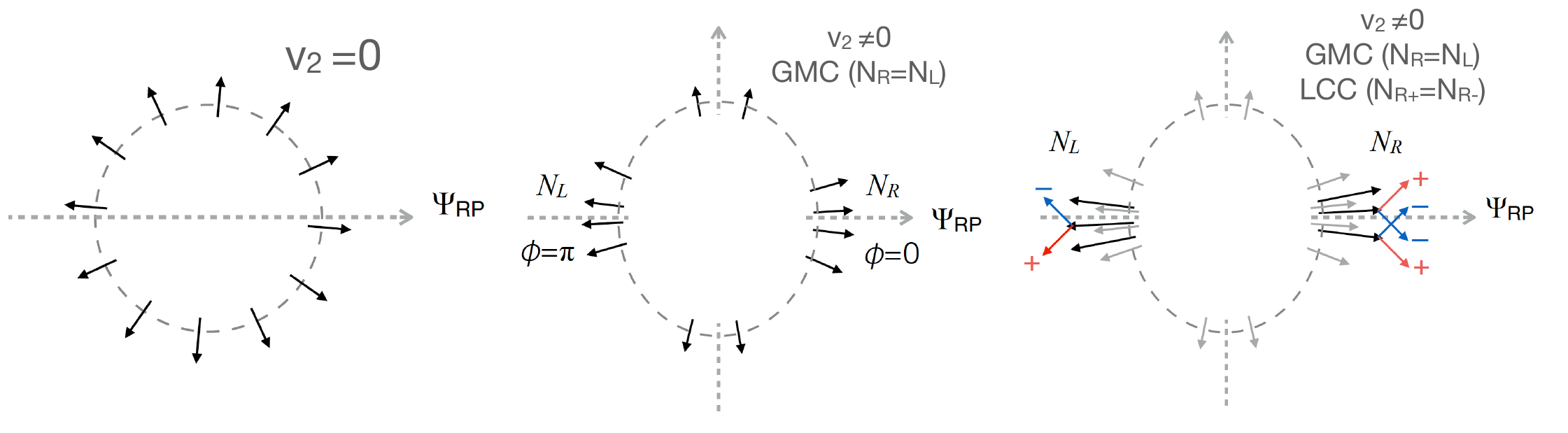}\qquad
\caption{The cartoon illustrates the azimuthal distribution of particles under different scenarios and their effect on charge separation across the reaction plane $\Psi_{RP}$. The left cartoon shows the case of isotropic emission of particles with respect to $\Psi_{RP}$, which has no elliptic anisotropy ($v_2=0$) and no charge separation. The middle and the right cartoons show the case of anisotropic emission of particles with respect to $\Psi_{RP}$, which has non-zero elliptic anisotropy ($v_2\ne0$) and charge separation. The middle cartoon shows the effect of global momentum conservation (GMC), which makes the number of left-going and right-going particles equal ($N_L=N_R$). The right cartoon shows the effect of local charge conservation (LCC), which makes the number of positive and negative particles equal ($N_+=N_-$) locally (in this example on left and right side separately). The combined effect of flow, GMC, and LCC can mimic charge separation across $\Psi_{RP}$.} 
\label{fig_cartoon_lcc_gmc_advanced_observables}
\end{center}
\end{figure}

Figure \ref{fig_cartoon_lcc_gmc_advanced_observables} shows a cartoon that summarizes the combined effect of flow, GMC, and LCC, which can produce an effect that mimics the charge separation across the reaction plane when studied using the $\gamma$-correlator. Note that this cartoon simplifies the picture by assuming scenarios such as all particles having identical momentum -- this is for demonstration only. Later on, we will compare it to a more realistic scenario of real-world phenomenology. With our simplistic picture in mind, the introduction of a background that mimics CME (nonzero $\Delta\gamma=\gamma^{os}-\gamma^{ss}$) can be summarized in the following steps. We first start with the left most panel of Fig.~\ref{fig_cartoon_lcc_gmc_advanced_observables} and we are talking about a scenario where we do not invoke CME. First, in the absence of any flow, one expects isotropic emission represented by a circular event topology, which does not lead to any nonzero $v_2$ and $\Delta\gamma$. Second, the particle emission topology changes when the system has anisotropic flow, elliptic flow in particular. The particle emission topology~\footnote{We will use the word topology here is a colloquial way to refer to the pattern of particle emission. For example, an event with large $v_2$ will have a topology of an ellipse.} changes to an elliptic shape, leading to more emission of particles along the reaction plane compared to its perpendicular direction. Once there is elliptic flow, it means $v_2\ne0$. You can even assume $v_2=1$ for a simplistic scenario, meaning particles are going left and right only -- still, you will not get any nonzero $\Delta\gamma$. The third effect is the impact of GMC, which ensures that the number of particles going left or right (and up and down) is conserved, as demonstrated by the middle panel of Figure \ref{fig_cartoon_lcc_gmc_advanced_observables}. This will lead to nonzero values of $\gamma$ that are the same for $\gamma^{OS}$ and $\gamma^{SS}$, but no nonzero $\Delta\gamma$. So far, we have not invoked the charge of the particles. This takes us to the final step, where LCC leads to the fact that charges only appear in pairs, locally, to enforce conservation. This means each right- or left-going particle is expected to split into a pair of positive and negative charges. This is demonstrated by the rightmost panel of Figure \ref{fig_cartoon_lcc_gmc_advanced_observables}. This leads to different values of $\gamma^{OS}$ and $\gamma^{SS}$, in other words, $\Delta\gamma\ne0$. Not only that, you can infer that $\gamma^{OS}>\gamma^{SS}$, in other words, $\Delta\gamma>0$, by considering the fact that the cartoon of the right panel of Figure \ref{fig_cartoon_lcc_gmc_advanced_observables} implies that somehow the correlation between a pair of opposite-sign particles is enhanced artificially due to this LCC combined with elliptic flow. For an experimentalist measuring a positive value of $\Delta\gamma$, there will be no way of distinguishing if CME is observed or if the effect is entirely due to flow and LCC.

As we noted before, in the LCC picture, a positive and a negative particle appear locally, and they can be assumed to be split from a neutral particle. A good way to think of LCC is to think of a neutral resonance decay, which we discuss in the next section. However, there is a common misconception about the rightmost panel of Fig.\ref{fig_cartoon_lcc_gmc_advanced_observables}. It is often drawn in a way that makes it look like the positive particles (shown by red color) are emitted along the upper half of the $\Psi_{RP}$ and the negative ones (shown by blue) go towards the lower half of $\Psi_{RP}$ -- thereby making it look like LCC leads to charge separation across event planes like CME. The follow-up argument is that LCC is a background for CME, as the event topology looks like CME.
Such an image in mind would be misleading, and we deliberately drew the cartoon in a way to emphasize the fact that there are no restrictions on positive and negative particles to go to the upper or lower half of the $\Psi_{RP}$. Although it is a bit counterintuitive, there is an important distinction of event topology here. The decay of a LCC or a flowing neutral resonance has a completely different event topology than a pair of charges separating across the reaction plane as CME. And yet, the former is able to mimic the latter when it comes to the measurement of $\Delta\gamma$ as a measure of charge separation. This has more to do with the design of the $\gamma-$ correlator than with the event topology. We follow this up in the next subsection.


\subsubsection{Neutral resonance decay and flowing clusters}
Some of the following content may repeat what was said in the previous subsection. As we mentioned before, the best way to understand the picture of LCC is to think about the decay of a flowing resonance (meaning one that goes preferentially along $\Psi_{RP}$). This is how it was introduced in the original paper where the $\gamma$ correlator was proposed~\cite{Voloshin:2004vk}. The idea in that paper was that the decay of a neutral resonance to a pair of positive and negative particles would mimic the effect of charge separation across the reaction plane. In a follow-up paper, this was referred to in terms of a more general terminology of neutral cluster flow~\cite{Wang:2009kd} and eventually LCC~\cite{Pratt:2010gy,Schlichting:2010qia}. At the most fundamental level, they are all the same effect. This happens in two steps, as discussed in the previous subsection. For a simple explanation, let us go back to the rightmost panel of Fig.\ref{fig_cartoon_lcc_gmc_advanced_observables}. A neutral resonance flows, meaning it has a higher probability to be emitted closer to the reaction plane than perpendicular to it. In the second step, the mother resonance decays into a pair of opposite daughter particles when it tries to emit closer to the reaction plane. This decay leads the pairs to go in either left or right direction together. Both the positive and negative particles going together in either left or right is important. The decay of the mother resonance ensures that this happens -- as the daughters will have smaller opening angles due to the boost from the momentum of the mother. If the mother has a sufficiently large momentum, one daughter cannot go to the right and another to the left. They will be forced to go together along left or right -- this is the source of stronger correlation among a pair of opposite sign. What then happens to the correlation among same sign pairs? Looking at the rightmost panel of Fig.\ref{fig_cartoon_lcc_gmc_advanced_observables}, there will be correlation between a pair of positive particles, but only due to combinatorics, which will be weaker than the LCC. This will lead to $\Delta\gamma = \gamma^{OS}-\gamma^{SS}$.

Let us consider that we have only positive and negative pions, a total of $N_\pi$ in the system and some of them come from the decay of neutral resonances. All the particles experience elliptic flow in the system. Let us define $v_{2,\rm res}$ and $\phi_{\rm res}$ as the elliptic flow and azimuthal angle of neutral resonance decaying to pions. We also define $ f_{\rm res}$ as the fraction of pions coming from decay of such neutral resonance. Then following the approach of Ref.~\cite{Voloshin:2004vk} we can write
\begin{eqnarray}
\gamma&=& \left< \cos(\phi_\alpha + \phi_\beta -2\Psi_{RP}) \right> \nonumber \\
&=& \left< \cos((\phi_\alpha + \phi_\beta - 2\phi_{\rm res}) + 2(\phi_{\rm res} -\Psi_{RP})) \right> \nonumber \\
&\sim& \frac{f_{\rm res}}{N_\pi} \left< \cos(\phi_\alpha + \phi_\beta - 2\phi_{\rm res}) \right> v_{2,\rm res}.
\label{eq_resonance_gamma}
\end{eqnarray}

Therefore, the effect of neutral resonance decay on the $\gamma-$correlator is inversely proportional to the number of pions and directly proportional to the elliptic flow of the neutral resonance. This effect is dominant for opposite-sign pairs ($\gamma^{OS}$), while the contribution from same-sign pairs ($\gamma^{SS}$) is lower due to combinatorics. Therefore, one will have $\gamma^{OS}>\gamma^{SS}$, leading to a nonzero $\Delta\gamma$. This $\Delta\gamma$ will act as a background for the measurement of charge separation across the reaction plane, which is attributed to the chiral magnetic effect (CME). However, the strength of this background depends on the fraction of resonances ($f_{\rm{res}}$) and the three-particle correlation ($\left< \cos(\phi_\alpha + \phi_\beta - 2\phi_{\rm res}) \right>$), which are unknown and need to be simulated using a realistic model. This requires some phenomenological input, which we will discuss in the next subsection. Historically, in the original paper where the $\gamma-$correlator was proposed~\cite{Voloshin:2004vk} and in the first measurement of $\Delta\gamma$ by the STAR collaboration~\cite{STAR:2009wot}, the total contribution from neutral resonance decay was estimated to be small and insufficient to explain the experimental observation. However, later on, the idea of LCC was introduced, which argued that the appearance of all charged particles happens by ensuring local charge conservation, which enhances the correlation between opposite-sign pairs. The addition of LCC on top of flowing neutral resonance was argued to be sufficient to explain the observed $\Delta\gamma$, thereby largely constraining the observability of CME signal in the measured charge separation by $\Delta\gamma$.

\subsubsection{Phenomenology of LCC and GMC}

\begin{figure}[!hbt]
\begin{center}
\includegraphics[height=0.35\textwidth]{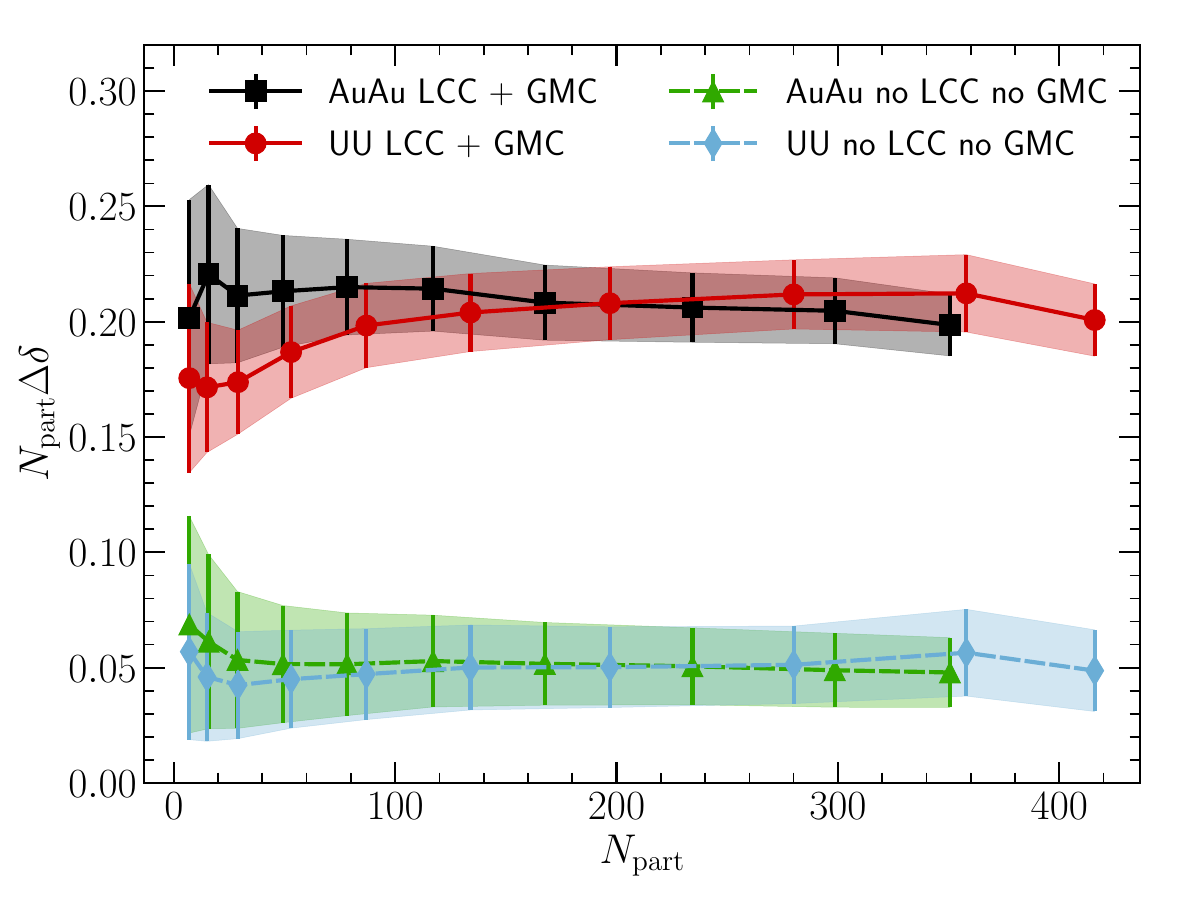}\qquad
\includegraphics[height=0.35\textwidth]{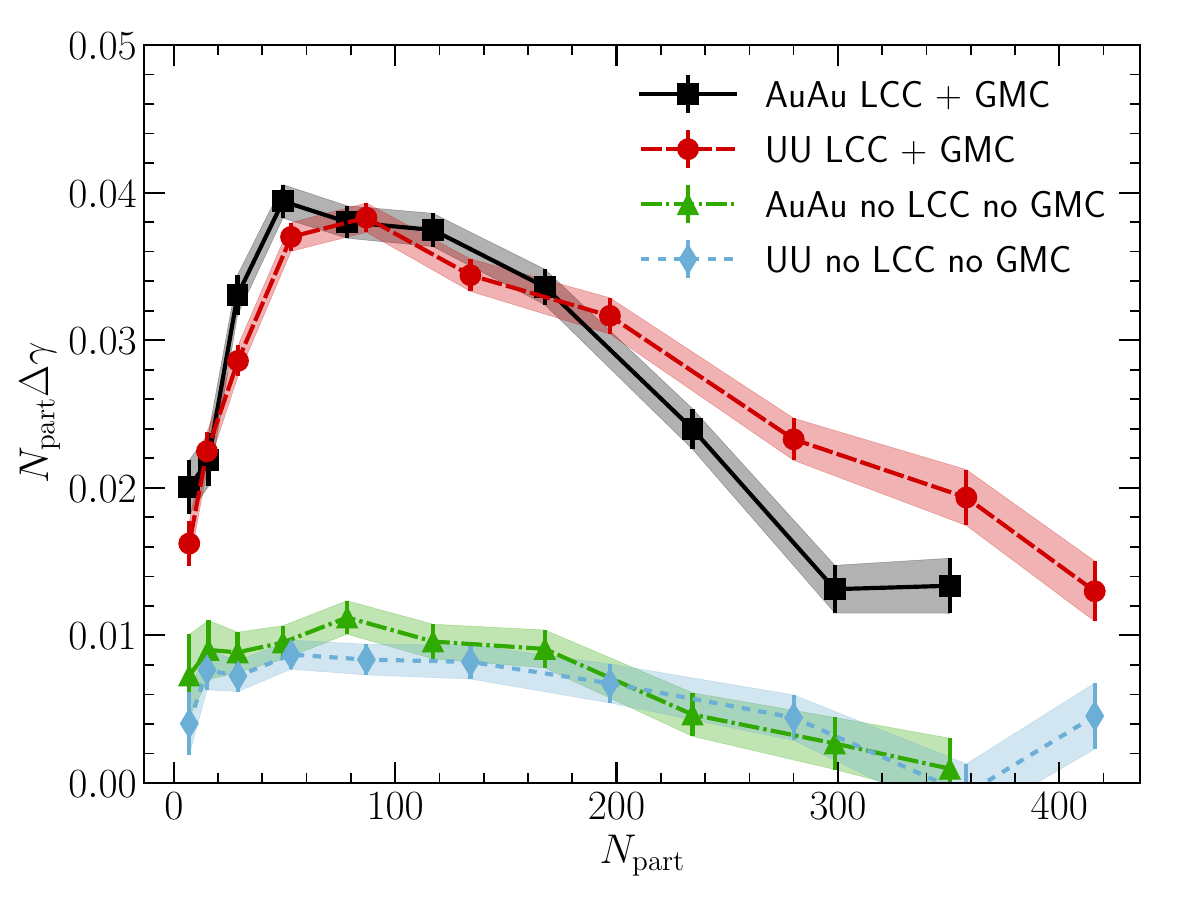}
\caption{(color online, reproduced from \protect{\cite{Schenke:2019ruo}} with permission) The difference between opposite-sign and same-sign $\delta$ (left) and $\gamma$ (right) correlation functions scaled by number of participants $N_\mathrm{part}$ in Au+Au and U+U collisions. Results with and without imposing local charge conservation (LCC) and global momentum conservation (GMC) are shown.
\label{fig.sec4_LCC}
} 
\end{center}
\end{figure}

The preceding discussion focused on a simplified scenario of background sources, particularly emphasizing the significant impact of GMC and LCC on the measurement of $\gamma$ and $\Delta\gamma$. The effects of GMC and LCC introduce non-vanishing multi-particle correlations, making it challenging to incorporate them into numerical simulations of the freeze-out process.

Early attempts to simulate these effects date back to Ref~\cite{Wang:2009kd,Pratt:2010gy,Pratt:2010zn,Schlichting:2010qia,Bzdak:2012ia}. In Ref~\cite{Schlichting:2010qia}, these effects were implemented using input from STAR data and the blast-wave model to account for LCC at freeze-out. For a comprehensive understanding, we direct the reader to the most recent implementation of GMC and LCC, which adopts a more principled approach by combining state-of-the-art models for the initial state of heavy-ion collisions, such as IP-Glasma, viscous hydrodynamic simulation model MUSIC, and UrQMD afterburner for hadronic rescattering~\cite{Schenke:2019ruo}. This approach is already known for providing a robust description of global data on charge-inclusive azimuthal correlations.

In~\cite{Schenke:2019ruo}, both the GMC and LCC effects are incorporated into the freeze-out process using the numerical implementation first proposed in~\cite{Bozek:2012en}. In this work, charged hadron-antihadron pairs are chosen to be produced at the same fluid cell, while their momenta are sampled independently in the local rest frame of the fluid cell. This procedure implicitly assumes that the correlation length is smaller than the size of the cell, providing an upper limit for the correlations between opposite-sign pairs. Furthermore, GMC is imposed by adjusting the momentum of final-state hadrons.

As illustrated in Fig.\ref{fig.sec4_LCC}, the LCC effect increases the $\Delta \gamma$ and $\Delta \delta$ correlators compared to the case with only resonance decay. Meanwhile, GMC alters the absolute values of same-sign and opposite-sign correlators but has a negligible influence on the difference between them. Subsequently, a more sophisticated particlization prescription was developed by the BEST collaboration\cite{Oliinychenko:2019zfk,Oliinychenko:2020cmr}, allowing for a more realistic estimation of GMC and LCC effects on the $\gamma$ and $\delta$ correlators. Discussed in detail in the next, this new particlization method employs the Markov Chain Monte Carlo algorithm to sample hadrons according to the desired distribution while respecting the conservation of energy, momentum, baryon number, electric charge, and strangeness within a localized batch of fluid cells at the freeze-out surface.

Based on the above discussion one can consider the following that observed experimental measurement of $\Delta\gamma$ can be expressed as 
\begin{equation}
    \Delta\gamma=\Delta\gamma^{\text{CME}}+ \Delta\gamma^{\text{bkg}}\,,
\end{equation}
where the second term $\Delta\gamma^{\text{bkg}}$ is dominantly from flow mediated background due to resonance decay and LCC. Phenomenological models indicate $\Delta\gamma^{\text{bkg}}$  can constitute a major part of the observed $\Delta\gamma$.  The major experimental challenge is to isolate the signal contribution $\Delta\gamma^{\text{CME}}$.



\subsubsection{Background sources from nonflow correlations}
The term ``nonflow" in the context of azimuthal correlations is often used to describe correlations that do not have a collective origin. Unlike sources for flow-mediated correlations, which depend on the reaction plane orientation and are distributed among many particles, nonflow correlations are independent of the reaction plane and primarily manifest in a few particles. They are known to originate from conservation processes such as charge and momentum, quantum processes like femtoscopic correlations, and final state effects such as Coulomb interactions~\cite{Tribedy:2017hwn}.
The main sources that strongly exhibit as nonflow include minijet production associated with charge conservation on the near side due to the fragmentation process and back-to-back correlations due to momentum conservation. All of these are examples of few-body correlations. In the conventional analysis of anisotropic flow, subtracting nonflow has been a significant endeavor. Specifically, for charge-inclusive anisotropic flow measurements, two-particle correlations contribute as a dominant source of nonflow. In the case of the CME, the relevant nonflow correlations that lead to charge-dependent azimuthal correlation are three-particle azimuthal correlations. For example, one can imagine a di-jet fragmenting as a pair of positive and negative pions with a narrow azimuthal angle, going in one direction, and a third particle, for which charge is not important, is going in the opposite direction. An analogy can be drawn between this process and the LCC and GMC discussed in the previous subsection. Here, no context of $\Psi_{RP}$ should be invoked; the axis of the di-jet serves as the equivalent of $\Psi_{RP}$. Such a process will result in a nonzero $\Delta\gamma$, following algebra similar to what was discussed in the previous section. We will not revisit them here.
Nonflow constitutes the second major source of non-CME background to $\Delta\gamma$, therefore one can write: \begin{equation}
\Delta\gamma=\Delta\gamma^{\text{CME}}+ \Delta\gamma^{flow-bkg} + \Delta\gamma^{nonflow-bkg}
\label{eqn_nonflow}
\end{equation}


The possibility of such nonflow background was discussed in the first publication of charge separation from STAR~\cite{STAR:2009wot}. It was argued that three-particle correlations induced by mini-jet fragmentation have two effects: 1) they influence the determination of the event plane, and 2) they introduce more opposite charge correlation than same charge correlation. The combination of these two effects is supposed to lead to non-zero $\Delta\gamma$ and mimic CME signals.

\begin{figure}[htb]
    \centering
    \includegraphics[width=0.8\textwidth]{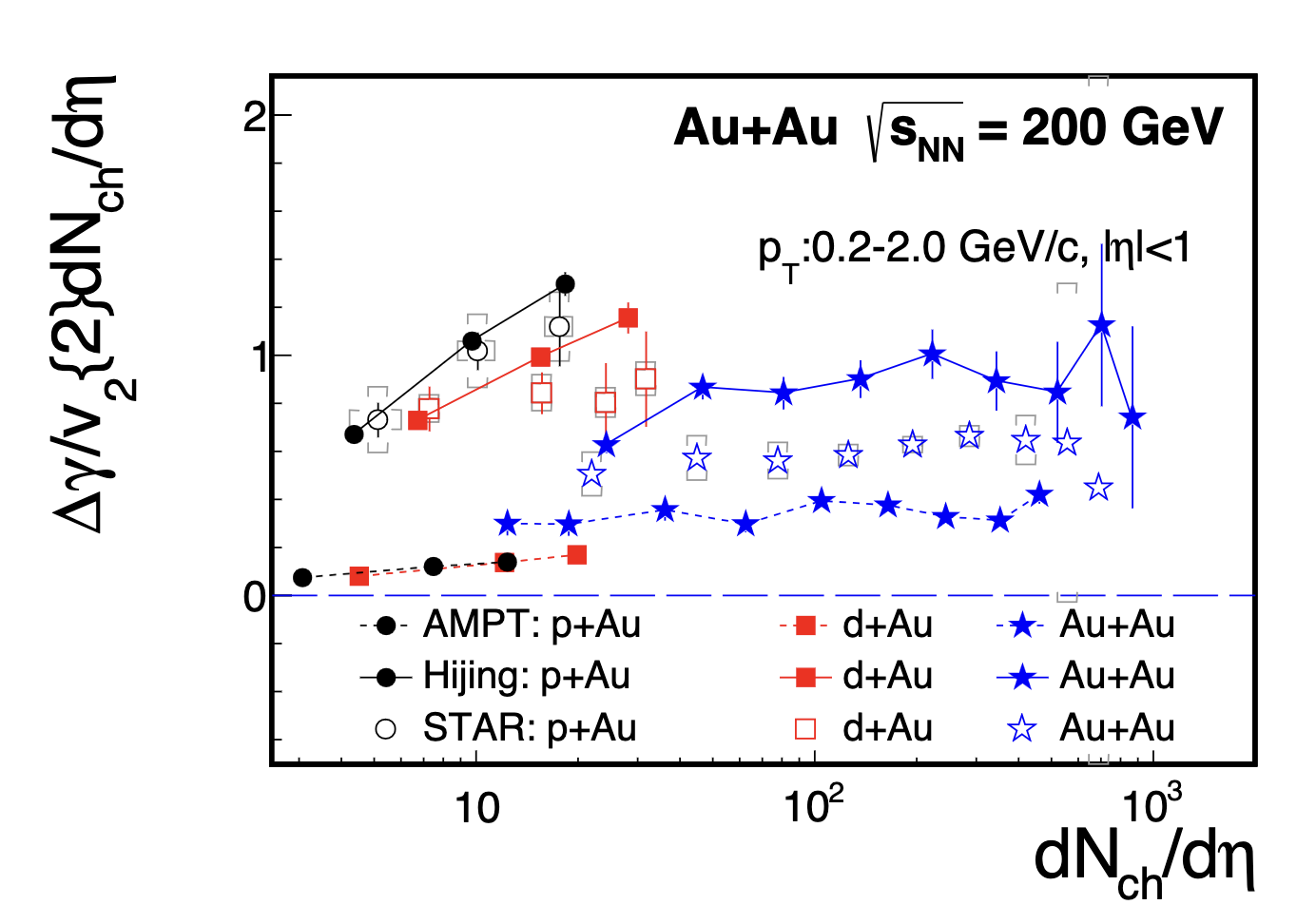}
    \caption{Model calculations showing that the HIJING model, which does not include the CME, can reproduce the experimental data if the elliptic anisotropy parameter is properly scaled. The paper also uses the AMPT model, which has a larger anisotropy parameter, to demonstrate that the data can be explained by models without the CME~\cite{Zhao:2019kyk}.}
    \label{fig:observables_hijing_nonflow}
\end{figure}



In Ref~\cite{STAR:2009wot}, an indication of larger contribution of reaction plane independent background can already be seen in: 1) the sharp increasing strength of $\Delta\gamma$ towards peripheral events and, 2) large $\Delta\gamma$ in Cu+Cu than in Au+Au system at the same centrality. Both observations can be supported by \textsc{hijing} calculation. In a recent calculation~\cite{Zhao:2019kyk} HIJING model estimations were revisited. It turns out in HIJING and AMPT models, one scaled by the elliptic anisotropy the charge sensitive correlators ($\Delta\gamma/v_2$) are comparable to data. The study indicates that the backgrounds in the CME-sensitive $\Delta\gamma$ observable arise from intrinsic particle correlations (nonflow), including resonance decays, cluster correlations, and (mini)jets.

The term $\Delta\gamma^{nonflow-bkg}$ in Eq.\ref{eqn_nonflow} can vary depending on the system size, centrality of collisions and kinematics of the measurement. For example, it can be smaller in smaller systems and peripheral events. In a later section, we revisit the idea of using small collision systems as a baseline to estimate the nonflow in heavy ion collisions. The CMS collaboration~\cite{Khachatryan:2016got} first demonstrated this approach by using p+Pb collisions as a baseline for peripheral Pb+Pb collisions, showing that the magnitude of $\Delta\gamma$ was equal, mainly due to nonflow. A similar measurement at RHIC, as shown in Fig.\ref{fig:observables_hijing_nonflow}, indicated that the $\Delta\gamma$ scaled by elliptic anisotropy $v_2$ was comparable in p+Au and d+Au collisions to that in Au+Au collisions, also because of nonflow. In the later section, we discuss the advantages and challenges of using a small collision system as a baseline for nonflow in a large system. This is one of the efforts to address nonflow, but it is not without difficulties.

In general, the removal of nonflow correlations poses a significant and potentially cumbersome challenge. Various ingenious experimental analyses have been devised, but so far, no single approach has been entirely successful in completely accounting for the nonflow background in a data-driven manner~\cite{Feng:2021pgf}. We delve into some of these approaches in the following section.

\subsection{Phenomenological modeling of CME in heavy ion collisions }

Critical to the success of the experimental program, is a precise and realistic characterization of the CME signals as well as background correlations in these collisions. To achieve this goal would require a  framework that addresses the main theoretical challenges discussed above: (1) dynamical CME transport in the relativistically expanding viscous QGP fluid; (2) initial conditions and subsequent relaxation for the axial charge; (3) co-evolution of the dynamical magnetic field with the medium; 
(4) proper implementation of major  background correlations such as resonance decays and local charge conservation (LCC).  
Such a framework, dubbed EBE-AVFD (Event-By-Event Anomalous-Viscous Fluid Dynamics)~\cite{Shi:2019wzi,Shi:2017cpu,Jiang:2016wve}, which addresses most of these effects, has been developed thorough  the BEST collaboration effort.

A number of different approaches have been developed  for modeling the CME transport in the heavy ion collision environment. For example, one could use the transport model such as the AMPT to simulate the bulk medium evoluton while implementing the CME-motivated charge distribution dipoles in the initial conditions~\cite{Ma:2011uma,Deng:2016knn,Zhao:2019crj,Wang:2021nvh,Li:2022bhl,Wu:2022fwz}.One could also utilize the so-called chiral kinetic theory as a weakly-coupled description to dynamically model the CME transport in the quark-gluon plasma phase, (see e.g.~\cite{Sun:2016nig,Sun:2016mvh,Sun:2018idn}).    The studies of CME were also performed within a strong-coupling, non-equilibrium  approach based on holography \cite{Ammon:2016fru, Ghosh:2021naw, Cartwright:2021maz}. 
In addition to simulation tools, advanced computing techniques such as deep learning based on convolutional neural network have been explored and applied to help identify the CME signal against background correlations~\cite{Zhao:2021yjo}. 

Within the hydrodynamic approach, the CME and CMW signals in heavy-ion collisions were investigated using ideal chiral hydrodynamics~\cite{Hongo:2013cqa,Hirono:2014oda} which describes the evolution of ideal, non-dissipative chiral currents on top a of viscous hydrodynamic background~\cite{Yee:2013cya,Yin:2015fca}. This approach suffers from a degree of inconsistency in the treatment of the chiral currents and that of the bulk fluid. The next step towards a more self-consistent treatment of anomalous transport, should take into account the non-equilibrium corrections both to the bulk background and to the vector as well as axial currents. 

\begin{figure}[!hbt]
\begin{center}
\includegraphics[height=0.5\textwidth]{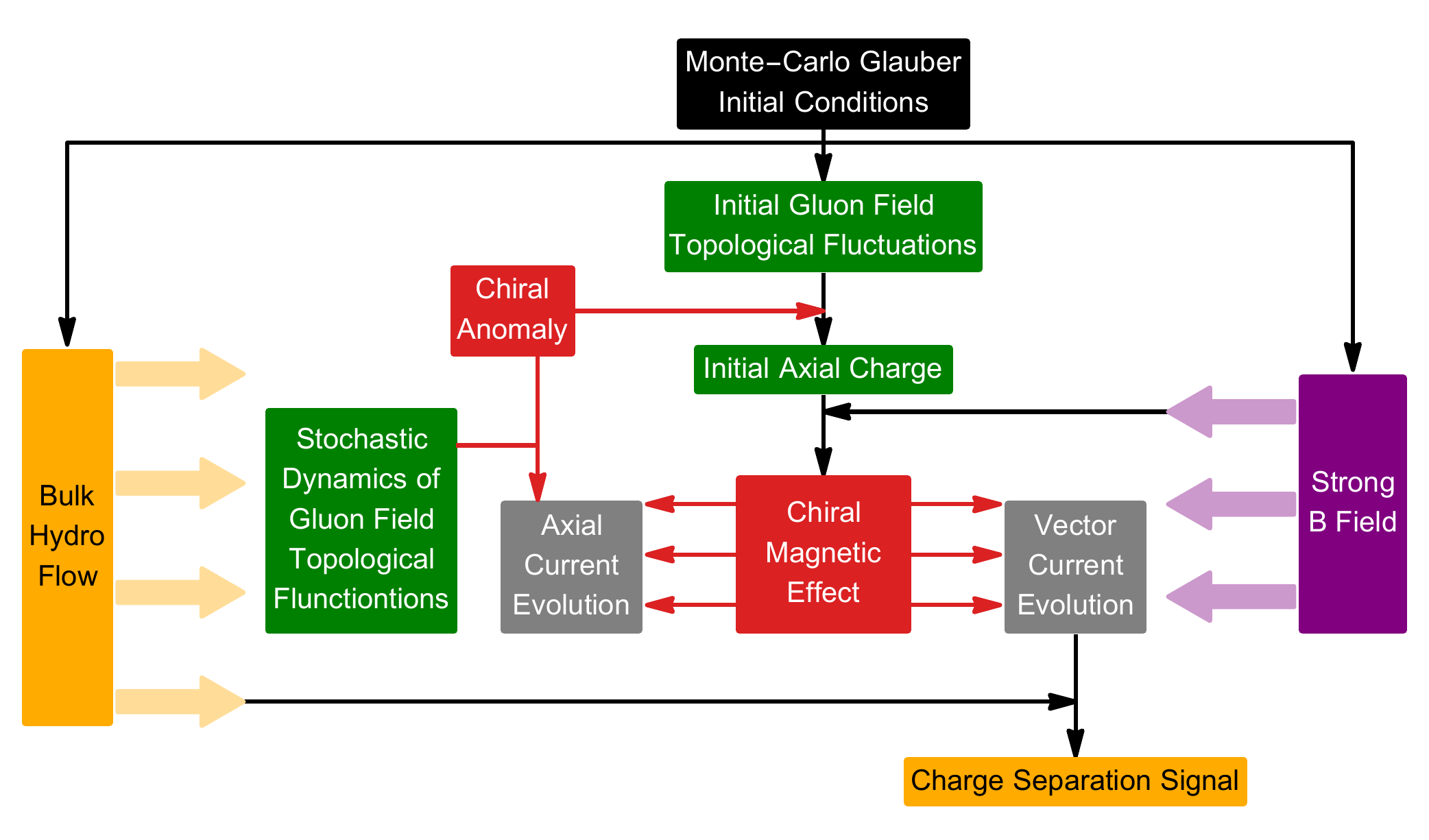} 
\caption{ An illustration of the components and structures of the event-by-event anomalous-viscous fluid dynamics (EBE-AVFD) framework with stochastic dynamics of gauge field topological fluctuations. See \cite{Jiang:2016wve,Shi:2017cpu,Shi:2019wzi,Huang:2021bhj} for details. 
\label{fig-avfd-chart}
}
\end{center}
\end{figure}

This has been  achieved by the Anomalous-Viscous Fluid Dynamics (AVFD) simulation framework~\cite{Jiang:2016wve,Shi:2017cpu}, which solves the evolution of vector and axial currents, including dissipation effects, as linear perturbation on top of the bulk viscous-hydrodynamic background. See Fig.~\ref{fig-avfd-chart} for a flowchart of this framework. Over the years the AVFD package has been continually developed and improved in essentially three major steps. 
1) In the first generation~\cite{Jiang:2016wve,Shi:2017cpu}, the simulations start with event-averaged initial condition, and allow systematically testing the sensitivity of the CME-induced charge separation with respect to various model ingredients such as   the axial charge imbalance and the magnetic field lifetime. 
2) Then, the second generation~\cite{Shi:2019wzi} was developed, which takes into account the fluctuating initial condition for hydro and magnetic field, and implements the LCC effect with prescription of Ref.~\cite{Schenke:2019ruo}. 
3) Finally through the  BEST Collaboration effort, the AVFD package is upgraded to its third generation, and implements the micro-canonical particle sampling~\cite{Oliinychenko:2019zfk,Oliinychenko:2020cmr} at freeze-out, followed by the updated hadron transport simulation package, SMASH~\cite{Weil:2016zrk}. It provides a global and quantitative  description of CME observables for different collision systems, including both the CME signal and the non-CME backgrounds from LCC and resonance decays.


 \begin{figure}[!hbt]
\begin{center}
\includegraphics[height=0.4\textwidth]{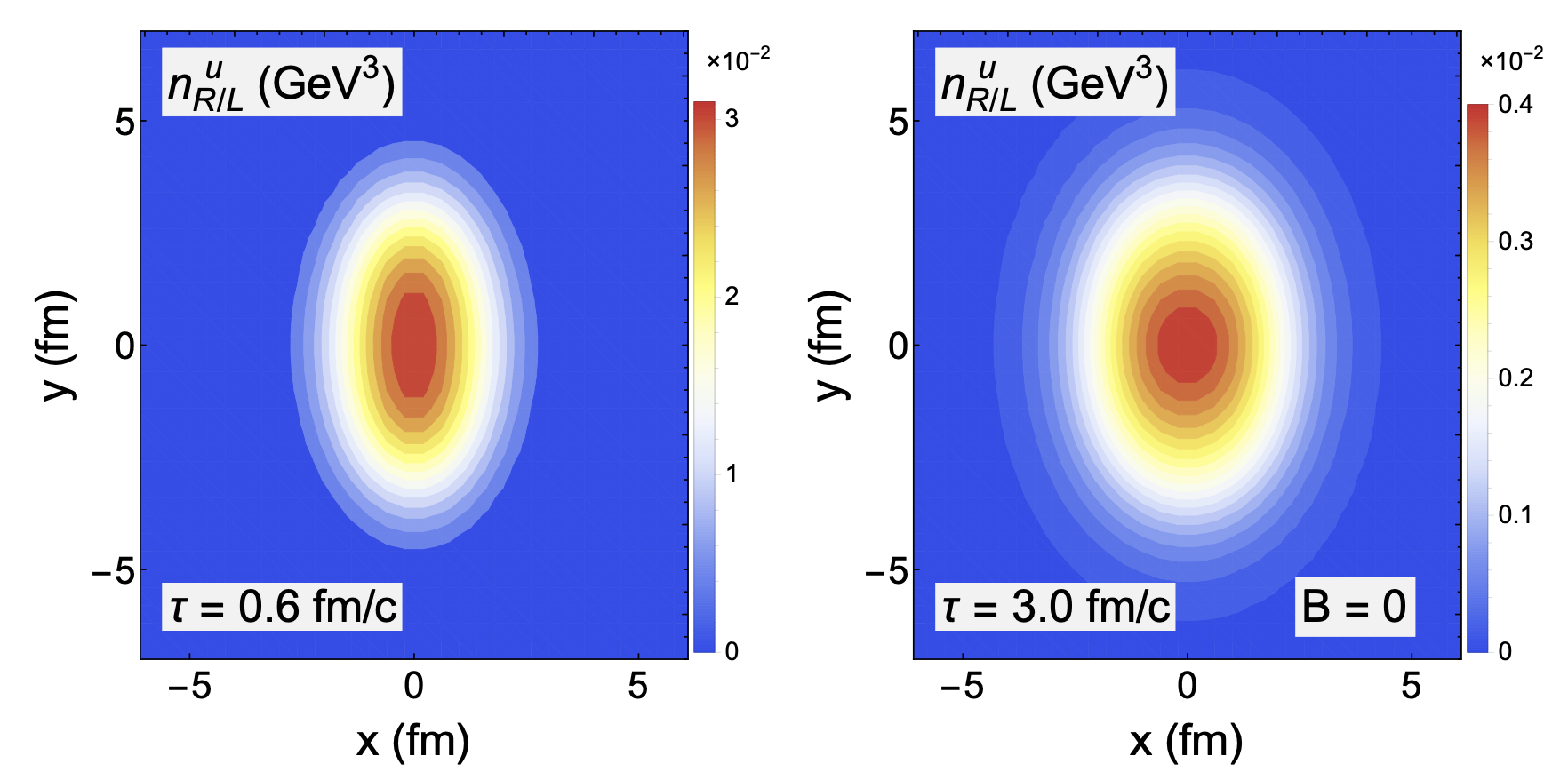}\\
\includegraphics[height=0.42\textwidth]{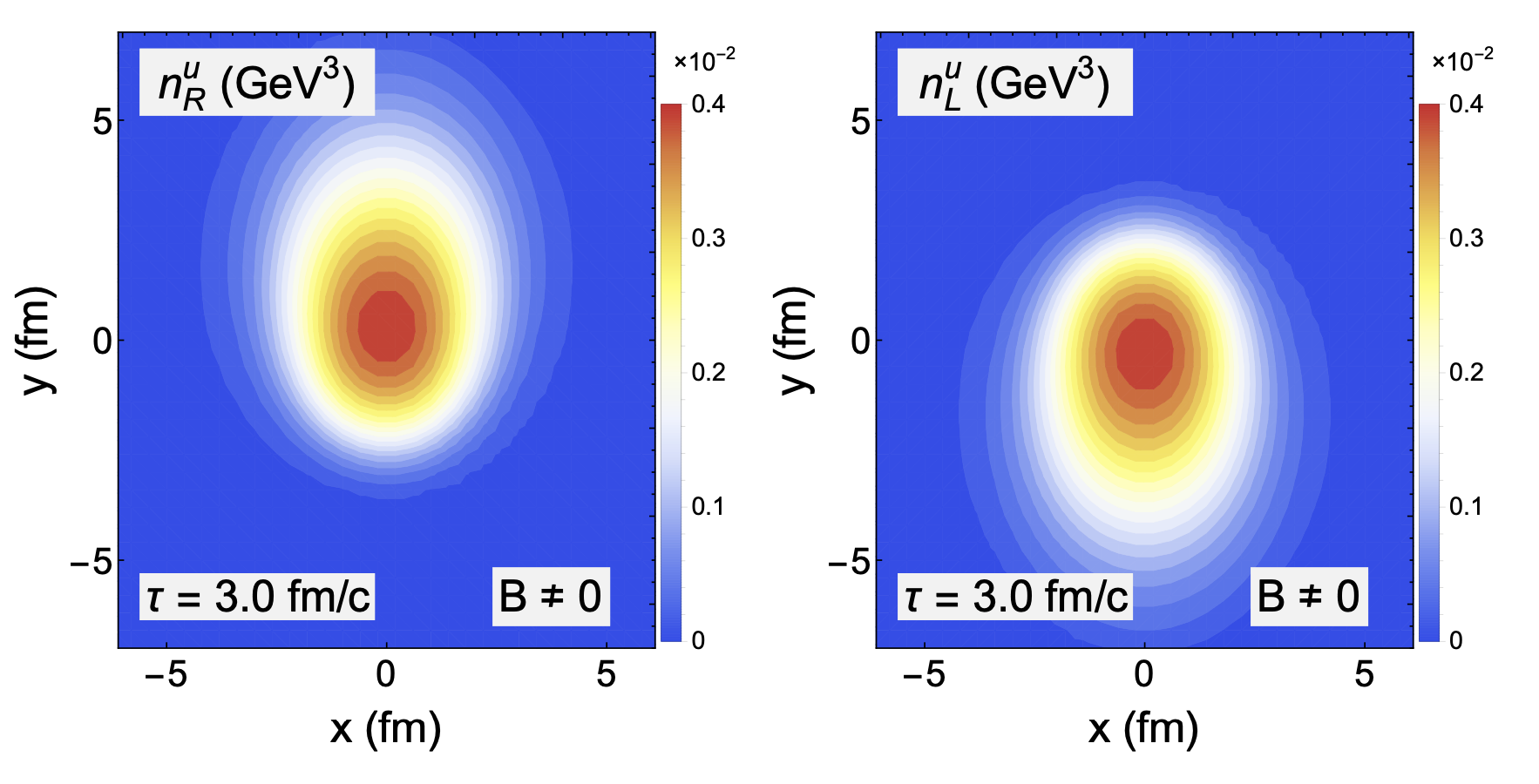} 
\end{center}
\caption{The evolution of $u-$flavor quark densities via solving AVFD equations from the same initial charge density distribution (for either RH or LH) at $\tau=0.60\rm fm/c$ (upper left panel) in three  cases: (a) (upper right panel) for either RH or LH density at $\tau=3.00\rm fm/c$ with magnetic field $B\to 0$ i.e. no anomalous transport; (b) (lower left panel) for RH density and (c)  (lower right panel) for LH density, both at $\tau=3.00\rm fm/c$ with nonzero $B$ field along positive y-axis. (Reproduced from  \cite{Shi:2017cpu}).
\label{fig.sec4_avfd_illustration}}
\end{figure}

To illustrate how the charge separation arises from the CME-induced anomalous transport within the AVFD framework, let us visualize how the quark densities evolve under normal and anomalous transport in Fig.~\ref{fig.sec4_avfd_illustration}.  When the hydrodynamic evolution starts (at proper time $0.6$ fm/c), we initialize the RH and LH u-quark number density as shown in the top left panel. If there is no external magnetic field applied, i.e. only normal transport, both RH and LH u-quarks expand with the fluid and also experience viscous transport like diffusion, in a symmetric fashion along x and y directions (shown in the top right panel). On the other hand, once an external magnetic field is turned on along the out-of-plane y-direction, the anomalous CME current propagates RH u-quarks toward the direction of $\mathbf{B}$ field and LH u-quarks toward the opposite direction, leading to an asymmetric pattern of the charge distribution along the out-of-plane direction (shown in the two bottom panels). As a result of the anomalous transport under the presence of chirality imbalance (i.e. either the
RH or LH pattern of the two lower panels would dominate), there will be accumulation of opposite charges on the
two poles above and below the reaction plane. Upon freeze-out, this pattern eventually translates into the charge separation signal of the measured final state hadrons.

 \begin{figure}[!hbt]
\begin{center}
\includegraphics[height=0.5\textwidth]{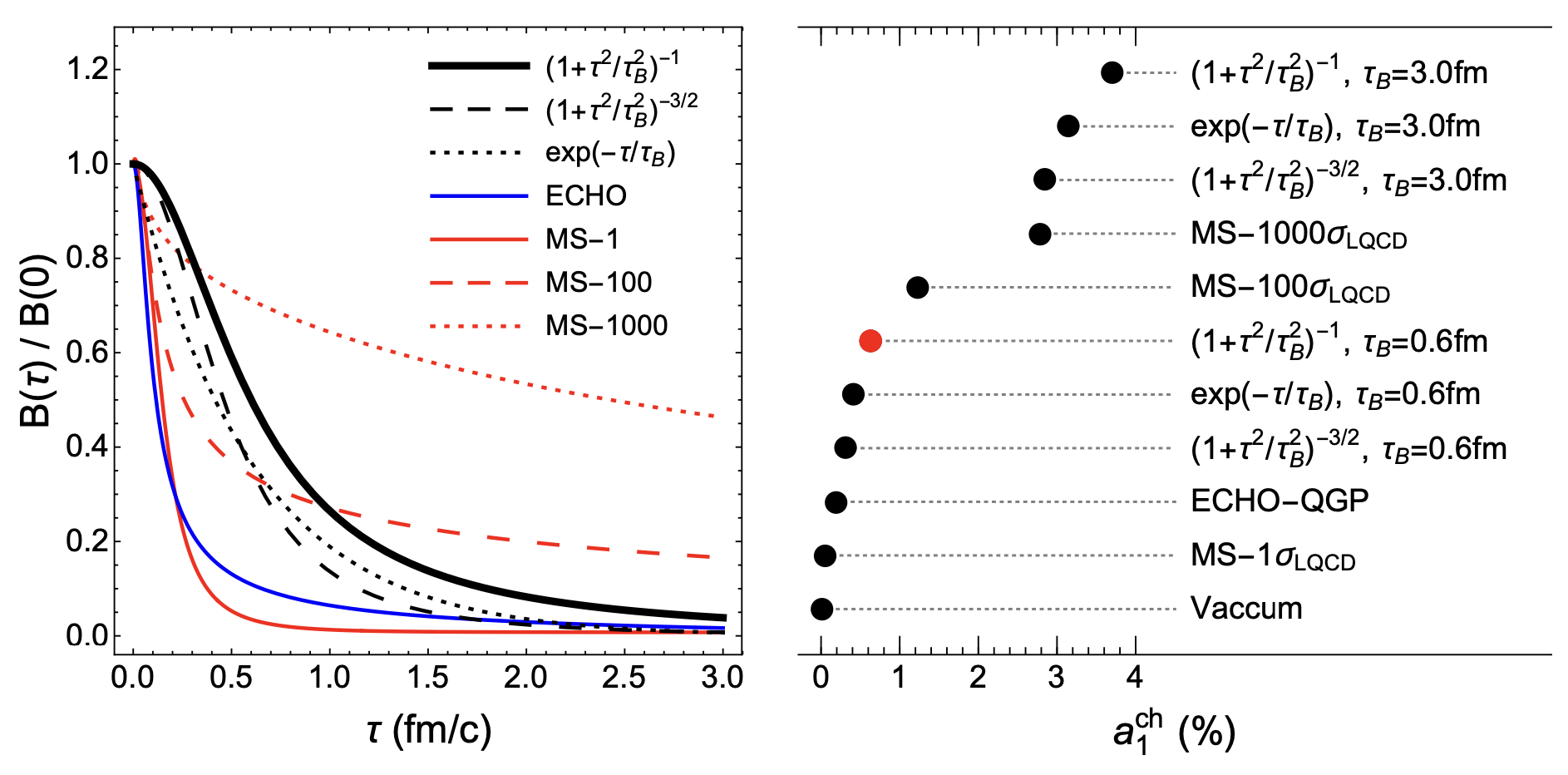} 
\caption{ 
(left) The time dependence of the magnetic field from different model studies: ECHO-QGP \cite{Inghirami:2016iru} (blue curve), 
McLerran-Skokov \cite{McLerran:2013hla} with different electric conductivity (solid red curve for $\sigma=\sigma_\text{LQCD}$, dashed red line for $\sigma=100\sigma_\text{LQCD}$, and dotted red line for $\sigma=1000\sigma_\text{LQCD}$.) Also, the thick, dashed, and dotted black curves represent the formulations $B \propto (1+\tau/\tau_B)^{-1}$, $(1+\tau^2/\tau_B^2)^{-3/2}$, and $\exp(-\tau^2/\tau_B^2)$ respectively, with $\tau_B=0.6$ fm/c.  
(right)  Comparison of charge separation signal computed from AVFD with different choices for the time dependence of magnetic field. 
(Reproduced from \cite{Shi:2017cpu}.)
\label{fig-avfd-B} 
}
\end{center}
\end{figure}

The AVFD framework as a hydrodynamic realization of anomalous transport in heavy ion collisions has allowed a systematic and quantitative understanding of the CME-induced charge separation. For example, Fig.~\ref{fig-avfd-B} shows  the AVFD results quantifying  the influence of the uncertainty in magnetic field lifetime on the predicted  CME signal. The charge separation signal $a^{ch}_1$ is  computed and compared with a wide variety of choices for the input magnetic field time dependence, given that all these calculations are done with the same initial axial charge condition $n_5/s=0.1$ and with the same peak value of $\vec{\bf B}$ field at time $\tau=0$. 
The comparison clearly demonstrates the strong sensitivity of CME signal to  the $\vec{\bf B}$ field lifetime. While this key information still needs to be better determined, the AVFD tool helps us understand and quantitatively calibrate its consequence on the output observables. In addition to magnetic field, the AVFD simulations have been performed to quantify the responses of CME signal to the initial axial charge and vector charge densities as well as to the bulk viscous transport parameters such as charge diffusion and second-order relaxation parameters:  see full details in \cite{Shi:2017cpu}.

 \begin{figure}[!hbt]
\begin{center}
\includegraphics[height=0.35\textwidth]{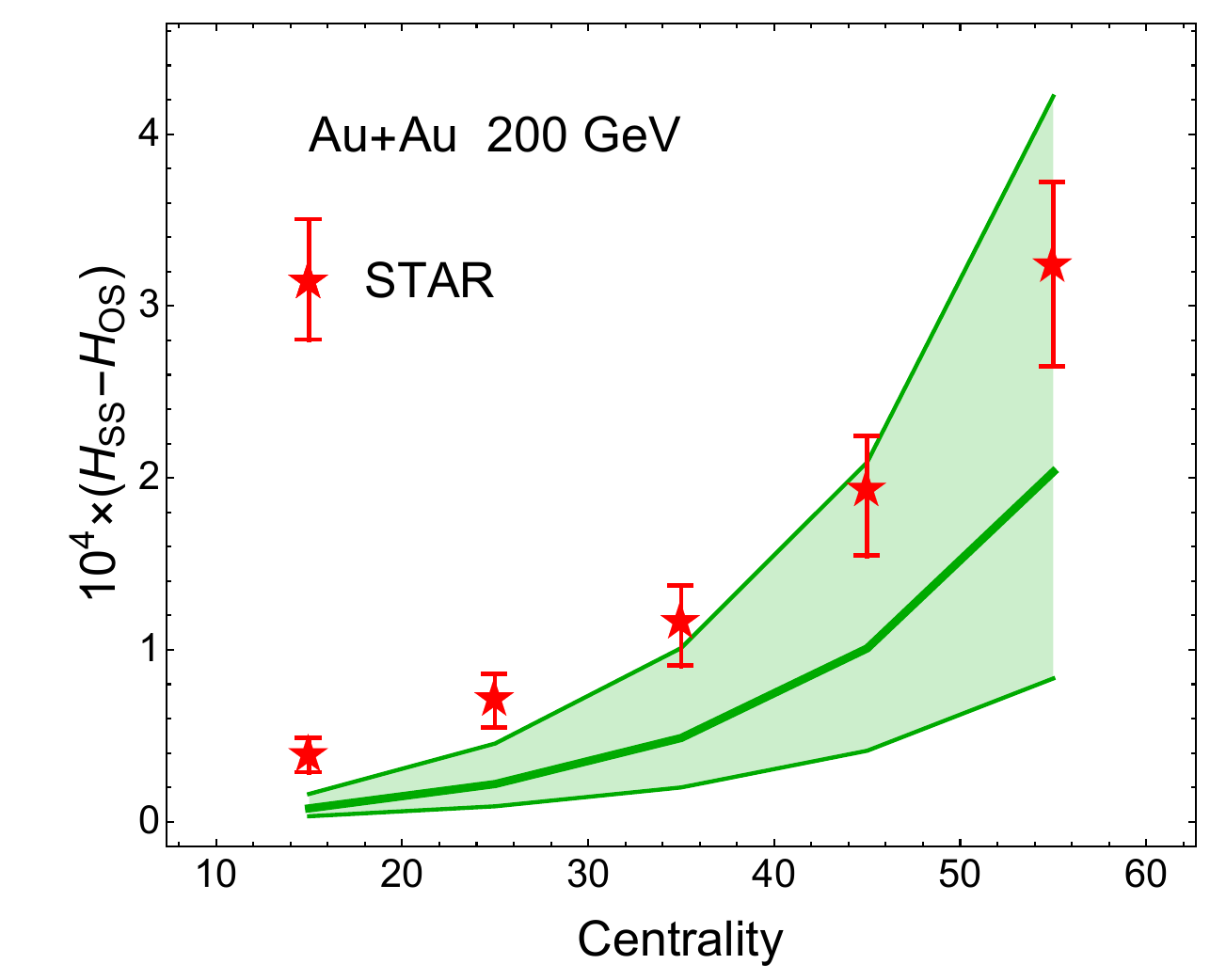}\quad
\includegraphics[height=0.5\textwidth]{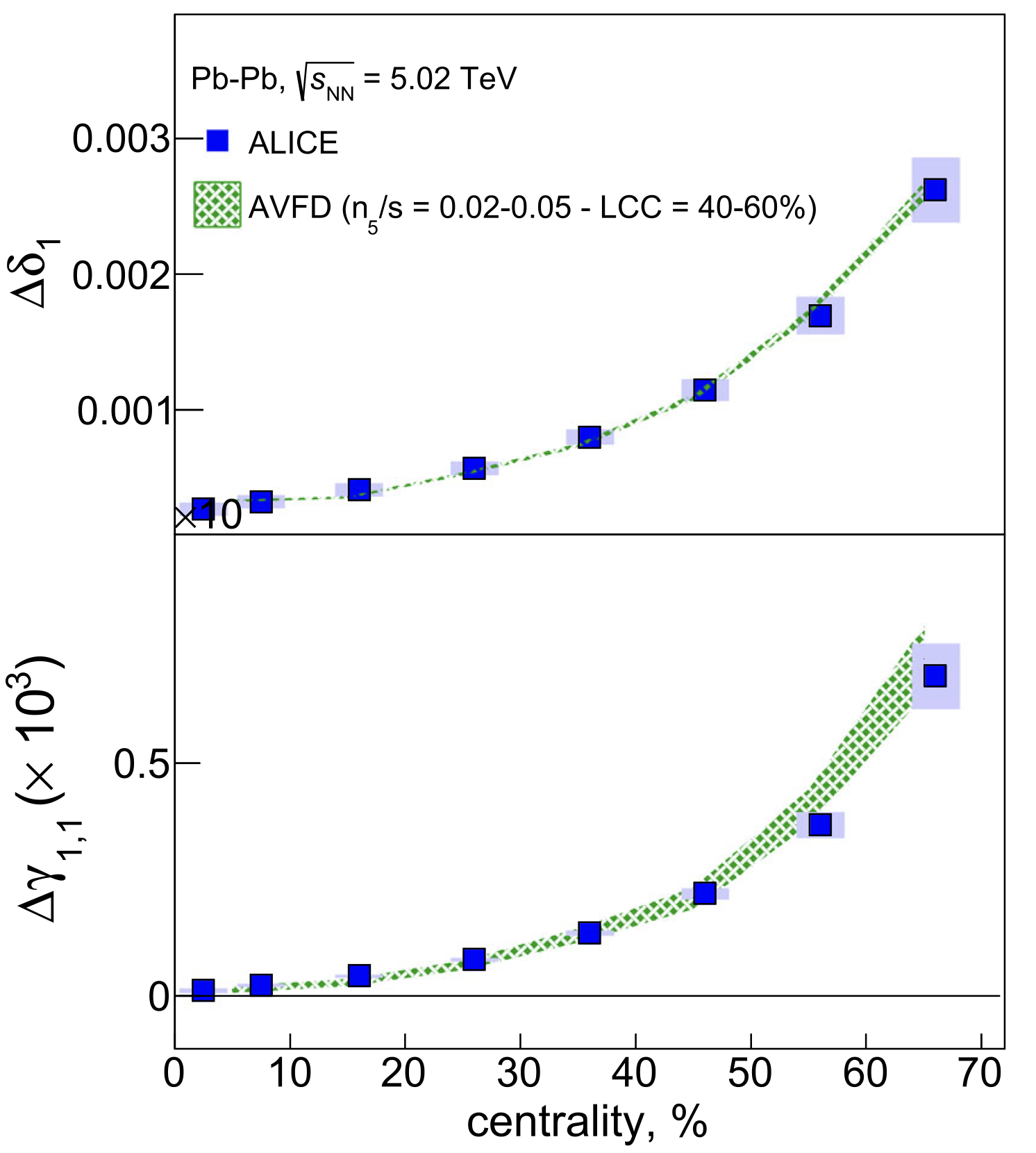} 
\end{center}
\caption{ The azimuthal correlation observable $\left(H_{SS}-H_{OS}\right) $ for various centrality, computed from AVFD simulations and compared with STAR measurement at RHIC (left, from \cite{Jiang:2016wve}) as well as ALICE measurement at LHC  (right, from \cite{Christakoglou:2021nhe}). 
\label{fig.sec4_avfd_1}}
\end{figure}

\begin{figure}[!hbt]
\begin{center}
\includegraphics[height=0.36\textwidth]{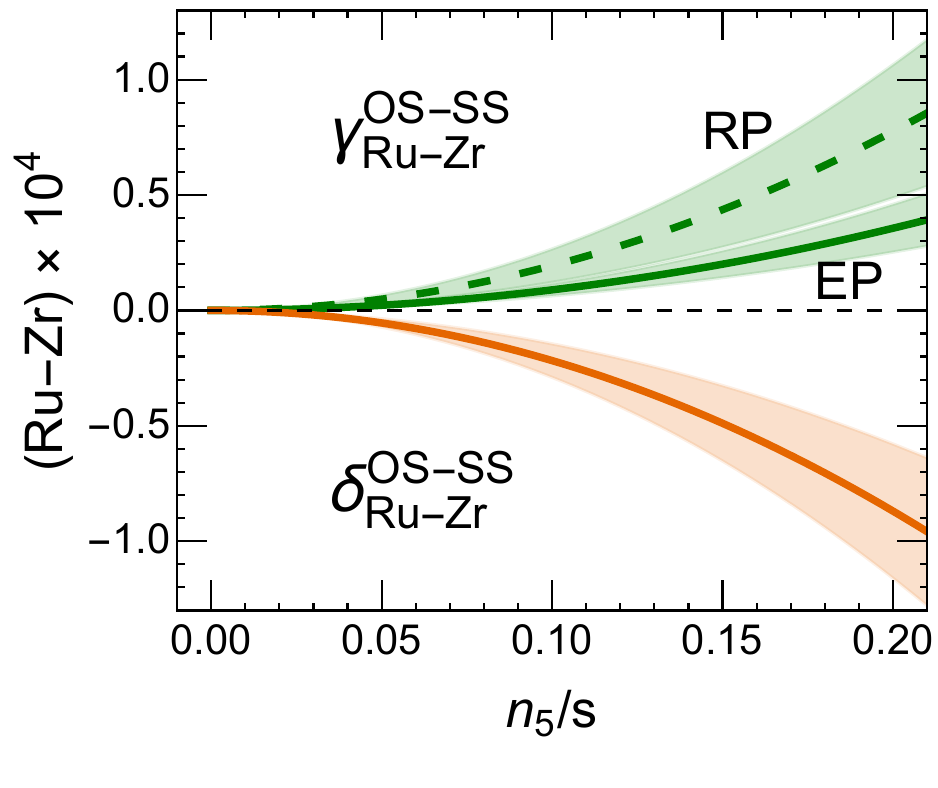} \qquad 
\includegraphics[height=0.4\textwidth]{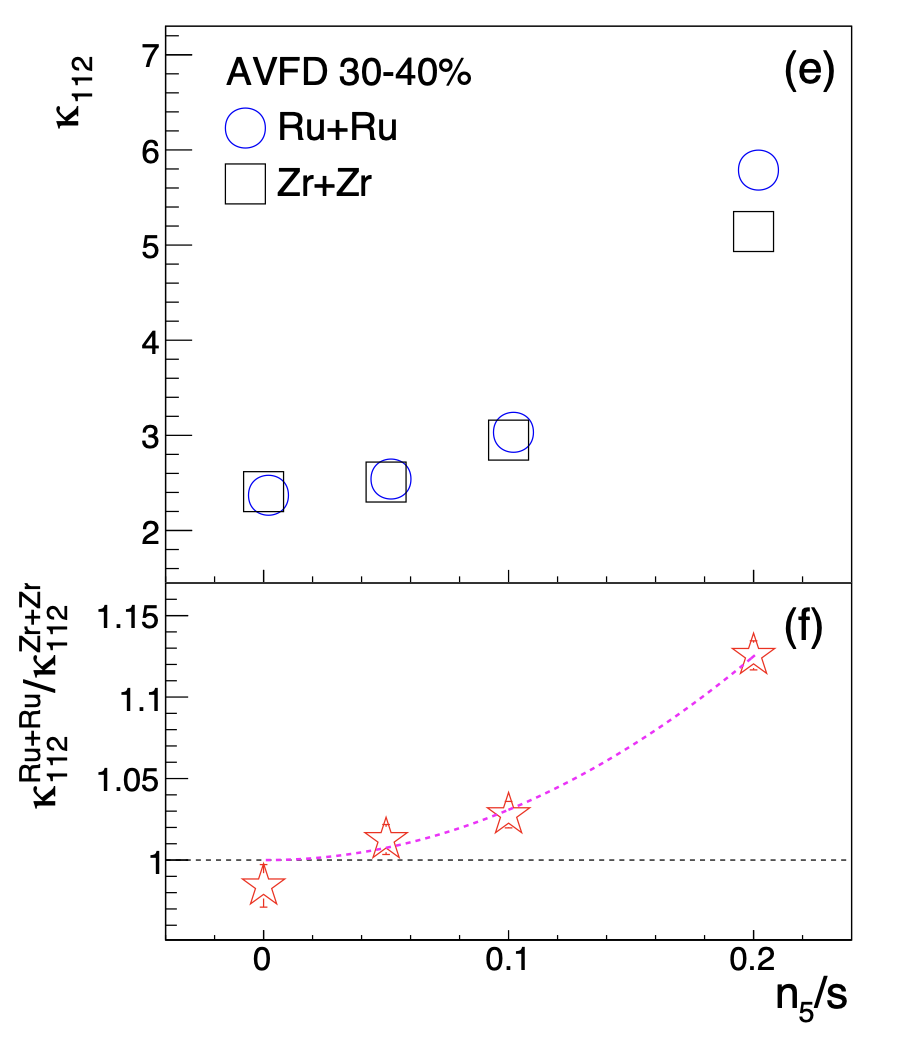}
\end{center}
\caption{
(left, from \cite{Shi:2019wzi})
EBE-AVFD predictions for $\gamma_{Ru-Zr}^{OS-SS}$ (green) and $\delta_{Ru-Zr}^{OS-SS}$ (orange) as functions of $n_5/s$ with respect to event-plane (EP) and reaction-plane (RP) respectively, where the error bands represent the statistical uncertainties from simulations. (right, from \cite{Milton:2021wku}) The expected difference based on EBE-AVFD simulations in the so-called $\kappa$ coefficient extracted via event-shape selection analysis method to remove flow-driven backgrounds.    } 
\label{fig.sec4_avfd_2}  
\end{figure}

The AVFD simulation results have been widely adopted to help interpret the experimental measurements of CME-motivated charge-dependent correlations. Shown in Fig.~\ref{fig.sec4_avfd_1} are two examples. In the left panel, the AVFD calculations with initial parameter range  $Q_s^2=(1\sim 1.5)\rm GeV^2$ (which is the saturation scale of the glasma stage and controls initial axial charge density) provide a good description of the so-called H-correlator (which is extracted from $\gamma$-correlator after flow background removal) across centrality. In the right panel, the comparison between AVFD results and ALICE data for both $\gamma$ and $\delta$ correlators suggests their interpretations in terms of a small but finite CME contributions along with a strong LCC backgrounds. Another important application of the AVFD framework is for developing various new CME observables and  understanding their sensitivities to both CME and backgrounds, see e.g.~\cite{Choudhury:2021jwd,Milton:2021wku,Tang:2019pbl,Magdy:2017yje}. Finally, the AVFD frame provides a powerful tool to make predictions for measurements in different colliding systems  such as the isobars of RuRu and ZrZr, as demonstrated in Fig.\ref{fig.sec4_avfd_2}. Shown in the left panel are predictions~\cite{Shi:2019wzi} for the difference between the isobar pairs in the $\gamma$-correlators measured with respect to both reaction plane (RP) and event plane (EP) as well as in the $\delta$ correlator based on identical event selections based on multiplicity and elliptic flow from both colliding systems. The righ panel shows the expected difference  between the isobar pairs in the so-called $\kappa$ coefficient extracted via event shape selection method~\cite{Milton:2021wku,Choudhury:2021jwd} for the removal of flow backgrounds.


\section{New theoretical developments}

\subsection{New CME results in lattice QCD}

Lattice QCQ is indispensable for obtaining first-principle results on strong interaction dynamics, including CME. The first CME studies on the lattice were performed shortly after the effect was proposed, in \cite{Buividovich:2009wi,Abramczyk:2009gb,Buividovich:2010tn,Yamamoto:2011gk,Braguta:2010ej}. Significant progress has been made since then. Let us mention some recent developments that highlight the non-equilibrium nature of CME.

The Bielefeld group \cite{Brandt:2024wlw} recently performed a study of an electric current in a static equilibrium setup, where the chiral chemical potential does not decay. This setup is analogous to the early formulation of Vilenkin \cite{vilenkin1980equilibrium}, and is expected to yield a zero CME current, as pointed out already in the original papers \cite{Kharzeev:2007jp,Fukushima:2008xe}, see also \cite{Kharzeev:2013ffa,Basar:2013iaa,Yamamoto:2015fxa} for a more detailed discussion of this crucial point. Perhaps the simplest way to see this is to observe that if the current $\Vec{J} \sim \vec{B}$ existed in equilibrium, then it would be possible to extract a non-zero power $P \sim \vec{J}\vec{E} \sim \vec{E}\vec{B}$ from the vacuum by choosing an appropriate direction of the electric field $\vec{E}$.

The results of \cite{Brandt:2024wlw} indeed indicate the expected absence of CME current in the Euclidian equilibrium formulation. The authors point out the importance of gauge-invariant lattice UV regularization in achieving this result, and discuss the future extension of their study to real-time Minkowski space by using the spectral densities of current correlators evaluated in Euclidean space.

Another recent lattice CME study has been performed by Buividovich \cite{Buividovich:2024bmu}. First, the author confirms the absence of CME in equilibrium. Second, he evaluates the Euclidean axial current correlator, and discusses in detail how to use it to obtain real-time, non-equilibrium CME results using lattice QCD.

In fact, CME in real time was already addressed in lattice QCQ some time ago \cite{Muller:2016}, and the presence of the effect was clealy established. The only limitation of this earlier study is the use of a classical approximation for the topological configurations of gauge fields. It would be very important to get real-time lattice QCD results on CME not restricted to classical configurations.

\subsection{Chiral magnetic effect in strongly coupled non-Abelian plasmas}

The chiral anomaly reflects the link between the topology of the gauge field and chirality of the fermions coupled to it. As a result, it has to be reproduced in quantum field theory at any value of the coupling constant, weak or strong. Coming from the weak coupling side, it is well known that the coefficient of the $AVV$ chiral anomaly (relating the {\em{operators}} of axial $A$ and vector $V$ currents) does not receive perturbative corrections. 

The topological nature of the chiral anomaly ensures that the same is true even when the coupling constant becomes large, and the perturbative expansion breaks down. Mathematically, the protection of the anomaly from the details of dynamics stems from the independence of the Chern-Simons term in the action on the metric. 

A very succinct description of this feature of the chiral anomaly is offered by the holographic correspondence between Conformal Field Theories (CFTs) and gravity in anti-de Sitter (AdS) space \cite{Maldacena:1997re,Gubser:1998bc}. As is well known, de Sitter space is a maximally symmetric vacuum solution of Einstein equations with a positive cosmological constant (i.e. with a positive vacuum energy density and a negative pressure). It possesses a positive scalar curvature, and describes an accelerating expansion of the Universe.  

Anti-de Sitter space is a vacuum solution of Einstein equations with a negative cosmological constant (negative energy density and a positive pressure), and possesses a negative scalar curvature. AdS space is known to be unstable - a perturbation around AdS metric leads to an instability leading to black hole formation. In the framework of AdS/CFT correspondence, the black hole in the AdS bulk corresponds to a finite temperature CFT, with the temperature given by the Hawking temperature of the black hole.

The common feature of all holographic realizations of CME is the presence of 5D Chern-Simons term for the $U_L(1)\times U_R(1)$ gauge field in the bulk that corresponds to the chiral anomaly in the boundary theory. There is an important subtlety that appears in the derivation of CME in holography that is intimately connected to the non-equilibrium nature of this effect. 
Namely, the variation of Chern-Simons term yields the so-called consistent anomaly and the CME current. However, once the action is corrected by the proper counter-term in accord with the covariant anomaly prescription, the new negative contribution to the CME current emerges that exactly cancels the ``consistent" one \cite{Rebhan:2009vc}. This cancellation reflects the absence of CME in equilibrium, i.e. at zero frequency. Once the magnetic field and/or the chiral imbalance become time-dependent, the cancellation no longer occurs \cite{Yee:2009vw} -- this clearly illustrates the non-equilibrium nature of the CME. 

Because of the non-equilibrium nature of CME, it is important to account for the dynamics of the axial charge relaxation. This requires including the non-Abelian chiral anomaly as well, that corresponds to a non-Abelian term in the holographic bulk. In the presence of dynamical Abelian and non-Abelian gauge fields, 
the axial current is no longer conserved:
\begin{equation*}
   \partial_\mu J_{5}^\mu =c_\text{strong}\, \text{tr}G\wedge G
   +c_\text{em}\!\left(3 F \wedge F+F^{(5)}\wedge F^{(5)}\right)
\end{equation*}

The  gravity degrees of freedom necessary to incorporate the non-Abelian chiral anomaly in holography have been described in the work of Klebanov, Ouyang, and Witten \cite{Klebanov:2002gr} who found that the anomaly emerges  from the form fields on the cycles in the internal part of the 10D background. 
A holographic $U(1)_A\times U(1)_V$ St\"uckelberg model was proposed in \cite{Jimenez-Alba:2014iia,Jimenez-Alba:2015awa}
\begin{eqnarray}
\label{eq:action}
S= \ && \frac{1}{2\kappa_5^2} \int_{\mathcal{M}} d^5 x\sqrt{-g}\left[R+\frac{12}{L^2}-\frac{1}{4}F^2-\frac{1}{4}F_{(5)}^2 +\frac{m_s^2}{2}(A_m-\partial_m\theta)^2 \right.\nonumber\\
&& \left.\quad+\frac{\alpha}{3} \epsilon^{mnklp} (A_m-\partial_m\theta)\left( 3 F_{nk}F_{lp}+F^{(5)}_{nk}F^{(5)}_{lp}\right) \right]+S_{bdy}+S_{ct} \,\, 
\end{eqnarray} 
with the axial field strength $F_{(5)}=\mathrm{d}A$, the vector field strength $F=\mathrm{d}V$ and the St\"uckelberg (pseudo)scalar $\theta$ which renders the axial gauge field massive while preserving gauge invariance. The strength of the Abelian $U(1)^3_A$ and $U(1)_A\times U(1)_V^2$ anomaly is governed by the parameter $\alpha$ in front of the mixed Chern-Simons term that couples the axial and vector gauge fields. Similarly, the strength of the non-Abelian anomaly is governed by the parameter $m_s$ that determines the mass of the axial gauge field and thus its anomalous dimension. Note that both couplings $\alpha$ and $m_s$ may be separately tuned to different values. 

Recently this model was used to consider dynamical fluctuations of the axial charge, and the resulting space-time correlations of the CME currents \cite{Grieninger:2023wuq}. 

\begin{figure*}
    \centering
\includegraphics[width=0.48\linewidth]{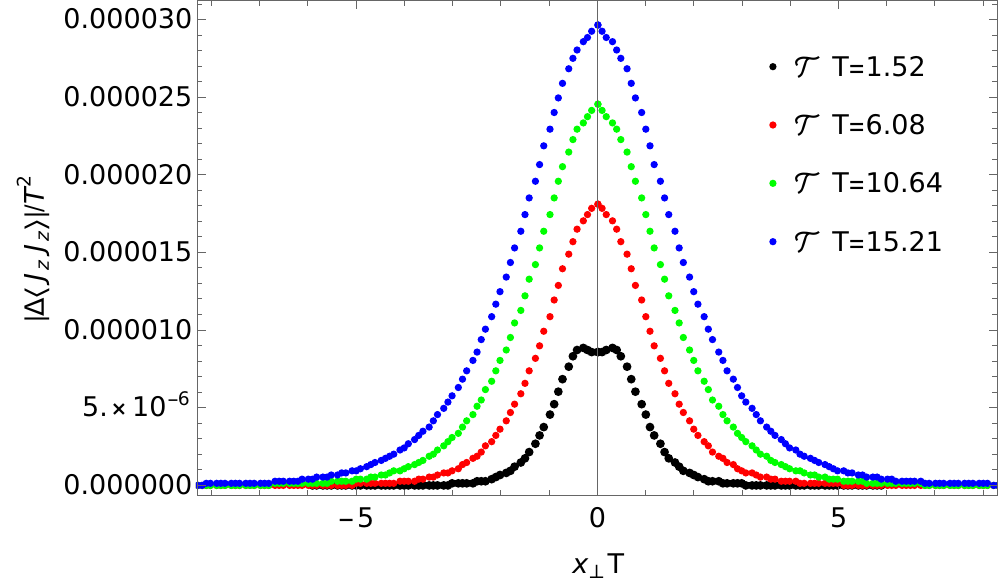}
\includegraphics[width=0.48\linewidth]{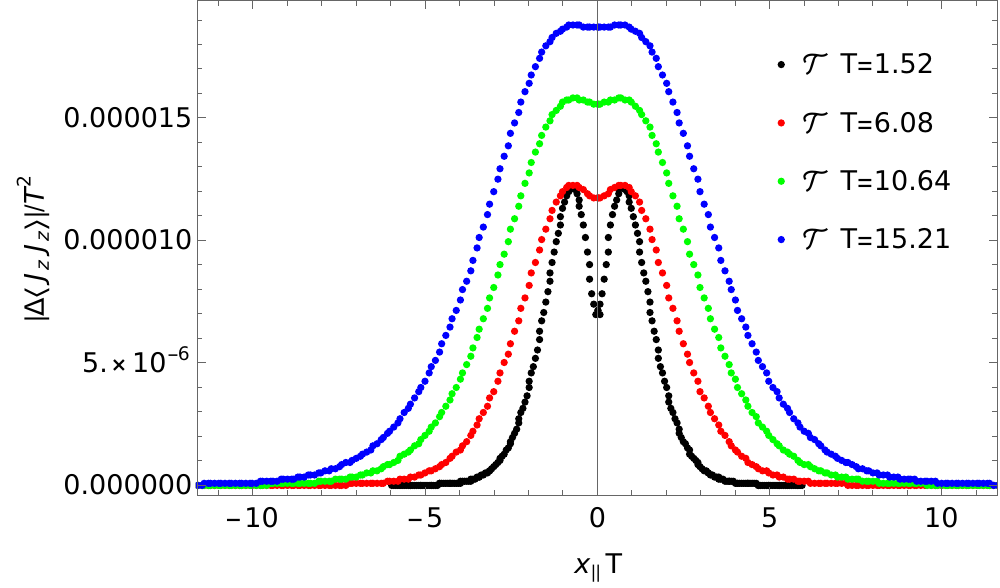}
\includegraphics[width=0.48\linewidth]{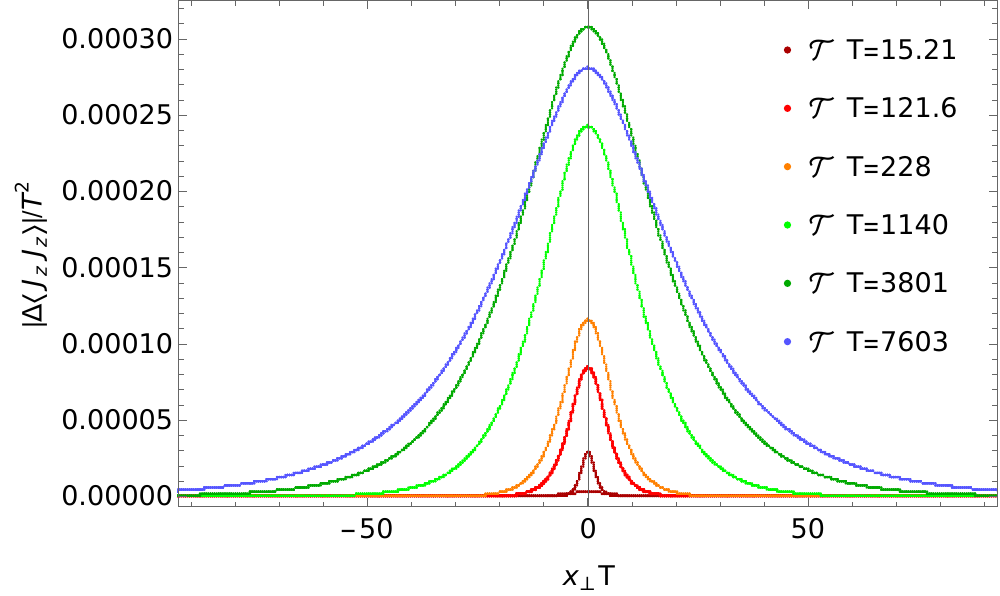}
\includegraphics[width=0.48\linewidth]{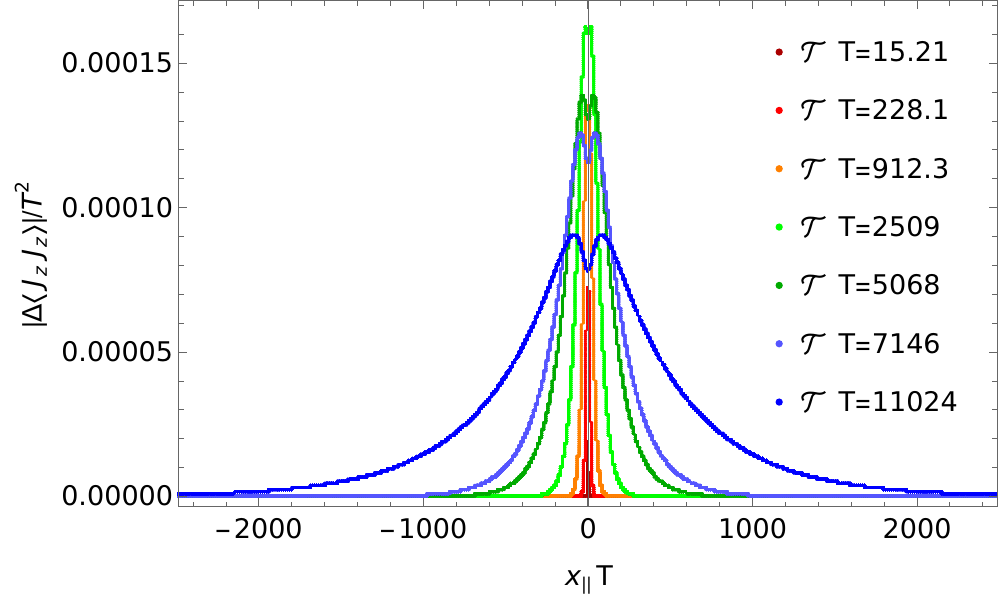}
    \caption{Initial spatial distribution (top panel) and late time spatial distribution (lower panel) of the subtracted electric current two point function. 
    We fixed $B\approx 0.22 T^2$, $\alpha=6/19$ and $m_sL=0.04$. The plots are shown in units of temperature.}
    \label{fig:01}
\end{figure*}
To illustrate the real-time dynamics as described in this paper, figure \ref{fig:01} depicts the spatial profile of the  two-point correlation function of the electric current at different moments in time. The top panel captures the build-up of the correlations at early times and the lower panel depicts the late-time dynamics. Transverse and longitudinal refer to the direction of the momentum (Fourier space) or Cartesian coordinate with respect to the direction of the magnetic field. For the earliest shown time interval (black curve, top panel) the transverse (left) and longitudinal distribution (right) show two peaks. This might be the strong coupling analogue of the two chiral fermions in the weak coupling picture. Increasing the length of the time interval, the distributions increase in spatial extent and magnitude. From the top panel it is evident that the area between the two off-axis peaks fills up, corresponding to filling up the sphaleron shell. After reaching the axial charge relaxation time $\tau_{5,\text{rel}} \simeq 2500/T$, the magnitude of the distributions starts to decrease, while their spatial extent continues to increase (lower panel). Furthermore, the two peaks start appearing again in the longitudinal distribution (lower right panel).

In order to give a realistic estimate of the spatial sizes of the electric current correlations, one can convert dimensionless quantities in these plots into dimensionful units.
For $T=300\ MeV,\, B=1\ m_\pi^2,\, {\cal T}= 10\, \text{fm}$ (for $m_sL=0.04$), we estimate the transverse and longitudinal sizes as \begin{equation*}\vspace{-0.18cm}
    x_\perp \simeq 1.25\, \text{fm}\qquad\text{and}\qquad x_\| \simeq 2\,\text{fm}.
\end{equation*}
At zero temperature and without a magnetic field, the average instanton size has been estimated to be $\simeq 0.3$ fm  \cite{Shuryak:2021iqu}. We see that the correlations between electric currents in our case have a significantly larger range. At weak coupling $g$, the size of the sphaleron at temperature $T$ can be estimated as $\sim 1/(g^2 T)$, as the sphaleron is a purely magnetic configuration at the top of the barrier, and its size should be determined by magnetic screening. So sphalerons are large objects (on thermal scale $1/T$) at weak coupling; the results of \cite{Grieninger:2023wuq} suggest that the sphalerons are  large at strong coupling as well. The effects of the plasma expansion on the dynamical fluctuations of axial charge were recently considered in \cite{Grieninger:2023myf}.

\subsection{Anomalous hydrodynamics} 








\subsubsection{Anomalous hydrodynamics and (lack of) entropy production}

Since in the strong coupling regime the plasma represents a fluid (for reviews, see \cite{Son:2007vk,Schafer:2009dj}), it is of great interest to study the effects 
of anomalies in relativistic hydrodynamics. 
A purely hydrodynamical derivation of the anomaly effects at first order in derivative expansion was given by Son and Surowka \cite{Son:2009tf}, motivated by earlier results in AdS/CFT correspondence  \cite{Erdmenger:2008rm,Banerjee:2008th,Torabian:2009qk} which found, among others, chiral vortical effect.
It has been generalized to anomalous superfluids \cite{Lublinsky:2009wr,Lin:2011mr,Bhattacharya:2011tr} and non-abelian symmetry \cite{Eling:2010hu,Neiman:2010zi}. 

The idea of Son and Surowka \cite{Son:2009tf} was to consider the local entropy production rate $\partial_\mu s^\mu$
and to impose on it the positivity constraint following from the second law of thermodynamics. The contributions from the anomaly to the entropy production 
were shown to be locally unbounded in either sign so that unless their coefficients identically vanished, they
could potentially violate the second law of thermodynamics. These arguments lead to a set of algebro-differential equations for the transport coefficients related to the anomaly; in many cases they can be solved.

\vskip0.3cm

The guiding principle proposed in \cite{Kharzeev:2011ds} for constructing anomalous hydrodynamics is that  the anomaly-induced ${\cal T}$-even terms should not contribute to the entropy production. To illustrate the significance of discrete symmetries, let us consider first the behavior of contributions to the entropy under  the spatial parity. The anomalous contributions to 
the entropy production are special in that
they change sign under spatial parity transformation $P$. Suppose there were a contribution to the entropy production from the anomalous terms we identify; then in the parity-flipped mirror world, this contribution
would become negative. Thinking of entropy production as originating from some dissipative work,
this is very unnatural. This consideration gives us the first hint that the anomalous terms should not contribute to the entropy production. 

\vskip0.3cm

The vanishing of the entropy production from the anomaly-induced terms has a simple physical meaning -- the corresponding {\it anomalous currents are non-dissipative}.
This rather unusual property originates from the fact that the  anomalous current is associated with the zero fermion modes, and the number of these zero modes is related to the topology of gauge fields by the Atiyah-Singer index theorem. Since the topology of gauge fields is determined at the boundary of the fluid, the processes in the bulk cannot lead to the dissipation of anomalous currents. This consideration can be 
made more formal by considering yet another discrete symmetry of the transport coefficients - time reversal ${\cal T}$. The ``usual" electric conductivity $\sigma$ is a ${\cal T}$-odd quantity, as can be easily seen from Ohm's law $J^i = \sigma E^i$: the electric field is ${\cal T}$-even, whereas the electric current  $J^i$ is ${\cal T}$-odd. On the other hand, let us consider the quantum Hall effect as an example of anomalous current in $(2+1)$ dimensions. The quantum Hall conductance is a ${\cal T}$-even quantity, as it is associated with a ${\cal T}$-odd magnetic field. 
The corresponding Hall current is non-dissipative, and the conductance of the integer quantum Hall effect is given by the Chern numbers of vector bundles associated with the energy bands of the Hamiltonian operator \cite{Thouless:1982zz}. 

In physical terms, the dissipative transport coefficients are described in terms of the response of the states near the Fermi energy, whereas the non-dissipative ones involve all of the states below the Fermi energy. The anomalous chiral magnetic current can be thought of as a quantum phenomenon that involves the entire Dirac sea \cite{Kharzeev:2010ym} (reflecting Gribov's view  of ``anomalies as a manifestation of the high momentum collective motion in the vacuum" \cite{Gribov:1981ku,Dokshitzer:2004ie}), and it is thus natural to expect that it is of non-dissipative, reversible nature. Indeed, the chiral magnetic conductivity $\sigma_\chi$ \cite{Kharzeev:2009pj} defined by $\vec{J} = \sigma_\chi \vec{B}$ is a manifestly ${\cal T}$-even quantity as it relates  magnetic field and electric current both of which are ${\cal T}$-odd. We note that this feature of anomalous currents makes them 
potentially important in various applications that include quantum computing, see e.g. \cite{Kharzeev:2019ceh}.

\vskip0.3cm

Therefore the terms originating from the anomaly, {\it i.e.} the terms
that are linear in $\kappa$, do not contribute to the net entropy production at all orders.
The first order result obeys this principle -- the first order contribution coming from
anomaly vanishes identically. The validity of this claim in the second order is supported by comparison of the anomalous  transport coefficient to the 
existing holographic results \cite{Erdmenger:2008rm, Banerjee:2008th, Ammon:2020rvg}.

\subsubsection{Chiral magnetic wave}\label{subsubsec_cmw}

It has been found that the CME current persists in hydrodynamics 
\cite{Sadofyev:2010pr} 
and is transferred by the sound-like gapless excitation -- ``the chiral magnetic wave" \cite{Kharzeev:2010gd,Burnier:2011bf}, see also \cite{Newman:2005hd} for an earlier study of collective excitations in anomalous hydrodynamics. 
Another line of development has been along the viewpoint of the effective field theory \cite{Sadofyev:2010is}.

For simplicity, let us consider single flavor ($N_F=1$) massless QCD with chiral symmetry $U(1)_L\times U(1)_R$, or equivalently $U(1)_V\times U(1)_A$ where $V(A)$ denotes vector(axial) respectively. The axial symmetry $U(1)_A$ suffers from both QCD anomaly with gluonic topological density and from the triangle anomaly of global chiral symmetry. The latter is in fact not harmful to the conservation of $U(1)_A$ as long as one does not elevate the global chiral symmetry to a gauged one, while the former indeed breaks the axial $U(1)_A$ symmetry by quantum fluctuations of topological density.

Our starting point is the anomalous generation of vector and axial currents along the applied magnetic field in the presence of axial (vector) chemical potential $\mu_A$ ($\mu_V$), as given by eqs. (\ref{eq_intro_cme}) and (\ref{eq_intro_cse}). We will now re-write these equations 
 in a more suggestive matrix form as
\begin{equation}
\left(\begin{array}{c} \vec j_V \\ \vec j_A\end{array}\right) = {N_c\ e \vec B\over 2\pi^2} \left(\begin{array}{cc}
0 & 1 \\ 1 & 0
\end{array}\right) \left(\begin{array}{c}
\mu_V \\ \mu_A
\end{array}\right)\quad.
\end{equation}
We are interested in small linearized fluctuations of the chiral currents $j_A$ and $j_V$ in the plasma; let us assume that this plasma is neutral, without any background charge density on average. We may then perform a linear expansion of the chemical potentials with respect to small charge densities $(j^0_V,j^0_A)$,
\begin{equation}
\left(\begin{array}{c}
 \mu_V \\ \mu_A
\end{array}\right)=\left(\begin{array}{cc}
\partial\mu_V\over \partial j^0_V & \partial\mu_V\over \partial j^0_A \\
 \partial\mu_A\over \partial j^0_V  & \partial\mu_A\over \partial j^0_A 
\end{array}\right) \left(\begin{array}{c}
 j^0_V \\ j^0_A 
\end{array}\right) + {\cal O}\left(\left(j^0\right)^2\right) \equiv
\left(\begin{array}{cc}
\alpha_{VV}& \alpha_{VA}\\
 \alpha_{AV}  & \alpha_{AA} 
\end{array}\right) \left(\begin{array}{c}
 j^0_V \\ j^0_A 
\end{array}\right)+ {\cal O}\left(\left(j^0\right)^2\right)
\quad.
\end{equation}
Remembering that 
\begin{equation}
    \mu_i = {\partial {\cal F}\over \partial j^0_i}\quad,\quad i=V,A
\end{equation}
where $\cal F$ is the Helmholtz free energy, the $\alpha$'s appearing above are nothing but the susceptibility matrices 
of vector/axial charge densities,
\begin{equation}
    \alpha_{ij} = {\partial^2 {\cal F}\over \partial j^0_i \partial j^0_j}\quad.
\end{equation}
Considering the parity $\cal{P}$ transformation $V\to - V$ and $A\to A$, one concludes that parity invariance of QCD implies that 
$\alpha_{VA}=\alpha_{AV}=0$ in the neutral plasma, $\mu_V=\mu_A=0$. Moreover, a simple large $N_c$ counting shows that
\begin{equation}
    \alpha_{VV}\sim \alpha_{AA} \sim {\cal O}\left(1\over N_c\right)\quad,
\end{equation}
while their difference in a deconfined and chirally symmetric phase is subleading
\begin{equation}
    \alpha_{VV}- \alpha_{AA} \sim {\cal O}\left(1\over N_c^2\right)\quad.
\end{equation}
Independently of this, the vanishing of the difference $\alpha_{VV}-\alpha_{AA}$ can be taken as
a signal of chiral symmetry restoration. Therefore, we expect it to be a good approximation to let $\alpha_{VV}=\alpha_{AA}\equiv \alpha$ in the chirally symmetric phase; this leads us to 
\begin{equation}
    \left(\begin{array}{c} \vec j_V \\ \vec j_A\end{array}\right) = {N_c\ e \vec B \alpha \over 2\pi^2} \left(\begin{array}{cc}
0 & 1 \\ 1 & 0
\end{array}\right) \left(\begin{array}{c}
j^0_V \\ j^0_A
\end{array}\right)\quad.\label{semifinal}
\end{equation}

It is natural to diagonalize the equation above by going to the chiral basis
\begin{equation}
    j^\mu_L\equiv {1\over 2}\left(j^\mu_V-j^\mu_A\right)\quad,\quad j^\mu_R\equiv
{1\over 2}\left(j^\mu_V+j^\mu_A\right)\quad.
\end{equation}
In terms of chiral currents, our previous assumptions and the definition of $\alpha$'s are easily translated to
\begin{equation}
    \alpha={1\over 2}\left(\partial \mu_L\over \partial j^0_L \right)={1\over 2}\left(\partial^2 {\cal F}\over \partial j^0_L \partial j^0_L\right)=
{1\over 2}\left(\partial \mu_R \over \partial j^0_R\right)={1\over 2}\left(\partial^2 {\cal F}\over \partial j^0_R \partial j^0_R\right)\quad.
\end{equation}
The (\ref{semifinal}) then leads to two decoupled relations
\begin{equation}
    \vec j_{L,R}= \mp \left(N_c e \vec B \alpha\over 2\pi^2\right) j^0_{L,R}\quad,\label{consti}
\end{equation}
where one should keep in mind the definite sign in front of the right-hand side depending on the chirality of the currents.

One can view the above expression as the leading constitutive equation for the currents in the long wavelength derivative expansion of hydrodynamics.
Indeed, our starting point (namely the CME and CSE) is strictly valid only when the variation of chemical potentials is sufficiently slow; for a finite frequency/momentum these expression gets modified resulting in frequency/momentum dependent chiral magnetic conductivity \cite{Kharzeev:2009pj,Yee:2009vw,Fukushima:2010vw,Fukushima:2009ft}. 
The equation (\ref{consti}) is the first leading term in the derivative expansion, while the next leading-order correction to the
chiral magnetic conductivity will be $\partial^2$ or $\omega^2\sim k^2$ in frequency/momentum space.
However, there is an important first-order derivative term in any constitutive equation of conserved current: a diffusion term  $-D \vec\nabla j^0$, with a diffusion constant $D$. In our case, we will be interested only in the waves propagating along the magnetic field direction which we call {\it longitudinal}; thus on general grounds, the constitutive relation including the next leading-order diffusion term reads as
\begin{equation}
\vec j_{L,R}= \mp \left(N_c\ e \vec B \alpha\over 2\pi^2\right) j^0_{L,R} - D_L {\vec B (\vec B\cdot \vec\nabla)\over B^2} j^0_{L,R} +\cdots \quad,\label{consti2}
\end{equation}
with a longitudinal diffusion constant $D_L$. Although we discuss only longitudinal dynamics in this paper,
it would also be interesting to study the transverse dynamics with the transverse diffusion constant $D_T$.

\subsubsection{Novel anomalous transport coefficients}

The CME and CVE are of first order in the hydrodynamic gradient expansion, and the corresponding transport coefficients can be derived in the framework of hydrodynamics by imposing the non-negativity of entropy production~\cite{son2009hydrodynamics}. At second order, there appear additional transport coefficients that have been classified in Ref.~\cite{Kharzeev:2011ds}. The relations between these transport coefficients have been derived from the {\it absence} of entropy production that stems from the time reversal invariance~\cite{Kharzeev:2011ds}. 
\vskip0.3cm

As a specific example of second-order anomalous transport effects, let us consider contributions to electric current that arise from the combination of shear and vorticity or magnetic field~\cite{Kharzeev:2011ds}:
\begin{equation}\label{cme2}
j^\mu_{(2)} = \xi_1 \sigma^{\mu\nu}\omega_\nu + \xi_2 Q \sigma^{\mu\nu}B_\nu,
\end{equation}
where $\sigma^{\mu\nu}= {1\over2} (\partial^\mu_\perp u^\nu + \partial^\nu_\perp u^\mu)$ is the transverse shear tensor ($u^\mu$ is the fluid velocity, and $\partial_\perp^\mu$ is the gradient perpendicular to $u^\mu$), $\omega^{\mu}={1\over 2}\epsilon^{\mu\nu\alpha\beta}u_\nu\partial_\alpha u_\beta$ is vorticity, and $B^\mu$ is magnetic field. The physical meaning of these contributions was discussed in \cite{buzzegoli2022shear}, where the first term in (\ref{cme2}) was called the shear-induced Chiral Vortical Effect (siCVE), and to the second term -- the shear-induced Chiral Magnetic Effect (siCME). 
\vskip0.3cm

What is the microscopic origin of the phenomena encoded in (\ref{cme2})? It is well known that the anomaly relation, and thus the expression for the CME current \ref{eq_intro_cme}
are exact \textit{at the operator level}.  Nevertheless, when the expectation value of this operator relation is taken over a physical state, there may well appear corrections arising from the renormalization of operator quantities that enter (\ref{eq_intro_cme}), see e.g.~\cite{Anselm:1989gi} and discussion in~\cite{Adler:2004qt}. In particular, the magnetic field in the medium can be renormalized by interactions. Moreover, if the shear (and/or vorticity) are present in the medium, they can rotate the orientation of an effective magnetic field by generating a component of the field in the direction perpendicular to initial ${\bf B}$.

To illustrate this argument, let us consider a vortex immersed in the flow and aligned initially along the axis $y$, with ${\vec \omega} \sim {\hat{\vec y}}$. The shear flow with $\sigma^{xy} \sim \omega_z$ will rotate the axis of the vortex in the $(x, y)$ plane, creating a component of an effective vorticity along the axis $x$. This ``tilting" of vorticity in shear flows has been extensively studied in hydrodynamics, see~\cite{kawahara1997wrap} and references therein. Perhaps the most spectacular manifestation of vorticity tilting in Nature is the emergence of tornadoes in ``supercell" thunderstorms, see~\cite{dahl2017tilting} for a review.

The ``conventional" first order chiral vortical effect will then create the current along the $x$ axis. Therefore, the second order anomalous transport phenomenon can be understood in terms of the modification of vorticity (or magnetic field) by the back-reaction of the medium. 

\vskip0.3cm
The values of the second-order transport coefficients $\xi_i$ had been evaluated at strong and weak coupling through holography and chiral kinetic theory, respectively. 

\subsubsection{Higher form anomalous hydrodynamics}

Recently, an interesting formulation of relativistic  magnetohydrodynamics (MHD) has been proposed \cite{Grozdanov:2016tdf}. In this approach, the conservation of magnetic flux in Maxwell electrodynamics is associated with a generalized global symmetry. Namely, the Bianchi identity (one of Maxwell's equations)
\begin{equation}
    \partial_\mu {\tilde F}^{\mu\nu} = 0
\end{equation}
is interpreted as a conservation of a 2-form current 
\begin{equation}\label{2form}
    J^{\mu\nu} \equiv \frac{1}{2} \epsilon^{\mu\nu\rho\sigma} F_{\rho\sigma}. 
\end{equation}
The integration of this 2-form current over a codimension-2 surface yields the number of magnetic lines that cross the surface. 
Relativistic MHD is then constructed as an effective field theory consistent with this generalized symmetry. 

The antisymmetric current (\ref{2form}) is coupled to an external 2-form gauge field $b_{\mu\nu}$, with an extra term in the action
\begin{equation}
    \Delta S = \int d^4x\ \sqrt{-g}\  b_{\mu\nu} J^{\mu\nu}.
\end{equation}
The chiral anomaly in terms of 2-form current acquires the form
\begin{equation}
   \partial_\mu J^\mu_A = \frac{1}{16 \pi^2}\ \epsilon^{\mu\nu\rho\sigma}\ J_{\mu\nu} J_{\rho\sigma} .
\end{equation}
This relation has been interpreted in terms of the so-called ``non-invertible symmetry"\cite{PhysRevLett.129.161601}, where the operator of the axial charge is forced by the anomaly to obey a composition law that is different from the one prescribed by the $U(1)$ symmetry. The relaxation of axial charge was recently studied in this formalism numerically using classical lattice simulations \cite{Das:2023nwl}.

\subsection{Quantum kinetic theory for chirality and spin transport }

An important aspect of the many-body theory for  anomalous chiral transport is to understand systems in the out-of-equilibrium situation. The natural framework is the quantum kinetic theory based on transport equations for the phase space distribution functions of such a system. Different from the classical kinetic theory, a proper description of the chiral fermions must account for intrinsic quantum and relativistic effects.  A lot of progress has been achieved in recent years to derive equations of such a theory and to understand their implications, see e.g.~\cite{Son:2012wh,Son:2012zy,Stephanov:2012ki,Chen:2014cla,Chen:2015gta,Kharzeev:2016sut,Chen:2012ca,Gao:2012ix,Gao:2017gfq,Hidaka:2016yjf,Hidaka:2017auj,Mueller:2017lzw,Mueller:2017arw,Gorbar:2017cwv,Wu:2016dam,Liu:2018xip,Gao:2022gqr,Liu:2020flb}.  There also exist phenomenological interests and active attempts to study anomalous chiral transport in the out-of-equilibrium setting~\cite{Mace:2016svc,Mace:2016shq,Mueller:2016ven,Fukushima:2015tza,Ebihara:2017suq,Sun:2016nig,Sun:2016mvh,Huang:2017tsq,Jiang:2016wve,Shi:2017cpu}.  Here we will briefly review these developments.

\subsubsection{Wigner function formalism}

We will start with the widely adopted approach to derive the quantum transport equations for chiral fermions in the  Wigner function formalism, which carries out a systematic semiclassical expansion in terms of $\hbar$~\cite{Vasak:1987um,Zhuang:1995pd,Zhuang:1995jb,Zhuang:1998kv,Ochs:1998qj,Guo:2017dzf} and provides the bridge connecting quantum field theory to relativistic kinetic theory~\cite{DeGroot:1980dk,Vasak:1987um}.

Let us consider a collisionless system in a background electromagnetic field $A^{\mu}$. The Winger function $W_{\alpha\beta}(x,p)$, defined as the expectation value of the Wigner operator for a Dirac fermion,  satisfies the quantum kinetic equation 
\cite{Vasak:1987um}
\begin{eqnarray}\label{eq:049}
\left(\slashed{\mathbf{K}}-m\right)W(x,p)=0 \,\, ,
\end{eqnarray}
where $\slashed{\mathbf{K}}=\gamma^{\mu} \mathbf{K}_{\mu}$, $\mathbf{K}_{\mu}=\pi_{\mu}+\frac{1}{2}i\hbar\triangledown_{\mu}$, and
\begin{eqnarray}
\pi^{\mu}&=&p^{\mu}-\frac{1}{2}Q\hbar \, j_{1}\left(\frac{1}{2}\hbar \triangle \right) \, F^{\mu\nu}  \partial^{p}_{\nu},\\
\triangledown^{\mu}&=&\partial^{\mu}-Q \, j_{0}\left(\frac{1}{2}\hbar \triangle \right) \, F^{\mu\nu} \partial^{p}_{\nu}.
\end{eqnarray}
Note that in the triangle operator $\triangle=\partial_{x}\cdot\partial_{p}$, $\partial_{x}$ acts only on electromagnetic tensor $F_{\mu\nu}=\partial^{\mu}A^{\nu}-\partial^{\nu}A^{\mu}$, while $\partial_{p}$ acts only on $W(x,p)$. In addition, $j_{0}(x)=x^{-1}\sin(x)$ and $j_{1}(x)=x^{-2}\sin(x)-x^{-1}\cos(x)$ are the spherical Bessel functions which are generated by the y-integrations. 
Combined with the Maxwell equation, the quantum kinetic equation of Wigner function Eq.(\ref{eq:049}) is equivalent to the QED field theory.

The Wigner function can be further expanded as
\begin{equation} 
W=\frac{1}{4}
\left(\mathscr{F}+i\gamma^5\mathscr{P}+\gamma^\mu\mathscr{V}_{\mu}+\gamma^\mu\gamma^5
\mathscr{A}_{\mu}+\frac{1}{2}\sigma^{\mu\nu}\mathscr{L}_{\mu\nu}\right),
\end{equation} 
where these sixteen real components are organized based on their Lorentz transformation properties, i.e. scalar, pseudo-scalar, vector, axial vector and antisymmetric tensor, respectively.   Each of these sixteen components is connected with a corresponding physical quantity\cite{BialynickiBirula:1991tx,Zhuang:1995pd}. 
In particular,  the vector component $\mathscr{V}_{\mu}(x,p) = \mathrm{tr}[\gamma_\mu W(x,p)]$ and axial vector component $\mathscr{A}_{\mu}(x,p) = \mathrm{tr}[\gamma_5 \gamma_\mu W(x,p)]$ 
can be used to construct the physical vector current density $J^{\mu}$ and  axial current density $J^{\mu}_{5}$,  
\begin{align}\begin{split}
&J^{\mu}(x)=\left<\bar{\psi}(x)\gamma^{\mu}\psi(x)\right>=\int d^4 p~\mathrm{tr}\left( \gamma^{\mu}W(x,p)\right)=\int d^4 p \mathscr{V}^{\mu}(x,p),\\
&J^{\mu}_{5}(x)=\left<\bar{\psi}(x)\gamma^{\mu}\gamma^{5}\psi(x)\right>=-\int d^4 p~\mathrm{tr}\left( \gamma^{5}\gamma^{\mu}W(x,p)\right)=-\int d^4 p \mathscr{A}^{\mu}(x,p), 
\end{split}\end{align} 
with the covariant derivative $D^{\mu}=\partial^{\mu}+iQA^{\mu}$ (where Q is the charge of fermion and $A^{\mu}$ is the vector potential).  

By plugging the decomposition back into the equation Eq.(\ref{eq:049}), one obtains a set of coupled equations for all sixteen components of the Wigner function. We will focus on the case of chiral fermions with $m=0$, where  the quantum kinetic equations for various components get partially decoupled. In this case, one obtains a closed set of equations for vector $\mathscr{V}_{\mu}$ and axial vector $\mathscr{A}_{\mu}$ components: 
\begin{align}\begin{split}\label{eq:051}
&\pi^{\mu}\mathscr{V}_{\mu}=0,\qquad\pi^{\mu}\mathscr{A}_{\mu}=0,\\
&\hbar\triangledown^{\mu}\mathscr{V}_{\mu}=0,\qquad\hbar\triangledown^{\mu}\mathscr{A}_{\mu}=0, \\
&\hbar\epsilon_{\mu\nu\rho\sigma}\triangledown^{\rho}\mathscr{V}^{\sigma}=2(\pi_{\mu}\mathscr{A}_{\nu}-\pi_{\nu}\mathscr{A}_{\mu}), \\
&\hbar\epsilon_{\mu\nu\rho\sigma}\triangledown^{\rho}\mathscr{A}^{\sigma}=2(\pi_{\mu}\mathscr{V}_{\nu}-\pi_{\nu}\mathscr{V}_{\mu}).
\end{split}\end{align}
Noting the specific patterns of vector (scalar) and axial-vector (pseudo-scalar) terms, one could further simplify the above two sets of equations by introducing the ``chiral basis''\cite{Ochs:1998qj, Gao:2017gfq} via
\begin{align}\begin{split}
\mathscr{T}_{\chi}&=\frac{1}{2}(\mathscr{F}+\chi\mathscr{P}),\\
\mathscr{S}^{\mu\nu}_{\chi}&=\frac{1}{2}\left(\mathscr{L}^{\mu\nu}+\chi\frac{1}{2}\epsilon^{\mu\nu\sigma\rho}\mathscr{L}_{\sigma\rho} \right),\\
\mathscr{J}^{\mu}_{\chi}&=\frac{1}{2}(\mathscr{V}^{\mu}-\chi\mathscr{A}^{\mu}),
\end{split}\end{align}
where $\chi=\pm1$ corresponds to the chirality of massless fermion. In such chiral basis, the equations for the right-handed(RH) and left-handed(LH) components get decoupled into:
\begin{eqnarray}
&&\pi_{\mu}\mathscr{T}_\chi+\frac{1}{2}\hbar\triangledown^{\nu}\mathscr{S}_{\mu\nu}^\chi=0,\\
&&\pi_{\mu}\mathscr{S}^{\mu\nu}_\chi+\frac{1}{2}\hbar\triangledown^{\nu}\mathscr{T}_\chi=0 \, .
\end{eqnarray}
Similarly  Eq.(\ref{eq:051}) can be recast into RH and LH sectors:  
\begin{eqnarray}
&&\hbar\epsilon_{\mu\nu\rho\sigma}\triangledown^{\rho}\mathscr{J}^{\sigma}_\chi=-2\chi(\pi_{\mu}\mathscr{J}_{\nu}^\chi-\pi_{\nu}\mathscr{J}_{\mu}^\chi),
\label{eq.CKE.1}\\
&&\pi^{\mu}\mathscr{J}_{\mu}^\chi=0, \label{eq.CKE.2}\\
&&\triangledown^{\mu}\mathscr{J}_{\mu}^\chi=0 \label{eq.CKE.3}.
\end{eqnarray}
The decoupling of the RH and LH components in these equations reflects a basic property of massless fermions: for the massless Dirac fermions, the RH and LH sectors can be completely separated in the Lagrangian. 
From now on, we will focus on the  equations for the chiral components $\mathscr{J}^{\mu}_{\chi}$, namely the Eqs. (\ref{eq.CKE.1}-\ref{eq.CKE.3}),  which  can be directly related to the  physical chiral currents: 
\begin{eqnarray}\label{eq:067}
& J^{\mu}_{\chi}=\left<\bar{\psi}_{\chi}\gamma^{\mu}\gamma^{5}\psi_{\chi}\right>=\int d^{4}p \mathscr{J}^{\mu}_{\chi}=\frac{1}{2}\left(J^{\mu}+\chi J^{\mu}_{5}\right).
\end{eqnarray}
In the above, $\psi_{\chi}=P_{\chi}\psi$ and  $\bar{\psi}_{\chi}=\bar{\psi}P_{-\chi}$, with $P_{\chi}=(1+\chi\gamma^{5})/2$ being the chirality projection operators.

 \subsubsection{Covariant chiral transport equations}

The set of Eqs. (\ref{eq.CKE.1}-\ref{eq.CKE.3}) are difficult to solve in general, and we apply the well-known $\hat{\mathbf{O}}(\hbar)$-expansion~\cite{Vasak:1987um}. By systematically expanding the spherical Bessel functions $\mathbf{j}_{0}$ and $\mathbf{j}_{1}$ (in the operators $\pi^{\mu}$ and $\triangledown^{\mu}$)  into a power series of $\hbar$, one then solves the $\mathscr{J}^\mu=\mathscr{J}^{(0)}_{\mu}+\hbar\mathscr{J}^{(1)}_{\mu}+\hat{\mathbf{O}}(\hbar^{2})$ order by order.  

The zeroth order results correspond to classical transport, equivalent to sending $\hbar\to 0$. In this limit the two constraint equations become simply:   
\begin{eqnarray}
p^{\mu}\mathscr{J}^{(0)}_{\mu}=0 \, , \,  \,  p_{\mu}\mathscr{J}^{(0)}_{\nu}-p_{\nu}\mathscr{J}^{(0)}_{\mu}=0 .
\end{eqnarray}
These constraints mandate the following general solution for the zeroth order, $\mathscr{J}_\chi^{(0)}=p_\mu f_\chi^{(0)} \delta(p^2)$. The $\delta$-function ensures precisely the classical on-shell condition for a massless particle: $p^2=p_0^2-\mathbf{p}^2=0$. Substituting this general solution into the evolution equation, one obtains the desired classical transport equation: 
\begin{eqnarray} \label{eq:Vlasov}
\delta(p^2) \, p^{\mu}(\partial_{\mu}-QF_{\mu\nu}\partial^{\nu}_{p}) f^{(0)}_{\chi}=0 \, .
\end{eqnarray}
This is of course the celebrated classical Vlasov equation. 

The general function $f^{(0)}_{\chi}(x,p)$ has the physical interpretation as the phase space distribution. It could be decomposed into the positive energy and negative energy components via $f^{(0)}_{\chi}=\sum_{\epsilon=\pm 1}\theta(\epsilon \, p^{0})f^{(0)}_{\chi ,\epsilon}(x,\epsilon p)$. 

Let us then move to the $\hat{\mathbf{O}}(\hbar)$ order where the first nontrivial quantum correction would emerge. At this order the Eqs. (\ref{eq.CKE.1}-\ref{eq.CKE.3}) become:  
 \begin{eqnarray}
 \label{eq:ohbar:2}
 p^{\mu}\mathscr{J}^{(1)}_{\mu} &=& 0 \, , \\ 
  \label{eq:ohbar:1}
 \epsilon_{\mu\nu\rho\sigma}\triangledown^{\rho}\mathscr{J}^{\sigma}_{(0)} &=& -2\chi(p_{\mu}\mathscr{J}^{(1)}_{\nu}-p_{\nu}\mathscr{J}^{(1)}_{\mu})  \, , \\
\label{eq:ohbar:3}
\triangledown^{\mu}\mathscr{J}^{(1)}_{\mu} &=& 0 \, . 
 \end{eqnarray} 

The first two equations (\ref{eq:ohbar:1},\ref{eq:ohbar:2}) strongly constrain the possible form of  
$\mathscr{J}^{(1)}_{\mu}$, for which the most general construction could be written down as:  
\begin{eqnarray} \label{eq:J1:hbar}
\mathscr{J}^{(1)}_{\mu} &=&  p_{\mu}f^{(1)}_{\chi} \delta(p^{2})+
\mathscr{K}_{\mu} \delta(p^{2}) \nonumber \\
&& + \chi Q \widetilde{F}_{\mu\nu}p^{\nu}f^{(0)}_{\chi}\delta^{'}(p^2) \, 
\end{eqnarray}
where $\delta^{'}(p^{2})=d\delta(p^2)/dp^2$, and we have used the relation $p^{2}\delta^{'}(p^{2})=-\delta(p^{2})$. $\widetilde{F}_{\mu\nu}=\frac{1}{2}\epsilon_{\mu\nu\rho\sigma}F^{\rho\sigma}$ is the dual tensor of $F_{\mu\nu}$. The $f^{(1)}_{\chi}$ in the first term is essentially the $\hat{\mathbf{O}}(\hbar)$ correction to the zeroth order distribution function $f^{(0)}_{\chi}$ in Eq.(\ref{eq:Vlasov}). The last term embeds the influence of the zeroth order solution to the subsequent order. 

The key element in (\ref{eq:J1:hbar}) is the $\mathscr{K}_{\mu}$, which needs to satisfy the following constraints followed from Eqs.(\ref{eq:ohbar:2},\ref{eq:ohbar:1}):   
\begin{eqnarray}
 p^{\mu}\mathscr{K}_{\mu}&=&0 \, ,    \\ 
 \epsilon_{\mu\nu\rho\sigma}p^{\sigma}\left(\triangledown^{\rho}f^{(0)}_{\chi}\right)\delta(p^2) &=& 2\chi(p_{\nu}\mathscr{K}_{\mu}-p_{\mu}\mathscr{K}_{\nu})\delta(p^{2})  . \,\,\, \quad 
\end{eqnarray} 
To solve the  $\mathscr{K}_{\mu}$, we introduce an {\em arbitrary} auxiliary time-like vector $n^\nu$   normalized to unity: $n^\mu n_\mu=1$. 
It is straightforward to prove mathematically that for each chosen $n^\mu$, there exists the unique  general solution of $\mathscr{K}_{\mu}$ satisfying these constraints, constructed as follows: 
 \begin{eqnarray} \label{eq:Kmu}
 \mathscr{K}_{\mu}= \frac{\chi}{2p\cdot n}\epsilon_{\mu\nu\lambda\rho}p^{\nu} n^{\lambda}  
 \left(\triangledown^{\rho}f^{(0)}_{\chi}\right) \, . 
 \end{eqnarray} 
 These also include all possible solutions, with each solution labelled by $n^\mu$. 
 At this point, we've consistently obtained the general $\mathscr{J}^{(1)}_{\mu}$, albeit with an arbitrary auxiliary quantity $n^\mu$ that appears to be a free choice at our disposal without clear physical meaning. 
Indeed, the solution $\mathscr{J}^{(1)}_{\mu}$ expressed through   (\ref{eq:J1:hbar}) and (\ref{eq:Kmu}) satisfies all the constraint equations for any choice of (even spacetime dependent) $n^\mu(x)$.  

Finally putting all the above analysis together, we can write down the general form of   $\mathscr{J}^{\mu}$ up to $\hat{\mathbf{O}}(\hbar)$ order: 
 \begin{eqnarray} \label{eq:Jall}
\mathscr{J}^{\mu}&=p^{\mu}f_{\chi}\delta(p^{2})+\hbar\chi Q\widetilde{F}^{\mu\nu}p_{\nu}f^{(0)}_{\chi}\delta^{'}(p^{2}) \nonumber \\
&-\hbar\frac{\chi}{2p\cdot u}\epsilon^{\mu\nu\lambda\rho}u_{\nu}p_{\lambda}\left(\triangledown_{\rho}f^{(0)}_{\chi}\right)\delta(p^2).
\end{eqnarray}
where we've introduced a total distribution function $f_{\chi}=f^{(0)}_{\chi}+\hbar f^{(1)}_{\chi}$ up to $\hat{\mathbf{O}}(\hbar)$ order. It can be similarly decomposed into positive/negative energy components via $f_{\chi}(x,p)=\sum_{\epsilon=\pm 1}\theta(\epsilon p^{0})f_{\chi, \, \epsilon}(x,\epsilon p)$. We emphasize that the above solution for the chiral component $\mathscr{J}^{\mu}$ of the Wigner function (up to $\hat{\mathbf{O}}(\hbar)$ order) is covariant, consistent and complete (i.e. without any additional term possible). In principle one could directly work with this covariant quantity for transport study. 

It must satisfy the transport equation $\triangledown^{\mu}\mathscr{J}_\mu=0$ for time evolution. After consistently keeping terms up to the $\hat{\mathbf{O}}(\hbar)$ order and using Taylor expansion of the delta function, one finally derives the following covariant chiral transport equation at $\hat{\mathbf{O}}(\hbar)$ order: 
\begin{eqnarray} \label{eq:CCTE}
&& \delta\left( p^{2}-\hbar\frac{\chi Q}{p\cdot n}p_{\lambda}\widetilde{F}^{\lambda\nu}n_{\nu} \right)  \times \nonumber \\
&& \Bigg\{ p\cdot\triangledown +\hbar\frac{\chi}{2\left(p\cdot n\right)^2}\left[(\partial_{\mu}n_{\sigma})p^{\sigma}-QF_{\mu\alpha}n^{\alpha}\right]\epsilon^{\mu\nu\lambda\rho}n_{\nu}p_{\lambda}\triangledown_{\rho}  \nonumber\\
&&  \quad -\hbar\frac{\chi}{2p\cdot n}\epsilon^{\mu\nu\lambda\rho}\left(\partial_{\mu}n_{\nu}\right) p_{\lambda}\triangledown_{\rho} \nonumber \\
&& \quad +\hbar\frac{\chi Q}{2p\cdot n}p_{\lambda}\left(\partial_{\sigma}\widetilde{F}^{\lambda\nu}\right)n_{\nu}\partial^{\sigma}_{p} \Bigg\} f_{\chi} = 0 \, .
\end{eqnarray}

\subsubsection{Lorentz invariance and frame dependence}

The transport theory of chiral fermions bears unusual subtlety and poses nontrivial challenges, particularly related to Lorentz invariance and frame dependence. A resolution was developed in the 3D formulation of chiral kinetic theory \cite{Chen:2014cla,Chen:2015gta,Hidaka:2016yjf}, but the origin of such issues remains  cloudy. As it turns out, the quantity $n^\mu$ plays a  crucial role in the chiral transport, especially in   the  confusing issues of Lorentz invariance and frame dependence. To understand the role of $n^\mu$, let us carefully examine the first two terms in Eq.(\ref{eq:J1:hbar}), which could be collectively written as $\mathscr{H}_{\mu}  \delta(p^{2})$ where $\mathscr{H}_{\mu}= p_{\mu}f^{(1)}_{\chi} + \mathscr{K}_{\mu}$ must satisfy $p^\mu \mathscr{H}_{\mu} =0$. In other words, one is trying to decompose a vector $\mathscr{H}_{\mu}$  
 orthogonal to $p^\mu$ into two parts. This decomposition is however subtle due to the light-like nature of $p^\mu$: $p^\mu p_\mu=0$ i.e. $p_\mu$ is ``self-orthogonal''. It deserves commenting that this light-like feature is of course ultimately because the chiral fermion is massless. 
 
 To {\em unambiguously} identify the first order correction to the distribution function, one must demand that  the part along $p_\mu$ should be attributed to the distribution term $f^{(1)}_{\chi}$ while the rest to the $\mathscr{K}_{\mu}$ term. In fact, such a requirement completely fixes the form of $\mathscr{K}_{\mu}$. For the light-like vector $p_\mu=(|\mathbf{p}|,\mathbf{p})$, there are three categories of orthogonal vectors: one parallel to $p_\mu$ itself, the other two taking the form $(0,\mathbf{K})$ with the spatial component satisfying $\mathbf{K}\cdot\mathbf{p}=0$. Therefore, for a uniquely defined $f^{(1)}_{\chi}$, the $\mathscr{K}_{\mu}$ must be in the latter two categories and thus $\mathscr{K}_0=0$ for any $\mathbf{p}$.  Combining this requirement with Eq.(\ref{eq:Kmu}), one arrives at the unique choice $n^\mu=(1,0,0,0)$ and the corresponding   $\mathscr{K}_{\mu}$  below: 
 \begin{eqnarray} \label{eq:Krest}
\mathscr{K}_{\mu} = \left ( 0,  \frac{\chi}{2|\mathbf{p}|}  \mathbf{p} \times 
(\vec{\triangledown}f^{(0)}_{\chi})   \right ) \,\, .
 \end{eqnarray}

This however is not the end of the story. While the above construction gives  well-defined $f^{(1)}_{\chi} p_\mu$ and $\mathscr{K}_{\mu}$  in the {\em current reference frame, this decomposition is actually  frame dependent}. To appreciate this less obvious subtlety,  suppose in the current frame there is a vector $\mathscr{K}_\mu=(0,\mathbf{K})$ which satisfies orthogonality to $p_\mu$ via $\mathbf{K}\cdot \mathbf{p}=0$. But upon boosting into a different frame with both $p_\mu$ and $\mathscr{K}_\mu$ transformed as Lorentz vectors into 
$p'$ and $\mathscr{K}'$, one finds that in general $\mathscr{K}'$ acquires a component along $p'$, despite that they still satisfy $\mathscr{K}' \cdot p'=0$. That means one has to redo the proper decomposition in the new reference frame and find a different $\mathscr{K}''=(0,\mathbf{K}'')$ satisfying $\mathbf{K}''\cdot p'=0$. This issue again arises from the light-like nature of $p_\mu$.    
 
A lengthy calculation proves that if one boosts from the current frame to a different frame of four-velocity $u^\mu$ (with respect to the current frame), then the  $\mathscr{K}_{\mu}$ from proper decomposition in this new frame should be precisely and uniquely given by Eq.(\ref{eq:Kmu}) with the identification $n^\mu \to u^\mu$ which leaves a well-defined $f^{(1)}_{\chi}$ in this new reference frame. Hence the role of $n^\mu$ now becomes clear. This result also explicates the fact that the distribution term $f^{(1)}_{\chi}$ becomes {\em frame-dependent} as well.  While the distribution function  in usual transport theory is a Lorentz scalar, 
our result clearly demonstrates that in chiral transport theory a nontrivial frame dependence of the distribution function arises precisely at the $\hat{\mathbf{O}}(\hbar)$ order correction and in the specific way discussed above. 
It may be mentioned that the structure of $\mathscr{K}_{\mu}$ in (\ref{eq:Kmu}) and the interpretation of $n^\mu$ naturally clarify
the  origin  of the side-jump effect and the frame dependence of spin antisymmetric tensor $S^{\mu\nu} \propto \frac{\epsilon^{\mu\nu\rho\sigma}p_\rho n_\sigma}{p\cdot n}$ discussed in \cite{Chen:2015gta,Hidaka:2016yjf}. In the approach presented here, however, one does not introduce the side-jump effect and it follows naturally and completely from the condition of a unique $f^{(1)}_{\chi} $ in the aforementioned decomposition for any given reference frame.

\subsubsection{Anomalous transport in chiral kinetic theory}

To investigate the anomalous transport phenomena with quantum transport equations, it is more convenient to utilize the 3D formulation of chiral kinetic theory, at the ``expense'' of losing  manifest Lorentz covariance. It can be shown that the 3D theory naturally and easily follows from the more general 4D covariant equation (\ref{eq:CCTE}). To do this, one essentially needs to take the series of moments of  the 4D equation by integrating it with $\int dp_0 p_0^q$. Specifically the first moment ($q=0$) leads to the desired 3D transport equation. 
 
 As a first step we adopt the rest-frame choice $n^\mu \to (1,0,0,0)$ in Eq.(\ref{eq:CCTE}) and integrate it over $p_0$ (which enforces the on-shell condition via the $\delta$ function). This immediately leads to the following dispersion relations for the chiral fermion (with $\hat{\mathbf{O}}(\hbar)$ correction):
 \begin{eqnarray}\label{eq:3Dshift}
&& p_{0}=\epsilon E_{\mathbf{p}} \, ,
\quad E_{\mathbf{p}}=|\mathbf{p}|\left( 1-\hbar\epsilon Q  \mathbf{B}\cdot \bf{b}_{\chi}\right) \, , \, \\ \label{eq:3Dshift:2}
&& \widetilde{\mathbf{v}}=\frac{\partial E_{\mathbf{p}}}{\partial\mathbf{p}}=\widehat{\mathbf{p}}\left(1+2\hbar\epsilon Q\mathbf{B}\cdot\mathbf{b}_{\chi} \right)-\hbar\epsilon Q b_{\chi}\mathbf{B} \, .
\end{eqnarray}
where $\mathbf{b}_{\chi}=\chi\frac{\mathbf{p}}{2|\mathbf{p}|^{3}}$ is the famous Berry curvature, 
$\widetilde{\mathbf{v}}$ is the fermion's group velocity, $\widehat{\mathbf{p}}=\mathbf{p}/|\mathbf{p}|$  and $\epsilon=\pm1$ corresponding to the particle/antiparticle. In the following we then  separate $f_\chi$ into particle/antiparticle sectors $f_{\chi, \, \epsilon}$.   
 
After carefully performing the energy integral and consistently keeping (up to) $\hat{\mathbf{O}}(\hbar)$ order, one finally arrives at the following 3D kinetic equation: 
\begin{eqnarray}\label{eq:CKT3D}
&& \Bigg\{ \partial_{t} + {\mathbf{G}}_{\mathbf{x}} \cdot\triangledown_{\mathbf{x}} + {\mathbf{G}}_{\mathbf{p}} \cdot\triangledown_{\mathbf{p}}  \Bigg \}  f_{\chi, \, \epsilon}(t,\mathbf{x},\mathbf{p}) = 0 \\
\label{eq:CKT3D:1}
&&  {\mathbf{G}}_{\mathbf{x}} = \frac{1}{\sqrt{G}} \left [ \widetilde{\mathbf{v}}+\hbar Q(\widetilde{\mathbf{v}}\cdot\mathbf{b}_{\chi})\mathbf{B}+\hbar Q\widetilde{\mathbf{E}}\times\mathbf{b}_{\chi} \right]\\
\label{eq:CKT3D:2}
&&  {\mathbf{G}}_{\mathbf{p}} = \frac{ \epsilon Q}{\sqrt{G}} \left[\widetilde{\mathbf{E}}+\widetilde{\mathbf{v}}\times\mathbf{B}+\hbar Q(\widetilde{\mathbf{E}}\cdot\mathbf{B})\mathbf{b}_{\chi} \right]
\end{eqnarray}
with the  Jacobian factor $\sqrt{G}=\left(1+\hbar Q\mathbf{b}_{\chi}\cdot\mathbf{B} \right)$ and   
$\widetilde{\mathbf{E}} =  \mathbf{E}-\frac{1}{\epsilon Q}\triangledown_{\mathbf{x}}E_{\mathbf{p}}$ where the $E_{\mathbf{p}}$ may vary in space due to space-dependence of the magnetic field. The set of  equations (\ref{eq:CKT3D},\ref{eq:CKT3D:1},\ref{eq:CKT3D:2})  precisely reproduce the 
3D chiral kinetic theory developed in \cite{Son:2012wh,Son:2012zy,Stephanov:2012ki} and widely discussed in the literature. 

The above equations are derived for free chiral fermions. Under the presence of interactions, the chiral kinetic theory could be generalized to take into account the collision term as well as anomalous magnetic moment, as follows: 
\begin{eqnarray}\label{eq:CKT3D:2}
&& \Bigg\{ \partial_{t} + {\mathbf{G}}_{\mathbf{x}} \cdot\triangledown_{\mathbf{x}} + {\mathbf{G}}_{\mathbf{p}} \cdot\triangledown_{\mathbf{p}}  \Bigg \}  f_{\chi, \, \epsilon}(t,\mathbf{x},\mathbf{p}) = \mathcal{C} \left [ f_{\chi, \, \epsilon} \right] 
\end{eqnarray}
along with the replacement $Q\mathbf{B} \to gQ\mathbf{B}$ for all relevant quantities in Eqs.(\ref{eq:3Dshift},\ref{eq:3Dshift:2},\ref{eq:CKT3D},\ref{eq:CKT3D:1},\ref{eq:CKT3D:2}), where the $g$ factor quantifies the anomalous magnetic moment with $g=2$ for free theory.

Let us now discuss how the above chiral kinetic equations lead to the   chiral anomaly relation, the Chiral Magnetic Effect (CME), as well as Chiral Vortical Effect (CVE). To do that, we examine the local current density given by the following:
\begin{eqnarray} \label{eq_ckt_j}
\vec{\bf j}  = \int_{\vec{\bf p}}  \, \left [ \widetilde{\mathbf{v}}+\hbar Q(\widetilde{\mathbf{v}}\cdot\mathbf{b}_{\chi})\mathbf{B}+\hbar Q\widetilde{\mathbf{E}}\times\mathbf{b}_{\chi} \right] \,  f_{\chi} 
\end{eqnarray}
with $\int_{\vec{\bf p}}\equiv \frac{d^3\vec{\bf p}}{(2\pi)^3}$ and $ f_{\chi} \equiv f_{\chi, \, +}-f_{\chi, \, -} $. 
 For simplicity, we focus on the collisionless case without the collision term.

Firstly, it is straightforward to verify the chiral anomaly that one would expect for the above chiral current. Indeed, after writing down the 4-current $j^\mu = (n, \vec{\bf j})$ with
$n= \int_{\vec{\bf p}}  \, \sqrt{G} \,  f_{\chi}$ and taking its full divergence, one obtains: 
\begin{eqnarray}
    \partial_\mu j^\mu   = \frac{\chi}{4\pi^2} \vec{\bf E} \cdot \vec{\bf B}
\end{eqnarray} 
where $\chi=\pm$ for RH or LH fermions respectively.

We next examine the CME current under the presence of external magnetic field and nonzero chiral chemical potential $\mu_5$. This can be computed by inserting the Fermi-Dirac distribution at a given temperature $T$ and $\mu_5$ into the above current, i.e. $f_\chi \to \frac{1}{e^{(|\vec{\bf p}|-  \mu_\chi)/T}}$. After performing the momentum integral, one arrives at the  following expected expression: 
\begin{eqnarray}
    \vec{\bf j}_{\rm CME}   = \frac{\chi}{4\pi^2}  \mu_\chi   \vec{\bf B} \, .
\end{eqnarray} 
The above results is for each specific chirality. Adding contributions from LH and RH sectors together will give the previously introduced CME current in Eq.(\ref{eq_intro_cme}). 

Finally one could also derive the CVE current within this formalism, by inserting an equilibrium distribution that generalizes the usual Fermi-Dirac to the case under finite vorticity field $\vec{\bf \omega}$. The final result could be obtained by simply making a replacement $\vec{\bf B} \to \epsilon_p \vec{\bf{\omega}}$ in the integral for $\vec{\bf j}$, leading to the result: 
\begin{eqnarray}
    \vec{\bf j}_{\rm CVE}   = \frac{\chi}{4\pi^2}  \mu_\chi^2   \vec{\bf \omega} \, .
\end{eqnarray} 
Again, adding contributions from LH and RH sectors together will give the previously introduced CVE current in Eq.(\ref{cve}).  It may be noted that additional terms of the form $\sim T^2$ will occur at finite temperature.

\subsubsection{New developments in quantum kinetic theory}

Many interesting new results, both in the theoretical framework and in the phenomenological application of quantum kinetic theory, have been obtained in the past several years. Here we briefly discuss some of these new developments.  For a recent in-depth review, see e.g. \cite{Hidaka:2022dmn,Gao:2020pfu,Gao:2020vbh}.

So far we have focused on the collisionless limit to emphasize the anomalous nature of chiral transport. However, it is important to include collision terms for describing real world systems. 
One convenient  approach is to use the relaxation time approximation(RTA), especially in near-equilibrium regime. Assuming the distribution function $f$ only differs from the equilibrium distribution function by a small quantity $f=f_{eq}+\delta f$, one can expand the collision term as 
$C_{coll}[f]=\frac{C}{\tau}\delta f+O(\delta f^2)$,  
where $\tau$ is the relaxation time quantifying how fast the system will evolve towards equilibrium, and $C$ can be a function of any scalars in the model. 
A more rigorous approach is to derive the chiral transport equations from interacting quantum field theory via Wigner functions, with the order $\hat{O}(\hbar)$ results obtained in \cite{Hidaka:2017auj}.  
The inclusion of collision terms has allowed studying the non-equilibrium response of the system to external disturbance such as the dynamical chiral magnetic conductivity under the presence of spacetime dependent temperature, density or external fields~\cite{Hidaka:2017auj,Kharzeev:2016sut,Horvath:2019dvl}.   

Another interesting direction is to understand the extension of quantum kinetic theory toward finite mass~\cite{Wang:2019moi,Hattori:2019ahi,Sheng:2020oqs,Manuel:2021oah,Das:2022azr}. Quantum corrections   up to $\hbar$ order were obtained in \cite{Wang:2019moi} for the classical  on-shell condition as well as for the transport equations. It was found that  to the linear order in the fermion mass, the mass correction does not change the structure of the chiral kinetic equations and behaves like an additional collision terms. Taking the massless limit of the quantum transport equations for massive fermions is a tricky issue and it was shown in \cite{Hattori:2019ahi,Weickgenannt:2019dks,Sheng:2020oqs} that with appropriate procedures the chiral kinetic equations can be smoothly connected with the more general quantum transport equations with finite mass.  

The studies of chiral transport equations have further  motivated strong interests in developing quantum transport theory that incorperates the spin degree of freedom in general, with many new results obtained --- see e.g. \cite{Gao:2019znl,Das:2022azr,Weickgenannt:2019dks,Sheng:2022ssd,Wagner:2022amr}.   This theoretical framework provides a powerful tool for deriving hydrodynamics with spin degrees of freedom, as shown in e.g. \cite{Florkowski:2017dyn,Bhadury:2020cop,Bhadury:2022ulr,Weickgenannt:2022qvh,Weickgenannt:2022zxs,She:2021lhe}. The formalism has also found important applications for  understanding the spin polarization effects observed in heavy ion collision~\cite{Wang:2020pej,Li:2020cwq,Li:2019qkf,Becattini:2021suc,Liu:2021uhn}.

The chiral transport theory provide an important tool for the phenomenological modeling of anomalous transport effects in heavy ion collisions. An early application of these equations was to compute the CME current generated during the early  stage of heavy ion collisions when the system is far from equilibrium whiles the magnetic fields is still strong~\cite{Huang:2017tsq}. Transport simulations based on chiral kinetic equations were also developed for describing the CME and CVE transport during the whole partonic stage~\cite{Sun:2016nig,Sun:2016mvh,Sun:2018idn,Yuan:2023skl}. A very recent work along this line \cite{Yuan:2023skl} has further included background effects and demonstrated nontrivial results for comparing with experimental measurements. In addition to anomalous transport effects, this formalism has also been used for computing the global and local spin polarization observables, see e.g. \cite{Sun:2017xhx,Liu:2019krs,Huang:2023tyn}.

Finally, it shall be noted that there are other theoretical approaches that have been developed for chiral transport theory, such as the Hamiltonian approach~\cite{Son:2012wh,Son:2012zy}, the path integral approach~\cite{Stephanov:2012ki}, the on-shell effective field theory(OSEFT)~\cite{Carignano:2018gqt}, the world-line formalism~\cite{Mueller:2017wom, Mueller:2017arw}, etc.


\section{Status of experimental search: Novel analysis techniques}

Given the  significance embodied in CME physics, it is of fundamental importance to search for its signatures in laboratory experiments. While analogous phenomena of anomalous transport were observed in condensed matter systems consisting of chiral quasi-particles~\cite{Li:2014bha,Xiong:2015X,Huang:2015X}, the search for CME and related effects in QCD systems produced by heavy ion collisions has proven to be extremely difficult. The main challenge is due to a very small signal embedded in substantial backgrounds in the relevant observables. It may be worth noting that a situation like this is not unfamiliar in physics, if one thinks of well-known searches for e.g. Higgs particle, gravitational waves, temperature fluctuations of cosmic microwave backgrounds, electric dipole moment pertaining to CP violations, dark matter particles, neutrino-less double-beta decays, etc. Many of these took decades of efforts.     

The initial theoretical idea~\cite{Kharzeev:2004ey} to look for CME in heavy ion collisions and the proposal of suitable experimental observable~\cite{Voloshin:2004vk}  both appeared in 2004. About fifteen years passed since the first CME measurement was published by STAR in 2009~\cite{STAR:2009wot}. While not reaching a final conclusion yet, the community has come a long way in fighting against relevant backgrounds and quantifying potential signal level. The progress came in three major steps: 1) recognizing the dominance of flow-driven backgrounds; 2) developing theoretical tools as well as experimental methods to quantify and separate the flow-driven backgrounds; 3) carefully addressing the residue non-flow backgrounds and getting ready for a final exaction of CME signal with anticipated high-statistics dataset.

\begin{figure}[!hbt]
\begin{center}
\includegraphics[height=0.9\textwidth]{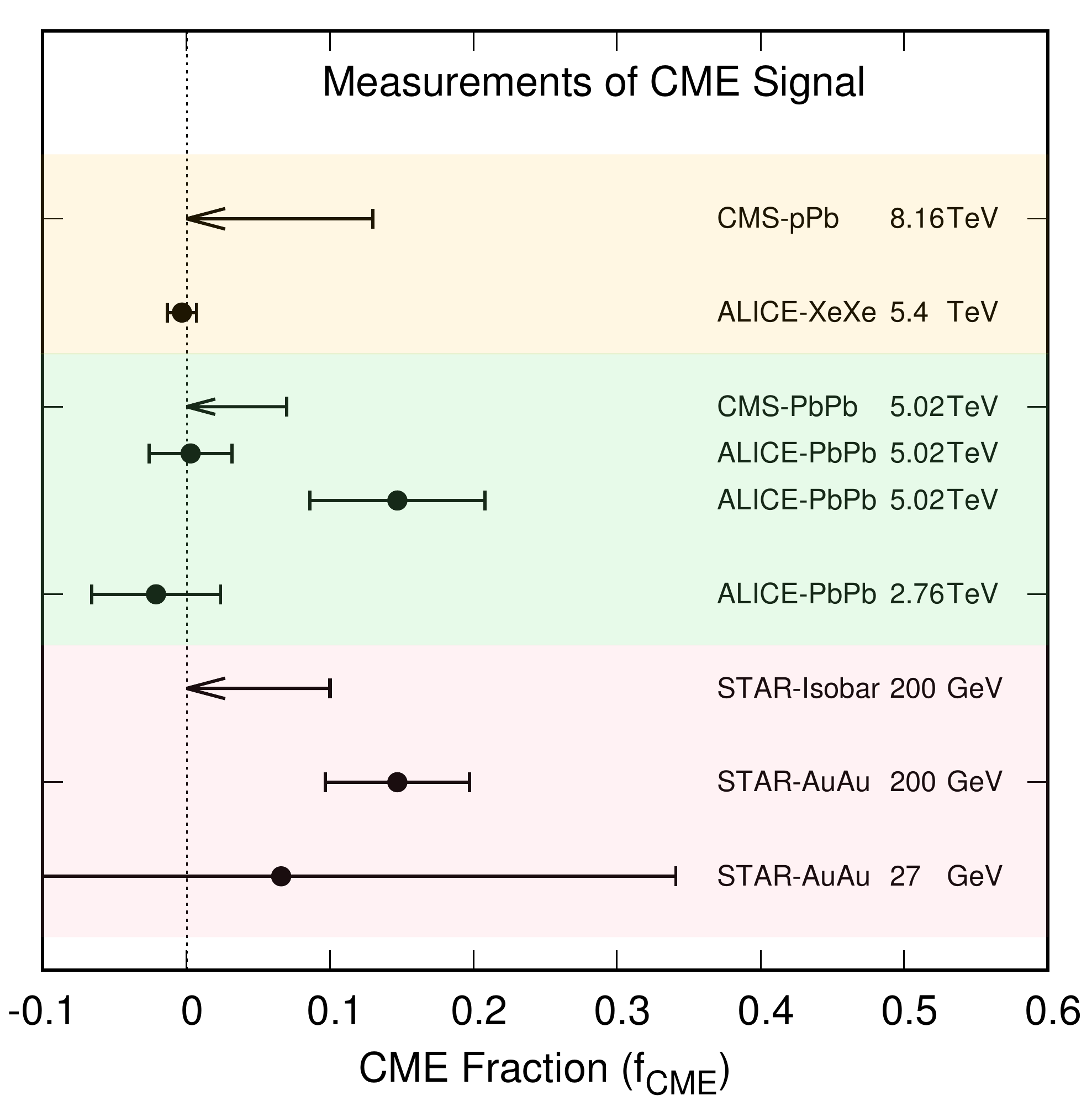} 
\caption{(color online)  
Compilation of measurement results for extracting the possible fraction of CME signal in the overall charge separation observable $\Delta\gamma$ by various experimental collaborations from different colliding systems and collision energies. Some measurements (markers) are presented in terms of absolute value with $1\sigma$ uncertainty while the others are presented (arrows) in terms of upper limits at 95\% confidence level. This compilation presents isobar collisions data from Ref.~\cite{STAR:2021mii,STAR:2023ioo}, 200 GeV Au+Au data from Ref.~\cite{STAR:2021pwb}, 27 GeV Au+Au data using input from Ref.~\cite{STAR:2022ahj}, p+Pb (8.16 TeV) and Pb+Pb (5.02 TeV) data from CMS from Ref.~\cite{Sirunyan:2017quh}, Xe+Xe 5.4 TeV data from ALICE from Ref.~\cite{ALICE:2022ljz}, Pb+Pb 2.76 GeV data from ALICE from Ref.~\cite{ALICE:2020siw}, two different measurements of Pb+Pb data at 5.02 TeV from ALICE from Ref.~\cite{ALICE:2020siw} and Ref.~\cite{ALICE:2022ljz}. \label{fig-exp-cme}}
\end{center}
\end{figure}
%
The rest of this section will present a thorough discussion of such progress, with an emphasis on the novel analysis techniques and the latest experimental results. But before getting down to all those details, it may be beneficial (especially for the impatient readers) to give a quick summary of all published extraction for the CME signal level from various experimental collaborations based on measurements across a variety of colliding systems over a wide span of collision beam energies: see Fig.~\ref{fig-exp-cme}. As one can see, there are fairly strong hints of a nonzero CME signal, especially for Au+Au collisions in the RHIC energy range.


\subsection{Novel analysis techniques}

\subsubsection{Mixed harmonics measurements with second and third order event planes}

In order to proceed in this section it is better to rewrite the 
conventional $\gamma$-correlator by a more general notation as $\gamma_{112}=\langle\cos(\phi_a^{\alpha}+\phi_b^{\beta}-2\Psi_{2})\rangle$. The idea is to measure charge separations across the third harmonic event plane by constructing a new correlator $\Delta\gamma_{123}=\gamma_{123}^{OS}-\gamma_{123}^{SS}$, where $\gamma_{123}=\langle\cos(\phi_a^{\alpha}+2\phi_b^{\beta}-3\Psi_{3})\rangle$ that was introduced by CMS collaboration in Ref~\cite{Sirunyan:2017quh}. Since the $\Psi_3$ plane is random and not correlated to B-field direction, $\gamma_{123}$ is purely driven by non-CME background, the contribution of which should go as $v_3/N$. This is very useful to contrast signal and background scenario by comparing the measurements in two isobaric collision systems. Since Ru+Ru has larger B-field than Zr+Zr but have comparable background, the case for CME would be as follows: $(\Delta\gamma_{112}/v_2)^{\rm Ru+Ru}/(\Delta\gamma_{112}/v_2)^{\rm Zr+Zr}>1$ and $(\Delta\gamma_{112}/v_2)^{\rm Ru+Ru}/(\Delta\gamma_{112}/v_2)^{\rm Zr+Zr}>(\Delta\gamma_{123}/v_3)^{\rm Ru+Ru}/(\Delta\gamma_{123}/v_3)^{\rm Zr+Zr}$. The quantities $\gamma_{112}$ and $\gamma_{123}$ have been measured in U+U, Au+Au collisions at the RHIC and in p+Pb and Pb+Pb collisions at the LHC. Within the uncertainties of the measurements, no significant difference in the trend of $\Delta\gamma_{112}/v_2$ and $\Delta\gamma_{123}/v_3$ is observed for the two collision systems except for the very central events. 

\subsubsection{Charge separation along participant and spectator planes}
\begin{figure}
    \centering
\includegraphics{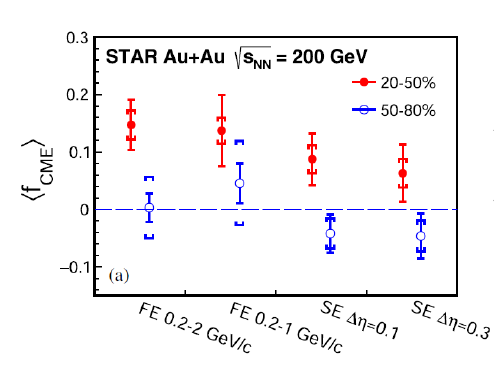}
    \caption{Measurements of the fraction of CME signal using the participant and spectator planes by the STAR collaboration from reference~\cite{STAR:2021pwb}. Various cases of acceptance selection are shown to lead to a dependence on the extracted signal. Overall about 2-3$\sigma$ deviation from zero were observed in the extracted fraction.}
    \label{fig_fcme_star_data}
\end{figure}
\begin{figure}
    \centering
    \includegraphics[width=0.9\textwidth]{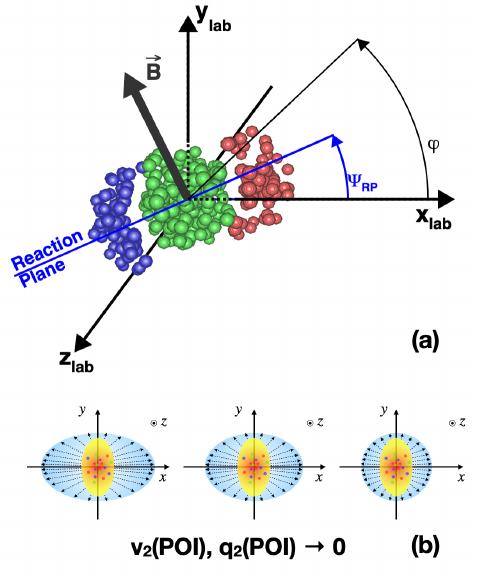}
    \caption{(a) Cartoon adapted from Ref~\cite{Xu:2023elq} showing the geometry of the collision zone with the participants (green) and the spectators (red and blue) nucleons. The reaction plane ($\Psi_{RP}$) is perpendicular to the magnetic field direction and strongly correlated with it. The reaction plane can be approximated by the plane of the spectators, which can be measured by the neutrons from the Zero-Degree-Calorimeters (not shown, but can be assumed to be defined by the lines connecting the centers of gravity of the spectator regions, $\Psi_{SP}$). The plane of the participant nucleons is also shown by $\phi$ (also denoted as $\Psi_{PP}$ here), which can be determined by the produced particles in the experiment. $\phi$ has a weaker correlation with the B-field direction. (b) Cartoon of the event shapes of the particle emission pattern in the transverse plane. The elliptic shape on the left shows the dominant elliptic anisotropy characterized by $v_2$ and the flow vector $q_2$ of particles. Moving from left to right shows the cases of events where the same initial distribution of participant nucleons produces less elliptic shapes due to dynamical fluctuations. In the experiment, events can be selected based on the distribution of $q_2$ and events with nearly spherical emission patterns can be chosen that correspond to $v_2\rightarrow0$. This technique is called Event Shape Engineering (ESE) or Event Shape Selection (ESS).}
    \label{fig_ese_pp_sp_planes}
\end{figure}

This analysis relies on the fact that we can measure the charge separation $\Delta\gamma$ with respect to two different planes: the participant plane ($\Psi_{PP}$) and the spectator plane ($\Psi_{SP}$). We try to depict this in Fig.\ref{fig_ese_pp_sp_planes}, where one can think $\Psi_{SP}$ is a close proxy for $\Psi_{RP}$. The signal driven by the magnetic field (B-field) is more correlated to $\Psi_{SP}$, as spectators mainly determine the B-field. In contrast, the background driven by the flow is dominant along the $\Psi_{PP}$ (depicted by the angle $\phi$ in Fig.\ref{fig_ese_pp_sp_planes}). In the experiment, we can use the elliptic flow anisotropy $\Psi_2$ of the produced particles at the central rapidity region of the collisions as a proxy for $\Psi_{PP}$. This is reasonable because $\Psi_2$ is the plane of elliptic anisotropy and the background to $\Delta\gamma$ is mainly due to elliptic flow. However, getting an experimental handle on $\Psi_{SP}$ is more challenging. We can use the planes of spectator neutrons $\Psi_1$ from a zero-degree-calorimeter at far-forward rapidity, or the forward directed flow plane of fragments $\Psi_1$. It is not easy to prove experimentally that the B-field is more correlated to these planes, but we can use model calculations to demonstrate it. Therefore, we have two measurements, one with respect to the participant plane:
 \begin{equation}
    \gamma^{\alpha\beta}_{PP} = \left<\cos(\phi_\alpha + \phi_\beta - 2 \Psi_{PP})\right>,     
 \end{equation}
 and also with spectator plane
  \begin{equation}
    \gamma^{\alpha\beta}_{SP} = \left<\cos(\phi_\alpha + \phi_\beta - 2 \Psi_{SP})\right>,     
 \end{equation}
 The idea was first introduced in Ref~\cite{Xu:2017qfs} and later on followed up in Ref~\cite{Voloshin:2018qsm}. 
The approach is based on three main assumptions: 1) measured $\Delta\gamma$ has contribution from signal and background that can be expressed as 
\begin{equation}
\Delta\gamma=\Delta\gamma^{\rm bkg}+\Delta\gamma^{\rm sig}, 
\end{equation}
which has been introduced multiple times. Given this equation one now tries to find the equation for the background contribution. 2) The background contribution to $\Delta\gamma$ should follow the scaling 
\begin{equation}
\Delta\gamma^{\rm bkg}_{\textsc{PP}}/\Delta\gamma^{\rm bkg}_{\textsc{SP}}=v_{2}(\textsc{PP})/v_{2}(\textsc{SP})
\end{equation}
 and, 3) the signal contribution to $\Delta\gamma$ should follow the scaling 
 \begin{equation}
\Delta\gamma^{\rm sig}_{\textsc{PP}}/\Delta\gamma^{\rm sig}_{\textsc{SP}}=v_{2}(\textsc{SP})/v_{2}(\textsc{PP}) 
\end{equation}
 The first two  have been known to be working assumptions, widely used for a long time and can be used to test the case of CME~\cite{Voloshin:2018qsm} if $\left(\Delta\gamma/v_2\right)(\textsc{SP})/\left(\Delta\gamma/v_2\right)(\textsc{PP})>1$. The validity of the last one was studied and demonstrated in Ref~\cite{Xu:2017qfs}. Using all three equations one can show that 
\begin{equation}
    \frac{(\Delta\gamma/v_2)_{\rm SP}}{(\Delta\gamma/v_2)_{\rm PP}}=1+\frac{\Delta\gamma^{\rm sig}_{\rm PP}}{\Delta\gamma_{\rm PP}}\left[\frac{(v_2)_{\rm SP}^2}{(v_2)_{\rm PP}^2} -1 \right], 
    \label{eq_double_ratio}
\end{equation}
using which one can extract~\cite{Zhao:2020utk} the fraction of possible CME signal as
\begin{equation}
f_{\textsc{cme}}=\frac{\Delta\gamma^{\rm sig}_{\rm PP}}{\Delta\gamma_{\rm PP}}.   
\label{eq_fcme}
\end{equation}
In this approach the extraction of $f_{\rm CME}$ is done in a fully data-driven as all the terms in Eq.\ref{eq_double_ratio} are measured experimentally. The results for such an approach using the Au+Au collision data at 200 GeV is shown in Fig.\ref{fig_fcme_star_data} by the STAR collaboration. The analysis is performed in various kinematic bins (referred to as full-event or FE and sub-event or SE). A hint of nonzero fraction of CME signal $f_{\rm CME}$ is observed with about $2-3\sigma$ significance~\cite{STAR:2021pwb}. 
%

\subsubsection{Event shape engineering}

As discussed in the previous sections, the main background contribution to the CME-sensitive $\Delta\gamma$ observable comes from the underlying elliptic flow $v_2$. We also discussed that in the scenario of isotropic particle emission (spherical pattern) shown in Fig.\ref{fig_cartoon_lcc_gmc_advanced_observables} (left), the background is absent. The idea of event-shape-engineering (ESE), which goes back to Ref~\cite{Schukraft:2012ah}, is to select events where $v_2\rightarrow0$ and achieve this spherical pattern. However, there is a major problem with this idea. The background to $\Delta\gamma$ comes from the $v_2^{res}$, i.e., the elliptic emission pattern of the mother resonances or neutral clusters that decay into a pair of particles. In other words 
\begin{equation}
    \Delta\gamma^{flow-bkg} = a v_2^{res}.
    \label{eq_ese_v2_res}
\end{equation}
Experimentally, there is no way to access all such mother resonances or track all the neutral clusters that decayed and measure $v_2^{res}$. Experimentally, one can only efficiently measure the $v_2$ of inclusive particles, which includes a fraction of decay daughters from neutral resonances or clusters, as well as some primordial particles that may not be associated with decaying resonances. So the challenge is to somehow use the $v_2$ of inclusive final state particles to make $v_2^{res}\rightarrow0$ -- this is really a challenge. The earliest attempt in this direction by the STAR collaboration was to directly measure $\Delta\gamma$ in various bins of $v_2$ of final state particles and then check if $\Delta\gamma\ne0$ when $v_2\rightarrow0$~\cite{STAR:2013zgu}, which resembles spherical particle emission patterns. However, limitations and complexities of such an approach were highlighted in several successive papers~\cite{Wen:2016zic, Wang:2016iov}. The main challenges are that requiring the $v_2$ of final state measured particles does not necessarily make the emission pattern of mother resonances spherical, as it seems to be effective only at the level of 50\% according to simulations~\cite{Wang:2016iov}. Also, the measured $v_2$ of final state particles is not a simple tool to map the emission pattern. Experimentally, measuring $v_2$ by event-plane method $v_2=\left<\cos(2\phi(A) -2\Psi_2(B))\right>$ or two-particle correlations $v_2^2 =\left<\cos(2\phi_1(A) - 2\phi_2(B))\right>$ involves two regions of acceptance $A$ and $B$ and many physics effects, such as de-correlation of initial state geometry or event-plane angles ($\Psi_2$). Selecting events with $v_2\rightarrow0$ may increase the de-correlation between the geometry in $A$ and $B$~\cite{Wen:2016zic}, which lowers the effectiveness of ESE. $\Psi_2$ planes are important for ESE because we also measure $\Delta\gamma$ using $\Psi_2$ as a proxy for $\Psi_{RP}$ or B-field direction -- one needs to keep track of this angle. Attempts to make $v_2\rightarrow0$ can also randomize $\Psi_2$ in $A$ relative to particles of interests in $B$. When $\Psi_2$ is random, both background and CME signal components in $\Delta\gamma$ disappears for trivial reasons. A challenge of ESE is to make $v_2^{res}\rightarrow0$ without randomizing $\Psi_2$.

How can we overcome this problem? How can we find an experimental quantity that can effectively make $v_2^{res}\rightarrow0$? There is no definitive answer to this question. One can try different strategies, such as choosing a quantity other than $v_2$, changing the acceptance of $\Psi_2$, or using a different harmonic $\Psi_1$ as a proxy for $\Psi_{RP}$. The idea is to start with a physics-motivated idea and then use ``engineering" tools to shape the event through various trials and errors, until we find the best approach. At the end, a closure test is needed to ensure the validity of the approach.

The idea of Ref~\cite{Schukraft:2012ah} is based on the fact that in a given centrality of collisions, which is largely determined by the overlap of two nuclei, there is a large variation of the geometry and the shape of the overlap zone due to fluctuations of nucleon positions. This leads to a large variation of eccentricities and therefore event-by-event fluctuations of elliptic flow coefficients. The main idea of ESE is that for a given centrality, events with different geometries and $v_2$ values can be classified by the variable called the flow vector, which has the components and the length given by
\begin{equation}
Q_{2,x} = \sum \limits_i^M \cos(n\phi_i); Q_{2,y} = \sum \limits_i^M \sin(n\phi_i); q_2 = Q_n/\sqrt{M},
\label{eq_flow_vectors}
\end{equation}
where $M$ is the particle multiplicity and $\phi_i$ are the particle angles. The flow vector is called a vector because it has both a magnitude, which reflects how elliptic the event is, and a direction, which is determined by the event plane $\Psi_2$. The distribution of $q_2$ can be regarded as a measure of the event-by-event magnitude of $v_2$~\cite{ALICE:2015lib, Schukraft:2012ah}. It follows that
\begin{equation}
\left<q_2^2 \right>  \simeq 1 + \left<(M-1)\right> \left<v_2^2 + \delta_2\right>,
\end{equation}
where $\delta_2$ is the non-flow correlation~\cite{Voloshin:2008dg}. In the limit of large $M$ and negligible $\delta_2$:
\begin{equation}
\left< v_2 \right> \approx \sqrt{\frac{ \left<q_2^2\right>}{\left<M\right>}}, 
\label{eq_v2_vs_q2}
\end{equation}
therefore $q_2$ can be used as a direct tool to measure the magnitude of $v_2$ and select event-by-event initial eccentricity ($\varepsilon$) that determines the elliptic geometry of the collisions~\cite{Schukraft:2012ah}. Since  $q_n\rightarrow0$, $v_2\rightarrow0$ and $\varepsilon\rightarrow0$, which means the events become more spherical.  Using $q_2$ instead of directly using $v_2$ avoids the complexities of event-planes that may reduce the effectiveness of ESE. Once events are classified by bins of $q_2$ distributions, one can estimate both $v_2$ of final state particles and CME-sensitive $\Delta\gamma$ and make a correlation plot. Extrapolating the correlation to $v_2\rightarrow0$ will lead to an intercept of $\Delta\gamma(v_2\rightarrow0)$, denoted as $\Delta\gamma^{ESE}$, in which the flow-driven background will be significantly minimized. In other words
\begin{equation}
    \Delta\gamma (q_2^2) = a v_2 (q_2^2) + \Delta \gamma^{ESE}\, , \, v_2 (q_2^2\rightarrow0) = 0.  
\end{equation}
where the quantity $\Delta\gamma^{ESE}$ can be used to extract an upper limit of the CME signal. 

\begin{figure}
    \centering
    \includegraphics[width=0.9\textwidth]{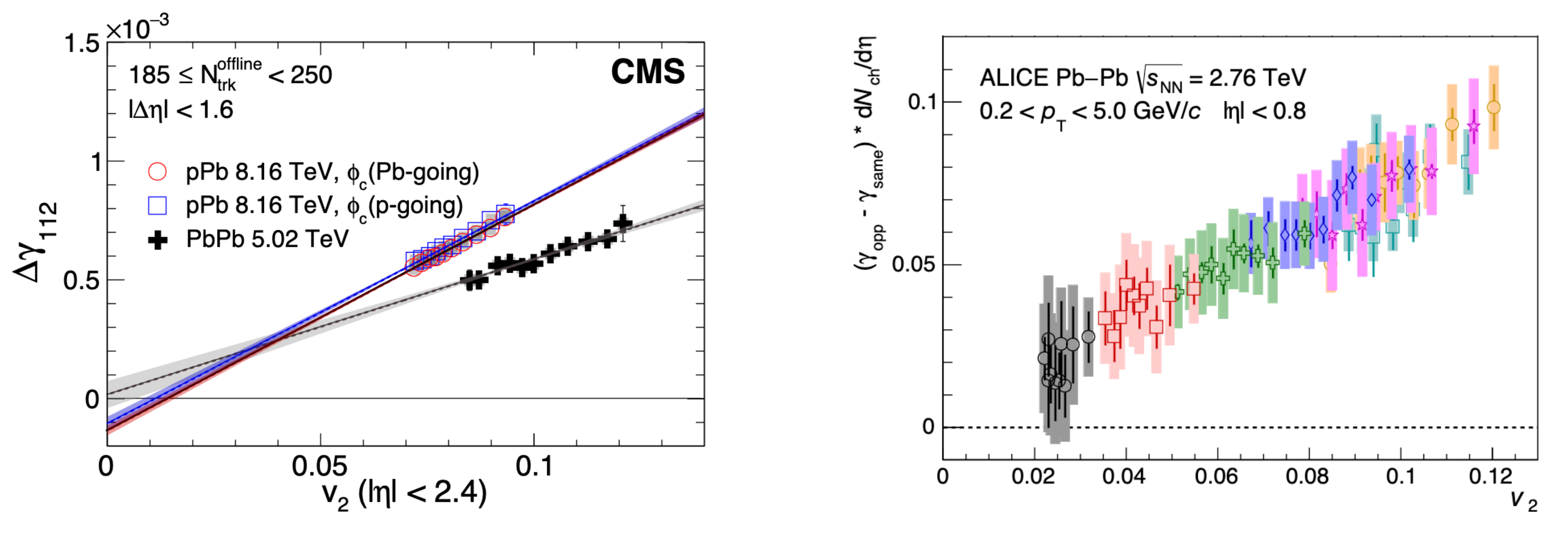}
    \caption{Results from CMS (left) and ALICE (right) Collaborations based on the event-shape-engineering approach. }
    \label{fig_alice_cms_ese}
\end{figure}

A more recent attempt by the STAR collaboration used ESE in Au+Au collisions at 200 GeV~\cite{STAR:2020gky}, addressing many of the difficulties we mentioned before. They used the distribution of $q_2$ instead of $v_2$, and ensured that making the event spherical did not randomize $\Psi_2$. They also analyzed pairs of pions and tracked the invariant mass spectrum of the possible mother resonances from the previous subsection. This approach gave an upper limit of 15\% for the CME signal in the inclusive $\Delta\gamma$ -- we discuss this approach in the next subsection. The CMS collaboration applied a similar ESE approach in Pb+Pb collisions at 5.02 TeV and p+Pb collisions at 8.16 TeV~\cite{CMS:2017lrw}. They fitted the correlation of $\Delta\gamma$ and $v_2$ with a linear function
\begin{equation}
\Delta\gamma = a v_2 + b,
\label{eq_ese_cms}
\end{equation}
and estimated the intercept $b$, assuming it was independent of $v_2$. They used forward rapidity acceptance to estimate $q_2$ for ESE and event-planes (equivalent in scalar-product method), while measuring $v_2$ and $\Delta\gamma$ at midrapidity. They obtained the CME fraction as $f_{\rm CME}=b/\Delta\gamma$, leading to an upper limit of 7\% and 13\% for the CME signal at 95\% confidence level. However, the assumption of $b$ being independent of $v_2$ is strong, as the CME signal strength $\Delta\gamma^{CME}$ may depend on the ellipticity. As we said before, $\Delta\gamma^{CME}$ depends on the correlation between $\Psi_2$ and $\vec{B}$, which can vary with $v_2$, as shown by the ALICE simulation~\cite{ALICE:2017sss}. Therefore, the ALICE search for CME signal used a slightly different approach, focusing on the slope parameter $a$ in the fit function of $\Delta\gamma = a v_2 + \Delta\gamma^{ESE}$ instead of the intercept $\Delta\gamma^{ESE}$. They used ESE to classify the events into bins of $q_2^2$ and estimate the correlation between $\Delta\gamma$ and $v_2$. They extracted the parameter $a$ from data and compared it to model calculations of the correlation between $v_2$ and projected magnetic field $\left<|B^2|\cos(\Phi_B-\Psi_2)\right>$ and the case for 100\% background. In this way, they estimated the upper limit of CME to be $f_{\rm CME} = 0.10\pm0.13$, $f_{\rm CME} = 0.08\pm 0.10$, and $f_{\rm CME} = 0.08\pm 0.11$ for the MC-Glauber, MC-KLN CGC and EKRT models respectively, in the centrality range 10-50\% in Pb+Pb  2.76 TeV. These results are consistent with zero CME fraction and correspond to upper limits on $f_{\rm CME}$ of 33\%, 26\%, and 29\%, respectively, at 95\% confidence level~\cite{ALICE:2017sss}.

Despite many attempts ESE approaches face some challenges, such as: 1) in Eq.\ref{eq_v2_vs_q2} if $v_2$ and $q_2$ are from different acceptances it becomes difficult for ESE to achieve $v_2\rightarrow0$, 2) if $v_2$ and $q_2$ are from the same acceptance, making $v_2\rightarrow0$ may randomize $\Psi_2$ and eliminate the signal component, and 3) there is no guarantee that making $v_2\rightarrow0$ also makes $v_2^{res}\rightarrow0$ and there are no quantitative estimates of how much background ($\propto v_2^{res}$) remains in the intercept $\Delta\gamma^{ESE}=\Delta\gamma(v_2\rightarrow0)$.

Recent approaches~\cite{Xu:2023elq,Milton:2021wku} try to address some of these issues. They propose a better tool for Event-Shape-Selection (ESS) by constructing flow-vectors using a pair of particles $q_{2, pair}^2$, which is defined similarly to Eq.\ref{eq_flow_vectors} except the angles correspond to a pair of particles $\phi_p =\tan^{-1} ((p_{1,y}+p_{2,y})/(p_{1,x}+p_{2,x}))$, where $p_1$ and $p_2$ are the momenta of two particles. This idea is based on the analogy that a resonance particle decays into a pair of daughters and a flow-vector from a pair might be more suitable to make $v_{2}^{res}\rightarrow0$. However, such pairs may not have real parents. The rest of the approach is similar to the previous ones, using the single particle $q_2$ vectors for measuring $v_2$ and $\Delta\gamma$. AVFD model calculations show that the ESS approach using the combination of single-particle $v_2$ and $q_{2, pair}^2\rightarrow0$ works best to recover the input signal used in AVFD simulation. An implementation of this ESS approach using the STAR data from the RHIC Beam Energy Scan program is ongoing and will be discussed in the following section~\cite{}. The STAR analysis also addresses the second issue. To avoid affecting the resolution of $\Psi_2$ when making $v_2\rightarrow0$, they use $\Psi_1$ plane as a proxy for $\Psi_{RP}$ in the ESS analysis from forward rapidity.

Similar to ESE, the ESS method also needs to address a number of challenges. Firstly, the two particles that make up $q_{2, pair}^2$ also appear in the $\gamma$-correlator. Whether this leads to any auto-correlation and consequently to $q_{2, pair}^2\rightarrow0$ when selecting statistical fluctuations requires further investigation. Secondly, the issue of quantifying the effectiveness of the ESS approach remains. To address this, an observable that is largely insensitive to CME, such as $\gamma^{132}=\cos(\phi^\alpha - 3\phi^\beta + 2\Psi_2)$, which is primarily determined by flow-driven background can be used as baseline. However, one needs to demonstrate that $\Delta\gamma^{132}\rightarrow0$ when $v_2\rightarrow0$ as a closure test. The analysis of STAR data based on ESS appaoch is ongoing and shows promise~\cite{Xu:2023wcy}. 



\subsubsection{Invariant mass dependence of charge separation}
\begin{figure}
    \centering
    \includegraphics[width=0.9\textwidth]{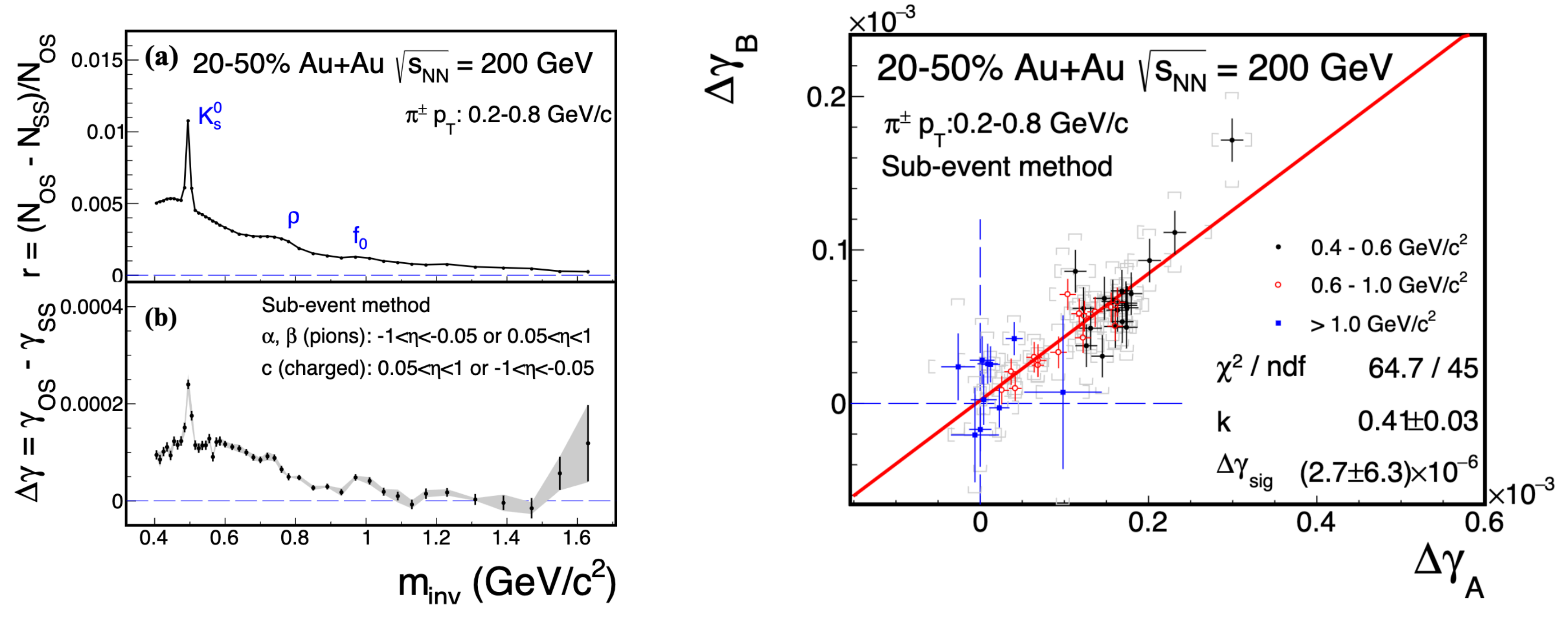}
    \caption{(Left) The top panel shows the invariant mass distribution ($m_{inv}$) of the excess yield of opposite-sign over same-sign pion pairs, which reveals peaks from various neutral resonances. The bottom panel shows the same for the CME-sensitive observable $\Delta\gamma$. The peaks appear at similar $m_{inv}$ values, suggesting that neutral resonance decay may be the main source of the measured $\Delta\gamma$. Even though the $m_{inv}$ distribution of the CME signal is unknown, the $m_{inv}$-integrated component can be obtained by using the top panel distribution and a two-component model. (Right) The correlation of $\Delta\gamma_A$ with $\Delta\gamma_B$ using events sorted by two different types of flow-vectors ($q_2$) that create different event-shapes (categorized in terms of labels $A$ and $B$). Various points are obtained from bins in invariant mass. This correlation plot is fitted with a function $\Delta\gamma_A = k \Delta\gamma_B + (1-k) \Delta\gamma_{sig}$. The combined approach of invariant mass and Event-Shape-Engineering led to an upper limit of CME component in $\Delta\gamma$ to be 15\% in Au+Au collisions at 200 GeV~\cite{STAR:2020gky}.}
\label{fig_inv_mass_cme_fraction}
\end{figure}

Differential measurements of $\Delta\gamma$ with invariant mass combined with ESE approach described above can be used to estimate the fraction of CME in the experiments~\cite{STAR:2020gky}. The idea to use invariant mass is simple and was first introduced in Ref~\cite{Zhao:2017nfq}. Resonances are widely identified by observing structures in the invariant mass spectra of their decay products. For example, a pair of opposite sign pions can come from the neutral resonances that show up in the invariant mass spectrum of $m_{inv}(\pi^+ + \pi^-)$. If we restrict the analysis to pairs of pions, differential measurements of $\Delta\gamma$ with $m_{inv}(\pi^+ + \pi^-)$ should also show similar peak-like structures if the background from neutral resonances dominates the charge separation. As shown in Fig.\ref{fig_inv_mass_cme_fraction} (left), indeed similar peak structures are observed in the excess yield of opposite-sign over same-sign pion pairs $r=(N_{OS}-N_{SS})/N_{SS}$, which reveals peaks from various neutral resonances. Assuming that the $\Delta\gamma$ data contain the flow-induced background and a possible CME signal, the inclusive $\Delta\gamma$ can be expressed as
\begin{equation}
\Delta\gamma(m_{inv}) = \Delta\gamma_{bkg} (m_{inv}) + \Delta\gamma_{sig}
\label{eq_delta_gamma_inv_mass}
\end{equation}
The first term is the background dependent on $m_{inv}$ and $v_2$ expressed as
\begin{equation}
\Delta\gamma_{bkg} (m_{inv}) = r (m_{inv}) \left< \cos(\phi_\alpha + \phi_\beta - 2 \phi_{res}) \right> v_{2,res}
\end{equation}
and the second term in Eq.\ref{eq_delta_gamma_inv_mass} is the possible CME signal that is assumed to be independent of $v_2$ and $m_{inv}$ (which may not be true~\cite{Muller:2016jod}). Clearly, the above equations have multiple unknowns to extract the term $\Delta\gamma_{sig}$. Therefore, the idea is to perform measurements in two different bins ($A$ and $B$) of flow-vectors $q_2$ that basically control the shape of particle emissions. This approach is same as the ESE that we discussed in the previous subsection. For now, the important thing to note is that we will have two such equations
\begin{eqnarray}
\Delta\gamma_A (m_{inv}, q_{2,A}) = \Delta\gamma_{bkg}(m_{inv}, q_{2,A}) + \Delta\gamma_{sig} \nonumber \\
\Delta\gamma_B (m_{inv}, q_{2,B}) = \Delta\gamma_{bkg}(m_{inv}, q_{2,B}) + \Delta\gamma_{sig}
\label{eq_inv_mass_ese}
\end{eqnarray}
where the assumption is that the signal component is not dependent on the event shape classifier $q_2$ or $m_{inv}$. Note here that we still have three unknowns in the right hand sides of Eq.\ref{eq_inv_mass_ese}. So more assumption will be needed to find $\Delta\gamma_{sig}$. Assuming that the invariant mass distribution does not change with $q_2$ selection
\begin{equation}
\Delta\gamma_{bkg}(m_{inv}, q_{2}) = F(m_{inv}) \times G(q_2).
\end{equation}
Where $F(m_{inv})$ is the universal function that includes the invariant mass dependence while $G$ is a function of $q_2$. How good this assumption is can be verified in a data-driven way by overlaying the measurements of $\Delta\gamma(m_{inv})$ in two regions of $q_2$ which is what was done by the STAR experiment and we refer the readers to Ref~\cite{STAR:2020gky}. For now, its straightforward to see that we can use this to express previous Eq.\ref{eq_inv_mass_ese} as
\begin{eqnarray}
\Delta\gamma_A(m_{inv}, q_{2,A}) = F(m_{inv}) \times G(q_{2,A}) + \Delta\gamma_{sig} \nonumber \\
\Delta\gamma_B(m_{inv}, q_{2,B}) = F(m_{inv}) \times G(q_{2,B}) + \Delta\gamma_{sig}. 
\label{eq_inv_mass_ese_cme_modified}
\end{eqnarray}
From this equations it is easy to eliminate the first terms involving background and rewrite them as
\begin{equation}
    \Delta\gamma_{A} (m_{inv}, q_{2,A}) = \frac{G(q_{2,A}) }{G(q_{2,B})} \Delta\gamma_{B} (m_{inv}, q_{2,B}) + \left(1-\frac{G(q_{2,A})}{G(q_{2,B})}\right) \Delta\gamma_{sig}
    \label{eq_inv_mass_subevent}
\end{equation}
 With this ansatz, a careful analysis is performed by the STAR collaboration to measure $\Delta\gamma_A$ and $\Delta\gamma_B$ in two class of events with different flow-vectors as a function of invariant mass and then make a correlation plot as shown in Fig.\ref{fig_inv_mass_cme_fraction} (right). Each point on this plot is obtained from a measurement at a small window of $m_{inv}$. One can now extract $\Delta\gamma_{sig}$ component by performig a fit to this correlation that is same as  Eq.\ref{eq_inv_mass_subevent} written as:
 \begin{equation}
     \Delta\gamma_A = k \Delta\gamma_B + (1-k) \Delta\gamma_{sig}
\label{eq_ese_inv_mass_fit_func}
 \end{equation}
where $k$ and $\Delta\gamma_{sig}$ are fit parameters. A couple of more steps remains to extract what is called the CME fraction defined as $f_{\rm CME}=\Delta\gamma_{sig}/\Delta\gamma$, where the term in the denominator is inclusive value of $\Delta\gamma$. In this above case the total class of event is divided into two categories of $q_2$ vectors labeled as $A$ and $B$. Therefore the inclusive measurement of $\Delta\gamma$ is the average of the same in this two event classes
\begin{equation}
    \Delta\gamma = \frac{1}{2} (\Delta\gamma_A + \Delta\gamma_B). 
    \label{eq_inclusive_gamma_ese_inv_mass}
\end{equation}
Combining Eq.\ref{eq_ese_inv_mass_fit_func} and Eq.\ref{eq_inclusive_gamma_ese_inv_mass} one can also show that
\begin{equation}
    \Delta\gamma = -(\Delta\gamma_A - \Delta\gamma_B) \left(\frac{1+k}{2(1-k)}\right) + \Delta\gamma_{sig}. 
\end{equation}
One can therefore express CME fraction as
\begin{equation}
    f_{\rm CME} = \frac{\Delta\gamma_{sig}}{\Delta\gamma} = 1 + 
    \frac{(\Delta\gamma_A - \Delta\gamma_B)}{\Delta\gamma}\left(\frac{1+k}{2(1-k)}\right). 
\end{equation}
In the above equation, one can measure $\Delta\gamma_A$, $\Delta\gamma_B$ and $\Delta\gamma$ over the inclusive range of invariant mass and the parameter $k$ from the fit in Fig.\ref{fig_inv_mass_cme_fraction} and extract $f_{\rm CME}$. The measurements from the STAR collaboration performed this analysis and found the upper limit of $f_{\rm CME}$ to be 15\% at 95\% confidence limit in Au+Au collisions at 200 GeV~\cite{STAR:2020gky}. 


%

The STAR collaboration recently performed a new ESE analysis on Au+Au data at 200 GeV~\cite{Xu:2023wcy}. The novelty of this method is to use different particle sets for ESE q-vector construction (e.g. in $|\eta|<0.3$) and for $\Delta\gamma$ and $v_2$ estimation ($0.3<|\eta|<1$). This aims to eliminate any potential auto-correlation in the ESE method, but it also requires a large extrapolation to obtain the ESE intercept of $\Delta\gamma$ at $v_2\rightarrow0$. Moreover, this method introduces a new feature by splitting the analysis into different invariant mass ($m_{inv}$) bins, which can reveal more information about the softness of the CME and its dependence on $m_{inv}$. A non-zero $\Delta\gamma_{ESE}$ (using intercept at $v_2\rightarrow0$) is indeed observed. Furthermore, the analysis employs spectator neutrons to enhance the correlation of magnetic field with spectator planes. The nonflow is naturally reduced as there is a large gap between the particles used for $\Delta\gamma$ and $v_2$. The first preliminary result from the STAR collaboration shows that the intercept is compatible with zero with a large uncertainty using Au+Au 200 GeV data. However, this analysis has the potential to achieve more precise measurements with future high statistics runs from RHIC in 2023 and 2025.

\subsubsection{Alternate measure: The novel R-observable}

The $R$-observable, introduced in Ref~\cite{Magdy:2017yje}, is a distribution that is defined as the ratio of two distribution functions of $\Delta S$, which is the quantity that measures the difference in the dipole moment of the positive and negative charge in an event (see Ref~\cite{Magdy:2017yje} for details). The ratio is taken for $\Delta S$ parallel and perpendicular to B-field direction, i.e. $R_{\Psi_m} (\Delta S)=C_{\Psi_m}(\Delta S)/C_{\perp}{\Psi_m}(\Delta S)$. The $R_{\Psi_2}(\Delta S)$ distribution is affected by both CME and non-CME background, while $R_{\Psi_3}(\Delta S)$ that is a variation of the observable where the third order harmonic is used, is only influenced by non-CME background and acts as a reference. This observable has several distinctive features according to model calculations: 1) CME signal will cause a concave shape of the $R_{\Psi_2}(\Delta S)$, 2) stronger CME signal will enhance the concavity of $R_{\Psi_2}(\Delta S)$, 3) with CME, the concavity of $R_{\Psi_2}(\Delta S)$ will be greater than that of $R_{\Psi_3}(\Delta S)$. In experimental analysis, a slightly modified quantity $\Delta S^{\prime\prime}$ is used instead of $(\Delta S)$, which includes correction for particle number fluctuations and event plane resolution. This observable has been mainly used in the analysis of the isobar data~\cite{STAR:2021mii}, which we will discuss later. For isobar collisions one expects stronger magnetic field in Ru+Ru than Zr+Zr, the expected features of R-observable for CME-like scenario are: 1) a concave shape is seen for the ratio of the observables  $R_{\Psi_2}(\Delta S)^{\rm Ru+Ru}/R_{\Psi_2}(\Delta S)^{\rm Zr+Zr}$ and 2) the concavity is weaker for $R_{\Psi_3}(\Delta S)^{\rm Ru+Ru}/R_{\Psi_3}(\Delta S)^{\rm Zr+Zr}$. The results agree with unity. The relation between the $\Delta\gamma$ and the $R$ observable was investigated in Ref~\cite{Choudhury:2021jwd}.


\subsubsection{Alternate measure: The signed Balance function}
An alternative method for detecting the Chiral Magnetic Effect (CME) is the signed balance function (SBF), as introduced by Tang et al. The concept involves considering the arrangement of momentum of charged particle pairs, as indicated by the width of the SBF. This width is expected to differ between measurements taken out-of-plane versus in-plane, as quantified by the ratio $r_{\rm lab}$. Additionally, the collective expansion of the system induces a boost, causing all particle pairs to move in the same direction, which can be accounted for by measuring the ratio in the pair rest frame, denoted as $r_{\rm rest}$. In the presence of CME, both individual ratios and the double ratio $R_B = r_{\rm rest}/r_{\rm lab}$ are anticipated to exceed unity. Currently, this observable is under investigation using isobar data in the STAR experiment, albeit not as part of a blind analysis. The expectation for CME in this context is twofold: 1) a higher value of $r$ for Ru+Ru collisions compared to Zr+Zr collisions, and 2) a greater $R_B$ for Ru+Ru collisions compared to Zr+Zr collisions. While this observable wasn't initially included in the isobar blind analysis, results from Au+Au collisions at 200 GeV were presented in a study from the STAR collaboration presented in Ref.~\cite{Lin:2020jcp}.

\subsection{Small colliding systems }
\begin{figure}
    \centering
\includegraphics[width=0.8\textwidth]{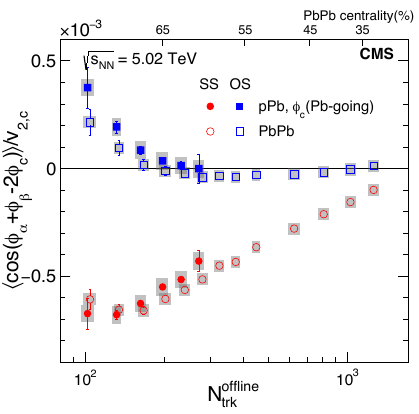}
    \caption{The results of the CME sensitive $\gamma$-correlator measuring charge separation for the same sign (SS) and opposite sign (OS) combinations as a function of the number of produced particles $N_{trk}^{offline}$ in p+Pb and Pb+Pb collisions at $\sqrt{s_{NN}} = 5.02$ TeV at the LHC (adopted from ~\cite{CMS:2016wfo}). The CME signal is expected to be negligible in the p+Pb system due to the lack of correlation between the direction of the magnetic field and the direction of the particle anisotropy, while in the Pb+Pb system it is expected to appear as a separation between $\gamma^{OS}$ and $\gamma^{SS}$. The results show that the measurements are consistent between the two systems within uncertainties in the overlapping range of $N_{trk}^{offline}$. This poses an apparent challenge to the CME interpretation in Pb+Pb collisions. However, upon closer inspection, we find that using p+Pb collisions as a baseline for Pb+Pb is not straightforward, as the relative contribution of signal and background are different in the two systems.}
    \label{fig_cme_collaboration_cme_search_experiment}
\end{figure}

Small collision systems such as p+Pb, p+Au, d+Au offer unique opportunities to study the background for charge separation measurements due to the weak correlation between the B-field direction and the elliptic anisotropy plane of produced particles with respect to which $\Delta\gamma$ is measured~\cite{Khachatryan:2016got,Belmont:2016oqp}. In low-multiplicity or minimum-bias collisions of small systems, these planes are primarily dominated by non-flow correlations arising from di-jets or momentum conservation. However, high-multiplicity events in small systems exhibit intriguing hints of collectivity, sparking significant discussion within the community. Several scenarios influence the elliptic anisotropy plane measured in experiments:

\begin{enumerate}
    \item Collectivity due to hydrodynamic flow: The plane is correlated with the geometric plane of participants.
    \item Collectivity due to non-hydrodynamic effects: The plane exhibits weak or no correlation with the geometric plane, arising from initial state momentum space correlations (e.g., CGC or escape mechanism).
    \item No collectivity: The plane is dominated by non-flow from di-jets and momentum conservation.
\end{enumerate}

These scenarios are crucial for CME searches as they determine the nature of the non-CME background dominating $\Delta\gamma$ measurements in small systems. Understanding the baseline behavior provided by small systems becomes vital for interpreting heavy-ion collision measurements. For instance, in scenario 1 (hydrodynamic flow), the dominant background in heavy-ion collisions involves hydrodynamic flow-driven background combined with local charge conservation. Conversely, scenarios 2 and 3 would result in reaction plane-independent backgrounds, relevant for peripheral and smaller heavy-ion collisions. Despite these considerations, the CME signal is expected to be weak in all scenarios due to the weak correlation between the B-field and the elliptic anisotropy plane in small systems, except for specific scenarios as discussed in Ref.~\cite{Kharzeev:2017uym}.

Experimental observations further emphasize the importance of studying the background in small systems. As shown on Fig.\ref{fig_cme_collaboration_cme_search_experiment}, the CMS measurement was the first to show that at overlapping multiplicities the $\gamma$-correlator measurements are quantitatively similar between p+Pb and Pb+Pb~\cite{Khachatryan:2016got}. STAR measurements performed in p+Au and d+Au systems show similar and in fact larger charge separation measured in terms of the scaled quantity $\Delta\gamma/v_{2} \times N_{\rm ch}$ than those in Au+Au measurements~\cite{STAR:2019xzd}. Such observations are striking as they tell us that a very large value of $\Delta\gamma$ is expected even for a 100\% background scenario. This highlights the need for a comprehensive understanding of the background contributions before drawing conclusions about the presence or absence of the CME in heavy-ion collisions.

However, after the appearance of the results from the CMS collaboration shown in Fig.\ref{fig_cme_collaboration_cme_search_experiment}, a possible notion was spread in the community. This notion states that the observation of similar $\gamma$ or $\Delta\gamma$ challenges the expectations of contributions to $\Delta\gamma$ from the CME. The argument can be presented as follows
\begin{equation}
\Delta\gamma_{\rm Pb+Pb}(N_{trk}^{offline})  = \Delta\gamma_{\rm p+Pb}(N_{trk}^{offline}) \Rightarrow \Delta\gamma_{\rm Pb+Pb}^{CME} = 0 (?).
\end{equation}

But is this really true? Does the observation of the same charge separation at overlapping multiplicities in large and a background dominated small collision systems challenge the CME in the large system?

The following describes why the overlapping values of $\Delta\gamma$ in p+Pb and Pb+Pb from LHC does not rule out the CME in Pb+Pb. The same is also true for the RHIC measurements. First, let us consider the LHC measurements. As shown in Fig. \ref{fig_cme_collaboration_cme_search_experiment}, the results of the CME sensitive $\gamma$-correlator measuring charge separation for the same sign (SS)
and opposite sign (OS) combinations as a function of the number of produced particles $N_{trk}^{offline}$ in p+Pb and Pb+Pb collisions at $\sqrt{s_{NN}} = 5.02$ TeV at the LHC~\cite{CMS:2016wfo} show that the measurements are consistent between the two systems within uncertainties in the overlapping range of $N_{trk}^{offline}$. From the previous section, we know that

\begin{equation}
\Delta\gamma=\Delta\gamma^{\text{CME}}+ \Delta\gamma^{flow-bkg} + \Delta\gamma^{nonflow-bkg}
\end{equation}

This means for Pb+Pb and p+Pb we have at the same value of multiplicity $N_{trk}^{offline}$:

\begin{equation}
\Delta\gamma_{\rm Pb+Pb}=\Delta\gamma^{CME}_{\rm Pb+Pb}+ \Delta\gamma^{flow-bkg}_{\rm Pb+Pb} + \Delta\gamma^{nonflow-bkg}_{\rm Pb+Pb}
\end{equation}

\begin{equation}
\Delta\gamma_{\rm p+Pb}=\Delta\gamma^{CME}_{\rm p+Pb}+ \Delta\gamma^{flow-bkg}_{\rm p+Pb} + \Delta\gamma^{nonflow-bkg}_{\rm p+Pb}
\end{equation}

The goal of the entire effort is to find $\Delta\gamma^{CME}_{\rm Pb+Pb}$. For example, if we can show that $\Delta\gamma^{CME}_{\rm Pb+Pb}=0$ then one can rule out CME in Pb+Pb collisions. First of all, it is easy to see that in the above two equations we have a total of six unknowns. However we have a few more equations to consider. From theoretical expectations, we have:
\begin{equation}
\Delta\gamma^{CME}_{\rm p+Pb} \approx 0,
\end{equation}

The CMS measurements also indicate:

\begin{equation}
\Delta\gamma_{\rm Pb+Pb} = \Delta\gamma_{\rm p+Pb}
\end{equation}

From this, we can conclude:
\begin{equation}
\Delta\gamma^{CME}_{\rm Pb+Pb} =  \left(\Delta\gamma^{flow-bkg}_{\rm p+Pb}- \Delta\gamma^{flow-bkg}_{\rm Pb+Pb}\right) + \left(\Delta\gamma^{nonflow-bkg}_{\rm p+Pb}- \Delta\gamma^{nonflow-bkg}_{\rm Pb+Pb}\right).
\end{equation}

Therefore, from the measurement one can not really conclude $\Delta\gamma^{CME}_{\rm Pb+Pb}=0$. Unless we can prove the right hand side of the above equation is zero. This would mean:

\begin{equation}
\left(\Delta\gamma^{flow-bkg}_{\rm p+Pb}- \Delta\gamma^{flow-bkg}_{\rm Pb+Pb}\right) + \left(\Delta\gamma^{nonflow-bkg}_{\rm p+Pb}- \Delta\gamma^{nonflow-bkg}_{\rm Pb+Pb}\right) = 0.
\end{equation}

So far, no one has established that this is true. For example, if we focus on the last two terms, it is very unlikely to have the nonflow correlations to be same in the two systems therefore:
\begin{equation}
\Delta\gamma^{nonflow-bkg}_{\rm p+Pb}- \Delta\gamma^{nonflow-bkg}_{\rm Pb+Pb} \ne 0.
\end{equation}
We know that nonflow correlations are generated from processes such as minijets production and how much they are quenched. Even if we consider the same value of multiplicity, due to the size differences between p+Pb and Pb+Pb, there is no reason that nonflow correlations will be the same. How about the flow-like background? Similar argument follows again, the origin of flow-like correlations in Pb+Pb and p+Pb may be different leading to differences in the magnitude of the flow-like background. Overall conclusion is that it is inconclusive to make either conclusion: 
\begin{equation}
     \Delta\gamma^{CME}_{\rm Pb+Pb} = 0,\,\,  \Delta\gamma^{CME}_{\rm Pb+Pb} \ne 0. 
\end{equation}

There is another complication. Heavy-ion measurements for CME search are performed where the system size, multiplicity do not necessarily overlap with that of small systems. For instance, one usually measures
\begin{equation}
\Delta\gamma_{\rm Pb+Pb} (N_{\rm trk}^{Pb+Pb}) \ne \Delta\gamma_{\rm p+Pb} (N_{\rm trk}^{p+Pb})\, , \, N_{\rm trk}^{Pb+Pb} \gg N_{\rm trk}^{p+Pb}.
\end{equation}

This introduces more uncertainties to constrain $\Delta\gamma_{\rm Pb+Pb}^{CME}$. Therefore, it is not straightforward to extrapolate the quantitative background baselines for $\Delta\gamma$ from small systems, such as p+Pb, to large systems, such as Pb+Pb, where the physics changes significantly. Namely, $\Delta\gamma$ measured for $N_{\rm trk}=10$ in p+Pb might not serve as a quantitative baseline for $\Delta\gamma$ in Pb+Pb at $N_{\rm trk}=100$. One could try to make a projection based on some assumptions, but that would result in a qualitative baseline and defeat the main purpose of using small systems as direct quantitative baselines. This is why isobar collisions are useful -- they ensure that measurements are compared in two systems with very similar sizes and shapes. The arguments used above for p+Pb and Pb+Pb at the LHC also apply to the similar measurements at RHIC for p/d+Au and Au+Au.

The overall conclusion is as follows. Observing similar $\Delta\gamma$ in p+A and A+A might seem to challenge the interpretation of CME in A+A. However, upon closer inspection, it becomes clear that it is difficult to rule out the presence of CME entirely based on $\Delta\gamma$(p+A)$=\Delta\gamma$(A+A), mainly because the flow and nonflow backgrounds are not the same in the two systems.



\subsection{Isobar collisions at RHIC}

In Ref.\cite{Voloshin:2010ut}, Voloshin proposed employing colliding isobars, such as Ru+Ru and Zr+Zr to discern Chiral Magnetic Effect (CME) signals from background noise, leveraging variations in magnetic field strength while the variations in flow-driven background, modulated by nuclear properties (Refs.\cite{Deng:2018dut,Xu:2017zcn,Hammelmann:2019vwd}) will be under control.

In the 2017-18 RHIC Beam-User-Request~\cite{starbur17}, the STAR collaboration proposed data collection for about three-and half weeks aiming for a five-sigma significance in detecting $\Delta\gamma$, assuming an 80\% non-CME background and minimal systematic uncertainties relative to statistical uncertainties. This initiative spurred a collaborative effort with the RHIC collider accelerator department to plan isobar running in the year 2018, identifying key sources of systematic uncertainties such as detector response variations, efficiency changes, and luminosity fluctuations affecting track reconstruction in the detectors. To mitigate these effects, a running proposal was devised, involving species switching between each store and maintaining consistent luminosity levels, aiming to minimize systematic uncertainties in observable rations and ensure balanced observations across different collision systems. 

Ahead of the 2018 isobar run, a blinding procedure was implemented to limit analyst data access, aiming to eliminate unconscious biases~\cite{Tribedy:2020npn}. STAR collected over 3 billion minimum-bias events per isobar species, with five institutional groups conducting blind analyses, focusing on specific tasks for result cross-validation. An Analysis Blinding Committee, in collaboration with STAR experts, ensured analysts had access to data with concealed species-specific information, maintaining integrity throughout the CME Isobar analyses (Ref.~\cite{VanBuren:2023mpn}).

The blind analyses of isobar data comprised four steps: a mock-data challenge to ensure blinding efficacy and analyst comprehension of data structures, followed by the isobar-mixed analysis, the most challenging step involving full QA and physics analysis, with documented and frozen procedures. Subsequently, in the isobar-blind analysis, analysts conducted run-by-run QA using blinded species data, leading to the final isobar-unblind analysis where species information was revealed for publication, though codes were run independently to ensure integrity.

\begin{figure}
    \centering
    \includegraphics[width=0.9\textwidth]{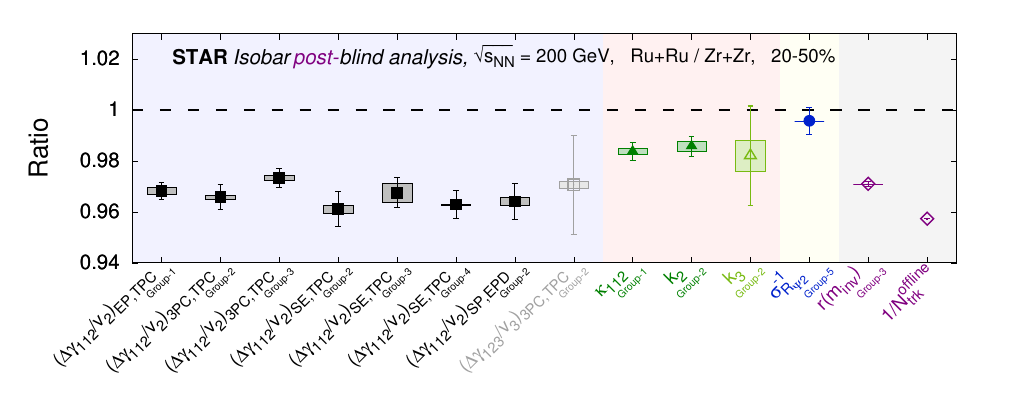}
    \caption{compilation of results obtained from the blind analysis referenced in ~\cite{STAR:2021mii}. The results depict the ratio of measurements in Ru+Ru collisions compared to those in Zr+Zr collisions. Solid dark symbols represent measures sensitive to CME, while open light symbols depict measures designed to be insensitive to CME. Statistical uncertainties are denoted by vertical lines, while systematic uncertainties are indicated by boxes. Background colors are used to differentiate between different types of measures. Additionally, two data points (open markers) on the right indicate the ratio of inverse multiplicities ($N_{\rm trk}^{\rm offline}$) and the ratio of relative pair multiplicity difference ($r$).}
    \label{fig_isobar_blind_analysis}
\end{figure}

As mentioned before, the isobars were collided to exploit the fact that the ruthenium collisions generate larger magnetic fields than the zirconium collisions by 5--9\%, leading to a 10--18\% larger CME correlation signal due to its $B^2$ dependence. Thus, the CME would make the ratio of CME-sensitive observables in Ru+Ru over Zr+Zr higher than one, assuming that the backgrounds are identical in the two systems. For instance, the double ratio of the main CME-sensitive correlator $\Delta \gamma$ scaled by ellipticity $v_2$ in ruthenium over zirconium should be higher than one if there is a non-zero CME fraction. In other words, one can write
\begin{equation}
R=\frac{(\Delta\gamma/v_2)_{\rm Ru+Ru}}{(\Delta\gamma/v_2)_{\rm Zr+Zr}} = 1 + f_{\rm CME}^{\rm Zr+Zr} \left[ \left(B_{\rm Ru+Ru}/B_{\rm Zr+Zr}\right)^2 -1 \right] > 1,
\label{eq_isobar_predefinition}
\end{equation}
where $f_{\rm CME}^{\rm Zr+Zr}$ is the fraction of CME in Zr+Zr collisions (one can rewrite Eq.\ref{eq_isobar_predefinition} using the same for Ru+Ru) and $B_{Ru+Ru,Zr+Zr}$ are the strength of the magnetic field in the isobar systems. It is interesting to note that the double ratio $R$ of the observable measured in the experiment depends on two dials. One dial is the ratio of the magnetic field square, which can be considered as an external dial and controlled by the experimentalists by colliding species like isobars with different atomic number ($Z$). The other dial that is hidden inside $f_{\rm CME}$ is more like an internal dial, which is not controlled by the experimentalists and more related to the physics of QCD vacuum. The combination of the two dials determines the outcome of the experiment, and hence the observability of $R>1$. Eq.\ref{eq_isobar_predefinition} is also called the pre-definition. To set up such a pre-definition, as required by the standard of the blind analysis, the multiplicity in Ru+Ru was assumed to be the same as Zr+Zr. The results from the isobar blind analysis showed that such an assumption was wrong~\cite{STAR:2021mii}. This unexpected result affected the CME baseline and necessitated further work to understand the analysis results, although the validity of the blind analysis was not compromised. The isobar run was specifically designed to minimize the systematics in this ratio $R$. To avoid the unconscious and pre-determined biases, a blind analysis was conducted with pre-defined criteria for the observation of a CME signal.

The measurements of the double ratio of $\Delta\gamma/v_2$ with various kinematic cuts from the isobar blind analysis are shown in Fig.\ref{fig_isobar_blind_analysis}. A precision in our measurement down to 0.4\% was observed in the measurement of the $\Delta\gamma/v_2$ ratio. However, no positive predefined signature of CME was observed.  The observation that the double ratio of $\Delta\gamma/v_2$ is significantly below unity can be attributed to the multiplicity difference between Ru+Ru and Zr+Zr as shown by the ratio of the inverse of uncorrected tracks $1/N_{\rm trk}^{\rm offline}$ measured within the acceptance of $|\eta|=0.5$. This ratio being less than one is explainable not by larger charge separation in Zr+Zr compared to Ru+Ru, rather $\Delta\gamma/v_2<1$ is explained by larger multiplicity dilution ($\propto 1/N_{\rm trk}^{\rm offline}$) in Ru+Ru. This argument is further demonstrated by the ratio of a similar quantity $r(m_{\rm inv})$ which measures the relative pair multiplicity difference between positive and negative pions-in a model in which the background for $\Delta\gamma$ is solely due to flowing clusters whose multiplicity scales with track multiplicity, $\Delta\gamma/v_2$ does scale simply as inverse multiplicity.

In Fig.\ref{fig_isobar_blind_analysis} a  number of other CME sensitive observables were also measured, such as the factorization coefficients $\kappa_{112}, k_2$, the inverse width of the $R$-variable as shown in Fig.\ref{fig_isobar_blind_analysis}. The ratios of these observables in Ru+Ru over Zr+Zr are also found to be less than unity, also not consistent with pre-defined CME signatures. In addition, CME-insensitive charge separation measures using third harmonic event planes such as $\Delta\gamma_{123}/v_3$ and $k_3$ were also measured to provide data-driven baselines. The utility of these baselines are not affected by multiplicity dilution although their constraining powers are limited by their larger uncertainties as compared to the equivalent observables involving second harmonics. 


\begin{figure}
    \centering
    \includegraphics[width=0.8\textwidth]{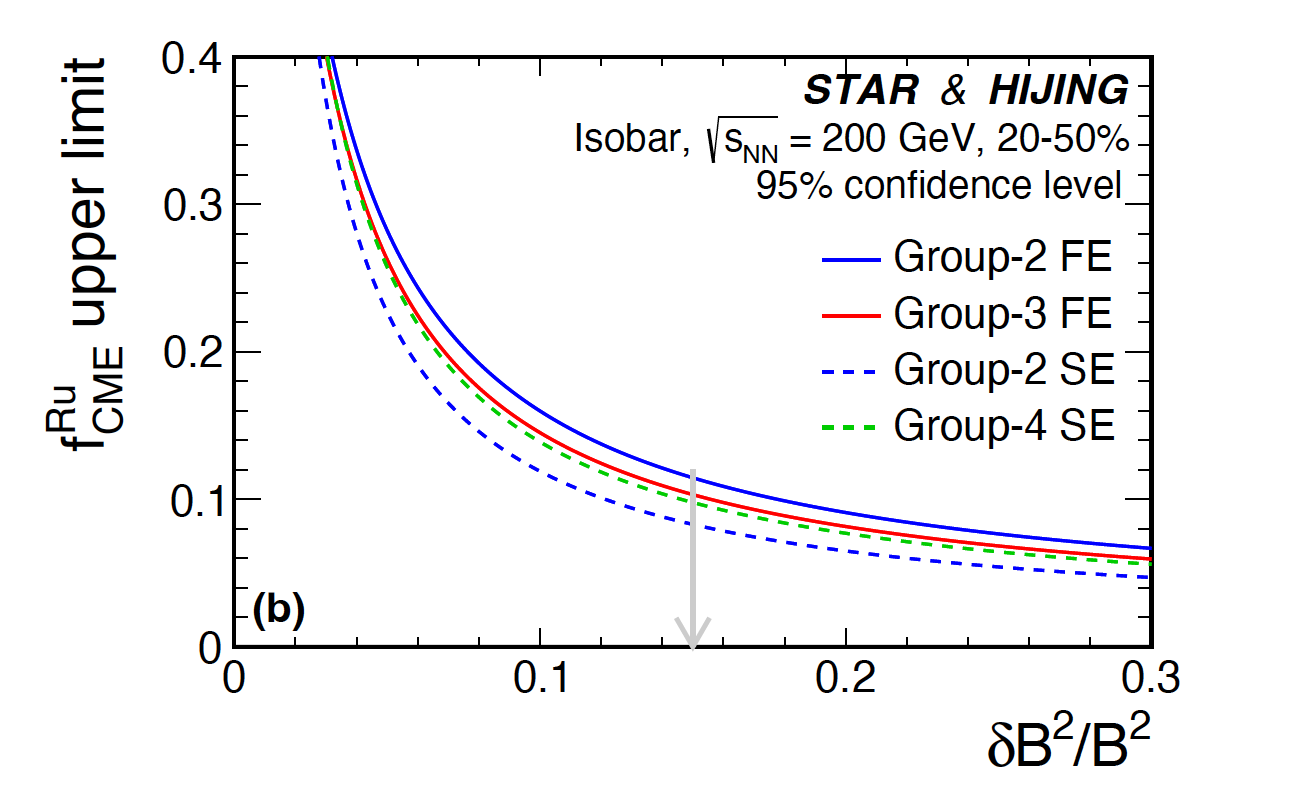}
    \caption{Figure take from the Ref.~\cite{STAR:2023ioo} with the permission from the STAR collaboration. Extracted upper limits as functions of relative difference in the magnetic field strength between the isobar pair.  The statistical and systematic uncertainties
are added in quadrature for obtaining the upper limits.}
    \label{fig:results_cme_limit}
\end{figure}

The overall conclusion from the blind analysis is that no predefined CME signature has been observed in the isobar data.  However, to extract a quantiative result utilizing the full sensitivity of the isobar run, careful consideration must be given to the baseline; the baseline of unity is expected to be affected by the multiplicity difference between the two isobars. We are in the process of careful study of the baseline and how it affects the physics conclusion from the isobar analysis. After the blind analysis, the STAR collaboration put effort towards quantifying possible remaining CME signal by incorporating the multiplicity difference between the two isobars and non-flow effects which are different between the isobars. As a first step, the estimates are made for the background contribution to the double ratio of the $\Delta\gamma/v_2$ by incorporating: 1) the difference in the multiplicity dilution ($\propto 1/N_{\rm trk}$) between the two isobars, 2)  data-driven estimates of various sources of two-particle non-flow correlations and, 3) sources of three-particle non-flow correlations estimated using a HIJING simulation.
The results from such analysis were present in \cite{STAR:2023ioo,STAR:2023gzg}. Here is one simple way to understand how to achieve the upper limit of CME in isobar collisions. Given the double ratio of $\Delta\gamma/v_2$ and $1/N_{\rm trk}$ known one can estimate CME baseline using the following approach discussed in Ref~\cite{Kharzeev:2022hqz} -- a very similar approach was used by the STAR collaboration in Ref~\cite{STAR:2023ioo,STAR:2023gzg}. From the blind analysis of isobar data shown in Fig.\ref{fig_isobar_blind_analysis} one has
\begin{equation}
    R = \frac{(\Delta\gamma/v_2)_{\rm Ru+Ru}}{(\Delta\gamma/v_2)_{\rm Zr+Zr}}\,\,,\,\, R_{\rm mult} = \frac{(1/N_{\rm trk})_{\rm Ru+Ru}}{(1/N_{\rm trk})_{\rm Zr+Zr}}.  
    \label{eq_isobar_ratio_result}
\end{equation}
By incorporating the multiplicity difference and various sources of nonflow in Ref~\cite{STAR:2023ioo,STAR:2023gzg}, the STAR collaboration estimates the background baseline for the double ratio as
\begin{equation}
    R_{\rm baseline} = \frac{(N_{\rm trk} \Delta\gamma/v_2)^{\rm bkg}_{\rm Ru+Ru}}{(N_{\rm trk} \Delta\gamma/v_2)^{\rm bkg}_{\rm Zr+Zr}}. 
    \label{eq_isobar_ratio_baseline_result}
\end{equation}
It turns our this modified baseline is much below the unity baseline that was used as a pre-condition for the STAR isobar blind analysis. Using Eq.\ref{eq_isobar_ratio_result} and Eq.\ref{eq_isobar_ratio_baseline_result} one therefore finds the background ratio
\begin{equation}
    R_{\rm bkg} = R_{\rm baseline} \times R_{\rm mult}. 
\end{equation}
Assuming the magnetic field square different between the two isobar is $\lambda_s=(B_{\rm Ru+Ru}/B_{\rm Zr+Zr})^2$ one can write the fraction of possible CME signal $f_{\rm CME}^{\rm Ru+Ru}$ in Ru+Ru collisions can be written as
\begin{equation}
    f_{\rm CME}^{\rm Ru+Ru} = \frac{1/R_{\rm bkg} -1/R}{\lambda_s + 1/R_{\rm bkg} - 1}.
\end{equation}
For assumptions of $\lambda_s=0.15$, i.e. about 15\% difference in the magnetic field square between the two isobar species the upper limit of CME fraction in Ru+Ru collisions from the RHIC isobar program has been reported to be 10\%~\cite{STAR:2023ioo,STAR:2023gzg}.




\subsubsection{Relative pseudorapidity and azimuth dependence: Quantification of nonflow}
It is possible to use the relative pseudorapidity dependence of azimuthal correlations to discern sources of long-range components, dominated by early time dynamics, from short-range components constrained by causality to appear at late times. This approach extends to charge-dependent correlations, prompting us to explore the dependence of the correlator, $\langle \cos(\phi_{a\alpha} + \phi_{b\beta} - 2\Psi_{\text{RP}}) \rangle$, on the pseudorapidity gap, $\Delta\eta_{ab} = |\eta_a - \eta_b|$, between charge-carrying particles. Measurements in STAR using Au+Au and U+U collisions reveal that short-range correlation sources, such as photon conversions ($e^+e^-$), Hanbury-Brown-Twiss (HBT), and Coulomb effects, manifest as Gaussian peaks at small $\Delta\eta_{ab}$. These peaks' width and magnitude strongly depend on collision centrality and system size. However, identifying these components becomes increasingly difficult at peripheral collision centralities due to overlap with di-jet fragmentation, which dominates both same-sign and opposite-sign correlations. While deconstructing the different components of the correlator based on $\Delta\eta_{ab}$ can be challenging, distinct changes in the shape of individual same-sign and opposite-sign measurements of the gamma-correlator offer a clear indication of the presence of different correlation sources, as demonstrated in Ref.~\cite{Tribedy:2017hwn}.

\subsection{Beam energy dependence}

We now discuss how the anomalous chiral transport may vary with the collision energy, which is an important question for understanding relevant measurements from RHIC Beam Energy Scan (BES) program as well as from LHC energies. The answer relies upon a number of key ingredients, to be discussed below. 

First, the necessary condition for anomalous transport to occur is a sufficiently hot environment such that  a substantial amount, in terms of both spatial volume and lifetime, of quark-gluon plasma with restored chiral symmetry is created.  One would then expect that with increasing collision energy this condition is satisfied better and better. Another relevant factor is the topological transition rate which controls the initial chirality imbalance for CME and which in general increases with the relevant medium scale, be it the pre-thermal saturation scale for the early stage or the temperature scale in the quark-gluon plasma. In either case, such a rate is expected to increase with the collision energy.
Thus it is natural to  expect a ``threshold energy'' below which a QGP with sufficient topological transitions would ceases to exist in the fireball.  A number of measurements from BES-I (such as the constituent quark scaling of elliptic flow, jet energy loss, directed flow, net proton fluctuations, light cluster production, etc) appear to hint at a qualitative change in the observed properties of the created bulk matter occurring around the beam energy range of $\sqrt{s_{NN}}\simeq (10\sim 20)$ GeV. 
Therefore, one should expect any signal from anomalous chiral transport to decrease towards the low enough beam energy regime and to eventually turn off when the collision energy drops below the production threshold for a substantial QGP with restored chiral symmetry. 

The other necessary condition is the presence of the ``driving forces'', i.e. the magnetic field $\mathbf{B}$, for which the situation is more complicated due to its strong time dependence. The peak strength of $\mathbf{B}$ scales with the nucleon Lorentz factor $\gamma \propto \sqrt{s_{NN}}$ and thus increases with increasing beam energy. On the other hand, without any medium feedback the time duration of this strong initial magnetic field scales inversely with $\gamma$  and thus decreases rapidly with increasing beam energy. This time scale can be estimated as $\tau_B \sim \frac{2R}{\gamma}\sim \frac{4R M_N}{\sqrt{s_{NN}}}$ with $R$ the nuclear radius (e.g. $R\sim 7$ fm for Au or Pb) and $M_N\simeq 0.939$ GeV the nucleon mass. Another important time scale is the formation time of the quarks after the initial collisions. Estimates of this timescale $\tau_f$ based on the glasma picture for the early stage in heavy ion collisions would suggest that 
$\tau_f \sim \frac{1}{Q_s} $, where $Q_s$ is the so-called saturation scale~\cite{Gelis:2010nm,Gelis:2012ri}.  $Q_s$ is expected to scale with beam energy in a specific way, $ {Q_s} \sim {Q_0} \left ( \frac{\sqrt{s_{NN}}}{E_0}  \right )^{\frac{0.3}{2}}$ with $Q_0$ the saturation scale at a reference energy scale $E_0$. We may use the RHIC values $Q_0\simeq 1.5$ GeV and $E_0\simeq 200$ GeV for a quick order-of-magnitude estimate~\cite{Gelis:2010nm,Gelis:2012ri}. Let us now compare the two time scales $\tau_B$ versus $\tau_f$, with the former decreasing rapidly with beam energy $\sqrt{s_{NN}}$ while the latter only decreasing mildly. Note that it  is the quarks that are needed for both the anomalous chiral transport and for the medium induction mechanism which has the potential of prolonging lifetime of the magnetic field. 
One may therefore expect that at certain high enough beam energy where $\tau_B\ll \tau_f$,  the initial magnetic field will exist only  like an extremely short pulse before the formation of any quark or antiquark medium and would not cause any anomalous transport. 
That is, the signals of anomalous chiral transport effects may be expected to eventually  {\em disappear at the high  beam energy end}. 
Let us give a concrete estimate for beam energies relevant to RHIC and the LHC. 
At RHIC top energy $\sqrt{s_{NN}}=200$ GeV, one has $\tau_B\sim 0.13$ fm/c and $\tau_f\sim 0.13$ fm/c, with the two scales  comparable $\tau_B\sim \tau_f$. At LHC energy e.g. $\sqrt{s_{NN}}=5020$ GeV,   $\tau_B \sim 0.005$ fm/c significantly decreases from RHIC while the $\tau_f\sim 0.08$ fm/c only slightly decreases, resulting in a situation where $\tau_B \ll \tau_f$.

To conclude this brief discussion, it is quite plausible that potential signals from anomalous chiral transport effects would have nontrivial  dependence on the collision beam energy, possibly  disappearing in collisions  both at very low energy end (e.g. below $\sim \hat{O}(10)\, \rm GeV$) and at very high energy end (e.g. beyond $\sim \hat{O}(1)\, \, \rm TeV$). From the estimate above, it appears 
quite likely that the  optimal  beam energy window   may be  in the  range of $\hat{O}(10\sim 100)\, \rm GeV$.  
Due to this non-monotonic trend, the beam energy scan program at RHIC provides the unique opportunity to look for the signals from anomalous chiral transport.  In fact, such a pattern of  beam energy dependence could in itself be considered as a characteristic signature  for the search of CME signal.

\subsubsection{Measurements from RHIC Beam Energy Scan (BES)}


The Beam Energy Scan program at RHIC aimed to investigate the chiral phase transition of Quantum Chromodynamics, the theory of strong interactions. Just to recall, to observe this effect in an experiment, three conditions must be met:

\begin{enumerate}
\item Chiral symmetry restoration must turn off the quark mass generated through the condensate.
\item Collisions must disturb the QCD vacuum to create an uneven distribution of right-handed and left-handed quarks through the phenomenon of quantum anomaly.
\item The collisions must produce a sufficiently strong electromagnetic field ($\Lambda_{\rm QCD}$) with a non-random orientation to the collision plane (reaction plane).
\end{enumerate}

As a result, a charge separation across the reaction plane is expected in heavy-ion collisions. The experimental search for CME involves varying these three prerequisites in innovative ways. CME has been extensively searched for in various collision systems, such as Au+Au and more recently isobars (Ru+Ru, Zr+Zr), with a particular focus on the highest collision energy accessible at RHIC. These searches aim to exploit the strength of the magnetic field by changing the systems or choosing two different planes as a proxy for the orientation of the collision plane, thereby leveraging the change in the signal of charge separation.

\begin{figure}
    \centering    
    \includegraphics[width=0.9\textwidth]{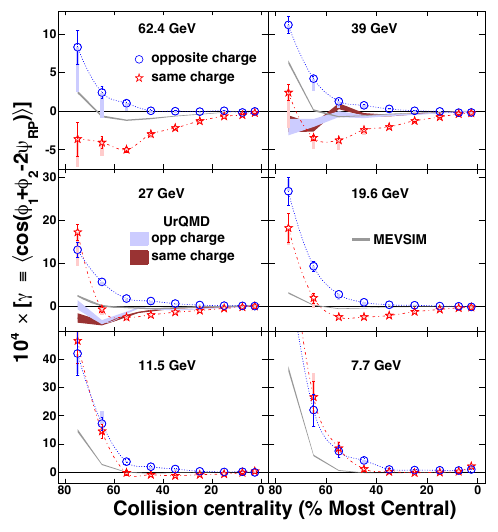}
    \caption{Energy dependence of charge separation in Au+Au collisions, as measured by the STAR detector using data from the RHIC beam energy scan program (adapted from reference \protect\cite{STAR:2014uiw}). The results illustrate the diminishing difference between same-sign and opposite-sign correlations as the collision energy approaches 7.7 GeV. This trend aligns with the expected disappearance of the Chiral Magnetic Effect (CME) signal, indicative of vanishing prerequisites such as chiral symmetry restoration and the deconfined phase. Additionally, it suggests a potential decrease in background contributions towards 7.7 GeV. Consequently, there is a need for new measurements employing innovative techniques and detectors to discern the vanishing component, thereby elucidating the underlying physics.}
    \label{fig_cme_bes1_star}
\end{figure}

\begin{figure}[h]
    \centering
    \includegraphics[width=0.9\textwidth]{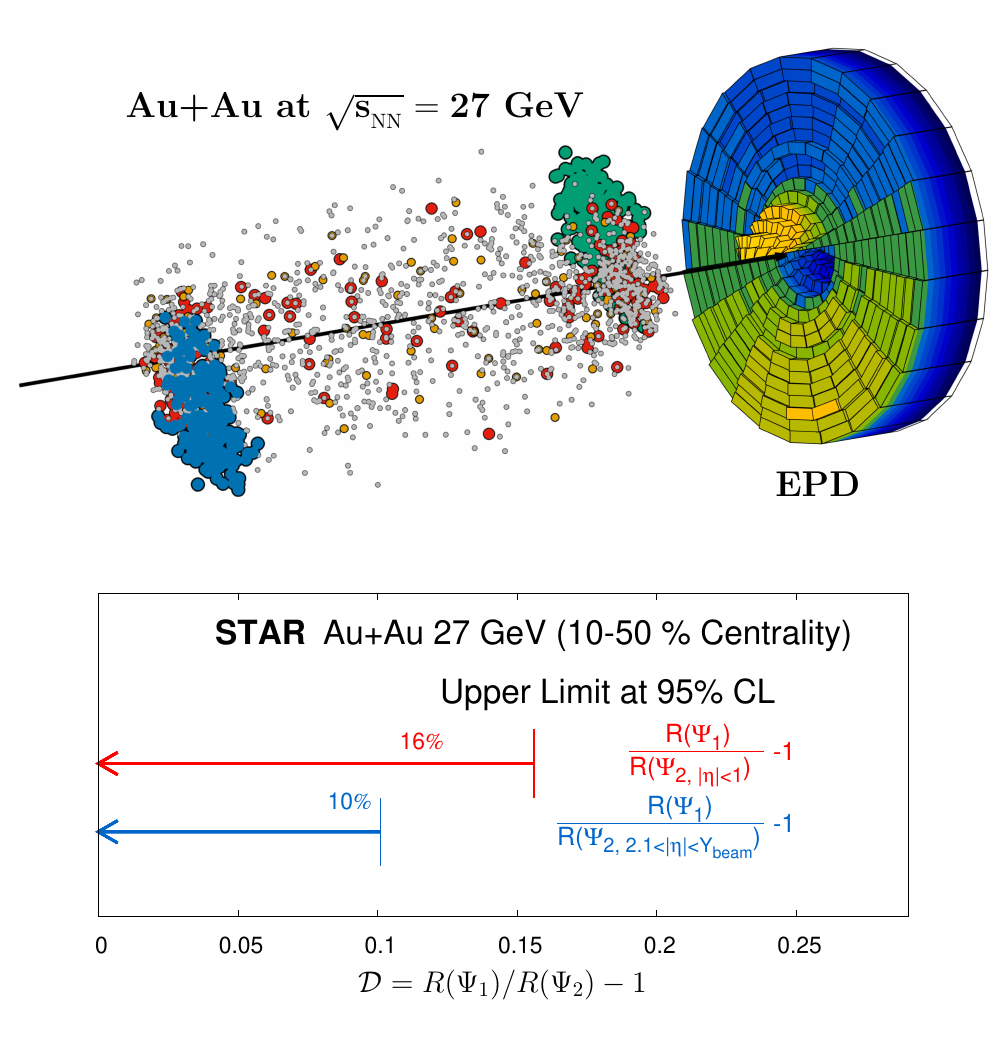}
    \caption{Top: Cartoon of the EPD detector acceptance and response to directed flow from both spectator protons and participant particles taken from ~\cite{STAR:2022ahj}. The directed flow from the participant particles is concentrated near the outer edge of the EPD, while the directed flow from the beam fragments, stopped and spectator protons are concentrated near the inner edge. Bottom: The upper limit at the 95\% CL calculated for the deviation quantity $\mathcal{D}$ in 10-50\% centrality. The deviation quantity $\mathcal{D}$ built to quantify the deviation from a flow-driven background scenario --  bigger $\mathcal{D}$ means the more confidence we have in the CME scenario. The results provide strong constraints on the observability of CME in Au+Au collisions at $\sqrt{s_{_{\rm{NN}}}}=27$ GeV.}
    \label{fig:my_label}
\end{figure}

The STAR Collaboration recently performed a search for the CME in Au+Au collisions at $\sqrt{s_{_{\rm{NN}}}}=27$ GeV with the STAR Forward Event Plane Detectors~\cite{STAR:2022ahj}. The aim of this work was to investigate the effect of lowering the collision energy on the CME. The change of collision energy affects the magnetic field lifetime, the domain size of left-right handed quark imbalance, and the presence of a medium where quarks and gluons are deconfined and the chiral symmetry of Quantum Chromodynamics (QCD) is restored. The observability of CME at lower collision energy is an outstanding question in the community. Previously, charge separation was measured with the BES-I data~\cite{STAR:2014uiw}, but the most recent development on disentangling the signal and background contribution was not known back then. 

 The STAR Collaboration address this issue by utilizing newly installed Event Plane Detectors (EPD) in the search for CME in Au+Au collisions with a center-of-mass energy of $\sqrt{s_{_{\rm{NN}}}}=27$ GeV, using data from 2018. Unlike the BES-I run, the combination of the EPD and time-projection chamber (TPC) enabled us to measure charge separation across both the directed flow plane ($\Psi_1$) at forward pseudorapidity ($Y_{\rm beam} < |\eta|<5.1$ ) and the elliptic flow plane ($\Psi_2$) at both central ($|\eta|<1.0$) and forward rapidity ($2.1 < |\eta|<Y_{\rm beam}$), where $Y_{\rm beam}=3.4$ is the beam rapidity. Model calculations show that the magnetic field will have a stronger correlation with $\Psi_1$ than with $\Psi_2$. Therefore, the anticipation is that the CME scenario will result in a larger charge separation across $\Psi_1$ than that of $\Psi_2$, when scaled by elliptic anisotropy along both planes. The flow-driven background scenario will lead to a consistent charge separation scaled by elliptic anisotropy across all the planes. 

  The observable $\mathcal{D}$ is built to quantify the deviation from a flow-driven background scenario. 
    Because the CME signal is expected to have different projections into different planes, $\mathcal{D}$ can tell us the room for signal. Besides, the $\mathcal{D}$ is not necessary to present as a percentage, it can be any number, and cannot directly tell us the fraction of the signal. But the bigger the number is, the more confidence we have in the CME scenario.

    If one performs the measurements of charge separation ($\Delta\gamma$) scaled by ellipticity ($v_2$) across two event planes $\Psi_1$ and $\Psi_2$, according to previous studies~\cite{Voloshin:2018qsm,Xu:2017qfs} one can obtain the following. Based on Eq.10 of ~\cite{Voloshin:2018qsm} one can think of the measured charge separation to be the sum of the signal and background term: 

    \begin{equation}
        \Delta\gamma (\Psi_1) = b \times v_2(\Psi_1) + (\Delta\gamma)^{\rm CME} \times \gamma_B(\Psi_1),
    \end{equation}
    \begin{equation}
        \Delta\gamma (\Psi_2) = b \times v_2(\Psi_2) + (\Delta\gamma)^{\rm CME} \times \gamma_B(\Psi_2),
    \end{equation}
     
   where $b$ is a constant related to background and $\Delta\gamma^{\rm CME}$ is charge separation correlated to the B-field direction. Also, $\gamma_B(\Psi_{1,2})=\left< \cos(2\Psi_B-2\Psi_{1,2}) \right> $ is the projection of the B-field direction on $\Psi_{1,2}$ planes. Therefore one can write:
    \begin{equation}
           \frac{(\Delta\gamma/v_{2})_{\Psi_1}}{(\Delta\gamma/v_{2})_{\Psi_2}} = 1 + f_{\rm CME}({\Psi_2}) \left(\frac{\gamma_{B}({\Psi_1})}{\gamma_{B}({\Psi_2})}\frac{v_{2}({\Psi_2})}{v_{2}({\Psi_1})}-1 \right),
    \end{equation}
    where $f_{CME}(\Psi_2)$ is the fraction of charge separation signal w.r.to $\Psi_2$ plane due to CME. Therefore, the observable $\mathcal{D}$ can be written as: 
     \begin{equation}
          {\cal D} = f_{\rm CME}({\Psi_2}) \left(\frac{\gamma_{B}({\Psi_1})}{\gamma_{B}({\Psi_2})}\frac{v_{2}({\Psi_2})}{v_{2}({\Psi_1})}-1 \right),
        \label{D_fcme}
    \end{equation}

   In the flow-driven background scenario, an expectation arises that $f_{CME}=0$, resulting in an expected value of $\mathcal{D}$ to be zero. Conversely, in the presence of CME, $\mathcal{D}$ is anticipated to be greater than zero. This anticipation arises due to the maximum elliptic flow concerning the $\Psi_2$ plane. Consequently, the ratio $v_{2}(\Psi_2)/v_{2}(\Psi_1)>1$ is consistently affirmed by STAR's measurement. Furthermore, since $\Psi_1$ is determined by the directed flow of forward protons (which also induce the B-field), it exhibits a stronger correlation with the B-field direction compared to the $\Psi_2$ plane. Consequently, $\gamma_{B}({\Psi_1})/\gamma_{B}({\Psi_2})>1$ is confirmed by UrQMD simulation. In the presence of CME, $f_{CME}>0$, hence $\mathcal{D}>0$. STAR finds that $\mathcal{D}$ remains consistent with zero within $2\sigma$, prompting the presentation of an upper limit for $\mathcal{D}$. In Ref~\cite{STAR:2022ahj} an upper limit on the deviation of $\mathcal{D}$ from a flow-driven background scenario at the 95\% confidence level was derived by STAR. This investigation offers a potential roadmap for future CME inquiries utilizing high-statistics data from the RHIC Beam Energy Scan Phase-II. Based on Eq.\ref{D_fcme} one can extract the CME fraction as
   \begin{equation}
          f_{\rm CME}({\Psi_2})= \frac{\cal D}{ \left(\frac{\gamma_{B}({\Psi_1})}{\gamma_{B}({\Psi_2})}\frac{v_{2}({\Psi_2})}{v_{2}({\Psi_1})}-1 \right)},
   \end{equation}
   However putting the numbers together from Ref~\cite{STAR:2022ahj}, we can estimate an $f_{\rm CME}=0.066\pm0.275$. 

As discussed in a previous section, the ESS method estimates the fraction of CME signal and has recently reported preliminary results in ~\cite{Xu:2023wcy}. The STAR collaboration has performed a CME search in Au+Au collisions at $\sqrt{s_{NN}}$ = 7.7, 14.6, 19.6, 27 from the RHIC Beam Energy scan phase two, which has improved detectors and a forward fragment-rich plane that is strongly correlated with the magnetic field direction. The preliminary results show a 3$\sigma$ significance for the ratio of the intercepts obtained using the ESS method $\Delta\gamma_{ESS}$ and the inclusive $\Delta\gamma$, which is an upper bound of the CME signal.

\subsubsection{CME measurements at LHC energies}

Going to the high end of the collision beam energy range, considerable efforts have been taken by ALICE and CMS experiments to perform measurements motivated by the search of CME at LHC energies. As already discussed at the beginning of this subsection, one would expect the CME signal to be rather limited at LHC (as compared with RHIC) due to extremely short magnetic field lifetime. Therefore, relevant measurements at LHC can help us achieve quantitative understanding of the backgrounds and develop methods for extracting/constraining CME signals. 

After the initial excitements as well as confusions triggered by the first CME results from STAR in 2009, it was natural to wonder what could happen to the charge asymmetry observables at the LHC. The first such measurement was reported by ALICE in 2012~\cite{ALICE:2012nhw} for Pb+Pb collisions at $\sqrt{s_{NN}}=2.76 \, \rm TeV$,  showing ``little or no collision energy dependence when compared to measurements at RHIC energies''. This result indeed provided a timely insight and strengthened the then-emerging understanding that the  charge-dependent azimuthal correlations observed at RHIC and LHC energies are dominated by non-CME background contributions. 

\begin{figure}[h]
    \centering
    \includegraphics[height=0.45\textwidth]{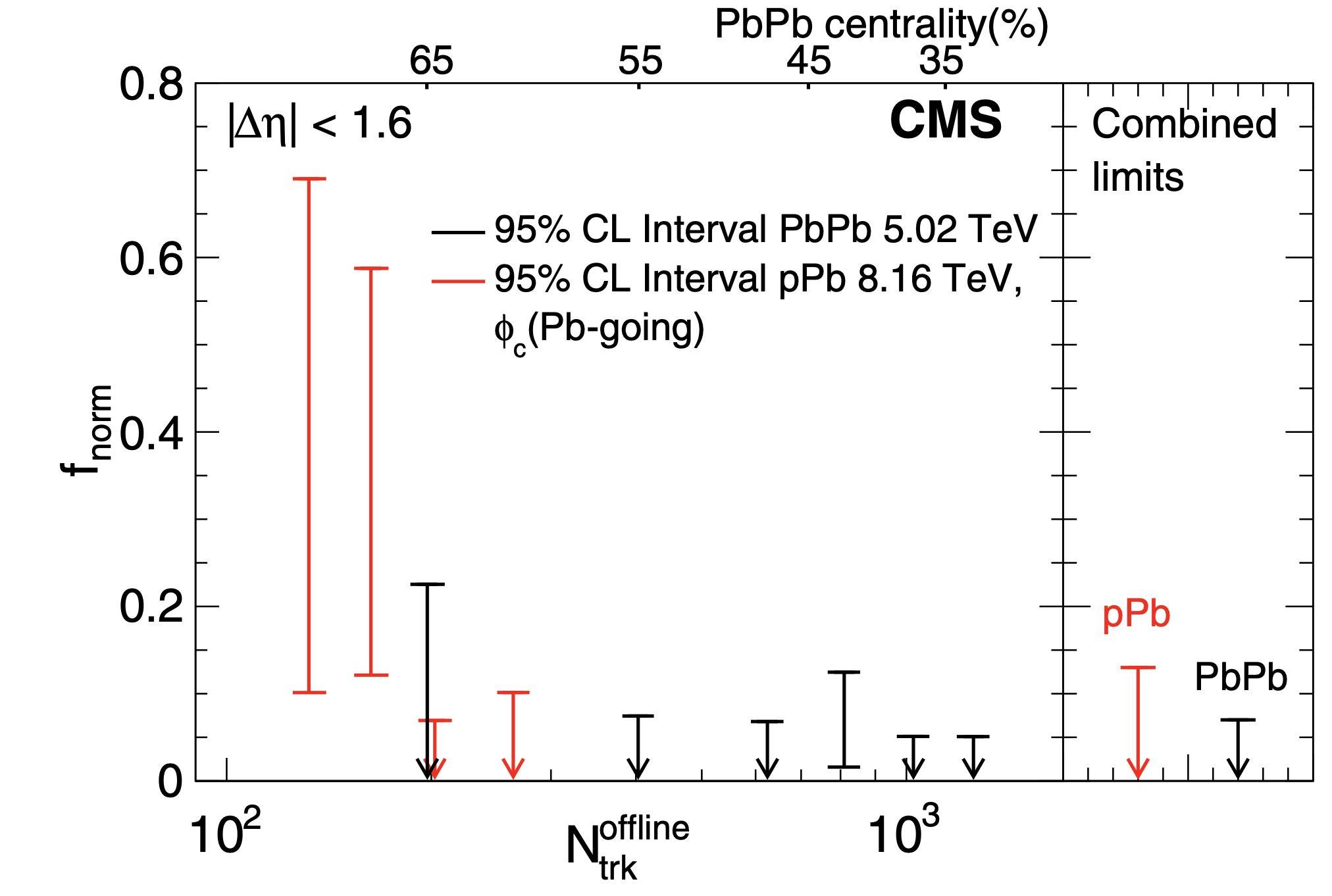}
    \caption{The extracted upper limit on possible CME signal fraction in the measured charge asymmetry azimuthal correlations by CMS Collaboration for pPb and Pb+Pb collisions at the LHC. (See \cite{CMS:2017lrw} for details.)}
    \label{fig:cme_CMS_limit}
\end{figure}

\begin{figure}[h]
    \centering
    \includegraphics[width=0.85\textwidth]{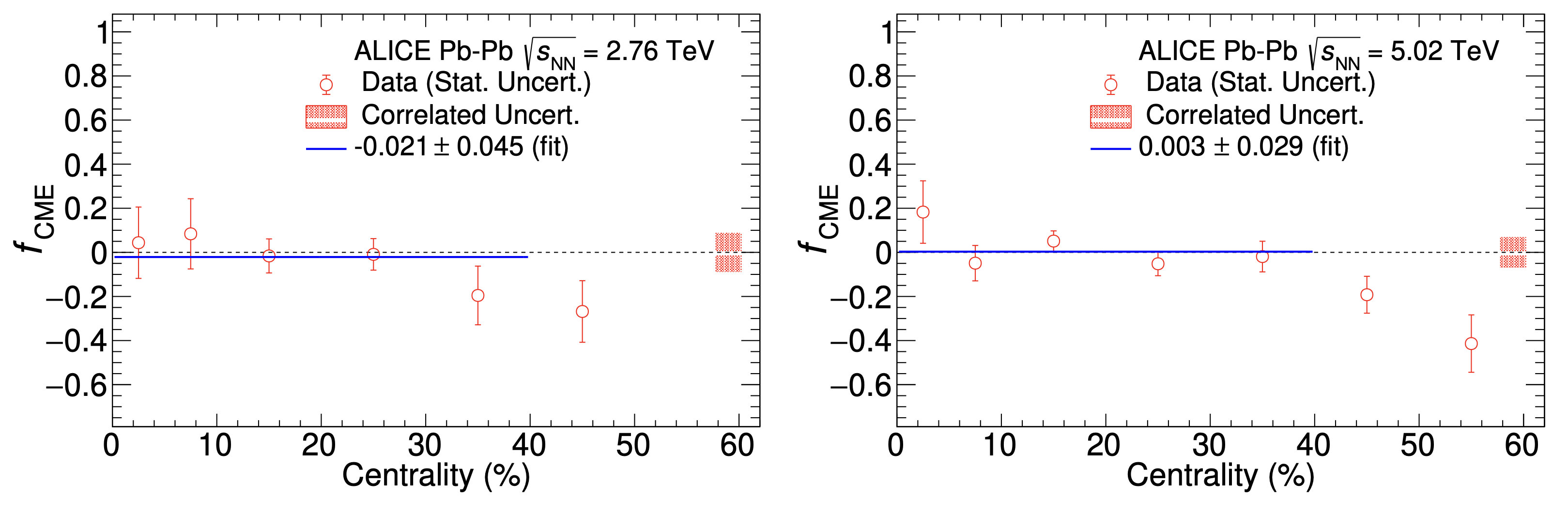}  
    \caption{ The extracted CME signal fraction in the measured charge asymmetry azimuthal correlations by ALICE Collaboration for Pb+Pb collisions at $2.76$ and $5.02\, \rm TeV$ respectively. (See \cite{ALICE:2020siw} for details.)}
    \label{fig:cme_ALICE_1}
\end{figure}

This crucial point was further reinforced by CMS measurements~\cite{CMS:2016wfo,CMS:2017lrw}, in particular, through the striking similarity of the charge-dependent azimuthal correlations between the small pPb (where CME is not expected to occur) and the large Pb+Pb colliding systems (--see Fig.~\ref{fig_cme_collaboration_cme_search_experiment}). The high precision data at LHC also allowed sophisticated analysis by both CMS and ALICE of the $\gamma$-correlator based on event-shape engineering~\cite{CMS:2017lrw,ALICE:2017sss} with respect to both the elliptic event plane and triangular event plane, with the latter providing a baseline estimate of  background effects (--see examples of such analysis in Fig.~\ref{fig_alice_cms_ese}). Based on this strategy, CMS went on to further extract an upper limit for the potential CME signal fraction within the total $\gamma$-correlator, shown in Fig.~\ref{fig:cme_CMS_limit} and estimated to be $13\%$ for pPb and $7\%$ for Pb+Pb at $95\%$ confidence level.

\begin{figure}[h]
    \centering
    \includegraphics[width=0.85\textwidth]{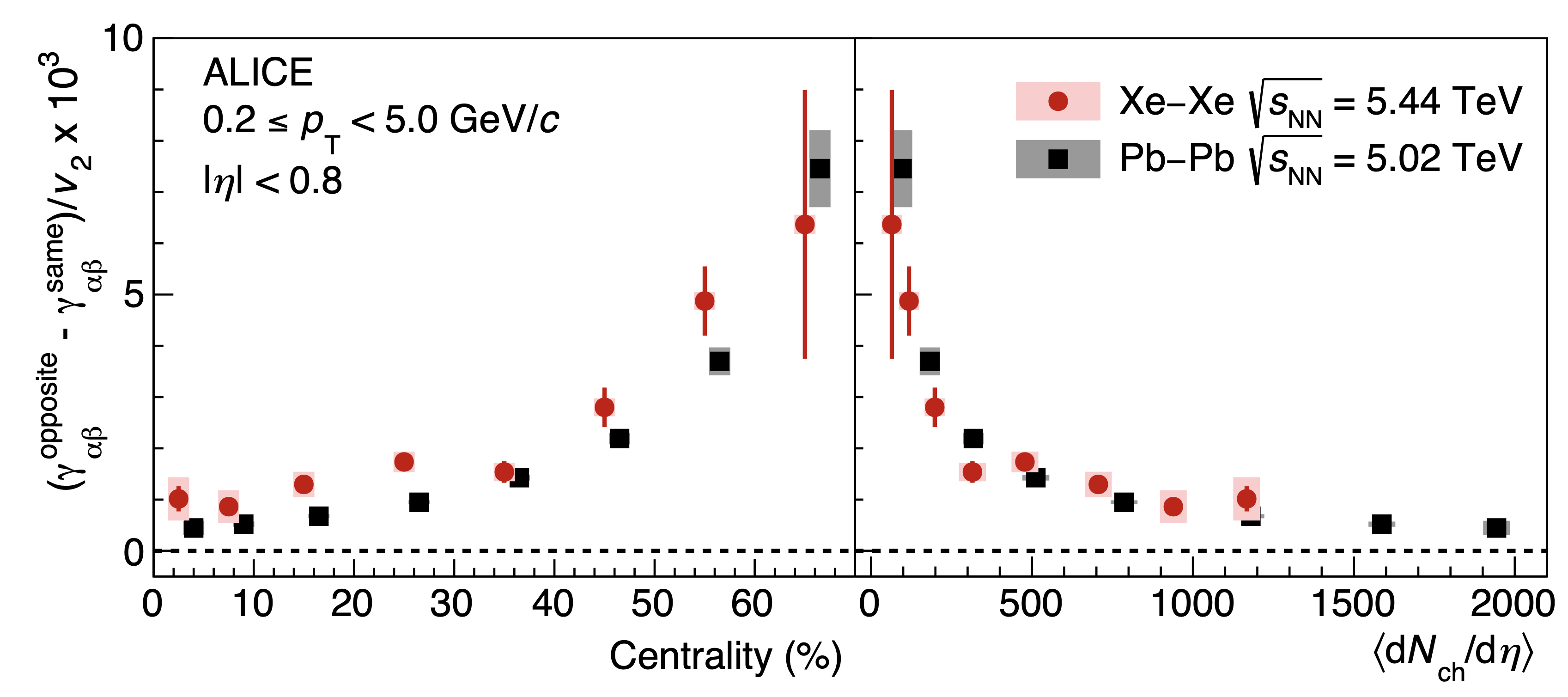}  
    \caption{
    The centrality (left) and multiplicity (right) dependence of the charge asymmetry azimuthal correlations, measured by ALICE Collaboration for Xe+Xe and Pb+Pb collisions at the LHC. (See \cite{ALICE:2022ljz} for details.) }
    \label{fig:cme_ALICE_2}
\end{figure}

\begin{figure}[h]
    \centering
    \includegraphics[width=0.85\textwidth]{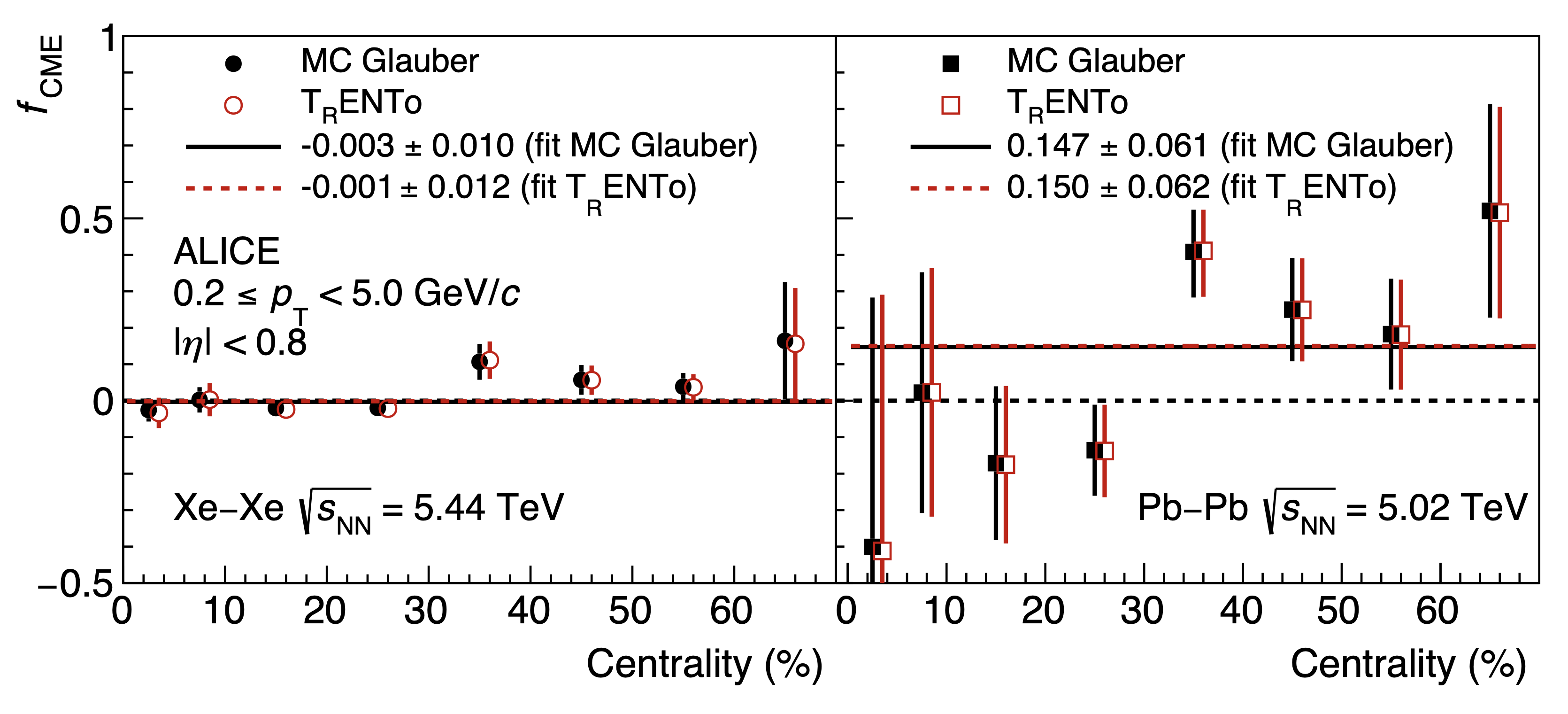}  
    \caption{  The extracted CME signal fraction in the measured charge asymmetry azimuthal correlations by ALICE Collaboration for Xe+Xe and Pb+Pb collisions at the LHC. (See \cite{ALICE:2022ljz} for details.)}
    \label{fig:cme_ALICE_3}
\end{figure}

More recently, ALICE collaboration reported in \cite{ALICE:2020siw} an extraction of the CME signal fraction in Pb+Pb collisions at both $2.76 \, \rm TeV$ and $5.02 \, \rm TeV$. The key issue is to quantify the background contributions from the major sources like the local charge conservation. The strategy by ALICE was to utilize the strength of charge-dependent azimuthal correlation with respect to the 3rd-harmonic plane as an estimator for the strength of background correlation with respect to the 2nd-harmonic plane. This assumption appears to be plausible based on their model study with blast-wave parametrization that incorporates local charge conservation.  The analysis suggests that the so-obtained background estimate 
is not able to fully describe the measured CME-sensitive $\gamma$-correlator, implying a potential room for some nonzero CME contribution. As shown in Fig.~\ref{fig:cme_ALICE_1}, this analysis suggests a CME signal fraction of $(-2.1\pm 4.5)\%$ at $2.76 \, \rm TeV$ and  $(0.3 \pm 2.9)\%$ at $5.02 \, \rm TeV$. As a caveat, it may be noted that there is unavoidable uncertainty involved in the aforementioned assumption about background estimate, the quantitative validation of which is difficult and model dependent.

The state-of-the-art LHC measurements by ALICE have extended the CME analysis to a new colliding system~\cite{ALICE:2022ljz}, the Xe+Xe collisions at $\sqrt{s_{NN}}=5.44 \, \rm TeV$. The results for the $\gamma$-correlator are shown in Fig.~\ref{fig:cme_ALICE_2}, in comparison with the Pb+Pb collisions at  $\sqrt{s_{NN}}=5.02 \, \rm TeV$. By adopting a two-component decomposition~\cite{Bzdak:2012ia}, with the signal scaling with magnetic field strength and the background scaling with elliptic anisotropy, ALICE extracted the CME signal fraction in both systems. The results  shown in Fig.~\ref{fig:cme_ALICE_3}, while demonstrating a mild dependence on the initial condition models used to estimate the magnetic field strength, suggest a negligible CME fraction in Xe+Xe collisions while a nonzero CME fraction of about $15\%$ in Pb+Pb collisions. Once again, there are uncertainties associated with the underlying assumptions, which may be part of the reason for the difference between the signal fraction extracted in this analysis and that in the previous analysis in \cite{ALICE:2020siw}.

Last but not least, anomalous transport effects like CME and CVE lead to not only charge separation but also baryon separation, as discussed in Sec.~\ref{sec:1:3}. Measurements of baryon separation as well as its correlation with charge separation could provide unique insights into the search of CME and CVE signals and the understanding of relevant backgrounds. ALICE recently took an important  step to measure the azimuthal correlations  between hyperon-proton pairs in Pb+Pb 5.02 TeV collisions, while using hypero-hadron and hadron-hadron pairs as reference~\cite{Wang:2023xhn}. The preliminary results demonstrate nontrivial baryon separation patterns specifically in the hyperon-proton pairs, which could have the potential to help study potential CVE transport as well as quantify local baryon conservation effect in heavy ion collisions.

\subsection{CMW measurements}

As previously discussed in \ref{subsubsec_cmw}, the chiral magnetic wave is a new type of gapless collective excitations in a system of chiral fermions under the presence of an external magnetic field  $\mathbf{B}$ that arises from the interplay between vector and axial density fluctuations coupled together by anomalous transport effects. Just like sound waves that can transport energy and momentum in the usual medium, the CMW can transport vector and axial charges   in chiral matter. 

Different from  the CME for which a chirality imbalance (i.e. nonzero axial density) is necessary, the existence of CMW does not require any particular background density and can be simply triggered by density fluctuations in chiral matter as long as there is a magnetic field.  
Therefore, CMW provides an independent way of manifesting anomalous transport, despite whether the CME is detectable or not.

\begin{figure}[h]
    \centering
    \includegraphics[width=0.4\textwidth]{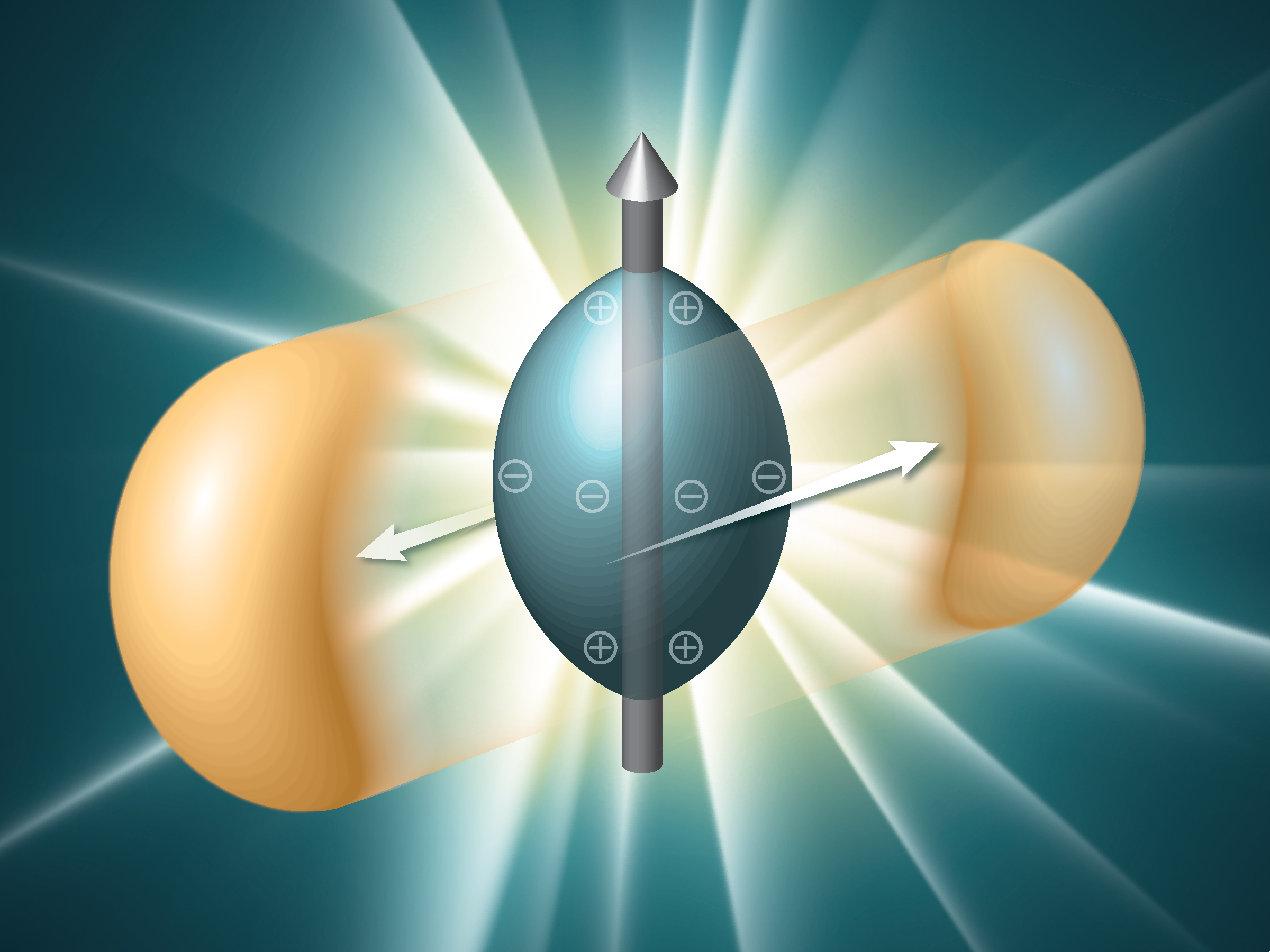} \hspace{0.2in}
    \includegraphics[width=0.45\textwidth]{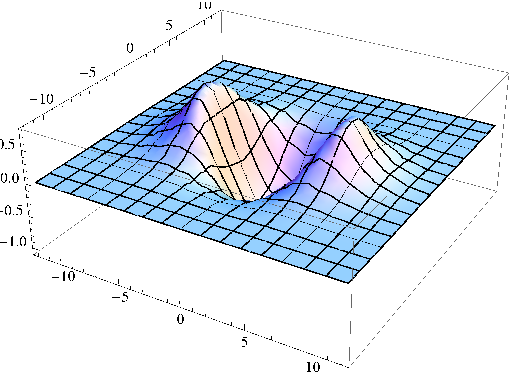}
    \caption{(left) An illustration of the charge quadruple moment arising from the CMW transport. (right) An example of computed charge density by solving CMW equations which shows a quadruple pattern with more positive charges on the two tips of the fireball while more negative charges around the equator --- see \cite{Burnier:2011bf} for details. }
    \label{fig:cmw}
\end{figure}

It is then natural to think about possible experimental observables for CMW. As first quantitatively demonstrated in \cite{Burnier:2011bf}, if the fireball has a nonzero electric charge density in the initial condition, then CMW transport leads to a quadruple pattern in the charge distribution along the magnetic field direction~\cite{Burnier:2011bf,Gorbar:2011ya}. This is illustrated in Fig.~\ref{fig:cmw} (left), with an example of the computed charge distribution from \cite{Burnier:2011bf} shown in  Fig.~\ref{fig:cmw} (right). When coupled with strong collective explosion of the fireball, the CMW-induced quadrupole results in  a specific splitting between the elliptic flow, $v_2$, of $\pi^-$ and $\pi^+$. Experimentally, the initial charge density is directly correlated with the charge asymmetry $A_{ch}= \frac{N_+ - N_-}{N_+ + N_-}$ of the final state charged hadrons. One would therefore expect the elliptic flow difference to be proportional to the charge asymmetry of a given event, i.e. 
\begin{eqnarray} \label{eq_cmw_r}
v_2^{\pi^-} - v_2^{\pi^+} = r A_{ch} \, ,  
\end{eqnarray}
with a positive slope parameter $r$ that is directly related to the CMW-induced quadruple moment of the charge distribution in the QGP. 
Such a feature was indeed observed in measurements both by STAR at RHIC and by ALICE and CMS at LHC, which we will discuss next.  On the theoretical side, a realistic description of those measurements is still lacking and requires more careful investigations of pertinent issues~\cite{Gorbar:2017cwv,Rybalka:2018uzh,Sukhachov:2018uuz,Wu:2022fwz,Wang:2021nvh} such as possible damping effects on CMW modes, interplay between CME and CMW induced charge transport patterns, bulk flow backgrounds, etc. We also note in passing that similar collective excitations could also arise in chrial fluids under the presence of vorticity, e.g. chiral vortical wave~\cite{Jiang:2015cva}.

\subsubsection{RHIC results} 

As stated earlier, signatures for CMW  and CME transport arise from different type of initial conditions and therefore the observation of CME is {\em not} a prerequisite for the  experimental search of CMW evidence. Furthermore, the vector charge density initial condition required for the CMW transport could be experimentally controlled by selecting event-wise charge asymmetry of final state hadrons. The first such measurement was reported by STAR in a 2015 paper \cite{STAR:2015wza}, showing a splitting pattern of $\pi^{\mp}$ elliptic flow $v_2$  exactly as predicted by CMW in Eq.(\ref{eq_cmw_r}). The magnitude of the slope parameter $r$, though, was found to be substantial and quantitatively difficult to explain. 
It was later understood that there exist two subtle issues pertaining to the CMW measurement. The first is a technicality related to the auto correlations in the analysis observable which need to be corrected. The second is a physical background due to the already familiar local charge conservation (LCC) that also contaminates the CMW observable~\cite{}. 

\begin{figure}[h]
    \centering
    \includegraphics[width=0.9\textwidth]{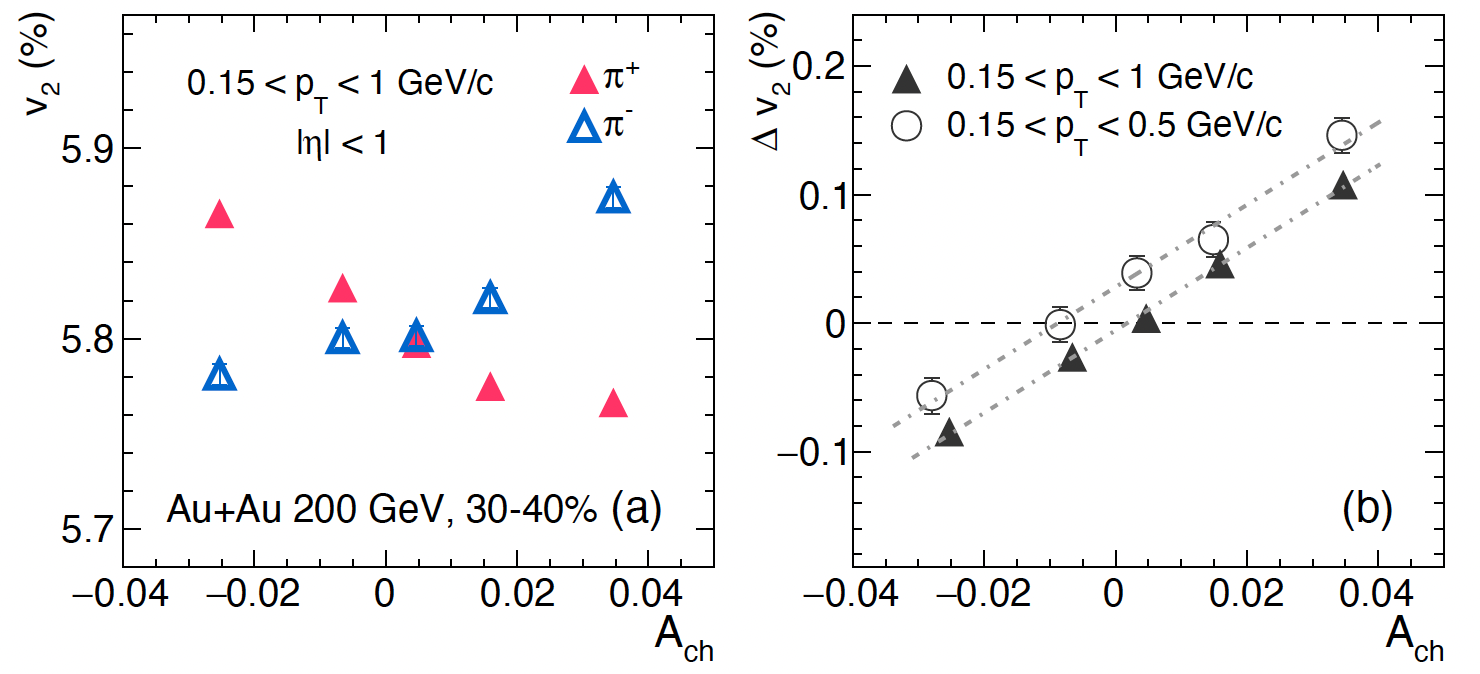} 
    \caption{
    The elliptic flow $v_2$ of $\pi^{\mp}$ individually (left) as well as their difference $\Delta v_2$ as functions of the event charge asymmetry $A_{ch}$, as measured by STAR Collaboration for Au+Au collisions at RHIC. 
     (See \cite{STAR:2022zpv} for details.)
    }
    \label{fig:cmw_ach}
\end{figure}

A careful new analysis was performed later by STAR, with the results reported in \cite{STAR:2022zpv}. In Fig.~\ref{fig:cmw_ach}, the left panel shows separately the $v_2$ of $\pi^{\mp}$ while the right panel shows the difference between them as functions of $A_{ch}$, where a linear dependence with a positive slope, as expected from CMW in Eq.(\ref{eq_cmw_r}), can be clearly identified. 

\begin{figure}[h]
    \centering
    \includegraphics[height=0.32\textwidth]{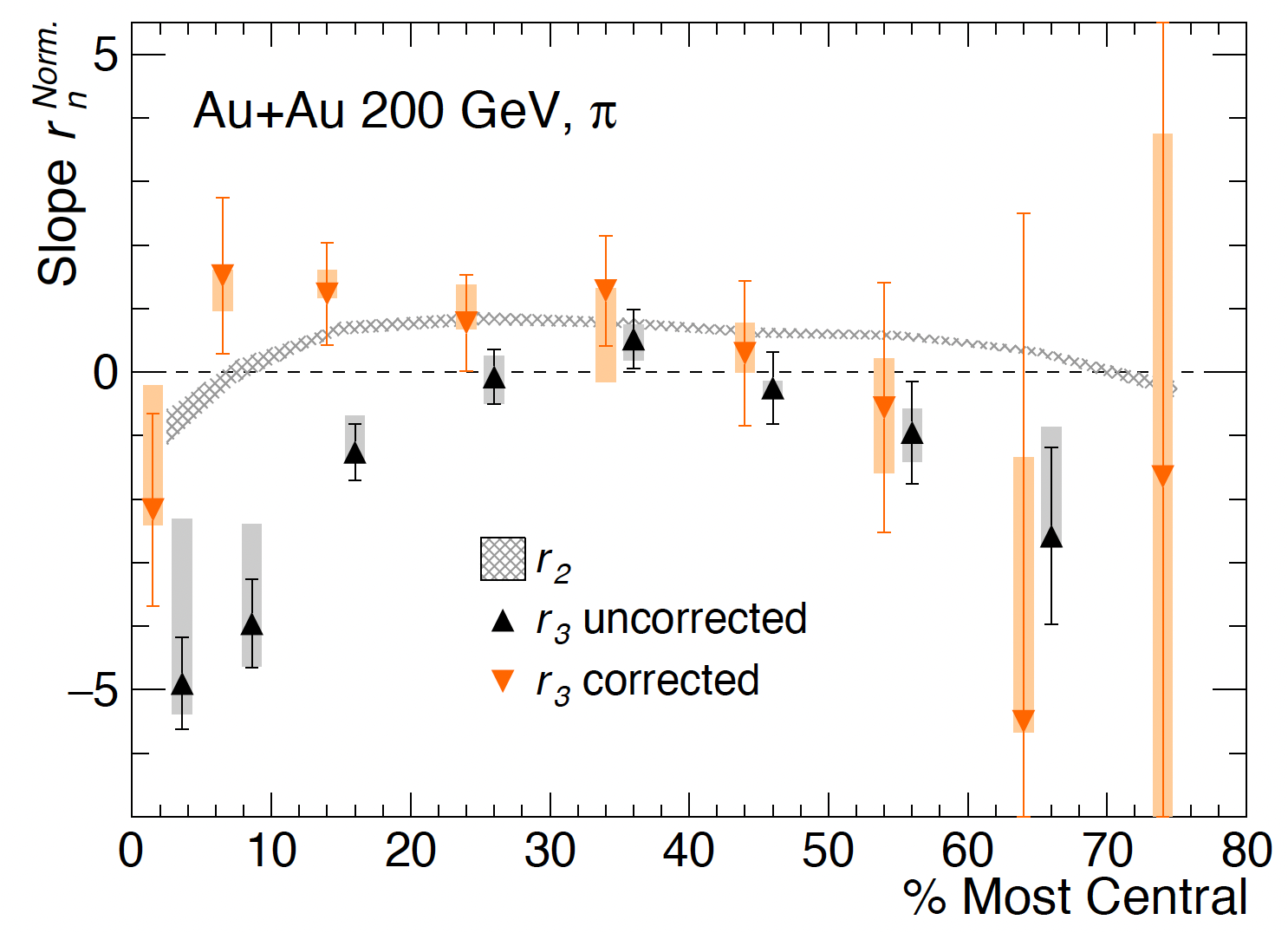} \hspace{0.1in}
    \includegraphics[height=0.32\textwidth]{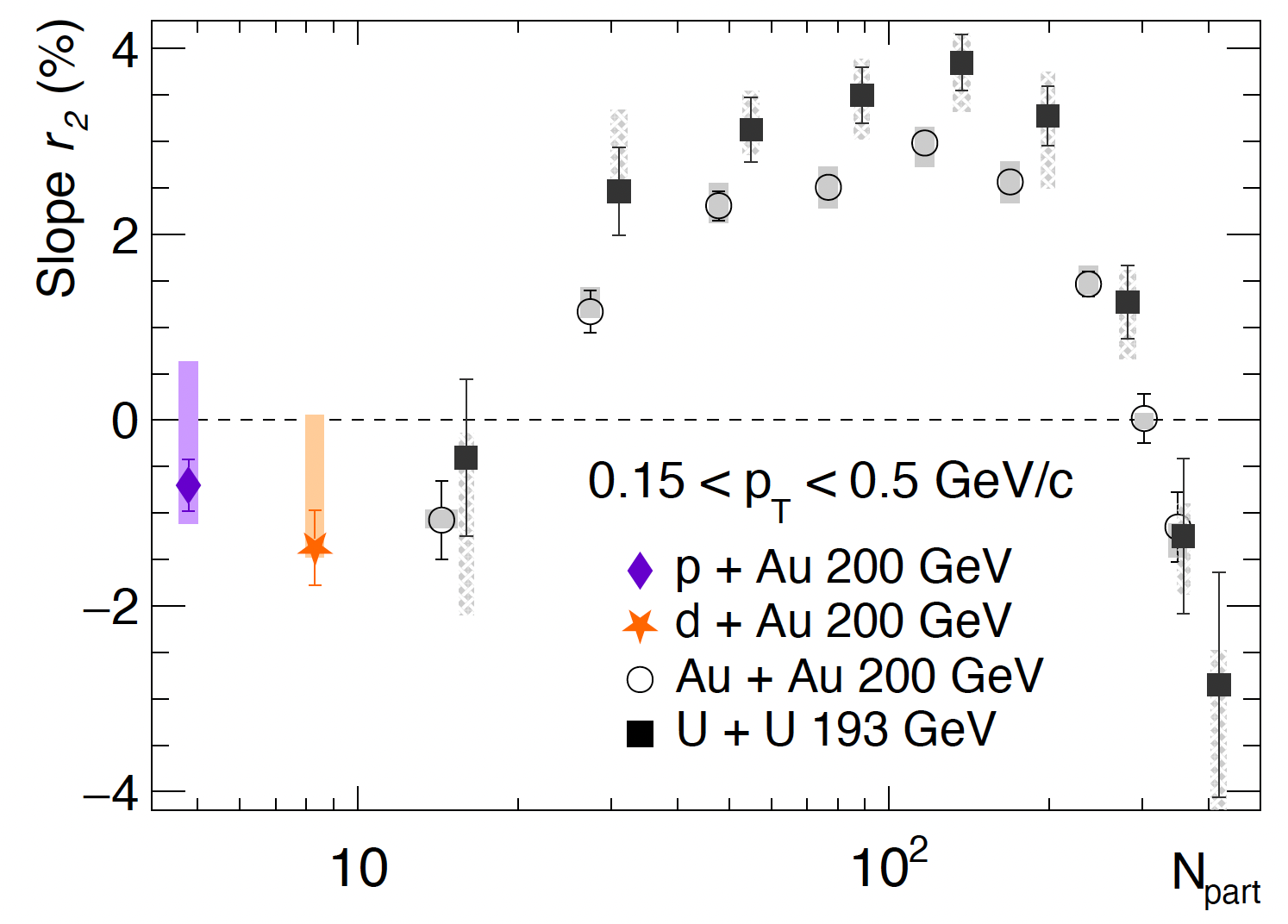}
    \caption{
(left) The slope parameters extracted from the elliptic as well as triangular flow differences between  $\pi^{\mp}$ versus   event charge asymmetry $A_{ch}$ as a function of centrality, measured by STAR Collaboration for Au+Au collisions at RHIC. (right) The slope parameter extracted from elliptic flow from small to large colliding systems measured by STAR Collaoration at RHIC. 
     (See \cite{STAR:2022zpv} for details.)
    }
    \label{fig:cmw_star}
\end{figure}

The slope parameter $r$ has been extracted for different centrality of Au+Au collisions, shown in Fig.~\ref{fig:cmw_star} (left panel) and compared with a similarly extracted slope parameter for the charge asymmetry dependence in the triangular flow $v_3$. The idea is that such a slope in the $v_3$ should come entirely from background correlations and therefore serve as a sort of baseline estimate for the background. As one can see from the comparison, while the error bars are still too large for a firm conclusion, the slope parameter in $v_2$ deviates from that in $v_3$ in the peripheral region where magnetic field could be more important. It should also be emphasized that whether the slope parameters in $v_2$ and $v_3$  contributed by  pure backgrounds should be quantitatively the same or not, is still an open question that calls for more investigations.   

Finally, the right panel of Fig.~\ref{fig:cmw_star} shows the extracted slope parameter in $v_2$ for a variety of colliding systems at RHIC, from p+Au and d+Au to Au+Au and U+U collisions. The slope parameter is consistent with zero for small colliding systems and very peripheral collisions of A+A systems, where magnetic field effects are expected to be absent. From the peripheral to mid-central regions, one observes a statistically significant positive slope which could be explained by CMW-induced charge transport. The slope approaches zero again toward the very central region which again could be consistent with the expected decrease of magnetic fields. The overall pattern appears to support an interpretation of this observable in terms of charge quadruple formation induced by CMW transport.

\subsubsection{LHC results}

Measurements of CMW-motivated observables were performed by CMS and ALICE at LHC energies. While one might expect that effects of magnetic field at LHC could be rather small due to its significantly shorter duration, it would be useful to examine and understand the behavior of pertinent observables under such circumstances.

\begin{figure}[h]
    \centering
    \includegraphics[height=0.38\textwidth]{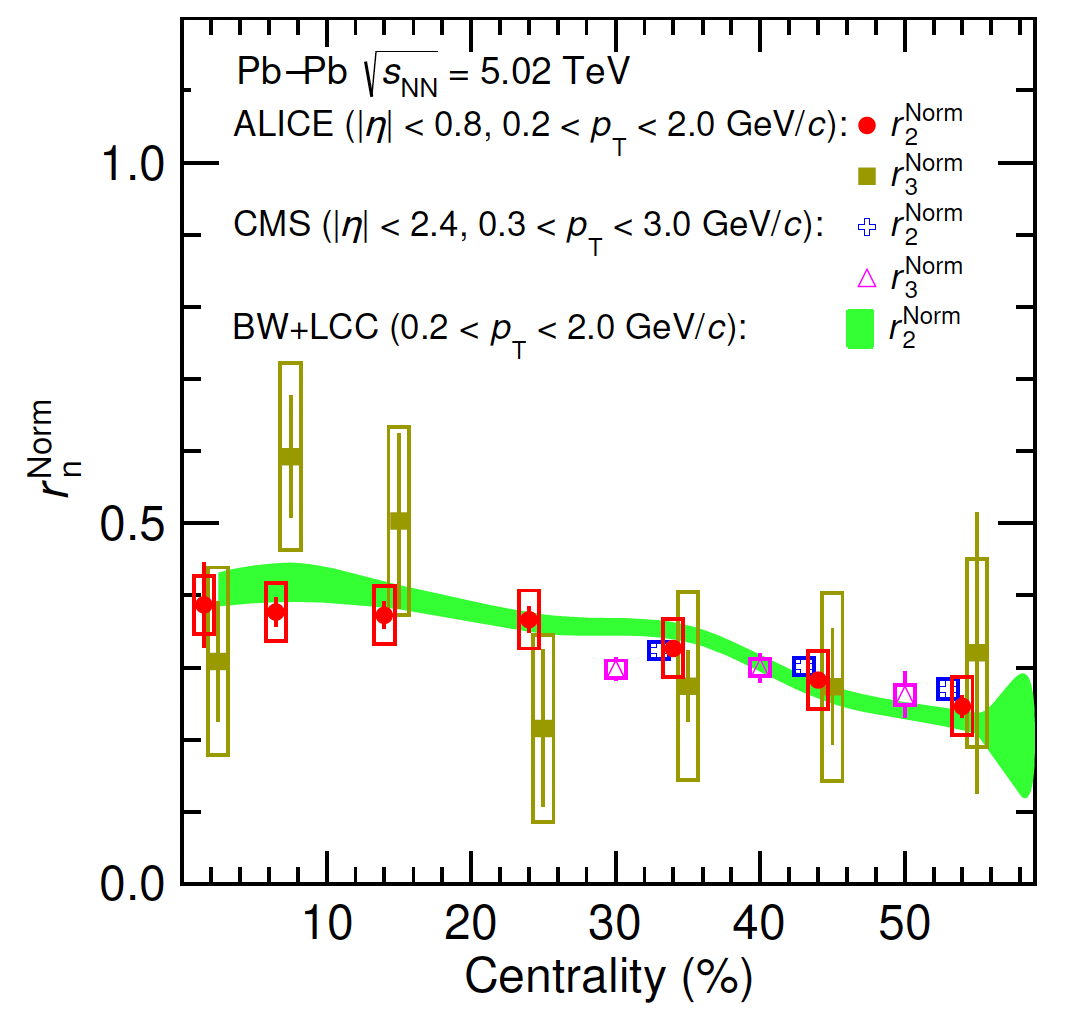} \hspace{0.1in}
    \includegraphics[height=0.38\textwidth]{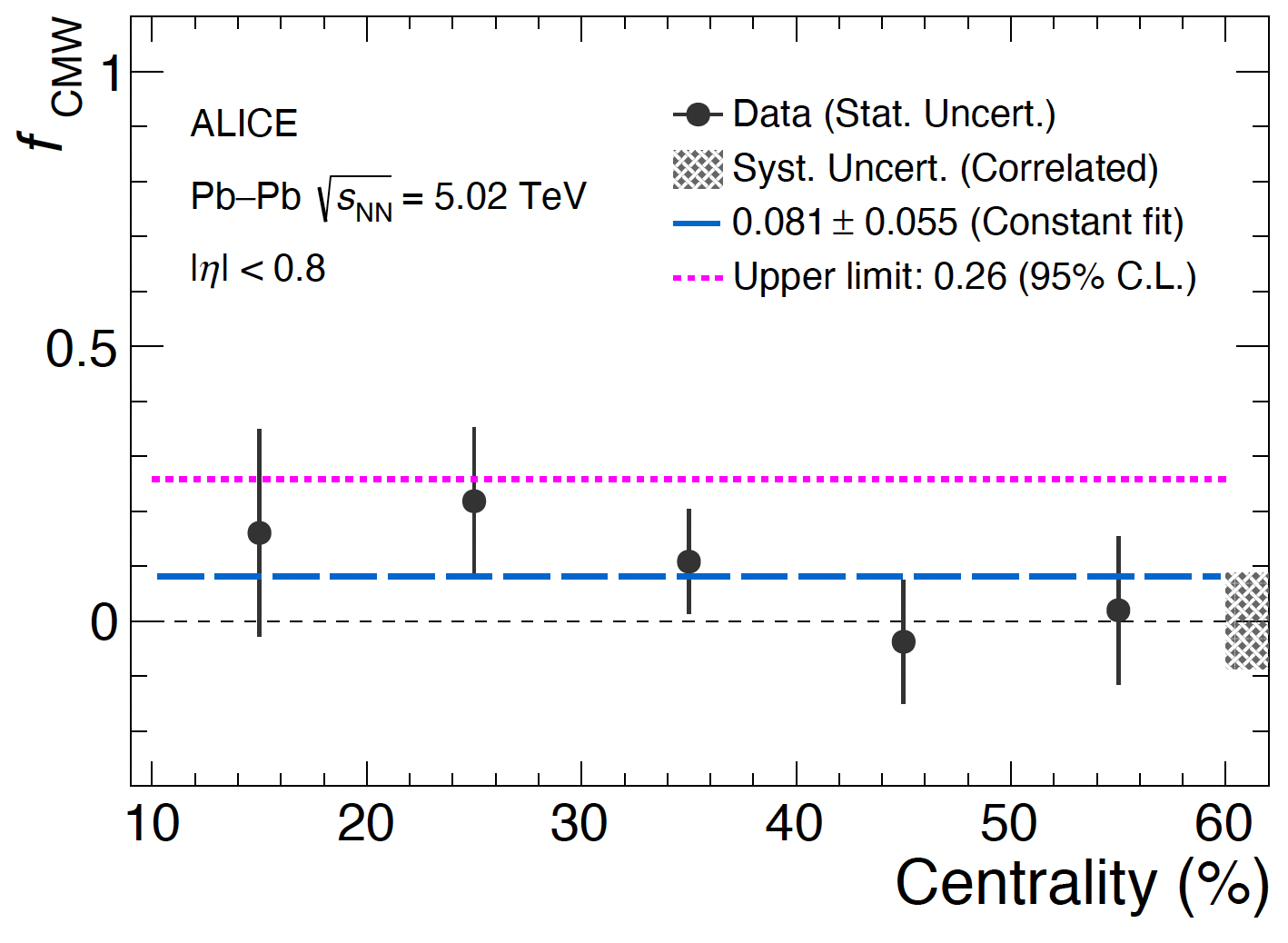}
    \caption{(left) The slope parameters extracted from the elliptic as well as triangular flow differences between  $\pi^{\mp}$ versus   event charge asymmetry $A_{ch}$ as a function of centrality, measured by ALICE Collaboration for Pb+Pb collisions at LHC and compared with measurements from CMS as well as with blast-wave model calculations. 
(right) The extracted upper limit for possible CMW signal fraction in the total slope parameter by ALICE Collaboration for Pb+Pb collisions at the LHC.  
(See details in \cite{ALICE:2023weh}).    
    }
    \label{fig:cmw_alice}
\end{figure}

The first results from ALICE in \cite{ALICE:2015cjr} for Pb+Pb collisions at 2.76~TeV, while  confirming a similar linear dependence of $v_2$ splitting on charge asymmetry with sizable slope parameter, also demonstrate a qualitatively similar behavior in higher harmonics like $v_3$ and $v_4$, thus pointing to potentially strong background contributions toward such slope parameter. Shortly after, CMS also reported their measurements for Pb+Pb as well as p+Pb collisions at 5.02~TeV  in \cite{CMS:2017pah}. The CMS results not only show a similarity between the $v_2$ and $v_3$ splitting patterns in Pb+Pb collisions but also suggest a similarity in $v_2$ splitting patterns between Pb+Pb and p+Pb collisions. These results collectively suggest a strong LCC background contribution to the charge asymmetry in collective flow patterns and could help the effort to achieve a quantitative understanding of such background.

The state-of-the-art results for the search of CMW at LHC energies, reported by ALICE in \cite{ALICE:2023weh}, are shown in Fig.~\ref{fig:cmw_alice}. The left panel shows a comparison among ALICE and CMS extractions of slope parameters in $v_2$ and $v_3$ as well as an estimate of background contribution based on blast-wave model with LCC. Based on such comparison, an extraction of potential CMW contribution to the flow splitting was obtained for the first time by ALICE, shown in  the right panel of Fig.~\ref{fig:cmw_alice}, suggesting a potential CMW signal level at $(8\pm 6)\%$ of the overall slope parameter for the flow splitting.

Finally, it is worth noting that a unified understanding of both CME and CMW observables, in which  anomalous chiral  transport as well as  background correlations (such as LCC) coexist and compete, would be very valuable. Efforts start to be made by utilizing  quantitative simulations to put global constraint on the magnitude of anomalous chiral effects~\cite{Wu:2022fwz}.

\subsection{Connections with other heavy ion measurements }


In addition to the charge-dependent azimuthal correlations for the direct search of CME signatures, there have been many interesting  measurements in heavy ion collision experiments that bear connection with and provide valuable insights for the CME search. Let us list a number of recent examples below, along with some brief discussions.

\begin{itemize}

\item {Ultraperipheral collisions (UPC) --- A fundamental component of CME transport is the presence of strong magnetic fields generated by the high-speed initial ions. The direct experimental confirmation of the existence and intensity of these initial vacuum electromagnetic fields has been demonstrated through intriguing UPC measurements, notably in processes such as $\gamma\gamma \rightarrow e^+e^-$ (sometimes called the Breit-Wheeler process) \cite{STAR:2019wlg}. The detection of two electrons from UPC $\gamma\gamma \rightarrow e^+e^-$ processes suggests the presence of electromagnetic fields (EM-field) on the order of the electron mass squared ($\sim m_e^2$). The most recent measurement of exclusive lepton pair cross sections at the LHC and $\tau$ lepton production ($\gamma\gamma\rightarrow \tau$) \cite{CMS:2022arf} indicate even larger ($\sim m_\tau^2>\sim m_e^2$) EM-field strength. It is comparable to the theoretical order of the magnetic field strength of the order of pion mass squared ($eB\sim m_\pi^2$)~\cite{Skokov:2009qp}. Alternatively, in $\gamma+A \rightarrow J/\psi$, also called exclusive vector meson production processes measured in isobar Ru+Ru and Zr+Zr collisions, the electromagnetic field is expected to have a larger cross-section in Ru+Ru due to a stronger magnetic field, as confirmed by preliminary data from STAR~\cite{Wang:2023kzx,Tribedy:2023pjj}. As discussed, despite using isobar collisions to search for CME, no significant CME difference has been found. However, $J/\psi$ measurements suggest that UPCs may serve as a means to verify the necessary prerequisites for CME were present. It must be noted that although UPCs serve as an independent way of confirming the initial EM-field is created, there are differences to be noted. In UPCs, the electromagnetic field is a combination of electric and magnetic fields existing in vacuum. For CME, we need strictly large initial magnetic fields. UPC processes provide us with knowledge of vacuum EM-fields. Once a medium of quark or gluon is formed, questions arise about what happens to these fields and how much of it is trapped by the medium, for which other alternative measurements may be important.} 

\item{In-medium presence of strong magnetic field -- In heavy-ion collisions, the search for direct evidence of EM-field effects in the medium continues. Theoretical predictions suggest that the Coulomb, Faraday and Hall effects would induce different slopes of directed flow ($d\Delta v_1/dy$) between positive and negative particles. The STAR Collaboration measured charge-dependent $v_1$ splitting between positively and negatively charged hadrons in Cu+Au and Au+Au collisions at $\sqrt{s_{\mathrm{NN}}}=200$~GeV~\cite{STAR:2016cio}. A significant $v_1$ splitting was observed in Cu+Au, attributed to the stronger Coulomb force in the mass-asymmetric system compared to Au+Au. The ALICE Collaboration reported a similar $v_1$ splitting in Pb+Pb collisions at $\sqrt{s_{\mathrm{NN}}}=5.02$~TeV~\cite{ALICE:2019sgg}. Moreover, both STAR and ALICE observed sizable $v_1$ values for $D^0$ and $\bar{D}^0$ mesons, as well as a non-zero $v_1$ splitting between them in Au+Au and Pb+Pb collisions at $\sqrt{s_{\mathrm{NN}}}=200$~GeV and $\sqrt{s_{\mathrm{NN}}}=5.02$~TeV, respectively~\cite{STAR:2019clv,ALICE:2019sgg}. A recent study by the STAR collaboration in Au+Au and isobar collisions, focusing on protons and anti-protons, revealed a sign change of the slope difference from central to peripheral events, indicating the influence of the EM field, especially the dominance of the Faraday effect~\cite{STAR:2023jdd}. Another study by the STAR collaboration in ref~\cite{STAR:2023wjl} highlighted a splitting in $v_1$ slope correlated with the electric charge and strangeness content differences of the constituent quarks, especially prominent in mid-central collisions, suggesting a possible presence of the dominant Hall effect in central collisions. The overall results of the directed flow splitting are consistent with the presence of a strong EM field in the medium. For the first time, dedicated measurements point us in this direction. The exact strength of the field requires input from models. An alternative explanation of the directed flow splitting based on baryon inhomogeneity has been discussed in a recent paper~\cite{Parida:2023ldu}, suggesting that no EM-field effect is needed to explain the observation of net-proton splitting and charge dependence on centrality. The effect of the directed flow splitting with electric charge can be interpreted as splitting with net-baryon or strangeness. Another observable that can be directly influenced by the presence of the magnetic field is the splitting between the global polarization of the $\bar{\Lambda}$ and $\Lambda$ in any collision systems. A recent measurement done in 27 GeV and 19 GeV Au+Au collisions~\cite{STAR:2023nvo} found no significant splitting at these collision energies and therefore sets an upper limit of $P_{\bar{\Lambda}}-P_{\Lambda}<0.24\%$ and $P_{\bar{\Lambda}}-P_{\Lambda}<0.35\%$, respectively, at a 95\% confidence level. This in turn leads to an upper limit on the naive extraction of the late-stage magnetic field of $B<9.4\times 10^{12}$ T and $B<1.4\times 10^{13}$ T at $\sqrt{s_{\mathrm{NN}}}=19.6$ GeV and $\sqrt{s_{\mathrm{NN}}}=27$ GeV, respectively. Although more thorough derivations are needed, this is one of the few direct methods to put constraints on the strength of magnetic fields. Overall, the direct evidence for the presence of a strong magnetic field is still under discussion. However, recent progress looks promising and quite encouraging for the search for CME.} 

\item {Magnetic field lifetime --- The lifetime of the in-medium magnetic field, a crucial parameter for CME transport, remains poorly constrained. This lifetime relates inversely to the medium's conductivity, as shown in~\cite{McLerran:2013hla}. Constraining this parameter could be achieved by searching for residual magnetic fields at later stages, such as freeze-out. As discussed earlier, directed flow splitting (the difference in directed flow between positive and negative particles) is expected to be affected by the in-medium electromagnetic field. Interestingly, recent measurements at lower collision energies have observed an increase in directed flow splitting~\cite{STAR:2023jdd,STAR:2023wjl,Dash:2024qcc}. Assuming this splitting is indeed due to the electromagnetic field, the observed energy dependence aligns with the expectation of a longer field lifetime at lower energies. However, no quantitative estimates on the upper limit of medium conductivity have been performed to date. Further data-model comparisons, particularly with measurements like the energy dependence of directed flow splitting, could be a promising avenue for achieving this goal.}

\item {Spin alignment and its implication to CME: Spin alignment of vector mesons is a phenomenon similar to that of spin polarization hyperon, but with differences in terms of their origin. The interest is the spin alignment coefficient $\rho_{00}$ as a function of the collision energy. For the $\phi$-mesons, we observe an 8.4$\sigma$ deviation from the baseline value of 1/3. The $K^{*0}$ results are consistent with 1/3 in Au+Au collisions at $\sqrt{s}=7.7-200$ GeV. An outstanding question is what causes the spin alignment of vector mesons and whether the source is the same as that of the hyperon global polarization that is due to the vortical structure and the global angular momentum of the system. It is hard to reconcile the fact that the sole source of spin alignment is the vortical structure. It seems that a model that includes a very strong vector meson field of the order of $m_\pi^2$ can provide an explanation for the $\phi$-mesons. What about the $K^*0$ mesons? For that, we perform new measurements of charged and neutral $K^{*0}$ mesons in isobar collisions to gain more insights \cite{Singha:2022syo}. Could there be a component from the magnetic field? The expected deviation of $\rho_{00}$ from 1/3 due to the vorticity and the magnetic field is $\sim 10^{-5}$ based on~\cite{Yang:2017sdk}, significantly less than its observed magnitude. It is not clear if spin alignment studies can put strong constraints on the magnetic field that would be informative for the CME search. However, a recent study indicates an interesting interplay between the vector meson spin alignment and the CME-induced charge separation, as suggested in e.g.~\cite{Shen:2022gtl}. The work demonstrates how the globally spin-aligned $\rho$ mesons affect the CME observables involving pions. More specifically, spin alignment leads to a background-like effect like elliptic anisotropy of neutral resonances such as the $\rho$ meson. As we discussed before, if a mother $\rho$ is preferentially emitted along the reaction plane due to its elliptic flow, its daughter pions will lead to a fake charge separation that will be indistinguishable from the CME. One can think of the spin alignment of a $\rho$ meson in the same way. When a $\rho$ meson is spin aligned, it can be shown that its daughters will give rise to a fake charge separation that would mimic the CME~\cite{Shen:2022gtl}. Unlike the flow background that is always positive, the global spin alignment of vector mesons can give a negative contribution if $\rho_{00}$ is smaller than 1/3, which is likely according to data. This further warrants the inclusion of the spin alignment effect in the background estimation, to avoid an over-subtraction of the background.}

\item {Correlation between charge separation and spin polarization --- A key ingredient of CME transport is the axial charges arising from QCD topological fluctuations, which nevertheless is hard to access. It is not easy to come up with an observable to measure the chirality of the quarks in collisions. A recent proposal to probe such fluctuations is to look for correlations between the CME-induced charge separation and the hyperon spin polarization on an event-by-event basis~\cite{Finch:2017xiz,Becattini:2020xbh}, with the first such measurement reported by STAR in Ref.~\cite{STAR:2023qyt} in Au+Au collisions at 27 GeV. The idea is to use $\Lambda$ hyperons that are self-analyzing in terms of their decay topology, meaning if one can reconstruct the decay $\Lambda\rightarrow p+\pi^-$ in the rest frame of the $\Lambda$, then the momentum direction of $p$ gives the orientation of the spin of the $\Lambda$. Given the spin orientation of the $\Lambda$ and its momentum, it is possible to obtain the helicity of the $\Lambda$. One can then in an event extract the number of left and right-handed $\Lambda$ hyperons as extract the normalized excess as $\Delta n = (N_L^\Lambda - N_R^\Lambda)/(N_L^\Lambda + N_R^\Lambda)$. Although $\Delta n$ measures the handedness of $\Lambda$ hyperons, it doesn't directly represent the chirality of the quarks. Yet, assuming that the quarks' chirality preference is transferred to the $\Lambda$ hyperons in the final state, one can analyze the event-by-event fluctuations in $\Delta n$ and correlate them with observables such as $\Delta a_1=a_1^+ - a_1^-$, where $a_1$ is defined in Eq.\ref{eq_a1_observables}. Since $\Delta a_1$ measures charge separation across the reaction plane, it is expected to be highly sensitive to the chirality imbalance of the quarks during collisions. Unlike average $\langle \Delta a_1 \rangle$ correlations, measures like $\langle \Delta n \, \Delta a_1 \rangle$ do not vanish after averaging, as both components are parity-odd, making the product parity-even. The first measurements from STAR~\cite{STAR:2023qyt} do not show any significant correlation between $\Delta n$ and $\Delta a_1$. But exploration in this direction with more precise data would be informative towards CME search.}
\end{itemize}




\section{Future opportunities}

\subsection{Open theoretical problems }

In this section, we list some of the open theoretical problems that need to be addressed in the near future.

\begin{itemize}

\item Chern-Simons fluctuations and CME near the critical point of the QCD phase diagram.
\vskip0.3cm

QCD phase diagram is expected to possess a critical endpoint, where the line of the first order phase transition, when moving from the domain of high baryon density to lower density, terminates \cite{Stephanov:1998dy,Stephanov:1999zu}. The fluctuations of the order parameter and the corresponding susceptibilities should grow near this critical point. What happens to the fluctuations of Chern-Simons number in this case? Since Chern-Simons number is coupled to the chiral charge density by the anomaly, it may be expected that the Chern-Simons fluctuations, and the corresponding fluctuations of the CME current, are also enhanced in the vicinity of the critical point. 

Indeed, within the Polyakov-Nambu-Jona-Lasinio model it has been found that the fluctuations of CME current are enhanced near the critical point \cite{Fukushima:2010fe}. Moreover, a study of the real-time dynamics of massive Schwinger model revealed a sharp peak in Cern-Simons diffusion rate near the critical point of the theory at $\theta = \pi$ \cite{Ikeda:2020agk}. This critical point closely resembles the one in QCD, and is also the endpoint of the first order phase transition at $m/g \simeq 0.33$ ($m$ is the fermion mass, and $g$ is the coupling). This growth of the real-time fluctuations of Chern-Simons number should lead to the enhancement of the CME current fluctuations in the vicinity of the critical point. 
\vskip0.3cm

It is very important to investigate this problem further theoretically and especially experimentally, by measuring the CME observables as a function of center-of-mass energy in heavy ion collisions. There is a preliminary STAR result \cite{Xu:2024xqc} indicating the presence of CME at $3\sigma$ level at energies where the critical point is expected to manifest itself.
\vskip0.3cm

    \item Self-consistent treatment of Chern-Simons fluctuations taking into account back-reaction from chirally imbalanced matter and the chiral magnetic current.
\vskip0.3cm

Usually, Chern-Simons diffusion rate is computed without taking into account the back-reaction from the chirally imbalanced matter. Nevertheless, this back-reaction can be very important. Indeed, suppose that a sphaleron process creates a configuration with a non-zero Chern-Simons number in the quark-gluon plasma. According to the anomaly equation, each sphaleron process changes chirality of $N_f$ quarks, and creates a local chirality imbalance. As a result, the local energy density increases, and this breaks the degeneracy between the probabilities of topological transitions by favoring the transition that reduces the chirality imbalance. The Cherrn-Simons diffusion therefore ceases to be a random walk once the back-reaction from the fermion sector is taken into account. 
\vskip0.3cm
    
    \item Development of dissipative chiral magnetohydrodynamics.
    \vskip0.3cm

Magnetohydrodynamics is usually considered in the limit of large electric conductivity. In this limit, the electric fields in the fluid are 
screened, and only magnetic fields appear in the equations of motion. From the viewpoint of chirality, this means that magnetic helicity of the fluid $\int d^3x\ {\bf A}\cdot {\bf B}$ (Abelian Chern-Simons number) is a conserved quantity, since its change in time necessarily involves the electric field ${\bf E} = -{\bf \dot{A}}$. Because of this, once the finite electric conductivity is taken into account and the electric field is not completely screened, magnetic helicity is no longer a conserved quantity in the fluid. This should significantly affect the dynamics of chiral charge relaxation in the fluid, and will modify the CME conductivity at finite frequency.
\vskip0.3cm

    \item Understanding the role of chiral anomaly in spin polarization in the quark-gluon plasma.
    \vskip0.3cm

    Observation of spin polarization of $\Lambda$ hyperons in heavy ion collisions at RHIC has been a very important advance in understanding the role of vorticity in quark-gluon plasma. Spin polarization of quarks can be a consequence of the chiral separation effect at finite baryon chemical potential, i.e. the generation of non-zero expectation value of the axial current expected as a consequence of anomaly both in an external magnetic field and at finite vorticity. It would be important to clarify what role these anomalous effects play in the observed $\Lambda$ polarization. On a more fundamental level, it is not yet clear whether the coupling of spin to vorticity can be understood as an effect of (mixed gauge-gravitational) chiral anomaly.

\vskip0.3cm
    
    \item Transition from chiral magnetohydrodynamics to spin magnetohydrodynamics as a function of fermion mass
    \vskip0.3cm

At zero fermion mass, chirality is a relevant quantum number, whereas for heavy fermions the relevant degree of freedom is spin, with arbitrary projection on the direction of momentum. Different theoretical methods have been used so far to construct chiral hydrodynamics of massless strongly interacting fermions and spin hydrodynamics of massive ones. It is clear that there must exist a unified approach that would allow to vary fermion mass and to obtain the chiral and spin hydrodynamics as different limits. It would be important to formulate such an approach.
    \vskip0.3cm
    
    \item Correlating charge asymmetry with other parity-odd observables, for example with baryon number asymmetry
    \vskip0.3cm
    
As we discussed at length throughout this review, the main difficulty in observing the charge separation predicted by the CME is separating this effect from mundane backgrounds. Because there is no global ${\mathcal{P}}$ and ${\mathcal{CP}}$ violation in QCD, the obserables that are first-order in charge separation vanish, and the second-order ones suffer from background contamination. One may however overcome this problem by correlating electric charge separation with the baryon number separation. A mixed electric charge -- baryon number correlator should not vanish even if it is first order in electric charge separation, because the chiral anomaly predicts a definite correlation between the separation of charge and baryon number \cite{Kharzeev:2010gr,Frenklakh:2024jff}. It would be interesting to study these correlators both experimentally and theoretically.

    \vskip0.3cm
    
    \item Quantitative predictions for CME, CVE, CMW and other anomaly-induced phenomena in heavy ion collisions
    \vskip0.3cm
A very significant progress has been made in recent years in quantitative modeling of CME and related phenomena. Solving the problems outlined above would allow making another step in this direction, and define a theoretical baseline that can be compared to experimental measurements.
    \vskip0.3cm
    \item New manifestations of CME in condensed matter physics: optics, Floquet dynamics, quantum sensing, quantum computing, ...
    \vskip0.3cm
By now, the existence of CME has been firmly established through magnetotransport in many Dirac and Weyl semimetals. In the limit of large chirality flipping rate, CME implies a new type of superconductivity \cite{Kharzeev:2016tvd}. However, in real materials chirality flipping rate is still significant, even though it can be several orders of magnitude smaller than the scattering rate. This limits the value of longitudinal DC magnetoconductivity. Nevertheless, when the frequency of external electric field is larger than chirality flipping rate, the material behaves as an almost perfect chiral medium with conserved chirality. This is why the optical experiments with chiral materials, including measurements of optical magnetoconductivity (optical CME) and CME current induced by circularly polarized light in an external magnetic field become the next frontier in the study of chiral anomaly effects in condensed matter. Chirally imbalanced states in Dirac and Weyl semimetals can be synthesized by periodic optical fields, and their properties can be theoretically addressed using Floquet dynamics. 
Applications of optical CME phenomena include quantum sensors, and perhaps even quantum computing with chiral qubits \cite{Kharzeev:2019ceh}, where the quantum superpositions of left- and roght-handed fermions are controlled by circularly polarized photons.

    \vskip0.3cm
    \item The role of CME and other anomaly-induced phenomena in the Early Universe: baryon asymmetry, magneto-genesis, generation of circularly-polarized gravitational waves, ...
\vskip0.3cm

Recently there has been a lot of interest in the possible role of CME in the Early Universe, see e.g. \cite{shao2023gravitational,schober2024chiral,brandenburg2024relic}. In particular, CME can be responsible for magneto-genesis -- the generation of large-scale magnetic fields characterized by non-zero magnetic helicity. This mechanism is powered by the chiral magnetic instability \cite{giovannini1998primordial}, which is the decay of a chirally imbalanced state of fermions into helical magnetic fields. The corresponding process is a self-similar inverse cascade \cite{hirono2015self}, and leads to the formation of large-scale helical magnetic fields, potentially explaining the current cosmological observations \cite{brandenburg2023chirality}. An interesting open question is the possibility of the emission of circularly polarized gravitational waves due to the mixed gauge-gravitational chiral anomaly. This emission could take place before and during the electroweak phase transition, for example due to electroweak sphaleron transitions (distorted by non-zero Weinberg angle) acting as local sources of chirality \cite{Kharzeev:2019rsy}. 
    
\end{itemize}

\subsection{Experimental quest for  CME  }

\begin{itemize}

\item {Prospects of CME Search Beyond Isobar-era: Experimental verification of the CME remains a key goal at RHIC. However, with the RHIC program approaching its final years, the window for dedicated CME searches is narrowing. Isobar collisions presented a unique opportunity to detect CME, revealing a notable contrast in $B^2$ and CME correlation signals between Ru+Ru and Zr+Zr collisions. Despite this, a statistically significant CME signal (with $5\sigma$  significance) was not established.  This outcome aligns with expectations, as isobar collisions have a smaller signal-to-background ratio~\cite{Feng:2021oub} compared to Au+Au collisions, where a positive CME signal of approximately 10\% with $2-3$ $\sigma$ significance has been observed ~\cite{STAR:2021pwb}. Efforts have been made to reevaluate the isobar data to extract the CME fraction, and further post-blind analysis may be conducted in the future. Notably, the observation of a 10\% signal in isobar collisions suggests an anticipation of a larger signal in larger collision systems. This underscores the importance of continued CME search with high-statistics Au+Au data from the upcoming RHIC runs or Pb+Pb collisions at the LHC in the future.}
    \vskip0.3cm
    
\item {Spectator/participant method: Analyzing Au+Au 200 GeV data during the 2023-25 run can achieve a significance level of $5 \sigma$ or higher  in CME-related measurements. Preliminary analysis of Au+Au 200 GeV data using participant vs spectator plane method shows a 4\% statistical uncertainty with 2.4 billion events (2-3$\sigma$ significance). With an anticipated 20 billion events during Run-23 and 25, achieving  significance beyond $5\sigma$ is plausible. Factors like enhanced B-field while going from isobar to Au+Au collisions and the acceptance increase due to iTPC upgrade will contribute to higher significance. The advantage of this method is that it has demonstrated sensitivity directly to the signal fraction in the overall measured correlations. The challenge at present is the large uncertainty, which is expected to be substantially improved with the anticipated large-statistics dataset to be taken.} 
    \vskip0.3cm
    
\item{Event Shape Engineering (ESE) Method:
The ESE method aims to suppress $v_2$-related background in CME-sensitive correlators. Using the flow vector, ESE selects events with minimal flow, effectively reducing background interference. Application to 30-40\% Au+Au collisions with RHIC BES-II demonstrates successful background reduction. ESE can be applied to future Au+Au 200 GeV data from the year 2023-25 run of RHIC, allowing more differential measurements. The advantage of ESE method is that it has a lot of precision. This however comes with a caveat that the efficacy of the method (how much background is removed) is an unknown parameter. Use of baseline observables or analysis in different invariant mass regions for which CME signal is expected to be vanishing as control measure could be possible ways to resolve this.}
    \vskip0.3cm

\item {Alternate Observables: Event-by-event correlations between CME charge separation and other parity-odd features are an intriguing avenue to explore. Correlations between charge separation and the net helicity of $\Lambda$($\bar{\Lambda}$) in each event can reveal unique insights into local parity violation and CME~\cite{Finch:2017xiz,Becattini:2020xbh}. A strong correlation would confirm the presence of CME. Event requirements for such correlations are being estimated, with possible implications for understanding strange quark contributions. A recent STAR analysis in this direction is an important step~\cite{STAR:2023qyt} , but no significant signal was observed in the data in Au+Au 27 GeV collisions. Higher energies may be a path forward, as they offer larger yields of $\Lambda$ although lower effective resolution for the proxy of reaction plane using spectators or fragments. The upcoming RHIC runs from 2023-25 with an anticipated statistics of 20 Billion Au+Au 200 GeV events still provide a unique opportunity.}
    \vskip0.3cm

\item {Opportunities at the LHC: As the RHIC program approaches its final years, the opportunity for dedicated CME searches is narrowing. However, the search can continue at the LHC, despite some apparent disadvantages such as higher energies leading to smaller domains of chiral imbalance and more transient magnetic fields. There are also advantages, such as larger initial strength of magnetic field and experimental benefits, such as higher particle yield, better event-plane capabilities, and more precise and differential measurements, such as with invariant mass. The spectator plane reconstruction using ZDCs~\cite{ALICE:2014sbx}
and the spectator-participant plane method to estimate CME signals are feasible at the LHC. The possibility of an isobar run similar to RHIC at the LHC has not been explored yet~\cite{Bhatta:2023cqf}, but high-statistics Pb+Pb data collection in upcoming LHC runs with novel techniques such as ESE and spectator/participant methods can be a promising way to proceed.} 
    \vskip0.3cm

\item {Exploring Chiral Vortical Effect: The CVE remains an intriguing yet underexplored phenomenon in heavy-ion collisions. Unlike the CME, which relies on the presence of a magnetic field, the CVE is contingent upon the existence of vortical fields~\cite{Kharzeev:2010gr}. While the direct manifestation of magnetic fields in mid-central heavy-ion collisions is still a subject of discussion~\cite{STAR:2023wjl,STAR:2023jdd}, the observation of vortical effects, such as through the measurement of global lambda polarization, is widely acknowledged~\cite{STAR:2017ckg,Pang:2016igs}. This underscores the significance of investigating the CVE further. Recent efforts, exemplified by the work of the STAR collaboration~\cite{Xu:2023wcy}, have attempted to probe the CVE using correlations between $\Lambda$ and $p$, employing the observable $\Delta\gamma$. However, thus far, no significant measurements have been reported. Nonetheless, the outlook for future investigations appears promising. Similarly, recent measurements at the LHC~\cite{Wang:2023xhn} have shed light on the CVE, particularly through the study of $\delta$ and $\gamma$ correlators involving baryon pairs like $\Lambda$–$p$. While preliminary observations suggest a notable separation in $\delta$ and $\gamma$ for $\Lambda$–$p$ pairs, further analysis is required to disentangle potential background contributions. These findings underscore the need for continued research to unravel the complexities surrounding the CVE and its implications in heavy-ion collisions. Further exploration with high-statistics upcoming RHIC and LHC data would be invaluable in advancing our understanding of the CVE and its role in heavy-ion collisions.}
    \vskip0.3cm
    
\item {Chiral Magnetic Wave: While extensive searches have been conducted for the Chiral Magnetic Effect (CME), the same level of attention has not been afforded to the exploration of the Chiral Magnetic Wave (CMW). It is crucial to note that CMW signatures can manifest even in the absence of detectable CME signals. This realization underscores the necessity to expand beyond the limited dedicated experimental searches conducted thus far~\cite{STAR:2015wza,ALICE:2015cjr,CMS:2017pah,STAR:2022zpv,ALICE:2023weh}. Novel methodologies have emerged to detect subtle CMW signals, such as employing three-particle correlators~\cite{Belmont:2014lta}, ESE~\cite{ALICE:2023weh} and devising strategies to mitigate auto-correlation effects~\cite{Xu:2020sln}. With the refinement of these tools, more future measurements are anticipated.}
    \vskip0.3cm
    
\item {The nonflow problem: Understanding flow and non-flow backgrounds is essential for CME searches. While the understanding of flow-driven background has improved over the years, satisfactory quantitative estimates of non-flow effects remain elusive.  Recent STAR analyses have made strides in addressing non-flow effects~\cite{STAR:2023ioo,STAR:2023gzg}. While challenges persist, novel data-driven methods offer promise, particularly in larger collision systems where traditional approximations may falter. Previous studies have established analysis procedures for non-flow contributions, utilizing techniques such as 2-dimensional distributions and variations in pseudorapidity. Further research is needed to refine these approaches and enhance our understanding of non-flow effects.}
    \vskip0.3cm
    

\end{itemize}



\section{Conclusion}

Over the past twenty years, there has been an impressive theoretical and experimental progress in understanding how the chiral anomaly induces the chiral magnetic effect and a variety of related macroscopic quantum transport phenomena. These phenomena appear in all systems possessing chiral fermions, whether fundamental (as in QCD) or emergent (as in Dirac and Weyl materials). 
As a result, the CME arises at different scales ranging from femtometers to parsecs. 

The experimental discovery of CME in condensed matter systems, and the confirmation of its existence in first principle through analytical and lattice computations, opens the path towards using it as a tool for extracting the information about the topological structure of QCD from measurements in heavy-ion collision experiments.  

This program, as reviewed above, requires accumulating high statistics heavy ion data and developing sophisticated analysis methods. As a result of a dedicated effort at RHIC and the LHC, extensive information has been accumulated about the charged hadron correlations, and an evidence for the CME in AuAu collisions has emerged in the RHIC energy region. The theoretical understanding has also improved significantly, and allows a quantitative interpretation of the expected CME signal. 

It is imperative now to reach a definitive conclusion on the presence or absence of CME in the final years of RHIC operation. This is a difficult task that will require a concerted effort by experimentalists and theorists together. However, given the fundamental importance of obtaining direct information about topology of non-Abelian gauge theories, and the broad impact of these findings, we believe that this is a task that has to be accomplished in the next several years. 



\section*{Acknowledgements}

The authors thank  
Francesco Becattini, 
Jinhui Chen, 
Kenji Fukushima, 
Xingyu Guo, 
Defu Hou, 
Anping Huang,
Huanzhong Huang, 
Mei Huang, 
Xu-Guang Huang, 
Yin Jiang, 
Dirk Rischke, 
Shuzhe Shi, 
Qiye Shou, 
Igor Shovkovy, 
Mikail Stephanov, 
Aihong Tang, 
Sergei Voloshin, 
Fuqiang Wang, 
Gang Wang,  
Zhangbu Xu, 
Ho-Ung Yee, 
Yi Yin, 
Pengfei Zhuang 
for useful discussions, communications and collaborations that have contributed to this review article. 
This work was supported in part by  
the U.S. Department of Energy (Grant No. DE-FG88ER41450 and DE-SC0012704)     
as well as
by the U.S. National Science Foundation (Grant No.~PHY-2209183).

\bibliographystyle{ws-rv-van}
\bibliography{cme.bib}

\begin{thebibliography}{325}
\providecommand{\natexlab}[1]{#1}
\providecommand{\url}[1]{\texttt{#1}}
\expandafter\ifx\csname urlstyle\endcsname\relax
  \providecommand{\doi}[1]{doi: #1}\else
  \providecommand{\doi}{doi: \begingroup \urlstyle{rm}\Url}\fi

\bibitem{Kharzeev:2004ey}
D.~Kharzeev, {Parity violation in hot QCD: Why it can happen, and how to look
  for it}, \emph{Phys. Lett. B}. {\bf 633}, \penalty0 260--264  (2006).
\newblock \doi{10.1016/j.physletb.2005.11.075}.

\bibitem{Kharzeev:2007jp}
D.~E. Kharzeev, L.~D. McLerran, and H.~J. Warringa, {The Effects of topological
  charge change in heavy ion collisions: 'Event by event P and CP violation'},
  \emph{Nucl. Phys. A}. {\bf 803}, \penalty0 227--253  (2008).
\newblock \doi{10.1016/j.nuclphysa.2008.02.298}.

\bibitem{Fukushima:2008xe}
K.~Fukushima, D.~E. Kharzeev, and H.~J. Warringa, {The Chiral Magnetic Effect},
  \emph{Phys. Rev. D}. {\bf 78}, \penalty0 074033  (2008).
\newblock \doi{10.1103/PhysRevD.78.074033}.

\bibitem{Kharzeev:2015znc}
D.~E. Kharzeev, J.~Liao, S.~A. Voloshin, and G.~Wang, {Chiral magnetic and
  vortical effects in high-energy nuclear collisions\textemdash{}A status
  report}, \emph{Prog. Part. Nucl. Phys.} {\bf 88}, \penalty0 1--28  (2016).
\newblock \doi{10.1016/j.ppnp.2016.01.001}.

\bibitem{Landsteiner:2016led}
K.~Landsteiner, {Notes on Anomaly Induced Transport}, \emph{Acta Phys. Polon.
  B}. {\bf 47}, \penalty0 2617  (2016).
\newblock \doi{10.5506/APhysPolB.47.2617}.

\bibitem{Kharzeev:2020jxw}
D.~E. Kharzeev and J.~Liao, {Chiral magnetic effect reveals the topology of
  gauge fields in heavy-ion collisions}, \emph{Nature Rev. Phys.} {\bf
  3}\penalty0 (1), \penalty0 55--63  (2021).
\newblock \doi{10.1038/s42254-020-00254-6}.

\bibitem{Zhao:2019hta}
J.~Zhao and F.~Wang, {Experimental searches for the chiral magnetic effect in
  heavy-ion collisions}, \emph{Prog. Part. Nucl. Phys.} {\bf 107}, \penalty0
  200--236  (2019).
\newblock \doi{10.1016/j.ppnp.2019.05.001}.

\bibitem{Li:2020dwr}
W.~Li and G.~Wang, {Chiral Magnetic Effects in Nuclear Collisions}, \emph{Ann.
  Rev. Nucl. Part. Sci.} {\bf 70}, \penalty0 293--321  (2020).
\newblock \doi{10.1146/annurev-nucl-030220-065203}.

\bibitem{Huang:2015oca}
X.-G. Huang, {Electromagnetic fields and anomalous transports in heavy-ion
  collisions --- A pedagogical review}, \emph{Rept. Prog. Phys.} {\bf
  79}\penalty0 (7), \penalty0 076302  (2016).
\newblock \doi{10.1088/0034-4885/79/7/076302}.

\bibitem{Hattori:2016emy}
K.~Hattori and X.-G. Huang, {Novel quantum phenomena induced by strong magnetic
  fields in heavy-ion collisions}, \emph{Nucl. Sci. Tech.} {\bf 28}\penalty0
  (2), \penalty0 26  (2017).
\newblock \doi{10.1007/s41365-016-0178-3}.

\bibitem{Liu:2020ymh}
Y.-C. Liu and X.-G. Huang, {Anomalous chiral transports and spin polarization
  in heavy-ion collisions}, \emph{Nucl. Sci. Tech.} {\bf 31}\penalty0 (6),
  \penalty0 56  (2020).
\newblock \doi{10.1007/s41365-020-00764-z}.

\bibitem{vilenkin1980equilibrium}
A.~Vilenkin, Equilibrium parity-violating current in a magnetic field,
  \emph{Physical Review D}. {\bf 22}\penalty0 (12), \penalty0 3080  (1980).

\bibitem{Atiyah:1968mp}
M.~F. Atiyah and I.~M. Singer, {The Index of elliptic operators. 1},
  \emph{Annals Math.} {\bf 87}, \penalty0 484--530  (1968).
\newblock \doi{10.2307/1970715}.

\bibitem{Aharonov:1978gb}
Y.~Aharonov and A.~Casher, {The Ground State of a Spin 1/2 Charged Particle in
  a Two-dimensional Magnetic Field}, \emph{Phys. Rev. A}. {\bf 19}, \penalty0
  2461--2462  (1979).
\newblock \doi{10.1103/PhysRevA.19.2461}.

\bibitem{Nielsen:1983rb}
H.~B. Nielsen and M.~Ninomiya, {ADLER-BELL-JACKIW ANOMALY AND WEYL FERMIONS IN
  CRYSTAL}, \emph{Phys. Lett. B}. {\bf 130}, \penalty0 389--396  (1983).
\newblock \doi{10.1016/0370-2693(83)91529-0}.

\bibitem{Adler:1969gk}
S.~L. Adler, {Axial vector vertex in spinor electrodynamics}, \emph{Phys. Rev.}
  {\bf 177}, \penalty0 2426--2438  (1969).
\newblock \doi{10.1103/PhysRev.177.2426}.

\bibitem{Bell:1969ts}
J.~S. Bell and R.~Jackiw, {A PCAC puzzle: $\pi^0 \to \gamma \gamma$ in the
  $\sigma$ model}, \emph{Nuovo Cim. A}. {\bf 60}, \penalty0 47--61  (1969).
\newblock \doi{10.1007/BF02823296}.

\bibitem{son2004quantum}
D.~Son and A.~R. Zhitnitsky, Quantum anomalies in dense matter, \emph{Physical
  Review D}. {\bf 70}\penalty0 (7), \penalty0 074018  (2004).

\bibitem{son2008axial}
D.~Son and M.~Stephanov, Axial anomaly and magnetism of nuclear and quark
  matter, \emph{Physical Review D}. {\bf 77}\penalty0 (1), \penalty0 014021
  (2008).

\bibitem{Basar:2013iaa}
G.~Basar, D.~E. Kharzeev, and H.-U. Yee, {Triangle anomaly in Weyl semimetals},
  \emph{Phys. Rev. B}. {\bf 89}\penalty0 (3), \penalty0 035142  (2014).
\newblock \doi{10.1103/PhysRevB.89.035142}.

\bibitem{Wilczek:1987mv}
F.~Wilczek, {Two Applications of Axion Electrodynamics}, \emph{Phys. Rev.
  Lett.} {\bf 58}, \penalty0 1799  (1987).
\newblock \doi{10.1103/PhysRevLett.58.1799}.

\bibitem{Kharzeev:2016mvi}
D.~Kharzeev, Y.~Kikuchi, and R.~Meyer, {Chiral magnetic effect without
  chirality source in asymmetric Weyl semimetals}, \emph{Eur. Phys. J. B}. {\bf
  91}\penalty0 (5), \penalty0 83  (2018).
\newblock \doi{10.1140/epjb/e2018-80418-1}.

\bibitem{li2016chiral}
Q.~Li, D.~E. Kharzeev, C.~Zhang, Y.~Huang, I.~Pletikosi{\'c}, A.~Fedorov,
  R.~Zhong, J.~Schneeloch, G.~Gu, and T.~Valla, Chiral magnetic effect in zrte
  5, \emph{Nature Physics}. {\bf 12}\penalty0 (6), \penalty0 550--554  (2016).

\bibitem{ong2021experimental}
N.~Ong and S.~Liang, Experimental signatures of the chiral anomaly in
  dirac--weyl semimetals, \emph{Nature Reviews Physics}. {\bf 3}\penalty0 (6),
  \penalty0 394--404  (2021).

\bibitem{Kharzeev:2007tn}
D.~Kharzeev and A.~Zhitnitsky, {Charge separation induced by P-odd bubbles in
  QCD matter}, \emph{Nucl. Phys. A}. {\bf 797}, \penalty0 67--79  (2007).
\newblock \doi{10.1016/j.nuclphysa.2007.10.001}.

\bibitem{Kharzeev:2010gr}
D.~E. Kharzeev and D.~T. Son, {Testing the chiral magnetic and chiral vortical
  effects in heavy ion collisions}, \emph{Phys. Rev. Lett.} {\bf 106},
  \penalty0 062301  (2011).
\newblock \doi{10.1103/PhysRevLett.106.062301}.

\bibitem{Kharzeev:2009fn}
D.~E. Kharzeev, {Topologically induced local P and CP violation in QCD x QED},
  \emph{Annals Phys.} {\bf 325}, \penalty0 205--218  (2010).
\newblock \doi{10.1016/j.aop.2009.11.002}.

\bibitem{Erdmenger:2008rm}
J.~Erdmenger, M.~Haack, M.~Kaminski, and A.~Yarom, {Fluid dynamics of R-charged
  black holes}, \emph{JHEP}. {\bf 01}, \penalty0 055  (2009).
\newblock \doi{10.1088/1126-6708/2009/01/055}.

\bibitem{wang2023search}
C.-Z. Wang.
\newblock Search for anomalous chiral effects in heavy-ion collisions with
  alice  (2023).

\bibitem{Frenklakh:2024jff}
D.~Frenklakh, D.~E. Kharzeev, and A.~Palermo, {Detecting anomalous CP violation
  in heavy ion collisions through baryon-electric charge correlations}  (2,
  2024).

\bibitem{Huang:2013iia}
X.-G. Huang and J.~Liao, {Axial Current Generation from Electric Field: Chiral
  Electric Separation Effect}, \emph{Phys. Rev. Lett.} {\bf 110}\penalty0 (23),
  \penalty0 232302  (2013).
\newblock \doi{10.1103/PhysRevLett.110.232302}.

\bibitem{Jiang:2014ura}
Y.~Jiang, X.-G. Huang, and J.~Liao, {Chiral electric separation effect in the
  quark-gluon plasma}, \emph{Phys. Rev. D}. {\bf 91}\penalty0 (4), \penalty0
  045001  (2015).
\newblock \doi{10.1103/PhysRevD.91.045001}.

\bibitem{Kharzeev:1998kz}
D.~Kharzeev, R.~D. Pisarski, and M.~H.~G. Tytgat, {Possibility of spontaneous
  parity violation in hot QCD}, \emph{Phys. Rev. Lett.} {\bf 81}, \penalty0
  512--515  (1998).
\newblock \doi{10.1103/PhysRevLett.81.512}.

\bibitem{Kharzeev:2001ev}
D.~Kharzeev, A.~Krasnitz, and R.~Venugopalan, {Anomalous chirality fluctuations
  in the initial stage of heavy ion collisions and parity odd bubbles},
  \emph{Phys. Lett. B}. {\bf 545}, \penalty0 298--306  (2002).
\newblock \doi{10.1016/S0370-2693(02)02630-8}.

\bibitem{Voloshin:2004vk}
S.~A. Voloshin, {Parity violation in hot QCD: How to detect it}, \emph{Phys.
  Rev. C}. {\bf 70}, \penalty0 057901  (2004).
\newblock \doi{10.1103/PhysRevC.70.057901}.

\bibitem{Mueller:2016ven}
N.~M\"uller, S.~Schlichting, and S.~Sharma, {Chiral magnetic effect and
  anomalous transport from real-time lattice simulations}, \emph{Phys. Rev.
  Lett.} {\bf 117}\penalty0 (14), \penalty0 142301  (2016).
\newblock \doi{10.1103/PhysRevLett.117.142301}.

\bibitem{Mace:2016svc}
M.~Mace, S.~Schlichting, and R.~Venugopalan, {Off-equilibrium sphaleron
  transitions in the Glasma}, \emph{Phys. Rev. D}. {\bf 93}\penalty0 (7),
  \penalty0 074036  (2016).
\newblock \doi{10.1103/PhysRevD.93.074036}.

\bibitem{Mace:2016shq}
M.~Mace, N.~Mueller, S.~Schlichting, and S.~Sharma, {Non-equilibrium study of
  the Chiral Magnetic Effect from real-time simulations with dynamical
  fermions}, \emph{Phys. Rev. D}. {\bf 95}\penalty0 (3), \penalty0 036023
  (2017).
\newblock \doi{10.1103/PhysRevD.95.036023}.

\bibitem{Mace:2017wcl}
M.~Mace, N.~Mueller, S.~Schlichting, and S.~Sharma, {Simulating chiral magnetic
  effect and anomalous transport phenomena in the pre-equilibrium stages of
  heavy-ion collisions}, \emph{Nucl. Phys. A}. {\bf 967}, \penalty0 752--755
  (2017).
\newblock \doi{10.1016/j.nuclphysa.2017.05.040}.

\bibitem{Lappi:2017skr}
T.~Lappi and S.~Schlichting, {Linearly polarized gluons and axial charge
  fluctuations in the Glasma}, \emph{Phys. Rev. D}. {\bf 97}\penalty0 (3),
  \penalty0 034034  (2018).
\newblock \doi{10.1103/PhysRevD.97.034034}.

\bibitem{Huang:2021bhj}
A.~Huang, S.~Shi, S.~Lin, X.~Guo, and J.~Liao, {Accessing topological
  fluctuations of gauge fields with the chiral magnetic effect}, \emph{Phys.
  Rev. D}. {\bf 107}\penalty0 (3), \penalty0 034012  (2023).
\newblock \doi{10.1103/PhysRevD.107.034012}.

\bibitem{Hirono:2014oda}
Y.~Hirono, T.~Hirano, and D.~E. Kharzeev, {The chiral magnetic effect in
  heavy-ion collisions from event-by-event anomalous hydrodynamics}  (11,
  2014).

\bibitem{Guo:2016nnq}
E.-d. Guo and S.~Lin, {Quark mass effect on axial charge dynamics}, \emph{Phys.
  Rev. D}. {\bf 93}\penalty0 (10), \penalty0 105001  (2016).
\newblock \doi{10.1103/PhysRevD.93.105001}.

\bibitem{Hou:2017szz}
D.-f. Hou and S.~Lin, {Fluctuation and Dissipation of Axial Charge from Massive
  Quarks}, \emph{Phys. Rev. D}. {\bf 98}\penalty0 (5), \penalty0 054014
  (2018).
\newblock \doi{10.1103/PhysRevD.98.054014}.

\bibitem{Lin:2018nxj}
S.~Lin, L.~Yan, and G.-R. Liang, {Axial Charge Fluctuation and Chiral Magnetic
  Effect from Stochastic Hydrodynamics}, \emph{Phys. Rev. C}. {\bf 98}\penalty0
  (1), \penalty0 014903  (2018).
\newblock \doi{10.1103/PhysRevC.98.014903}.

\bibitem{Liang:2020sgr}
G.-R. Liang, J.~Liao, S.~Lin, L.~Yan, and M.~Li, {Chiral Magnetic Effect in
  Isobar Collisions from Stochastic Hydrodynamics}, \emph{Chin. Phys. C}. {\bf
  44}\penalty0 (9), \penalty0 094103  (2020).
\newblock \doi{10.1088/1674-1137/44/9/094103}.

\bibitem{Shi:2019wzi}
S.~Shi, H.~Zhang, D.~Hou, and J.~Liao, {Signatures of Chiral Magnetic Effect in
  the Collisions of Isobars}, \emph{Phys. Rev. Lett.} {\bf 125}, \penalty0
  242301  (2020).
\newblock \doi{10.1103/PhysRevLett.125.242301}.

\bibitem{Bloczynski:2012en}
J.~Bloczynski, X.-G. Huang, X.~Zhang, and J.~Liao, {Azimuthally fluctuating
  magnetic field and its impacts on observables in heavy-ion collisions},
  \emph{Phys. Lett. B}. {\bf 718}, \penalty0 1529--1535  (2013).
\newblock \doi{10.1016/j.physletb.2012.12.030}.

\bibitem{Shi:2017cpu}
S.~Shi, Y.~Jiang, E.~Lilleskov, and J.~Liao, {Anomalous Chiral Transport in
  Heavy Ion Collisions from Anomalous-Viscous Fluid Dynamics}, \emph{Annals
  Phys.} {\bf 394}, \penalty0 50--72  (2018).
\newblock \doi{10.1016/j.aop.2018.04.026}.

\bibitem{Bzdak:2019pkr}
A.~Bzdak, S.~Esumi, V.~Koch, J.~Liao, M.~Stephanov, and N.~Xu, {Mapping the
  Phases of Quantum Chromodynamics with Beam Energy Scan}, \emph{Phys. Rept.}
  {\bf 853}, \penalty0 1--87  (2020).
\newblock \doi{10.1016/j.physrep.2020.01.005}.

\bibitem{Wang:2018ygc}
F.-Q. Wang and J.~Zhao, {Search for the chiral magnetic effect in heavy ion
  collisions}, \emph{Nucl. Sci. Tech.} {\bf 29}\penalty0 (12), \penalty0 179
  (2018).
\newblock \doi{10.1007/s41365-018-0520-z}.

\bibitem{Gursoy:2018yai}
U.~G\"ursoy, D.~Kharzeev, E.~Marcus, K.~Rajagopal, and C.~Shen,
  {Charge-dependent Flow Induced by Magnetic and Electric Fields in Heavy Ion
  Collisions}, \emph{Phys. Rev. C}. {\bf 98}\penalty0 (5), \penalty0 055201
  (2018).
\newblock \doi{10.1103/PhysRevC.98.055201}.

\bibitem{Huang:2022qdn}
A.~Huang, D.~She, S.~Shi, M.~Huang, and J.~Liao, {Dynamical magnetic fields in
  heavy-ion collisions}, \emph{Phys. Rev. C}. {\bf 107}\penalty0 (3), \penalty0
  034901  (2023).
\newblock \doi{10.1103/PhysRevC.107.034901}.

\bibitem{McLerran:2013hla}
L.~McLerran and V.~Skokov, {Comments About the Electromagnetic Field in
  Heavy-Ion Collisions}, \emph{Nucl. Phys. A}. {\bf 929}, \penalty0 184--190
  (2014).
\newblock \doi{10.1016/j.nuclphysa.2014.05.008}.

\bibitem{Tuchin:2015oka}
K.~Tuchin, {Initial value problem for magnetic fields in heavy ion collisions},
  \emph{Phys. Rev. C}. {\bf 93}\penalty0 (1), \penalty0 014905  (2016).
\newblock \doi{10.1103/PhysRevC.93.014905}.

\bibitem{Inghirami:2016iru}
G.~Inghirami, L.~Del~Zanna, A.~Beraudo, M.~H. Moghaddam, F.~Becattini, and
  M.~Bleicher, {Numerical magneto-hydrodynamics for relativistic nuclear
  collisions}, \emph{Eur. Phys. J. C}. {\bf 76}\penalty0 (12), \penalty0 659
  (2016).
\newblock \doi{10.1140/epjc/s10052-016-4516-8}.

\bibitem{Inghirami:2019mkc}
G.~Inghirami, M.~Mace, Y.~Hirono, L.~Del~Zanna, D.~E. Kharzeev, and
  M.~Bleicher, {Magnetic fields in heavy ion collisions: flow and charge
  transport}, \emph{Eur. Phys. J. C}. {\bf 80}\penalty0 (3), \penalty0 293
  (2020).
\newblock \doi{10.1140/epjc/s10052-020-7847-4}.

\bibitem{Roy:2017yvg}
V.~Roy, S.~Pu, L.~Rezzolla, and D.~H. Rischke, {Effect of intense magnetic
  fields on reduced-MHD evolution in $\sqrt{s_{\rm NN}}$ = 200 GeV Au+Au
  collisions}, \emph{Phys. Rev. C}. {\bf 96}\penalty0 (5), \penalty0 054909
  (2017).
\newblock \doi{10.1103/PhysRevC.96.054909}.

\bibitem{Pu:2016ayh}
S.~Pu, V.~Roy, L.~Rezzolla, and D.~H. Rischke, {Bjorken flow in one-dimensional
  relativistic magnetohydrodynamics with magnetization}, \emph{Phys. Rev. D}.
  {\bf 93}\penalty0 (7), \penalty0 074022  (2016).
\newblock \doi{10.1103/PhysRevD.93.074022}.

\bibitem{Muller:2018ibh}
B.~M\"uller and A.~Sch\"afer, {Chiral magnetic effect and an experimental bound
  on the late time magnetic field strength}, \emph{Phys. Rev. D}. {\bf
  98}\penalty0 (7), \penalty0 071902  (2018).
\newblock \doi{10.1103/PhysRevD.98.071902}.

\bibitem{Guo:2019joy}
Y.~Guo, S.~Shi, S.~Feng, and J.~Liao, {Magnetic Field Induced Polarization
  Difference between Hyperons and Anti-hyperons}, \emph{Phys. Lett. B}. {\bf
  798}, \penalty0 134929  (2019).
\newblock \doi{10.1016/j.physletb.2019.134929}.

\bibitem{Guo:2019mgh}
X.~Guo, J.~Liao, and E.~Wang, {Spin Hydrodynamic Generation in the Charged
  Subatomic Swirl}, \emph{Sci. Rep.} {\bf 10}\penalty0 (1), \penalty0 2196
  (2020).
\newblock \doi{10.1038/s41598-020-59129-6}.

\bibitem{Hernandez:2017mch}
J.~Hernandez and P.~Kovtun, {Relativistic magnetohydrodynamics}, \emph{JHEP}.
  {\bf 05}, \penalty0 001  (2017).
\newblock \doi{10.1007/JHEP05(2017)001}.

\bibitem{Denicol:2018rbw}
G.~S. Denicol, X.-G. Huang, E.~Moln\'ar, G.~M. Monteiro, H.~Niemi, J.~Noronha,
  D.~H. Rischke, and Q.~Wang, {Nonresistive dissipative magnetohydrodynamics
  from the Boltzmann equation in the 14-moment approximation}, \emph{Phys. Rev.
  D}. {\bf 98}\penalty0 (7), \penalty0 076009  (2018).
\newblock \doi{10.1103/PhysRevD.98.076009}.

\bibitem{Denicol:2019iyh}
G.~S. Denicol, E.~Moln\'ar, H.~Niemi, and D.~H. Rischke, {Resistive dissipative
  magnetohydrodynamics from the Boltzmann-Vlasov equation}, \emph{Phys. Rev.
  D}. {\bf 99}\penalty0 (5), \penalty0 056017  (2019).
\newblock \doi{10.1103/PhysRevD.99.056017}.

\bibitem{Shokri:2018qcu}
M.~Shokri and N.~Sadooghi, {Evolution of magnetic fields from the 3 + 1
  dimensional self-similar and Gubser flows in ideal relativistic
  magnetohydrodynamics}, \emph{JHEP}. {\bf 11}, \penalty0 181  (2018).
\newblock \doi{10.1007/JHEP11(2018)181}.

\bibitem{Siddique:2019gqh}
I.~Siddique, R.-j. Wang, S.~Pu, and Q.~Wang, {Anomalous magnetohydrodynamics
  with longitudinal boost invariance and chiral magnetic effect}, \emph{Phys.
  Rev. D}. {\bf 99}\penalty0 (11), \penalty0 114029  (2019).
\newblock \doi{10.1103/PhysRevD.99.114029}.

\bibitem{Hattori:2022hyo}
K.~Hattori, M.~Hongo, and X.-G. Huang, {New Developments in Relativistic
  Magnetohydrodynamics}, \emph{Symmetry}. {\bf 14}\penalty0 (9), \penalty0 1851
   (2022).
\newblock \doi{10.3390/sym14091851}.

\bibitem{Wang:2023imu}
S.~Wang and X.-G. Huang, {Chiral magnetovortical instability}, \emph{Phys. Rev.
  D}. {\bf 109}\penalty0 (12), \penalty0 L121302  (2024).
\newblock \doi{10.1103/PhysRevD.109.L121302}.

\bibitem{Guo:2015nsa}
X.~Guo, S.~Shi, N.~Xu, Z.~Xu, and P.~Zhuang, {Magnetic Field Effect on
  Charmonium Production in High Energy Nuclear Collisions}, \emph{Phys. Lett.
  B}. {\bf 751}, \penalty0 215--219  (2015).
\newblock \doi{10.1016/j.physletb.2015.10.038}.

\bibitem{Xu:2020sui}
K.~Xu, S.~Shi, H.~Zhang, D.~Hou, J.~Liao, and M.~Huang, {Extracting the
  magnitude of magnetic field at freeze-out in heavy-ion collisions},
  \emph{Phys. Lett. B}. {\bf 809}, \penalty0 135706  (2020).
\newblock \doi{10.1016/j.physletb.2020.135706}.

\bibitem{Aaboud:2018eph}
M.~Aaboud et~al., {Observation of centrality-dependent acoplanarity for muon
  pairs produced via two-photon scattering in Pb+Pb collisions at
  $\sqrt{s_{\mathrm{NN}}}=5.02$ TeV with the ATLAS detector}, \emph{Phys. Rev.
  Lett.} {\bf 121}\penalty0 (21), \penalty0 212301  (2018).
\newblock \doi{10.1103/PhysRevLett.121.212301}.

\bibitem{Adam:2018tdm}
J.~Adam et~al., {Low-$p_T$ $e^{+}e^{-}$ pair production in Au$+$Au collisions
  at $\sqrt{s_{NN}}$ = 200 GeV and U$+$U collisions at $\sqrt{s_{NN}}$ = 193
  GeV at STAR}, \emph{Phys. Rev. Lett.} {\bf 121}\penalty0 (13), \penalty0
  132301  (2018).
\newblock \doi{10.1103/PhysRevLett.121.132301}.

\bibitem{Zha:2018tlq}
W.~Zha, J.~D. Brandenburg, Z.~Tang, and Z.~Xu, {Initial transverse-momentum
  broadening of Breit-Wheeler process in relativistic heavy-ion collisions},
  \emph{Phys. Lett. B}. {\bf 800}, \penalty0 135089  (2020).
\newblock \doi{10.1016/j.physletb.2019.135089}.

\bibitem{Zha:2018ywo}
W.~Zha, L.~Ruan, Z.~Tang, Z.~Xu, and S.~Yang, {Coherent lepton pair production
  in hadronic heavy ion collisions}, \emph{Phys. Lett. B}. {\bf 781}, \penalty0
  182--186  (2018).
\newblock \doi{10.1016/j.physletb.2018.04.006}.

\bibitem{Klein:2018fmp}
S.~Klein, A.~H. Mueller, B.-W. Xiao, and F.~Yuan, {Acoplanarity of a Lepton
  Pair to Probe the Electromagnetic Property of Quark Matter}, \emph{Phys. Rev.
  Lett.} {\bf 122}\penalty0 (13), \penalty0 132301  (2019).
\newblock \doi{10.1103/PhysRevLett.122.132301}.

\bibitem{Xu:2017qfs}
H.-j. Xu, J.~Zhao, X.~Wang, H.~Li, Z.-W. Lin, C.~Shen, and F.~Wang, {Varying
  the chiral magnetic effect relative to flow in a single nucleus-nucleus
  collision}, \emph{Chin. Phys. C}. {\bf 42}\penalty0 (8), \penalty0 084103
  (2018).
\newblock \doi{10.1088/1674-1137/42/8/084103}.

\bibitem{Zhao:2017nfq}
J.~Zhao, H.~Li, and F.~Wang, {Isolating the chiral magnetic effect from
  backgrounds by pair invariant mass}, \emph{Eur. Phys. J. C}. {\bf
  79}\penalty0 (2), \penalty0 168  (2019).
\newblock \doi{10.1140/epjc/s10052-019-6671-1}.

\bibitem{Voloshin:2018qsm}
S.~A. Voloshin, {Estimate of the signal from the chiral magnetic effect in
  heavy-ion collisions from measurements relative to the participant and
  spectator flow planes}, \emph{Phys. Rev. C}. {\bf 98}\penalty0 (5), \penalty0
  054911  (2018).
\newblock \doi{10.1103/PhysRevC.98.054911}.

\bibitem{Magdy:2017yje}
N.~Magdy, S.~Shi, J.~Liao, N.~Ajitanand, and R.~A. Lacey, {New correlator to
  detect and characterize the chiral magnetic effect}, \emph{Phys. Rev. C}.
  {\bf 97}\penalty0 (6), \penalty0 061901  (2018).
\newblock \doi{10.1103/PhysRevC.97.061901}.

\bibitem{Magdy:2018lwk}
N.~Magdy, S.~Shi, J.~Liao, P.~Liu, and R.~A. Lacey, {Examination of the
  observability of a chiral magnetically driven charge-separation difference in
  collisions of the $\mathrm{^{96}_{44}Ru +\, ^{96}_{44}Ru}$ and
  $\mathrm{^{96}_{40}Zr +\, ^{96}_{40}Zr}$ isobars at energies available at the
  BNL Relativistic Heavy Ion Collider}, \emph{Phys. Rev. C}. {\bf 98}\penalty0
  (6), \penalty0 061902  (2018).
\newblock \doi{10.1103/PhysRevC.98.061902}.

\bibitem{Tang:2019pbl}
A.~H. Tang, {Probe chiral magnetic effect with signed balance function},
  \emph{Chin. Phys. C}. {\bf 44}\penalty0 (5), \penalty0 054101  (2020).
\newblock \doi{10.1088/1674-1137/44/5/054101}.

\bibitem{Kharzeev:2019zgg}
D.~E. Kharzeev and J.~Liao, {Isobar Collisions at RHIC to Test Local Parity
  Violation in Strong Interactions}, \emph{Nucl. Phys. News}. {\bf 29}\penalty0
  (1), \penalty0 26--31  (2019).
\newblock \doi{10.1080/10619127.2018.1495479}.

\bibitem{Skokov:2016yrj}
V.~Koch, S.~Schlichting, V.~Skokov, P.~Sorensen, J.~Thomas, S.~Voloshin,
  G.~Wang, and H.-U. Yee, {Status of the chiral magnetic effect and collisions
  of isobars}, \emph{Chin. Phys. C}. {\bf 41}\penalty0 (7), \penalty0 072001
  (2017).
\newblock \doi{10.1088/1674-1137/41/7/072001}.

\bibitem{Voloshin:2010ut}
S.~A. Voloshin, {Testing the Chiral Magnetic Effect with Central U+U
  collisions}, \emph{Phys. Rev. Lett.} {\bf 105}, \penalty0 172301  (2010).
\newblock \doi{10.1103/PhysRevLett.105.172301}.

\bibitem{Adam:2019fbq}
J.~Adam et~al., {Methods for a blind analysis of isobar data collected by the
  STAR collaboration}, \emph{Nucl. Sci. Tech.} {\bf 32}\penalty0 (5), \penalty0
  48  (2021).
\newblock \doi{10.1007/s41365-021-00878-y}.

\bibitem{Bzdak:2012ia}
A.~Bzdak, V.~Koch, and J.~Liao, {Charge-Dependent Correlations in Relativistic
  Heavy Ion Collisions and the Chiral Magnetic Effect}, \emph{Lect. Notes
  Phys.} {\bf 871}, \penalty0 503--536  (2013).
\newblock \doi{10.1007/978-3-642-37305-3_19}.

\bibitem{STAR:2013ksd}
L.~Adamczyk et~al., {Fluctuations of charge separation perpendicular to the
  event plane and local parity violation in $\sqrt{s_{NN}}=200$ GeV Au+Au
  collisions at the BNL Relativistic Heavy Ion Collider}, \emph{Phys. Rev. C}.
  {\bf 88}\penalty0 (6), \penalty0 064911  (2013).
\newblock \doi{10.1103/PhysRevC.88.064911}.

\bibitem{STAR:2005btp}
J.~Adams et~al., {Directed flow in Au+Au collisions at s(NN)**(1/2) = 62-GeV},
  \emph{Phys. Rev. C}. {\bf 73}, \penalty0 034903  (2006).
\newblock \doi{10.1103/PhysRevC.73.034903}.

\bibitem{STAR:2009wot}
B.~I. Abelev et~al., {Azimuthal Charged-Particle Correlations and Possible
  Local Strong Parity Violation}, \emph{Phys. Rev. Lett.} {\bf 103}, \penalty0
  251601  (2009).
\newblock \doi{10.1103/PhysRevLett.103.251601}.

\bibitem{Wang:2009kd}
F.~Wang, {Effects of Cluster Particle Correlations on Local Parity Violation
  Observables}, \emph{Phys. Rev. C}. {\bf 81}, \penalty0 064902  (2010).
\newblock \doi{10.1103/PhysRevC.81.064902}.

\bibitem{Pratt:2010gy}
S.~Pratt, {Alternative Contributions to the Angular Correlations Observed at
  RHIC Associated with Parity Fluctuations}  (2, 2010).

\bibitem{Schlichting:2010qia}
S.~Schlichting and S.~Pratt, {Charge conservation at energies available at the
  BNL Relativistic Heavy Ion Collider and contributions to local parity
  violation observables}, \emph{Phys. Rev. C}. {\bf 83}, \penalty0 014913
  (2011).
\newblock \doi{10.1103/PhysRevC.83.014913}.

\bibitem{Schenke:2019ruo}
B.~Schenke, C.~Shen, and P.~Tribedy, {Multi-particle and charge-dependent
  azimuthal correlations in heavy-ion collisions at the Relativistic Heavy-Ion
  Collider}, \emph{Phys. Rev. C}. {\bf 99}\penalty0 (4), \penalty0 044908
  (2019).
\newblock \doi{10.1103/PhysRevC.99.044908}.

\bibitem{Pratt:2010zn}
S.~Pratt, S.~Schlichting, and S.~Gavin, {Effects of Momentum Conservation and
  Flow on Angular Correlations at RHIC}, \emph{Phys. Rev. C}. {\bf 84},
  \penalty0 024909  (2011).
\newblock \doi{10.1103/PhysRevC.84.024909}.

\bibitem{Bozek:2012en}
P.~Bozek and W.~Broniowski, {Charge conservation and the shape of the ridge of
  two-particle correlations in relativistic heavy-ion collisions}, \emph{Phys.
  Rev. Lett.} {\bf 109}, \penalty0 062301  (2012).
\newblock \doi{10.1103/PhysRevLett.109.062301}.

\bibitem{Oliinychenko:2019zfk}
D.~Oliinychenko and V.~Koch, {Microcanonical Particlization with Local
  Conservation Laws}, \emph{Phys. Rev. Lett.} {\bf 123}\penalty0 (18),
  \penalty0 182302  (2019).
\newblock \doi{10.1103/PhysRevLett.123.182302}.

\bibitem{Oliinychenko:2020cmr}
D.~Oliinychenko, S.~Shi, and V.~Koch, {Effects of local event-by-event
  conservation laws in ultrarelativistic heavy-ion collisions at
  particlization}, \emph{Phys. Rev. C}. {\bf 102}\penalty0 (3), \penalty0
  034904  (2020).
\newblock \doi{10.1103/PhysRevC.102.034904}.

\bibitem{Tribedy:2017hwn}
P.~Tribedy, {Disentangling flow and signals of Chiral Magnetic Effect in U+U,
  Au+Au and p+Au collisions}, \emph{Nucl. Phys. A}. {\bf 967}, \penalty0
  740--743  (2017).
\newblock \doi{10.1016/j.nuclphysa.2017.05.078}.

\bibitem{Zhao:2019kyk}
J.~Zhao, Y.~Feng, H.~Li, and F.~Wang, {HIJING can describe the
  anisotropy-scaled charge-dependent correlations at the BNL Relativistic Heavy
  Ion Collider}, \emph{Phys. Rev. C}. {\bf 101}\penalty0 (3), \penalty0 034912
  (2020).
\newblock \doi{10.1103/PhysRevC.101.034912}.

\bibitem{Khachatryan:2016got}
V.~Khachatryan et~al., {Observation of charge-dependent azimuthal correlations
  in $p$-Pb collisions and its implication for the search for the chiral
  magnetic effect}, \emph{Phys. Rev. Lett.} {\bf 118}\penalty0 (12), \penalty0
  122301  (2017).
\newblock \doi{10.1103/PhysRevLett.118.122301}.

\bibitem{Feng:2021pgf}
Y.~Feng, J.~Zhao, H.~Li, H.-j. Xu, and F.~Wang, {Two- and three-particle
  nonflow contributions to the chiral magnetic effect measurement by spectator
  and participant planes in relativistic heavy ion collisions}, \emph{Phys.
  Rev. C}. {\bf 105}\penalty0 (2), \penalty0 024913  (2022).
\newblock \doi{10.1103/PhysRevC.105.024913}.

\bibitem{Jiang:2016wve}
Y.~Jiang, S.~Shi, Y.~Yin, and J.~Liao, {Quantifying the chiral magnetic effect
  from anomalous-viscous fluid dynamics}, \emph{Chin. Phys. C}. {\bf
  42}\penalty0 (1), \penalty0 011001  (2018).
\newblock \doi{10.1088/1674-1137/42/1/011001}.

\bibitem{Ma:2011uma}
G.-L. Ma and B.~Zhang, {Effects of final state interactions on charge
  separation in relativistic heavy ion collisions}, \emph{Phys. Lett. B}. {\bf
  700}, \penalty0 39--43  (2011).
\newblock \doi{10.1016/j.physletb.2011.04.057}.

\bibitem{Deng:2016knn}
W.-T. Deng, X.-G. Huang, G.-L. Ma, and G.~Wang, {Test the chiral magnetic
  effect with isobaric collisions}, \emph{Phys. Rev. C}. {\bf 94}, \penalty0
  041901  (2016).
\newblock \doi{10.1103/PhysRevC.94.041901}.

\bibitem{Zhao:2019crj}
X.-L. Zhao, G.-L. Ma, and Y.-G. Ma, {Impact of magnetic-field fluctuations on
  measurements of the chiral magnetic effect in collisions of isobaric nuclei},
  \emph{Phys. Rev. C}. {\bf 99}\penalty0 (3), \penalty0 034903  (2019).
\newblock \doi{10.1103/PhysRevC.99.034903}.

\bibitem{Wang:2021nvh}
C.-Z. Wang, W.-Y. Wu, Q.-Y. Shou, G.-L. Ma, Y.-G. Ma, and S.~Zhang,
  {Interpreting the charge-dependent flow and constraining the chiral magnetic
  wave with event shape engineering}, \emph{Phys. Lett. B}. {\bf 820},
  \penalty0 136580  (2021).
\newblock \doi{10.1016/j.physletb.2021.136580}.

\bibitem{Li:2022bhl}
F.~Li, Y.-G. Ma, S.~Zhang, G.-L. Ma, Q.~Shou, and Q.-Y. Shou, {Impact of
  nuclear structure on the background in the chiral magnetic effect in
  $^{96}_{44}$Ru + $^{96}_{44}$Ru and $^{96}_{40}$Zr + $^{96}_{40}$Zr
  collisions at $\sqrt{s_{NN}}$ = 7.7 $\sim$ 200 GeV from a multiphase
  transport model}, \emph{Phys. Rev. C}. {\bf 106}\penalty0 (1), \penalty0
  014906  (2022).
\newblock \doi{10.1103/PhysRevC.106.014906}.

\bibitem{Wu:2022fwz}
W.-Y. Wu et~al., {Global constraint on the magnitude of anomalous chiral
  effects in heavy-ion collisions}, \emph{Phys. Rev. C}. {\bf 107}\penalty0
  (3), \penalty0 L031902  (2023).
\newblock \doi{10.1103/PhysRevC.107.L031902}.

\bibitem{Sun:2016nig}
Y.~Sun, C.~M. Ko, and F.~Li, {Anomalous transport model study of chiral
  magnetic effects in heavy ion collisions}, \emph{Phys. Rev. C}. {\bf
  94}\penalty0 (4), \penalty0 045204  (2016).
\newblock \doi{10.1103/PhysRevC.94.045204}.

\bibitem{Sun:2016mvh}
Y.~Sun and C.~M. Ko, {Chiral vortical and magnetic effects in the anomalous
  transport model}, \emph{Phys. Rev. C}. {\bf 95}\penalty0 (3), \penalty0
  034909  (2017).
\newblock \doi{10.1103/PhysRevC.95.034909}.

\bibitem{Sun:2018idn}
Y.~Sun and C.~M. Ko, {Chiral kinetic approach to the chiral magnetic effect in
  isobaric collisions}, \emph{Phys. Rev. C}. {\bf 98}\penalty0 (1), \penalty0
  014911  (2018).
\newblock \doi{10.1103/PhysRevC.98.014911}.

\bibitem{Ammon:2016fru}
M.~Ammon, S.~Grieninger, A.~Jimenez-Alba, R.~P. Macedo, and L.~Melgar,
  {Holographic quenches and anomalous transport}, \emph{JHEP}. {\bf 09},
  \penalty0 131  (2016).
\newblock \doi{10.1007/JHEP09(2016)131}.

\bibitem{Ghosh:2021naw}
J.~K. Ghosh, S.~Grieninger, K.~Landsteiner, and S.~Morales-Tejera, {Is the
  chiral magnetic effect fast enough?}, \emph{Phys. Rev. D}. {\bf 104}\penalty0
  (4), \penalty0 046009  (2021).
\newblock \doi{10.1103/PhysRevD.104.046009}.

\bibitem{Cartwright:2021maz}
C.~Cartwright, M.~Kaminski, and B.~Schenke, {Energy dependence of the chiral
  magnetic effect in expanding holographic plasma}, \emph{Phys. Rev. C}. {\bf
  105}\penalty0 (3), \penalty0 034903  (2022).
\newblock \doi{10.1103/PhysRevC.105.034903}.

\bibitem{Zhao:2021yjo}
Y.-S. Zhao, L.~Wang, K.~Zhou, and X.-G. Huang, {Detecting the chiral magnetic
  effect via deep learning}, \emph{Phys. Rev. C}. {\bf 106}\penalty0 (5),
  \penalty0 L051901  (2022).
\newblock \doi{10.1103/PhysRevC.106.L051901}.

\bibitem{Hongo:2013cqa}
M.~Hongo, Y.~Hirono, and T.~Hirano, {Anomalous-hydrodynamic analysis of
  charge-dependent elliptic flow in heavy-ion collisions}, \emph{Phys. Lett.
  B}. {\bf 775}, \penalty0 266--270  (2017).
\newblock \doi{10.1016/j.physletb.2017.10.028}.

\bibitem{Yee:2013cya}
H.-U. Yee and Y.~Yin, {Realistic Implementation of Chiral Magnetic Wave in
  Heavy Ion Collisions}, \emph{Phys. Rev. C}. {\bf 89}\penalty0 (4), \penalty0
  044909  (2014).
\newblock \doi{10.1103/PhysRevC.89.044909}.

\bibitem{Yin:2015fca}
Y.~Yin and J.~Liao, {Hydrodynamics with chiral anomaly and charge separation in
  relativistic heavy ion collisions}, \emph{Phys. Lett. B}. {\bf 756},
  \penalty0 42--46  (2016).
\newblock \doi{10.1016/j.physletb.2016.02.065}.

\bibitem{Weil:2016zrk}
J.~Weil et~al., {Particle production and equilibrium properties within a new
  hadron transport approach for heavy-ion collisions}, \emph{Phys. Rev. C}.
  {\bf 94}\penalty0 (5), \penalty0 054905  (2016).
\newblock \doi{10.1103/PhysRevC.94.054905}.

\bibitem{Christakoglou:2021nhe}
P.~Christakoglou, S.~Qiu, and J.~Staa, {Systematic study of the chiral magnetic
  effect with the AVFD model at LHC energies}, \emph{Eur. Phys. J. C}. {\bf
  81}\penalty0 (8), \penalty0 717  (2021).
\newblock \doi{10.1140/epjc/s10052-021-09498-7}.

\bibitem{Milton:2021wku}
R.~Milton, G.~Wang, M.~Sergeeva, S.~Shi, J.~Liao, and H.~Z. Huang, {Utilization
  of event shape in search of the chiral magnetic effect in heavy-ion
  collisions}, \emph{Phys. Rev. C}. {\bf 104}\penalty0 (6), \penalty0 064906
  (2021).
\newblock \doi{10.1103/PhysRevC.104.064906}.

\bibitem{Choudhury:2021jwd}
S.~Choudhury et~al., {Investigation of experimental observables in search of
  the chiral magnetic effect in heavy-ion collisions in the STAR experiment *},
  \emph{Chin. Phys. C}. {\bf 46}\penalty0 (1), \penalty0 014101  (2022).
\newblock \doi{10.1088/1674-1137/ac2a1f}.

\bibitem{Buividovich:2009wi}
P.~V. Buividovich, M.~N. Chernodub, E.~V. Luschevskaya, and M.~I. Polikarpov,
  {Numerical evidence of chiral magnetic effect in lattice gauge theory},
  \emph{Phys. Rev. D}. {\bf 80}, \penalty0 054503  (2009).
\newblock \doi{10.1103/PhysRevD.80.054503}.

\bibitem{Abramczyk:2009gb}
M.~Abramczyk, T.~Blum, G.~Petropoulos, and R.~Zhou, {Chiral magnetic effect in
  2+1 flavor QCD+QED}, \emph{PoS}. {\bf LAT2009}, \penalty0 181  (2009).
\newblock \doi{10.22323/1.091.0181}.

\bibitem{Buividovich:2010tn}
P.~V. Buividovich, M.~N. Chernodub, D.~E. Kharzeev, T.~Kalaydzhyan, E.~V.
  Luschevskaya, and M.~I. Polikarpov, {Magnetic-Field-Induced
  insulator-conductor transition in SU(2) quenched lattice gauge theory},
  \emph{Phys. Rev. Lett.} {\bf 105}, \penalty0 132001  (2010).
\newblock \doi{10.1103/PhysRevLett.105.132001}.

\bibitem{Yamamoto:2011gk}
A.~Yamamoto, {Chiral magnetic effect in lattice QCD with a chiral chemical
  potential}, \emph{Phys. Rev. Lett.} {\bf 107}, \penalty0 031601  (2011).
\newblock \doi{10.1103/PhysRevLett.107.031601}.

\bibitem{Braguta:2010ej}
V.~V. Braguta, P.~V. Buividovich, T.~Kalaydzhyan, S.~V. Kuznetsov, and M.~I.
  Polikarpov, {The Chiral Magnetic Effect and chiral symmetry breaking in SU(3)
  quenched lattice gauge theory}, \emph{Phys. Atom. Nucl.} {\bf 75}, \penalty0
  488--492  (2012).
\newblock \doi{10.1134/S1063778812030052}.

\bibitem{Brandt:2024wlw}
B.~B. Brandt, G.~Endr\H{o}di, E.~Garnacho-Velasco, and G.~Mark\'o, {On the
  absence of the Chiral Magnetic Effect in equilibrium QCD}  (5, 2024).

\bibitem{Kharzeev:2013ffa}
D.~E. Kharzeev, {The Chiral Magnetic Effect and Anomaly-Induced Transport},
  \emph{Prog. Part. Nucl. Phys.} {\bf 75}, \penalty0 133--151  (2014).
\newblock \doi{10.1016/j.ppnp.2014.01.002}.

\bibitem{Yamamoto:2015fxa}
N.~Yamamoto, {Generalized Bloch theorem and chiral transport phenomena},
  \emph{Phys. Rev. D}. {\bf 92}\penalty0 (8), \penalty0 085011  (2015).
\newblock \doi{10.1103/PhysRevD.92.085011}.

\bibitem{Buividovich:2024bmu}
P.~V. Buividovich, {Out-of-equilibrium Chiral Magnetic Effect from simulations
  on Euclidean lattices}  (4, 2024).

\bibitem{Muller:2016}
N.~Müller, S.~Schlichting, and S.~Sharma, Chiral magnetic effect and anomalous
  transport from real-time lattice simulations, \emph{Physical Review Letters}.
  {\bf 117}\penalty0 (14)  (Sept., 2016).
\newblock ISSN 1079-7114.
\newblock \doi{10.1103/physrevlett.117.142301}.
\newblock URL \url{http://dx.doi.org/10.1103/PhysRevLett.117.142301}.

\bibitem{Maldacena:1997re}
J.~M. Maldacena, {The Large N limit of superconformal field theories and
  supergravity}, \emph{Adv. Theor. Math. Phys.} {\bf 2}, \penalty0 231--252
  (1998).
\newblock \doi{10.4310/ATMP.1998.v2.n2.a1}.

\bibitem{Gubser:1998bc}
S.~S. Gubser, I.~R. Klebanov, and A.~M. Polyakov, {Gauge theory correlators
  from noncritical string theory}, \emph{Phys. Lett. B}. {\bf 428}, \penalty0
  105--114  (1998).
\newblock \doi{10.1016/S0370-2693(98)00377-3}.

\bibitem{Rebhan:2009vc}
A.~Rebhan, A.~Schmitt, and S.~A. Stricker, {Anomalies and the chiral magnetic
  effect in the Sakai-Sugimoto model}, \emph{JHEP}. {\bf 01}, \penalty0 026
  (2010).
\newblock \doi{10.1007/JHEP01(2010)026}.

\bibitem{Yee:2009vw}
H.-U. Yee, {Holographic Chiral Magnetic Conductivity}, \emph{JHEP}. {\bf 11},
  \penalty0 085  (2009).
\newblock \doi{10.1088/1126-6708/2009/11/085}.

\bibitem{Klebanov:2002gr}
I.~R. Klebanov, P.~Ouyang, and E.~Witten, {A Gravity dual of the chiral
  anomaly}, \emph{Phys. Rev. D}. {\bf 65}, \penalty0 105007  (2002).
\newblock \doi{10.1103/PhysRevD.65.105007}.

\bibitem{Jimenez-Alba:2014iia}
A.~Jimenez-Alba, K.~Landsteiner, and L.~Melgar, {Anomalous magnetoresponse and
  the St\"uckelberg axion in holography}, \emph{Phys. Rev. D}. {\bf 90},
  \penalty0 126004  (2014).
\newblock \doi{10.1103/PhysRevD.90.126004}.

\bibitem{Jimenez-Alba:2015awa}
A.~Jimenez-Alba, K.~Landsteiner, Y.~Liu, and Y.-W. Sun, {Anomalous
  magnetoconductivity and relaxation times in holography}, \emph{JHEP}. {\bf
  07}, \penalty0 117  (2015).
\newblock \doi{10.1007/JHEP07(2015)117}.

\bibitem{Grieninger:2023wuq}
S.~Grieninger and D.~E. Kharzeev, {Spacetime dynamics of chiral magnetic
  currents in a hot non-Abelian plasma}, \emph{Phys. Rev. D}. {\bf
  108}\penalty0 (12), \penalty0 126004  (2023).
\newblock \doi{10.1103/PhysRevD.108.126004}.

\bibitem{Shuryak:2021iqu}
E.~Shuryak and I.~Zahed, {How to observe the QCD instanton/sphaleron processes
  at hadron colliders?}  (1, 2021).

\bibitem{Grieninger:2023myf}
S.~Grieninger and S.~Morales-Tejera, {Real-time dynamics of axial charge and
  chiral magnetic current in a non-Abelian expanding plasma}, \emph{Phys. Rev.
  D}. {\bf 108}\penalty0 (12), \penalty0 126010  (2023).
\newblock \doi{10.1103/PhysRevD.108.126010}.

\bibitem{Son:2007vk}
D.~T. Son and A.~O. Starinets, {Viscosity, Black Holes, and Quantum Field
  Theory}, \emph{Ann. Rev. Nucl. Part. Sci.} {\bf 57}, \penalty0 95--118
  (2007).
\newblock \doi{10.1146/annurev.nucl.57.090506.123120}.

\bibitem{Schafer:2009dj}
T.~Sch\"afer and D.~Teaney, {Nearly Perfect Fluidity: From Cold Atomic Gases to
  Hot Quark Gluon Plasmas}, \emph{Rept. Prog. Phys.} {\bf 72}, \penalty0 126001
   (2009).
\newblock \doi{10.1088/0034-4885/72/12/126001}.

\bibitem{Son:2009tf}
D.~T. Son and P.~Surowka, {Hydrodynamics with Triangle Anomalies}, \emph{Phys.
  Rev. Lett.} {\bf 103}, \penalty0 191601  (2009).
\newblock \doi{10.1103/PhysRevLett.103.191601}.

\bibitem{Banerjee:2008th}
N.~Banerjee, J.~Bhattacharya, S.~Bhattacharyya, S.~Dutta, R.~Loganayagam, and
  P.~Surowka, {Hydrodynamics from charged black branes}, \emph{JHEP}. {\bf 01},
  \penalty0 094  (2011).
\newblock \doi{10.1007/JHEP01(2011)094}.

\bibitem{Torabian:2009qk}
M.~Torabian and H.-U. Yee, {Holographic nonlinear hydrodynamics from AdS/CFT
  with multiple/non-Abelian symmetries}, \emph{JHEP}. {\bf 08}, \penalty0 020
  (2009).
\newblock \doi{10.1088/1126-6708/2009/08/020}.

\bibitem{Lublinsky:2009wr}
M.~Lublinsky and I.~Zahed, {Anomalous Chiral Superfluidity}, \emph{Phys. Lett.
  B}. {\bf 684}, \penalty0 119--122  (2010).
\newblock \doi{10.1016/j.physletb.2010.01.015}.

\bibitem{Lin:2011mr}
S.~Lin, {On the anomalous superfluid hydrodynamics}, \emph{Nucl. Phys. A}. {\bf
  873}, \penalty0 28--46  (2012).
\newblock \doi{10.1016/j.nuclphysa.2011.10.001}.

\bibitem{Bhattacharya:2011tr}
S.~Bhattacharya, J.~L. Diaz-Cruz, E.~Ma, and D.~Wegman, {Dark
  Vector-Gauge-Boson Model}, \emph{Phys. Rev. D}. {\bf 85}, \penalty0 055008
  (2012).
\newblock \doi{10.1103/PhysRevD.85.055008}.

\bibitem{Eling:2010hu}
C.~Eling, Y.~Neiman, and Y.~Oz, {Holographic Non-Abelian Charged Hydrodynamics
  from the Dynamics of Null Horizons}, \emph{JHEP}. {\bf 12}, \penalty0 086
  (2010).
\newblock \doi{10.1007/JHEP12(2010)086}.

\bibitem{Neiman:2010zi}
Y.~Neiman and Y.~Oz, {Relativistic Hydrodynamics with General Anomalous
  Charges}, \emph{JHEP}. {\bf 03}, \penalty0 023  (2011).
\newblock \doi{10.1007/JHEP03(2011)023}.

\bibitem{Kharzeev:2011ds}
D.~E. Kharzeev and H.-U. Yee, {Anomalies and time reversal invariance in
  relativistic hydrodynamics: the second order and higher dimensional
  formulations}, \emph{Phys. Rev. D}. {\bf 84}, \penalty0 045025  (2011).
\newblock \doi{10.1103/PhysRevD.84.045025}.

\bibitem{Thouless:1982zz}
D.~J. Thouless, M.~Kohmoto, M.~P. Nightingale, and M.~den Nijs, {Quantized Hall
  Conductance in a Two-Dimensional Periodic Potential}, \emph{Phys. Rev. Lett.}
  {\bf 49}, \penalty0 405--408  (1982).
\newblock \doi{10.1103/PhysRevLett.49.405}.

\bibitem{Kharzeev:2010ym}
D.~E. Kharzeev.
\newblock {Axial anomaly, Dirac sea, and the chiral magnetic effect}.
\newblock In \emph{{Gribov-80 Memorial Workshop on Quantum Chromodynamics and
  Beyond}}, pp. 293--306  (2011).
\newblock \doi{10.1142/9789814350198_0028}.

\bibitem{Gribov:1981ku}
V.~N. Gribov.
\newblock {ANOMALIES, AS A MANIFESTATION OF THE HIGH MOMENTUM COLLECTIVE MOTION
  IN THE VACUUM}.
\newblock pp. 74--91  (8, 1981).

\bibitem{Dokshitzer:2004ie}
Y.~L. Dokshitzer and D.~E. Kharzeev, {The Gribov conception of quantum
  chromodynamics}, \emph{Ann. Rev. Nucl. Part. Sci.} {\bf 54}, \penalty0
  487--524  (2004).
\newblock \doi{10.1146/annurev.nucl.54.070103.181224}.

\bibitem{Kharzeev:2009pj}
D.~E. Kharzeev and H.~J. Warringa, {Chiral Magnetic conductivity}, \emph{Phys.
  Rev. D}. {\bf 80}, \penalty0 034028  (2009).
\newblock \doi{10.1103/PhysRevD.80.034028}.

\bibitem{Kharzeev:2019ceh}
D.~E. Kharzeev and Q.~Li, {The Chiral Qubit: quantum computing with chiral
  anomaly}  (3, 2019).

\bibitem{Ammon:2020rvg}
M.~Ammon, S.~Grieninger, J.~Hernandez, M.~Kaminski, R.~Koirala, J.~Leiber, and
  J.~Wu, {Chiral hydrodynamics in strong external magnetic fields},
  \emph{JHEP}. {\bf 04}, \penalty0 078  (2021).
\newblock \doi{10.1007/JHEP04(2021)078}.

\bibitem{Sadofyev:2010pr}
A.~V. Sadofyev and M.~V. Isachenkov, {The Chiral magnetic effect in
  hydrodynamical approach}, \emph{Phys. Lett. B}. {\bf 697}, \penalty0 404--406
   (2011).
\newblock \doi{10.1016/j.physletb.2011.02.041}.

\bibitem{Kharzeev:2010gd}
D.~E. Kharzeev and H.-U. Yee, {Chiral Magnetic Wave}, \emph{Phys. Rev. D}. {\bf
  83}, \penalty0 085007  (2011).
\newblock \doi{10.1103/PhysRevD.83.085007}.

\bibitem{Burnier:2011bf}
Y.~Burnier, D.~E. Kharzeev, J.~Liao, and H.-U. Yee, {Chiral magnetic wave at
  finite baryon density and the electric quadrupole moment of quark-gluon
  plasma in heavy ion collisions}, \emph{Phys. Rev. Lett.} {\bf 107}, \penalty0
  052303  (2011).
\newblock \doi{10.1103/PhysRevLett.107.052303}.

\bibitem{Newman:2005hd}
G.~M. Newman, {Anomalous hydrodynamics}, \emph{JHEP}. {\bf 01}, \penalty0 158
  (2006).
\newblock \doi{10.1088/1126-6708/2006/01/158}.

\bibitem{Sadofyev:2010is}
A.~V. Sadofyev, V.~I. Shevchenko, and V.~I. Zakharov, {Notes on chiral
  hydrodynamics within effective theory approach}, \emph{Phys. Rev. D}. {\bf
  83}, \penalty0 105025  (2011).
\newblock \doi{10.1103/PhysRevD.83.105025}.

\bibitem{Fukushima:2010vw}
K.~Fukushima, D.~E. Kharzeev, and H.~J. Warringa, {Real-time dynamics of the
  Chiral Magnetic Effect}, \emph{Phys. Rev. Lett.} {\bf 104}, \penalty0 212001
  (2010).
\newblock \doi{10.1103/PhysRevLett.104.212001}.

\bibitem{Fukushima:2009ft}
K.~Fukushima, D.~E. Kharzeev, and H.~J. Warringa, {Electric-current
  Susceptibility and the Chiral Magnetic Effect}, \emph{Nucl. Phys. A}. {\bf
  836}, \penalty0 311--336  (2010).
\newblock \doi{10.1016/j.nuclphysa.2010.02.003}.

\bibitem{son2009hydrodynamics}
D.~T. Son and P.~Surowka, Hydrodynamics with triangle anomalies, \emph{Physical
  Review Letters}. {\bf 103}\penalty0 (19), \penalty0 191601  (2009).

\bibitem{buzzegoli2022shear}
M.~Buzzegoli, D.~E. Kharzeev, Y.-C. Liu, S.~Shi, S.~A. Voloshin, and H.-U. Yee,
  Shear-induced anomalous transport and charge asymmetry of triangular flow in
  heavy-ion collisions, \emph{Physical Review C}. {\bf 106}\penalty0 (5),
  \penalty0 L051902  (2022).

\bibitem{Anselm:1989gi}
A.~A. Anselm and A.~A. Johansen, {Radiative Corrections to the Axial Anomaly},
  \emph{JETP Lett.} {\bf 49}, \penalty0 214--218  (1989).

\bibitem{Adler:2004qt}
S.~L. Adler, \emph{{Anomalies to all orders}}, In ed. G.~'t~Hooft, \emph{{50
  years of Yang-Mills theory}}, pp. 187--228.
\newblock  (2005).
\newblock \doi{10.1142/9789812567147_0009}.

\bibitem{kawahara1997wrap}
G.~Kawahara, S.~Kida, M.~Tanaka, and S.~Yanase, Wrap, tilt and stretch of
  vorticity lines around a strong thin straight vortex tube in a simple shear
  flow, \emph{Journal of Fluid Mechanics}. {\bf 353}, \penalty0 115--162
  (1997).

\bibitem{dahl2017tilting}
J.~M.~L. Dahl, Tilting of horizontal shear vorticity and the development of
  updraft rotation in supercell thunderstorms, \emph{Journal of the Atmospheric
  Sciences}. {\bf 74}\penalty0 (9), \penalty0 2997 -- 3020  (2017).
\newblock \doi{10.1175/JAS-D-17-0091.1}.
\newblock URL
  \url{https://journals.ametsoc.org/view/journals/atsc/74/9/jas-d-17-0091.1.xml}.

\bibitem{Grozdanov:2016tdf}
S.~Grozdanov, D.~M. Hofman, and N.~Iqbal, {Generalized global symmetries and
  dissipative magnetohydrodynamics}, \emph{Phys. Rev. D}. {\bf 95}\penalty0
  (9), \penalty0 096003  (2017).
\newblock \doi{10.1103/PhysRevD.95.096003}.

\bibitem{PhysRevLett.129.161601}
Y.~Choi, H.~T. Lam, and S.-H. Shao, Noninvertible global symmetries in the
  standard model, \emph{Phys. Rev. Lett.} {\bf 129}, \penalty0 161601  (Oct,
  2022).
\newblock \doi{10.1103/PhysRevLett.129.161601}.
\newblock URL \url{https://link.aps.org/doi/10.1103/PhysRevLett.129.161601}.

\bibitem{Das:2023nwl}
A.~Das, A.~Florio, N.~Iqbal, and N.~Poovuttikul, {Higher-form symmetry and
  chiral transport in real-time lattice $U(1)$ gauge theory}  (9, 2023).

\bibitem{Son:2012wh}
D.~T. Son and N.~Yamamoto, {Berry Curvature, Triangle Anomalies, and the Chiral
  Magnetic Effect in Fermi Liquids}, \emph{Phys. Rev. Lett.} {\bf 109},
  \penalty0 181602  (2012).
\newblock \doi{10.1103/PhysRevLett.109.181602}.

\bibitem{Son:2012zy}
D.~T. Son and N.~Yamamoto, {Kinetic theory with Berry curvature from quantum
  field theories}, \emph{Phys. Rev. D}. {\bf 87}\penalty0 (8), \penalty0 085016
   (2013).
\newblock \doi{10.1103/PhysRevD.87.085016}.

\bibitem{Stephanov:2012ki}
M.~A. Stephanov and Y.~Yin, {Chiral Kinetic Theory}, \emph{Phys. Rev. Lett.}
  {\bf 109}, \penalty0 162001  (2012).
\newblock \doi{10.1103/PhysRevLett.109.162001}.

\bibitem{Chen:2014cla}
J.-Y. Chen, D.~T. Son, M.~A. Stephanov, H.-U. Yee, and Y.~Yin, {Lorentz
  Invariance in Chiral Kinetic Theory}, \emph{Phys. Rev. Lett.} {\bf
  113}\penalty0 (18), \penalty0 182302  (2014).
\newblock \doi{10.1103/PhysRevLett.113.182302}.

\bibitem{Chen:2015gta}
J.-Y. Chen, D.~T. Son, and M.~A. Stephanov, {Collisions in Chiral Kinetic
  Theory}, \emph{Phys. Rev. Lett.} {\bf 115}\penalty0 (2), \penalty0 021601
  (2015).
\newblock \doi{10.1103/PhysRevLett.115.021601}.

\bibitem{Kharzeev:2016sut}
D.~E. Kharzeev, M.~A. Stephanov, and H.-U. Yee, {Anatomy of chiral magnetic
  effect in and out of equilibrium}, \emph{Phys. Rev. D}. {\bf 95}\penalty0
  (5), \penalty0 051901  (2017).
\newblock \doi{10.1103/PhysRevD.95.051901}.

\bibitem{Chen:2012ca}
J.-W. Chen, S.~Pu, Q.~Wang, and X.-N. Wang, {Berry Curvature and
  Four-Dimensional Monopoles in the Relativistic Chiral Kinetic Equation},
  \emph{Phys. Rev. Lett.} {\bf 110}\penalty0 (26), \penalty0 262301  (2013).
\newblock \doi{10.1103/PhysRevLett.110.262301}.

\bibitem{Gao:2012ix}
J.-H. Gao, Z.-T. Liang, S.~Pu, Q.~Wang, and X.-N. Wang, {Chiral Anomaly and
  Local Polarization Effect from Quantum Kinetic Approach}, \emph{Phys. Rev.
  Lett.} {\bf 109}, \penalty0 232301  (2012).
\newblock \doi{10.1103/PhysRevLett.109.232301}.

\bibitem{Gao:2017gfq}
J.-h. Gao, S.~Pu, and Q.~Wang, {Covariant chiral kinetic equation in the Wigner
  function approach}, \emph{Phys. Rev. D}. {\bf 96}\penalty0 (1), \penalty0
  016002  (2017).
\newblock \doi{10.1103/PhysRevD.96.016002}.

\bibitem{Hidaka:2016yjf}
Y.~Hidaka, S.~Pu, and D.-L. Yang, {Relativistic Chiral Kinetic Theory from
  Quantum Field Theories}, \emph{Phys. Rev. D}. {\bf 95}\penalty0 (9),
  \penalty0 091901  (2017).
\newblock \doi{10.1103/PhysRevD.95.091901}.

\bibitem{Hidaka:2017auj}
Y.~Hidaka, S.~Pu, and D.-L. Yang, {Nonlinear Responses of Chiral Fluids from
  Kinetic Theory}, \emph{Phys. Rev. D}. {\bf 97}\penalty0 (1), \penalty0 016004
   (2018).
\newblock \doi{10.1103/PhysRevD.97.016004}.

\bibitem{Mueller:2017lzw}
N.~Mueller and R.~Venugopalan, {The chiral anomaly, Berry's phase and chiral
  kinetic theory, from world-lines in quantum field theory}, \emph{Phys. Rev.
  D}. {\bf 97}\penalty0 (5), \penalty0 051901  (2018).
\newblock \doi{10.1103/PhysRevD.97.051901}.

\bibitem{Mueller:2017arw}
N.~Mueller and R.~Venugopalan, {Worldline construction of a covariant chiral
  kinetic theory}, \emph{Phys. Rev. D}. {\bf 96}\penalty0 (1), \penalty0 016023
   (2017).
\newblock \doi{10.1103/PhysRevD.96.016023}.

\bibitem{Gorbar:2017cwv}
E.~V. Gorbar, V.~A. Miransky, I.~A. Shovkovy, and P.~O. Sukhachov,
  {Second-order chiral kinetic theory: Chiral magnetic and pseudomagnetic
  waves}, \emph{Phys. Rev. B}. {\bf 95}\penalty0 (20), \penalty0 205141
  (2017).
\newblock \doi{10.1103/PhysRevB.95.205141}.

\bibitem{Wu:2016dam}
Y.~Wu, D.~Hou, and H.-c. Ren, {Field theoretic perspectives of the Wigner
  function formulation of the chiral magnetic effect}, \emph{Phys. Rev. D}.
  {\bf 96}\penalty0 (9), \penalty0 096015  (2017).
\newblock \doi{10.1103/PhysRevD.96.096015}.

\bibitem{Liu:2018xip}
Y.-C. Liu, L.-L. Gao, K.~Mameda, and X.-G. Huang, {Chiral kinetic theory in
  curved spacetime}, \emph{Phys. Rev. D}. {\bf 99}\penalty0 (8), \penalty0
  085014  (2019).
\newblock \doi{10.1103/PhysRevD.99.085014}.

\bibitem{Gao:2022gqr}
L.-L. Gao and X.-G. Huang, {Chiral Anomaly in Non-Relativistic Systems: Berry
  Curvature and Chiral Kinetic Theory}, \emph{Chin. Phys. Lett.} {\bf
  39}\penalty0 (2), \penalty0 021101  (2022).
\newblock \doi{10.1088/0256-307X/39/2/021101}.

\bibitem{Liu:2020flb}
Y.-C. Liu, K.~Mameda, and X.-G. Huang, {Covariant Spin Kinetic Theory I:
  Collisionless Limit}, \emph{Chin. Phys. C}. {\bf 44}\penalty0 (9), \penalty0
  094101  (2020).
\newblock \doi{10.1088/1674-1137/ac009b}.
\newblock [Erratum: Chin.Phys.C 45, 089001 (2021)].

\bibitem{Fukushima:2015tza}
K.~Fukushima, {Simulating net particle production and chiral magnetic current
  in a $CP$-odd domain}, \emph{Phys. Rev. D}. {\bf 92}\penalty0 (5), \penalty0
  054009  (2015).
\newblock \doi{10.1103/PhysRevD.92.054009}.

\bibitem{Ebihara:2017suq}
S.~Ebihara, K.~Fukushima, and S.~Pu, {Boost invariant formulation of the chiral
  kinetic theory}, \emph{Phys. Rev. D}. {\bf 96}\penalty0 (1), \penalty0 016016
   (2017).
\newblock \doi{10.1103/PhysRevD.96.016016}.

\bibitem{Huang:2017tsq}
A.~Huang, Y.~Jiang, S.~Shi, J.~Liao, and P.~Zhuang, {Out-of-equilibrium chiral
  magnetic effect from chiral kinetic theory}, \emph{Phys. Lett. B}. {\bf 777},
  \penalty0 177--183  (2018).
\newblock \doi{10.1016/j.physletb.2017.12.025}.

\bibitem{Vasak:1987um}
D.~Vasak, M.~Gyulassy, and H.~T. Elze, {Quantum Transport Theory for Abelian
  Plasmas}, \emph{Annals Phys.} {\bf 173}, \penalty0 462--492  (1987).
\newblock \doi{10.1016/0003-4916(87)90169-2}.

\bibitem{Zhuang:1995pd}
P.~Zhuang and U.~W. Heinz, {Relativistic quantum transport theory for
  electrodynamics}, \emph{Annals Phys.} {\bf 245}, \penalty0 311--338  (1996).
\newblock \doi{10.1006/aphy.1996.0011}.

\bibitem{Zhuang:1995jb}
P.-f. Zhuang and U.~W. Heinz, {Relativistic kinetic equations for
  electromagnetic, scalar and pseudoscalar interactions}, \emph{Phys. Rev. D}.
  {\bf 53}, \penalty0 2096--2101  (1996).
\newblock \doi{10.1103/PhysRevD.53.2096}.

\bibitem{Zhuang:1998kv}
P.-f. Zhuang and U.~W. Heinz, {Equal-Time Hierarchies in Quantum Transport
  Theory}, \emph{Phys. Rev. D}. {\bf 57}, \penalty0 6525--6543  (1998).
\newblock \doi{10.1103/PhysRevD.57.6525}.

\bibitem{Ochs:1998qj}
S.~Ochs and U.~W. Heinz, {Wigner functions in covariant and single time
  formulations}, \emph{Annals Phys.} {\bf 266}, \penalty0 351--416  (1998).
\newblock \doi{10.1006/aphy.1998.5796}.

\bibitem{Guo:2017dzf}
X.~Guo and P.~Zhuang, {Out-of-equilibrium $U_A$(1) symmetry breaking in
  electromagnetic fields}, \emph{Phys. Rev. D}. {\bf 98}\penalty0 (1),
  \penalty0 016007  (2018).
\newblock \doi{10.1103/PhysRevD.98.016007}.

\bibitem{DeGroot:1980dk}
S.~R. De~Groot, \emph{{Relativistic Kinetic Theory. Principles and
  Applications}}  (1980).

\bibitem{BialynickiBirula:1991tx}
I.~Bialynicki-Birula, P.~Gornicki, and J.~Rafelski, {Phase space structure of
  the Dirac vacuum}, \emph{Phys. Rev. D}. {\bf 44}, \penalty0 1825--1835
  (1991).
\newblock \doi{10.1103/PhysRevD.44.1825}.

\bibitem{Hidaka:2022dmn}
Y.~Hidaka, S.~Pu, Q.~Wang, and D.-L. Yang, {Foundations and applications of
  quantum kinetic theory}, \emph{Prog. Part. Nucl. Phys.} {\bf 127}, \penalty0
  103989  (2022).
\newblock \doi{10.1016/j.ppnp.2022.103989}.

\bibitem{Gao:2020pfu}
J.-H. Gao, Z.-T. Liang, and Q.~Wang, {Quantum kinetic theory for spin-1/2
  fermions in Wigner function formalism}, \emph{Int. J. Mod. Phys. A}. {\bf
  36}\penalty0 (01), \penalty0 2130001  (2021).
\newblock \doi{10.1142/S0217751X21300015}.

\bibitem{Gao:2020vbh}
J.-H. Gao, G.-L. Ma, S.~Pu, and Q.~Wang, {Recent developments in chiral and
  spin polarization effects in heavy-ion collisions}, \emph{Nucl. Sci. Tech.}
  {\bf 31}\penalty0 (9), \penalty0 90  (2020).
\newblock \doi{10.1007/s41365-020-00801-x}.

\bibitem{Horvath:2019dvl}
M.~Horvath, D.~Hou, J.~Liao, and H.-c. Ren, {Chiral magnetic response to
  arbitrary axial imbalance}, \emph{Phys. Rev. D}. {\bf 101}\penalty0 (7),
  \penalty0 076026  (2020).
\newblock \doi{10.1103/PhysRevD.101.076026}.

\bibitem{Wang:2019moi}
Z.~Wang, X.~Guo, S.~Shi, and P.~Zhuang, {Mass Correction to Chiral Kinetic
  Equations}, \emph{Phys. Rev. D}. {\bf 100}\penalty0 (1), \penalty0 014015
  (2019).
\newblock \doi{10.1103/PhysRevD.100.014015}.

\bibitem{Hattori:2019ahi}
K.~Hattori, Y.~Hidaka, and D.-L. Yang, {Axial Kinetic Theory and Spin Transport
  for Fermions with Arbitrary Mass}, \emph{Phys. Rev. D}. {\bf 100}\penalty0
  (9), \penalty0 096011  (2019).
\newblock \doi{10.1103/PhysRevD.100.096011}.

\bibitem{Sheng:2020oqs}
X.-L. Sheng, Q.~Wang, and X.-G. Huang, {Kinetic theory with spin: From massive
  to massless fermions}, \emph{Phys. Rev. D}. {\bf 102}\penalty0 (2), \penalty0
  025019  (2020).
\newblock \doi{10.1103/PhysRevD.102.025019}.

\bibitem{Manuel:2021oah}
C.~Manuel and J.~M. Torres-Rincon, {Chiral kinetic theory with small mass
  corrections and quantum coherent states}, \emph{Phys. Rev. D}. {\bf
  103}\penalty0 (9), \penalty0 096022  (2021).
\newblock \doi{10.1103/PhysRevD.103.096022}.

\bibitem{Das:2022azr}
A.~Das, W.~Florkowski, A.~Kumar, R.~Ryblewski, and R.~Singh, {Semi-classical
  Kinetic Theory for Massive Spin-half Fermions with Leading-order Spin
  Effects}, \emph{Acta Phys. Polon. B}. {\bf 54}\penalty0 (8), \penalty0 8--A4
  (2023).
\newblock \doi{10.5506/APhysPolB.54.8-A4}.

\bibitem{Weickgenannt:2019dks}
N.~Weickgenannt, X.-L. Sheng, E.~Speranza, Q.~Wang, and D.~H. Rischke, {Kinetic
  theory for massive spin-1/2 particles from the Wigner-function formalism},
  \emph{Phys. Rev. D}. {\bf 100}\penalty0 (5), \penalty0 056018  (2019).
\newblock \doi{10.1103/PhysRevD.100.056018}.

\bibitem{Gao:2019znl}
J.-H. Gao and Z.-T. Liang, {Relativistic Quantum Kinetic Theory for Massive
  Fermions and Spin Effects}, \emph{Phys. Rev. D}. {\bf 100}\penalty0 (5),
  \penalty0 056021  (2019).
\newblock \doi{10.1103/PhysRevD.100.056021}.

\bibitem{Sheng:2022ssd}
X.-L. Sheng, Q.~Wang, and D.~H. Rischke, {Lorentz-covariant kinetic theory for
  massive spin-1/2 particles}, \emph{Phys. Rev. D}. {\bf 106}\penalty0 (11),
  \penalty0 L111901  (2022).
\newblock \doi{10.1103/PhysRevD.106.L111901}.

\bibitem{Wagner:2022amr}
D.~Wagner, N.~Weickgenannt, and D.~H. Rischke, {Lorentz-covariant nonlocal
  collision term for spin-1/2 particles}, \emph{Phys. Rev. D}. {\bf
  106}\penalty0 (11), \penalty0 116021  (2022).
\newblock \doi{10.1103/PhysRevD.106.116021}.

\bibitem{Florkowski:2017dyn}
W.~Florkowski, B.~Friman, A.~Jaiswal, R.~Ryblewski, and E.~Speranza,
  {Spin-dependent distribution functions for relativistic hydrodynamics of
  spin-1/2 particles}, \emph{Phys. Rev. D}. {\bf 97}\penalty0 (11), \penalty0
  116017  (2018).
\newblock \doi{10.1103/PhysRevD.97.116017}.

\bibitem{Bhadury:2020cop}
S.~Bhadury, W.~Florkowski, A.~Jaiswal, A.~Kumar, and R.~Ryblewski, {Dissipative
  Spin Dynamics in Relativistic Matter}, \emph{Phys. Rev. D}. {\bf
  103}\penalty0 (1), \penalty0 014030  (2021).
\newblock \doi{10.1103/PhysRevD.103.014030}.

\bibitem{Bhadury:2022ulr}
S.~Bhadury, W.~Florkowski, A.~Jaiswal, A.~Kumar, and R.~Ryblewski,
  {Relativistic Spin Magnetohydrodynamics}, \emph{Phys. Rev. Lett.} {\bf
  129}\penalty0 (19), \penalty0 192301  (2022).
\newblock \doi{10.1103/PhysRevLett.129.192301}.

\bibitem{Weickgenannt:2022qvh}
N.~Weickgenannt, D.~Wagner, E.~Speranza, and D.~H. Rischke, {Relativistic
  dissipative spin hydrodynamics from kinetic theory with a nonlocal collision
  term}, \emph{Phys. Rev. D}. {\bf 106}\penalty0 (9), \penalty0 L091901
  (2022).
\newblock \doi{10.1103/PhysRevD.106.L091901}.

\bibitem{Weickgenannt:2022zxs}
N.~Weickgenannt, D.~Wagner, E.~Speranza, and D.~H. Rischke, {Relativistic
  second-order dissipative spin hydrodynamics from the method of moments},
  \emph{Phys. Rev. D}. {\bf 106}\penalty0 (9), \penalty0 096014  (2022).
\newblock \doi{10.1103/PhysRevD.106.096014}.

\bibitem{She:2021lhe}
D.~She, A.~Huang, D.~Hou, and J.~Liao, {Relativistic viscous hydrodynamics with
  angular momentum}, \emph{Sci. Bull.} {\bf 67}, \penalty0 2265--2268  (2022).
\newblock \doi{10.1016/j.scib.2022.10.020}.

\bibitem{Wang:2020pej}
Z.~Wang, X.~Guo, and P.~Zhuang, {Equilibrium Spin Distribution From Detailed
  Balance}, \emph{Eur. Phys. J. C}. {\bf 81}\penalty0 (9), \penalty0 799
  (2021).
\newblock \doi{10.1140/epjc/s10052-021-09586-8}.

\bibitem{Li:2020cwq}
S.~Li, {Quantum Kinetic Equation for spin polarization of massive quarks from
  pQCD}, \emph{Nucl. Phys. A}. {\bf 1005}, \penalty0 121977  (2021).
\newblock \doi{10.1016/j.nuclphysa.2020.121977}.

\bibitem{Li:2019qkf}
S.~Li and H.-U. Yee, {Quantum Kinetic Theory of Spin Polarization of Massive
  Quarks in Perturbative QCD: Leading Log}, \emph{Phys. Rev. D}. {\bf
  100}\penalty0 (5), \penalty0 056022  (2019).
\newblock \doi{10.1103/PhysRevD.100.056022}.

\bibitem{Becattini:2021suc}
F.~Becattini, M.~Buzzegoli, and A.~Palermo, {Spin-thermal shear coupling in a
  relativistic fluid}, \emph{Phys. Lett. B}. {\bf 820}, \penalty0 136519
  (2021).
\newblock \doi{10.1016/j.physletb.2021.136519}.

\bibitem{Liu:2021uhn}
S.~Y.~F. Liu and Y.~Yin, {Spin polarization induced by the hydrodynamic
  gradients}, \emph{JHEP}. {\bf 07}, \penalty0 188  (2021).
\newblock \doi{10.1007/JHEP07(2021)188}.

\bibitem{Yuan:2023skl}
Z.~Yuan, A.~Huang, W.-H. Zhou, G.-L. Ma, and M.~Huang, {Evolution of
  topological charge through chiral anomaly transport}, \emph{Phys. Rev. C}.
  {\bf 109}\penalty0 (3), \penalty0 L031903  (2024).
\newblock \doi{10.1103/PhysRevC.109.L031903}.

\bibitem{Sun:2017xhx}
Y.~Sun and C.~M. Ko, {$\Lambda$ hyperon polarization in relativistic heavy ion
  collisions from a chiral kinetic approach}, \emph{Phys. Rev. C}. {\bf
  96}\penalty0 (2), \penalty0 024906  (2017).
\newblock \doi{10.1103/PhysRevC.96.024906}.

\bibitem{Liu:2019krs}
S.~Y.~F. Liu, Y.~Sun, and C.~M. Ko, {Spin Polarizations in a Covariant
  Angular-Momentum-Conserved Chiral Transport Model}, \emph{Phys. Rev. Lett.}
  {\bf 125}\penalty0 (6), \penalty0 062301  (2020).
\newblock \doi{10.1103/PhysRevLett.125.062301}.

\bibitem{Huang:2023tyn}
A.~Huang, Z.~Yuan, and M.~Huang, {The dynamical magnetic-induced quark spin
  polarization at the early stage in heavy-ion collisions}  (7, 2023).

\bibitem{Carignano:2018gqt}
S.~Carignano, C.~Manuel, and J.~M. Torres-Rincon, {Consistent relativistic
  chiral kinetic theory: A derivation from on-shell effective field theory},
  \emph{Phys. Rev. D}. {\bf 98}\penalty0 (7), \penalty0 076005  (2018).
\newblock \doi{10.1103/PhysRevD.98.076005}.

\bibitem{Mueller:2017wom}
N.~Mueller and R.~Venugopalan, {World-line approach to chiral kinetic theory in
  topological background gauge fields}, \emph{PoS}. {\bf CPOD2017}, \penalty0
  047  (2018).
\newblock \doi{10.22323/1.311.0047}.

\bibitem{Li:2014bha}
Q.~Li, D.~E. Kharzeev, C.~Zhang, Y.~Huang, I.~Pletikosic, A.~V. Fedorov, R.~D.
  Zhong, J.~A. Schneeloch, G.~D. Gu, and T.~Valla, {Observation of the chiral
  magnetic effect in ZrTe5}, \emph{Nature Phys.} {\bf 12}, \penalty0 550--554
  (2016).
\newblock \doi{10.1038/nphys3648}.

\bibitem{Xiong:2015X}
J.~{Xiong}, S.~K. {Kushwaha}, T.~{Liang}, J.~W. {Krizan}, M.~{Hirschberger},
  W.~{Wang}, R.~J. {Cava}, and N.~P. {Ong}, {Evidence for the chiral anomaly in
  the Dirac semimetal Na$_{3}$Bi}, \emph{Science}. {\bf 350}\penalty0 (6259),
  \penalty0 413--416  (Oct., 2015).
\newblock \doi{10.1126/science.aac6089}.

\bibitem{Huang:2015X}
X.~{Huang}, L.~{Zhao}, Y.~{Long}, P.~{Wang}, D.~{Chen}, Z.~{Yang}, H.~{Liang},
  M.~{Xue}, H.~{Weng}, Z.~{Fang}, X.~{Dai}, and G.~{Chen}, {Observation of the
  Chiral-Anomaly-Induced Negative Magnetoresistance in 3D Weyl Semimetal TaAs},
  \emph{Physical Review X}. 5\penalty0 (3):\penalty0 031023  (July, 2015).
\newblock \doi{10.1103/PhysRevX.5.031023}.

\bibitem{STAR:2021mii}
M.~Abdallah et~al., {Search for the chiral magnetic effect with isobar
  collisions at $\sqrt {s_{NN}}$=200 GeV by the STAR Collaboration at the BNL
  Relativistic Heavy Ion Collider}, \emph{Phys. Rev. C}. {\bf 105}\penalty0
  (1), \penalty0 014901  (2022).
\newblock \doi{10.1103/PhysRevC.105.014901}.

\bibitem{STAR:2023ioo}
{Estimate of Background Baseline and Upper Limit on the Chiral Magnetic Effect
  in Isobar Collisions at $\sqrt{s_{\text{NN}}}=200$ GeV at the Relativistic
  Heavy-Ion Collider}, \emph{Accepted by Phys. Rev. C} .

\bibitem{STAR:2021pwb}
M.~S. Abdallah et~al., {Search for the Chiral Magnetic Effect via
  Charge-Dependent Azimuthal Correlations Relative to Spectator and Participant
  Planes in Au+Au Collisions at $\sqrt{s_{NN}}$ =\, 200\,GeV}, \emph{Phys. Rev.
  Lett.} {\bf 128}\penalty0 (9), \penalty0 092301  (2022).
\newblock \doi{10.1103/PhysRevLett.128.092301}.

\bibitem{STAR:2022ahj}
B.~Aboona et~al., {Search for the Chiral Magnetic Effect in Au+Au collisions at
  $\sqrt{s_{_{\rm{NN}}}}=27$ GeV with the STAR forward Event Plane Detectors},
  \emph{Phys. Lett. B}. {\bf 839}, \penalty0 137779  (2023).
\newblock \doi{10.1016/j.physletb.2023.137779}.

\bibitem{Sirunyan:2017quh}
A.~M. Sirunyan et~al., {Constraints on the chiral magnetic effect using
  charge-dependent azimuthal correlations in $p\mathrm{Pb}$ and PbPb collisions
  at the CERN Large Hadron Collider}, \emph{Phys. Rev. C}. {\bf 97}\penalty0
  (4), \penalty0 044912  (2018).
\newblock \doi{10.1103/PhysRevC.97.044912}.

\bibitem{ALICE:2022ljz}
{Search for the Chiral Magnetic Effect with charge-dependent azimuthal
  correlations in Xe-Xe collisions at $\sqrt{s_{\mathrm{NN}}} = 5.44$ TeV}
  (10, 2022).

\bibitem{ALICE:2020siw}
S.~Acharya et~al., {Constraining the Chiral Magnetic Effect with
  charge-dependent azimuthal correlations in Pb-Pb collisions at $
  \sqrt{s_{\mathrm{NN}}} $ = 2.76 and 5.02 TeV}, \emph{JHEP}. {\bf 09},
  \penalty0 160  (2020).
\newblock \doi{10.1007/JHEP09(2020)160}.

\bibitem{Zhao:2020utk}
J.~Zhao, {Search for CME in U+U and Au+Au collisions in STAR with different
  approaches of handling backgrounds}, \emph{Nucl. Phys. A}. {\bf 1005},
  \penalty0 121766  (2021).
\newblock \doi{10.1016/j.nuclphysa.2020.121766}.

\bibitem{Xu:2023elq}
Z.~Xu, B.~Chan, G.~Wang, A.~Tang, and H.~Z. Huang, {Event shape selection
  method in search of the chiral magnetic effect in heavy-ion collisions},
  \emph{Phys. Lett. B}. {\bf 848}, \penalty0 138367  (2024).
\newblock \doi{10.1016/j.physletb.2023.138367}.

\bibitem{Schukraft:2012ah}
J.~Schukraft, A.~Timmins, and S.~A. Voloshin, {Ultra-relativistic nuclear
  collisions: event shape engineering}, \emph{Phys. Lett. B}. {\bf 719},
  \penalty0 394--398  (2013).
\newblock \doi{10.1016/j.physletb.2013.01.045}.

\bibitem{STAR:2013zgu}
L.~Adamczyk et~al., {Measurement of charge multiplicity asymmetry correlations
  in high-energy nucleus-nucleus collisions at $\sqrt{{s}_{NN}} =$ 200 GeV},
  \emph{Phys. Rev. C}. {\bf 89}\penalty0 (4), \penalty0 044908  (2014).
\newblock \doi{10.1103/PhysRevC.89.044908}.

\bibitem{Wen:2016zic}
F.~Wen, J.~Bryon, L.~Wen, and G.~Wang, {Event-shape-engineering study of charge
  separation in heavy-ion collisions}, \emph{Chin. Phys. C}. {\bf 42}\penalty0
  (1), \penalty0 014001  (2018).
\newblock \doi{10.1088/1674-1137/42/1/014001}.

\bibitem{Wang:2016iov}
F.~Wang and J.~Zhao, {Challenges in flow background removal in search for the
  chiral magnetic effect}, \emph{Phys. Rev. C}. {\bf 95}\penalty0 (5),
  \penalty0 051901  (2017).
\newblock \doi{10.1103/PhysRevC.95.051901}.

\bibitem{ALICE:2015lib}
J.~Adam et~al., {Event shape engineering for inclusive spectra and elliptic
  flow in Pb-Pb collisions at 2.76 TeV}, \emph{Phys. Rev.
  C}. {\bf 93}\penalty0 (3), \penalty0 034916  (2016).
\newblock \doi{10.1103/PhysRevC.93.034916}.

\bibitem{Voloshin:2008dg}
S.~A. Voloshin, A.~M. Poskanzer, and R.~Snellings, {Collective phenomena in
  non-central nuclear collisions}, \emph{Landolt-Bornstein}. {\bf 23},
  \penalty0 293--333  (2010).
\newblock \doi{10.1007/978-3-642-01539-7_10}.

\bibitem{STAR:2020gky}
M.~S. Abdallah et~al., {Pair invariant mass to isolate background in the search
  for the chiral magnetic effect in Au~+~Au collisions at sNN=200~GeV},
  \emph{Phys. Rev. C}. {\bf 106}\penalty0 (3), \penalty0 034908  (2022).
\newblock \doi{10.1103/PhysRevC.106.034908}.

\bibitem{CMS:2017lrw}
A.~M. Sirunyan et~al., {Constraints on the chiral magnetic effect using
  charge-dependent azimuthal correlations in $p\mathrm{Pb}$ and PbPb collisions
  at the CERN Large Hadron Collider}, \emph{Phys. Rev. C}. {\bf 97}\penalty0
  (4), \penalty0 044912  (2018).
\newblock \doi{10.1103/PhysRevC.97.044912}.

\bibitem{ALICE:2017sss}
S.~Acharya et~al., {Constraining the magnitude of the Chiral Magnetic Effect
  with Event Shape Engineering in Pb-Pb collisions at $\sqrt{s_\mathrm{NN}}$ =
  2.76 TeV}, \emph{Phys. Lett. B}. {\bf 777}, \penalty0 151--162  (2018).
\newblock \doi{10.1016/j.physletb.2017.12.021}.

\bibitem{Xu:2023wcy}
Z.~Xu.
\newblock {Search for the Chiral Magnetic and Vortical Effects Using Event
  Shape Variables in Au+Au Collisions at STAR}.
\newblock In \emph{{30th International Conference on Ultrarelativstic
  Nucleus-Nucleus Collisions}}  (12, 2023).

\bibitem{Muller:2016jod}
N.~M\"uller, S.~Schlichting, and S.~Sharma, {Chiral magnetic effect and
  anomalous transport from real-time lattice simulations}, \emph{Phys. Rev.
  Lett.} {\bf 117}\penalty0 (14), \penalty0 142301  (2016).
\newblock \doi{10.1103/PhysRevLett.117.142301}.

\bibitem{Lin:2020jcp}
Y.~Lin, {Measurement of the charge separation along the magnetic field with
  Signed Balance Function in 200 GeV Au + Au collisions at STAR}, \emph{Nucl.
  Phys. A}. {\bf 1005}, \penalty0 121828  (2021).
\newblock \doi{10.1016/j.nuclphysa.2020.121828}.

\bibitem{CMS:2016wfo}
V.~Khachatryan et~al., {Observation of charge-dependent azimuthal correlations
  in $p$-Pb collisions and its implication for the search for the chiral
  magnetic effect}, \emph{Phys. Rev. Lett.} {\bf 118}\penalty0 (12), \penalty0
  122301  (2017).
\newblock \doi{10.1103/PhysRevLett.118.122301}.

\bibitem{Belmont:2016oqp}
R.~Belmont and J.~L. Nagle, {To CME or not to CME? Implications of p+Pb
  measurements of the chiral magnetic effect in heavy ion collisions},
  \emph{Phys. Rev. C}. {\bf 96}\penalty0 (2), \penalty0 024901  (2017).
\newblock \doi{10.1103/PhysRevC.96.024901}.

\bibitem{Kharzeev:2017uym}
D.~Kharzeev, Z.~Tu, A.~Zhang, and W.~Li, {Effect of the fluctuating proton size
  on the study of the chiral magnetic effect in proton-nucleus collisions},
  \emph{Phys. Rev. C}. {\bf 97}\penalty0 (2), \penalty0 024905  (2018).
\newblock \doi{10.1103/PhysRevC.97.024905}.

\bibitem{STAR:2019xzd}
J.~Adam et~al., {Charge-dependent pair correlations relative to a third
  particle in $p$ + Au and $d$+ Au collisions at RHIC}, \emph{Phys. Lett. B}.
  {\bf 798}, \penalty0 134975  (2019).
\newblock \doi{10.1016/j.physletb.2019.134975}.

\bibitem{Deng:2018dut}
W.-T. Deng, X.-G. Huang, G.-L. Ma, and G.~Wang, {Predictions for isobaric
  collisions at $\sqrt{s_{_{\rm NN}}}$ = 200 GeV from a multiphase transport
  model}, \emph{Phys. Rev. C}. {\bf 97}\penalty0 (4), \penalty0 044901  (2018).
\newblock \doi{10.1103/PhysRevC.97.044901}.

\bibitem{Xu:2017zcn}
H.-J. Xu, X.~Wang, H.~Li, J.~Zhao, Z.-W. Lin, C.~Shen, and F.~Wang, {Importance
  of isobar density distributions on the chiral magnetic effect search},
  \emph{Phys. Rev. Lett.} {\bf 121}\penalty0 (2), \penalty0 022301  (2018).
\newblock \doi{10.1103/PhysRevLett.121.022301}.

\bibitem{Hammelmann:2019vwd}
J.~Hammelmann, A.~Soto-Ontoso, M.~Alvioli, H.~Elfner, and M.~Strikman,
  {Influence of the neutron-skin effect on nuclear isobar collisions at
  energies available at the BNL Relativistic Heavy Ion Collider}, \emph{Phys.
  Rev. C}. {\bf 101}\penalty0 (6), \penalty0 061901  (2020).
\newblock \doi{10.1103/PhysRevC.101.061901}.

\bibitem{starbur17}
STAR BUR 2019 STAR 2017 BUR
  \url{https://drupal.star.bnl.gov/STAR/system/files/STAR_BUR_Run1718_v22_0.pdf}.

\bibitem{Tribedy:2020npn}
P.~Tribedy, {Status of CME Search Before Isobar Collisions and Methods of Blind
  Analysis From STAR}, \emph{J. Phys. Conf. Ser.} {\bf 1602}\penalty0 (1),
  \penalty0 1  (2020).
\newblock \doi{10.1088/1742-6596/1602/1/012002}.

\bibitem{VanBuren:2023mpn}
G.~Van~Buren and J.~Lauret, {The unseen: revealing the blind production
  procedure and experience for NP data}, \emph{J. Phys. Conf. Ser.} {\bf
  2438}\penalty0 (1), \penalty0 012065  (2023).
\newblock \doi{10.1088/1742-6596/2438/1/012065}.

\bibitem{STAR:2023gzg}
{Upper Limit on the Chiral Magnetic Effect in Isobar Collisions at the
  Relativistic Heavy-Ion Collider}  (8, 2023).

\bibitem{Kharzeev:2022hqz}
D.~E. Kharzeev, J.~Liao, and S.~Shi, {Implications of the isobar-run results
  for the chiral magnetic effect in heavy-ion collisions}, \emph{Phys. Rev. C}.
  {\bf 106}\penalty0 (5), \penalty0 L051903  (2022).
\newblock \doi{10.1103/PhysRevC.106.L051903}.

\bibitem{Gelis:2010nm}
F.~Gelis, E.~Iancu, J.~Jalilian-Marian, and R.~Venugopalan, {The Color Glass
  Condensate}, \emph{Ann. Rev. Nucl. Part. Sci.} {\bf 60}, \penalty0 463--489
  (2010).
\newblock \doi{10.1146/annurev.nucl.010909.083629}.

\bibitem{Gelis:2012ri}
F.~Gelis, {Color Glass Condensate and Glasma}, \emph{Int. J. Mod. Phys. A}.
  {\bf 28}, \penalty0 1330001  (2013).
\newblock \doi{10.1142/S0217751X13300019}.

\bibitem{STAR:2014uiw}
L.~Adamczyk et~al., {Beam-energy dependence of charge separation along the
  magnetic field in Au+Au collisions at RHIC}, \emph{Phys. Rev. Lett.} {\bf
  113}, \penalty0 052302  (2014).
\newblock \doi{10.1103/PhysRevLett.113.052302}.

\bibitem{ALICE:2012nhw}
B.~Abelev et~al., {Charge separation relative to the reaction plane in Pb-Pb
  collisions at $\sqrt{s_{NN}}= 2.76$ TeV}, \emph{Phys. Rev. Lett.} {\bf
  110}\penalty0 (1), \penalty0 012301  (2013).
\newblock \doi{10.1103/PhysRevLett.110.012301}.

\bibitem{Wang:2023xhn}
C.-Z. Wang, {Search for anomalous chiral effects in heavy-ion collisions with
  ALICE}  (12, 2023).

\bibitem{Gorbar:2011ya}
E.~V. Gorbar, V.~A. Miransky, and I.~A. Shovkovy, {Normal ground state of dense
  relativistic matter in a magnetic field}, \emph{Phys. Rev. D}. {\bf 83},
  \penalty0 085003  (2011).
\newblock \doi{10.1103/PhysRevD.83.085003}.

\bibitem{Rybalka:2018uzh}
D.~O. Rybalka, E.~V. Gorbar, and I.~A. Shovkovy, {Hydrodynamic modes in a
  magnetized chiral plasma with vorticity}, \emph{Phys. Rev. D}. {\bf
  99}\penalty0 (1), \penalty0 016017  (2019).
\newblock \doi{10.1103/PhysRevD.99.016017}.

\bibitem{Sukhachov:2018uuz}
P.~O. Sukhachov, V.~A. Miransky, I.~A. Shovkovy, and E.~V. Gorbar, {Collective
  excitations in Weyl semimetals in the hydrodynamic regime}, \emph{J. Phys.
  Condens. Matter}. {\bf 30}\penalty0 (27), \penalty0 275601  (2018).
\newblock \doi{10.1088/1361-648X/aac500}.

\bibitem{Jiang:2015cva}
Y.~Jiang, X.-G. Huang, and J.~Liao, {Chiral vortical wave and induced flavor
  charge transport in a rotating quark-gluon plasma}, \emph{Phys. Rev. D}. {\bf
  92}\penalty0 (7), \penalty0 071501  (2015).
\newblock \doi{10.1103/PhysRevD.92.071501}.

\bibitem{STAR:2015wza}
L.~Adamczyk et~al., {Observation of charge asymmetry dependence of pion
  elliptic flow and the possible chiral magnetic wave in heavy-ion collisions},
  \emph{Phys. Rev. Lett.} {\bf 114}\penalty0 (25), \penalty0 252302  (2015).
\newblock \doi{10.1103/PhysRevLett.114.252302}.

\bibitem{STAR:2022zpv}
M.~I. Abdulhamid et~al., {Search for the chiral magnetic wave using anisotropic
  flow of identified particles at energies available at the BNL Relativistic
  Heavy Ion Collider}, \emph{Phys. Rev. C}. {\bf 108}\penalty0 (1), \penalty0
  014908  (2023).
\newblock \doi{10.1103/PhysRevC.108.014908}.

\bibitem{ALICE:2023weh}
S.~Acharya et~al., {Probing the chiral magnetic wave with charge-dependent flow
  measurements in Pb-Pb collisions at the LHC}, \emph{JHEP}. {\bf 12},
  \penalty0 067  (2023).
\newblock \doi{10.1007/JHEP12(2023)067}.

\bibitem{ALICE:2015cjr}
J.~Adam et~al., {Charge-dependent flow and the search for the chiral magnetic
  wave in Pb-Pb collisions at $\sqrt{s_{\rm NN}} =$ 2.76 TeV}, \emph{Phys. Rev.
  C}. {\bf 93}\penalty0 (4), \penalty0 044903  (2016).
\newblock \doi{10.1103/PhysRevC.93.044903}.

\bibitem{CMS:2017pah}
A.~M. Sirunyan et~al., {Probing the chiral magnetic wave in $pPb$ and PbPb
  collisions at $\sqrt {s_{NN}}$ =5.02TeV using charge-dependent azimuthal
  anisotropies}, \emph{Phys. Rev. C}. {\bf 100}\penalty0 (6), \penalty0 064908
  (2019).
\newblock \doi{10.1103/PhysRevC.100.064908}.

\bibitem{STAR:2019wlg}
J.~Adam et~al., {Measurement of $e^+e^-$ Momentum and Angular Distributions
  from Linearly Polarized Photon Collisions}, \emph{Phys. Rev. Lett.} {\bf
  127}\penalty0 (5), \penalty0 052302  (2021).
\newblock \doi{10.1103/PhysRevLett.127.052302}.

\bibitem{CMS:2022arf}
A.~Tumasyan et~al., {Observation of $\tau$ lepton pair production in
  ultraperipheral lead-lead collisions at $\sqrt{s_\mathrm{NN}}$ = 5.02 TeV},
  \emph{Phys. Rev. Lett.} {\bf 131}, \penalty0 151803  (2023).
\newblock \doi{10.1103/PhysRevLett.131.151803}.

\bibitem{Skokov:2009qp}
V.~Skokov, A.~Y. Illarionov, and V.~Toneev, {Estimate of the magnetic field
  strength in heavy-ion collisions}, \emph{Int. J. Mod. Phys. A}. {\bf 24},
  \penalty0 5925--5932  (2009).
\newblock \doi{10.1142/S0217751X09047570}.

\bibitem{Wang:2023kzx}
X.~Wang, {Beam Energy and Collision Species Dependences of Photon-induced
  Lepton Pair Production at STAR}, \emph{Acta Phys. Polon. Supp.} {\bf
  16}\penalty0 (1), \penalty0 1--A96  (2023).
\newblock \doi{10.5506/APhysPolBSupp.16.1-A96}.

\bibitem{Tribedy:2023pjj}
P.~Tribedy, {Highlights from the STAR Experiment}, \emph{Acta Phys. Polon.
  Supp.} {\bf 16}\penalty0 (1), \penalty0 1--A6  (2023).
\newblock \doi{10.5506/APhysPolBSupp.16.1-A6}.

\bibitem{STAR:2016cio}
L.~Adamczyk et~al., {Charge-dependent directed flow in Cu+Au collisions at
  $\sqrt{s_{_{NN}}}$ = 200 GeV}, \emph{Phys. Rev. Lett.} {\bf 118}\penalty0
  (1), \penalty0 012301  (2017).
\newblock \doi{10.1103/PhysRevLett.118.012301}.

\bibitem{ALICE:2019sgg}
S.~Acharya et~al., {Probing the effects of strong electromagnetic fields with
  charge-dependent directed flow in Pb-Pb collisions at the LHC}, \emph{Phys.
  Rev. Lett.} {\bf 125}\penalty0 (2), \penalty0 022301  (2020).
\newblock \doi{10.1103/PhysRevLett.125.022301}.

\bibitem{STAR:2019clv}
J.~Adam et~al., {First Observation of the Directed Flow of $D^{0}$ and
  $\overline{D^0}$ in Au+Au Collisions at $\sqrt{s_{\rm NN}}$ = 200 GeV},
  \emph{Phys. Rev. Lett.} {\bf 123}\penalty0 (16), \penalty0 162301  (2019).
\newblock \doi{10.1103/PhysRevLett.123.162301}.

\bibitem{STAR:2023jdd}
M.~I. Abdulhamid et~al., {Observation of the electromagnetic field effect via
  charge-dependent directed flow in heavy-ion collisions at the Relativistic
  Heavy Ion Collider}, \emph{Phys. Rev. X}. {\bf 14}\penalty0 (1), \penalty0
  011028  (2024).
\newblock \doi{10.1103/PhysRevX.14.011028}.

\bibitem{STAR:2023wjl}
{Electric charge and strangeness-dependent directed flow splitting of produced
  quarks in Au+Au collisions}  (4, 2023).

\bibitem{Parida:2023ldu}
T.~Parida and S.~Chatterjee, {Baryon inhomogeneities driven charge dependent
  directed flow in heavy ion collisions}  (5, 2023).

\bibitem{STAR:2023nvo}
M.~I. Abdulhamid et~al., {Global polarization of \ensuremath{\Lambda} and
  \ensuremath{\Lambda}\textasciimacron{} hyperons in Au+Au collisions at
  sNN=19.6 and 27 GeV}, \emph{Phys. Rev. C}. {\bf 108}\penalty0 (1), \penalty0
  014910  (2023).
\newblock \doi{10.1103/PhysRevC.108.014910}.

\bibitem{Dash:2024qcc}
A.~P. Dash, {Exploring Electromagnetic Field Effects and Constraining Transport
  Parameters of QGP using STAR BES-II data}  (1, 2024).

\bibitem{Singha:2022syo}
S.~Singha, {Measurement of Global Spin Alignment of Vector Mesons at RHIC},
  \emph{Acta Phys. Polon. Supp.} {\bf 16}\penalty0 (1), \penalty0 1--A34
  (2023).
\newblock \doi{10.5506/APhysPolBSupp.16.1-A34}.

\bibitem{Yang:2017sdk}
Y.-G. Yang, R.-H. Fang, Q.~Wang, and X.-N. Wang, {Quark coalescence model for
  polarized vector mesons and baryons}, \emph{Phys. Rev. C}. {\bf 97}\penalty0
  (3), \penalty0 034917  (2018).
\newblock \doi{10.1103/PhysRevC.97.034917}.

\bibitem{Shen:2022gtl}
D.~Shen, J.~Chen, A.~Tang, and G.~Wang, {Impact of globally spin-aligned vector
  mesons on the search for the chiral magnetic effect in heavy-ion collisions},
  \emph{Phys. Lett. B}. {\bf 839}, \penalty0 137777  (2023).
\newblock \doi{10.1016/j.physletb.2023.137777}.

\bibitem{Finch:2017xiz}
L.~E. Finch and S.~J. Murray, {Investigating local parity violation in
  heavy-ion collisions using \ensuremath{\Lambda} helicity}, \emph{Phys. Rev.
  C}. {\bf 96}\penalty0 (4), \penalty0 044911  (2017).
\newblock \doi{10.1103/PhysRevC.96.044911}.

\bibitem{Becattini:2020xbh}
F.~Becattini, M.~Buzzegoli, A.~Palermo, and G.~Prokhorov, {Polarization as a
  signature of local parity violation in hot QCD matter}, \emph{Phys. Lett. B}.
  {\bf 822}, \penalty0 136706  (2021).
\newblock \doi{10.1016/j.physletb.2021.136706}.
\newblock [Erratum: Phys.Lett.B 826, 136909 (2022)].

\bibitem{STAR:2023qyt}
M.~I. Abdulhamid et~al., {Event-by-event correlations between
  \ensuremath{\Lambda}~(\ensuremath{\Lambda}\textasciimacron{}) hyperon global
  polarization and handedness with charged hadron azimuthal separation in Au+Au
  collisions at sNN=27 GeV from STAR}, \emph{Phys. Rev. C}. {\bf 108}\penalty0
  (1), \penalty0 014909  (2023).
\newblock \doi{10.1103/PhysRevC.108.014909}.

\bibitem{Stephanov:1998dy}
M.~A. Stephanov, K.~Rajagopal, and E.~V. Shuryak, {Signatures of the
  tricritical point in QCD}, \emph{Phys. Rev. Lett.} {\bf 81}, \penalty0
  4816--4819  (1998).
\newblock \doi{10.1103/PhysRevLett.81.4816}.

\bibitem{Stephanov:1999zu}
M.~A. Stephanov, K.~Rajagopal, and E.~V. Shuryak, {Event-by-event fluctuations
  in heavy ion collisions and the QCD critical point}, \emph{Phys. Rev. D}.
  {\bf 60}, \penalty0 114028  (1999).
\newblock \doi{10.1103/PhysRevD.60.114028}.

\bibitem{Fukushima:2010fe}
K.~Fukushima, M.~Ruggieri, and R.~Gatto, {Chiral magnetic effect in the PNJL
  model}, \emph{Phys. Rev. D}. {\bf 81}, \penalty0 114031  (2010).
\newblock \doi{10.1103/PhysRevD.81.114031}.

\bibitem{Ikeda:2020agk}
K.~Ikeda, D.~E. Kharzeev, and Y.~Kikuchi, {Real-time dynamics of Chern-Simons
  fluctuations near a critical point}, \emph{Phys. Rev. D}. {\bf 103}\penalty0
  (7), \penalty0 L071502  (2021).
\newblock \doi{10.1103/PhysRevD.103.L071502}.

\bibitem{Xu:2024xqc}
Z.~Xu.
\newblock \emph{{Search for the Chiral Magnetic Effect from RHIC Beam Energy
  Scan II data with STAR}}.
\newblock PhD thesis, UCLA, Los Angeles (main)  (2024).

\bibitem{Kharzeev:2016tvd}
D.~E. Kharzeev, {Chiral magnetic superconductivity}, \emph{EPJ Web Conf.} {\bf
  137}, \penalty0 01011  (2017).
\newblock \doi{10.1051/epjconf/201713701011}.

\bibitem{shao2023gravitational}
J.~Shao and M.~Huang, Gravitational waves and primordial black holes from
  chirality imbalanced qcd first-order phase transition with p and c p
  violation, \emph{Physical Review D}. {\bf 107}\penalty0 (4), \penalty0 043011
   (2023).

\bibitem{schober2024chiral}
J.~Schober, I.~Rogachevskii, and A.~Brandenburg, Chiral anomaly and dynamos
  from inhomogeneous chemical potential fluctuations, \emph{Physical Review
  Letters}. {\bf 132}\penalty0 (6), \penalty0 065101  (2024).

\bibitem{brandenburg2024relic}
A.~Brandenburg, E.~Clarke, T.~Kahniashvili, A.~J. Long, and G.~Sun, Relic
  gravitational waves from the chiral plasma instability in the standard
  cosmological model, \emph{Physical Review D}. {\bf 109}\penalty0 (4),
  \penalty0 043534  (2024).

\bibitem{giovannini1998primordial}
M.~Giovannini and M.~E. Shaposhnikov, Primordial hypermagnetic fields and the
  triangle anomaly, \emph{Physical Review D}. {\bf 57}\penalty0 (4), \penalty0
  2186  (1998).

\bibitem{hirono2015self}
Y.~Hirono, D.~E. Kharzeev, and Y.~Yin, Self-similar inverse cascade of magnetic
  helicity driven by the chiral anomaly, \emph{Physical Review D}. {\bf
  92}\penalty0 (12), \penalty0 125031  (2015).

\bibitem{brandenburg2023chirality}
A.~Brandenburg.
\newblock Chirality in astrophysics.
\newblock In \emph{CHIRAL MATTER: Proceedings of the Nobel Symposium 167}, pp.
  15--35  (2023).

\bibitem{Kharzeev:2019rsy}
D.~Kharzeev, E.~Shuryak, and I.~Zahed, {Sphalerons, baryogenesis, and helical
  magnetogenesis in the electroweak transition of the minimal standard model},
  \emph{Phys. Rev. D}. {\bf 102}\penalty0 (7), \penalty0 073003  (2020).
\newblock \doi{10.1103/PhysRevD.102.073003}.

\bibitem{Feng:2021oub}
Y.~Feng, Y.~Lin, J.~Zhao, and F.~Wang, {Revisit the chiral magnetic effect
  expectation in isobaric collisions at the relativistic heavy ion collider},
  \emph{Phys. Lett. B}. {\bf 820}, \penalty0 136549  (2021).
\newblock \doi{10.1016/j.physletb.2021.136549}.

\bibitem{ALICE:2014sbx}
B.~B. Abelev et~al., {Performance of the ALICE Experiment at the CERN LHC},
  \emph{Int. J. Mod. Phys. A}. {\bf 29}, \penalty0 1430044  (2014).
\newblock \doi{10.1142/S0217751X14300440}.

\bibitem{Bhatta:2023cqf}
S.~Bhatta, C.~Zhang, and J.~Jia, {Energy dependence of heavy-ion initial
  condition in isobar collisions}  (1, 2023).

\bibitem{STAR:2017ckg}
L.~Adamczyk et~al., {Global $\Lambda$ hyperon polarization in nuclear
  collisions: evidence for the most vortical fluid}, \emph{Nature}. {\bf 548},
  \penalty0 62--65  (2017).
\newblock \doi{10.1038/nature23004}.

\bibitem{Pang:2016igs}
L.-G. Pang, H.~Petersen, Q.~Wang, and X.-N. Wang, {Vortical Fluid and $\Lambda$
  Spin Correlations in High-Energy Heavy-Ion Collisions}, \emph{Phys. Rev.
  Lett.} {\bf 117}\penalty0 (19), \penalty0 192301  (2016).
\newblock \doi{10.1103/PhysRevLett.117.192301}.

\bibitem{Belmont:2014lta}
R.~Belmont, {Charge-dependent anisotropic flow studies and the search for the
  Chiral Magnetic Wave in ALICE}, \emph{Nucl. Phys. A}. {\bf 931}, \penalty0
  981--985  (2014).
\newblock \doi{10.1016/j.nuclphysa.2014.09.070}.

\bibitem{Xu:2020sln}
H.-j. Xu, J.~Zhao, Y.~Feng, and F.~Wang, {Importance of non-flow background on
  the chiral magnetic wave search}, \emph{Nucl. Phys. A}. {\bf 1005}, \penalty0
  121770  (2021).
\newblock \doi{10.1016/j.nuclphysa.2020.121770}.

\end{thebibliography}


\end{document}